\documentclass[aps,prd,notitlepage,showpacs,nofootinbib,superscriptaddress]{revtex4-1}
 \pdfoutput=1
\usepackage[usenames,dvipsnames]{color}  
\usepackage{graphicx}
\usepackage{setspace}
\usepackage{caption}
\usepackage{amsmath}
\usepackage{slashed}
\usepackage{amssymb}
\usepackage[colorlinks=true,citecolor=darkred,urlcolor=darkred, pdfborder={0 0 0}]{hyperref}
\usepackage[normalem]{ulem}
\usepackage[table,dvipsnames]{xcolor}
\usepackage{gensymb}
\usepackage{braket}
\usepackage{multirow}
\usepackage[export]{adjustbox}
\usepackage{array}
\usepackage{soul}
\usepackage{bm}
\usepackage{subfig}
\usepackage{hhline}

\makeatletter
\def\p@subsection{}
\makeatother

\definecolor{darkred}{rgb}{0.6,0,0}
\definecolor{dgr}{rgb}{0,0.5,0}

\definecolor{linkcolor}{rgb}{0,0,0.5}

\colorlet{gr}{gray!20!White}
\colorlet{grc}{gray!40!White}
\colorlet{cyn}{Cyan!20!White}
\colorlet{cync}{Cyan!40!White}
\colorlet{lmg}{LimeGreen!20!White}
\colorlet{lmgc}{LimeGreen!50!White}
\colorlet{brw}{brown!20!white!95!red}
\colorlet{brwc}{brown!50!white}
\colorlet{ylwc}{Dandelion!60!yellow!80!white}
\colorlet{mgn}{magenta!20!white}

\def\gsim{\raise0.3ex\hbox{$\;>$\kern-0.75em\raise-1.1ex\hbox{$\sim\;$}}}
\def\lsim{\raise0.3ex\hbox{$\;<$\kern-0.75em\raise-1.1ex\hbox{$\sim\;$}}}
\def \znbb {$0\nu\beta\beta$ }
\def \zn4b {$0\nu4\beta$ } 
\def\beqn#1{\begin{equation}\label{#1}}
\def\eeqn{\end{equation}}

\def\beqa#1{\begin{eqnarray}\label{#1}}
\def\eeqa{\end{eqnarray}}

\definecolor{a:green}{cmyk}{0.25, 0, 0.25, 0.20}
\definecolor{a:blue}{cmyk}{0.59, 0.16, 0.0, 0.31}
\definecolor{a:red}{cmyk}{0.0, 0.64, 0.68, 0.46}
\definecolor{a:orange}{cmyk}{0.0, 0.14, 0.60, 0.0}
\definecolor{a:purple}{cmyk}{0.0, 0.25, 0.0, 0.20}
\definecolor{a:turquoise}{cmyk}{0.26, 0, 0.02, 0.05}

\newcommand {\ignore}[1]{}

\newcommand{\sm}{{Standard Model }}
\def\lfv{lepton flavor violation }
\def\lnv{lepton number violation }

\def\lfv{lepton flavor violation }

\def\SM{$\mathrm{SU(3)_c \otimes SU(2)_L \otimes U(1)_Y}$ }

\def\321{$\mathrm{SU(3) \otimes SU(2) \otimes U(1)}$ }

\def\dsf{$\Delta m_{\Sigma F}^{}$ } 
\def\depf{$\Delta m_{\eta^+ F}^{}$ } 
\def\deip{$\Delta m^2_{\eta_I^0\eta^+}$ }
\def\mdm{$m_{\text{DM}}^{}$ }

\def\red{\color{red}{}}

\newcommand{\AddrIvania}{Departamento de Ciencias, Facultad de Artes Liberales, Universidad Adolfo Ib\'a\~{n}ez, Diagonal Las Torres 2640, Santiago, Chile}

\begin{document}

\title{\boldmath 
\color{BrickRed} Dark matter as the source of neutrino mass:\\
theory overview and experimental prospects
}

\author{Ivania M. \'Avila}\email{ivania.maturana@edu.uai.cl}
\affiliation{~\AddrIvania}

\author{Anirban Karan}\email{anirban.karan@unical.it}
\affiliation{Institut de F\'{i}sica Corpuscular
, CSIC/Universitat de Valencia, Parc Cientific de Paterna, C/ Catedratico Jose Beltran, 2, E-46980 Paterna, (Valencia), Spain}
\affiliation{
INFN Cosenza \& Dipartimento di Fisica, Università della Calabria, I-87036 Arcavacata di Rende, Cosenza, Italy}

\author{Sanjoy Mandal}\email{smandal@kias.re.kr}
\affiliation{Korea Institute for Advanced Study, Seoul 02455, Korea}

\author{Soumya Sadhukhan}\email{soumya.sadhukhan@rkmrc.in}
\affiliation{Ramakrishna Mission Residential College (Autonomous), Vivekananda Centre for Research, Narendrapur, Kolkata, India-700103} 


\author{Jos\'{e} W. F. Valle}\email{valle@ific.uv.es}
\affiliation{Institut de F\'{i}sica Corpuscular
, CSIC/Universitat de Valencia, Parc Cientific de Paterna, C/ Catedratico Jose Beltran, 2, E-46980 Paterna, (Valencia), Spain}

\begin{abstract}

We review theoretical frameworks in which small neutrino masses arise radiatively through interactions with a dark sector that also accounts for cosmological dark matter (DM). A prototype is provided by \textit{scotogenic} schemes, that extend the inert Higgs doublet model to include dark fermions. We outline their key features and limitations, discussing
the advantages of the \textit{revamped scotogenic} extension. The phenomenological signatures of fermionic and bosonic scotogenic dark matter are discussed, along with \textit{scoto-seesaw} models that merge scotogenic and seesaw mechanisms. We also consider scenarios where the dark sector seeds a low-scale seesaw. These frameworks can accommodate dark matter as Weakly or Feebly Interacting Massive Particles (WIMPs or FIMPs). 
While \textit{hidden} dark sector models are inherently difficult to exclude, \textit{visible} dark sector schemes should be confirmed—or ruled out—by forthcoming dark matter, collider, and lepton flavor violation studies.

\end{abstract}

 
\maketitle
\vskip -.6cm
{
  \hypersetup{linkcolor=blue}
  \small 
  \tableofcontents
}

\newpage

\section{Introduction}
\label{sec:intro}

 Advances in experimental particle physics and observational cosmology have led to two major frameworks—the Standard Model (SM) of particle physics and the Standard $\Lambda$CDM cosmological model—that together form the foundation of our current understanding of the universe, from the smallest scales of particles to the largest structures in space.
Major milestones have been the discovery of the Higgs boson at CERN~\cite{ATLAS:2012yve,CMS:2012qbp}
and the observation of Gravitational Waves (GWs)~\cite{LIGOScientific:2016aoc}. While the former provides a confirmation of the SM of particle physics and the Brout-Englert-Higgs spontaneous symmetry-breaking  mechanism~\cite{Higgs:1964pj,Englert:1964et,Guralnik:1964eu}, the latter provides a long-awaited confirmation of Einstein's General Relativity~\cite{Einstein:1916vd} with profound implications for our understanding of the universe.
Specially due to the fact that GWs can arise from first order phase transitions taking place in the early universe~\cite{Witten:1984rs,Kamionkowski:1993fg,Addazi:2019dqt,Roshan:2024qnv}.

Beyond these important breakthroughs in particle and astrophysics, there are important indications for the existence of new physics.
The discovery of neutrino
oscillations~\cite{McDonald:2016ixn,Kajita:2016cak} has opened a new chapter in particle physics that may hold the key to understanding major drawbacks in our current description of nature~\cite{Valle:2015pba}, such as understanding the pattern of quark and lepton masses from first principles~\cite{Ding:2024ozt}. Underpinning neutrino parameters and unveiling the nature of the mechanism responsible for neutrino mass generation constitute major open challenges.
On the other hand the
evidence for dark matter constitutes one of the most solid indications of new physics. 
Indeed, there is a crack in our basic understanding of nature coming, e.g., from a combination of studies of the cosmic microwave background, 
distant supernovae, galaxy clusters, and baryon acoustic oscillations. 
Altogether, these have firmly established~\cite{Planck:2015fie,Planck:2018vyg} a picture of cosmology~\cite{Dolgov:1981hv,Dodelson:2003ft,Kolb:1990vq} in which there must exist a new form of matter, dubbed \textit{dark matter} (DM), accounting for 26\% of the cosmic energy budget,
and about 85\% of the matter content of the Universe~\cite{Planck:2018vyg}. The quest for such dark matter is now a basic science challenge in the agenda of many upcoming explorations.

Many lines of evidence suggest that most of the cosmological dark matter is non-baryonic~\cite{Bergstrom:2000pn},
required to adequately account for the formation of large-scale structure in the Universe~\cite{Davis:1985rj,Blumenthal:1984bp}.
Such Cold Dark Matter (CDM), 
non-relativistic at the matter-radiation equality epoch, seems to be a key ingredient, missing in the SM of particle physics~\cite{Bertone:2004pz,Feng:2010gw,Bergstrom:2012fi}. 
Therefore, if cosmological dark matter is a fundamental particle, its existence provides indication for new physics, beyond the SM. 
Out of the many candidates postulated to make up the dark matter, WIMPs produced thermally in the early Universe, appear to be one of the most compelling ones. 
Despite an extensive experimental search effort, combining direct, indirect, as well as collider probes, there has been so far no conclusive WIMP DM detection. 
Nonetheless, the search continues and WIMPs remain as a leading CDM candidate. 

 Here we review the idea that the two major shortcomings of the Standard Model -- the lack of neutrino mass and of a viable WIMP DM candidate -- have a common origin. 
This is the essence of the scotogenic approach, that postulates a radiative origin for neutrino mass generation~\cite{Boucenna:2014zba,Cai:2017jrq}, in which a dark sector acts as the mediator~\cite{Ma:2006km,Tao:1996vb}.
In its simplest form the idea requires the presence of a $\mathbb{Z} _{2}$ symmetry to stabilize dark matter and ensure that neutrino masses arise radiatively.
We examine the original scotogenic model and its simplest generalizations, in which Majorana neutrinos get one-loop-induced masses. 

 Within scotogenic frameworks, the \textit{lightest scotogenic particle} (LSP) is stable and typically serves as the WIMP dark matter candidate. It can be either a dark fermion or a dark scalar boson, both of which provide a consistent scotogenic CDM picture.
 A simple realization was proposed in~\cite{Hirsch:2013ola} to overcome limitations of the original model and harbor a richer phenomenology~\cite{ 
Merle:2016scw,Rocha-Moran:2016enp,Diaz:2016udz,Restrepo:2019ilz,
Avila:2019hhv,Karan:2023adm,Lozano:2025tst}. 
 In contrast to the simplest model, the unwanted spontaneous breaking of the $\mathbb{Z}_{2}$ parity symmetry can be naturally avoided due to the presence of triplets~\cite{Merle:2016scw}. \\[-.3cm]

 {Notice that the required dark matter stabilizing-symmetry in scotogenic schemes may have a deeper theoretical origin. As an interesting example, this symmetry could be interpreted as a matter-parity (the non-supersymmetric analogue of R-parity) arising as a residual symmetry within extended gauge setups~\cite{Leite:2019grf,Kang:2019sab,CarcamoHernandez:2020ehn,Hernandez:2021zje,Garnica:2024wur,VanDong:2023xmf,Leite:2023gzl}. 
 The stabilizing symmetry could also arise from a global symmetry~\cite{Bonilla:2023egs}, a higher $\mathbb{Z}_n$ symmetry~\cite{Bonilla:2018ynb} or 
 accidentally, as a result of the imposition of some family symmetries ~\cite{Lavoura:2012cv}. } \\[-.4cm]

{It has been noted that dark matter stability may be related to the possible Dirac nature of neutrinos~\cite{CentellesChulia:2016rms}. Indeed, one can envisage WIMP-mediated scotogenic schemes with Dirac neutrino masses~\cite{Farzan:2012sa,CentellesChulia:2016rms,Bonilla:2016diq,Wang:2017mcy,CentellesChulia:2019xky,Li:2022chc}. 
Some radiative models involve leptoquarks~\cite{Nomura:2025ovm, AristizabalSierra:2007nf, Chua:1999si, Dorsner:2017wwn, Cheung:2016fjo,Batra:2023erw,Li:2022chc,Hati:2024ppg} as a way to address neutrino mass generation and other issues, such as possible anomalies that may also be present in the muon (g-2) and the B-meson sector.} \\[-.3cm]
 
{Scotogenic schemes may also incorporate the idea of scale invariance~\cite{Coleman:1973jx} which can help explain the smallness of the electroweak breaking scale compared to the Planck scale (the so-called hierarchy problem)~\cite{Bardeen:1995kv,Holthausen:2013ota}. Specific attempts at formulating scale-invariant scotogenic models have been given in Refs.~\cite{Ahriche:2016cio,Guo:2018iix,Soualah:2021xbn}.} \\[-.3cm]
 
 Concerning phenomenology, for the case of fermionic DM, the revamped scotogenic picture~\cite{Hirsch:2013ola} recovers the physics of supersymmetric neutralino dark matter~\cite{PhysRevLett.50.1419,Ellis:1983ew,Jungman:1995df}, but within a simpler theory framework and with a strong connection to neutrino physics and \lfv processes.
 For the bosonic DM case, there is close similarity, but also differences, with the inert Higgs doublet picture.
In both cases there are strong synergies between 
cosmological and laboratory searches~\cite{Gerbino:2022nvz}, as well as between dark matter, neutrino oscillations, collider and flavor violation studies.\\[-.3cm]
 
 This review is organized as follows.
 In Sec.~\ref{sec:dark-preliminaries} we briefly describe the observational evidence for dark matter and discuss possible DM candidates, as well as the evidence for neutrino masses and mixing from oscillation experiments. 
 In Sec.~\ref{sec:Inert} we briefly review the inert Higgs doublet model (IHDM), one of the simplest Standard Model extensions incorporating dark matter.
 In Sec.~\ref{sec:simpl-scot-setup} we describe the simplest scotogenic setup, 
 introducing its main ingredients,  such as its new fields and interactions, and discuss the associated phenomenological prospects.
In Sec.~\ref{sec:singlet-triplet-scoto} we describe a revamped version of the simplest scenario, which bypasses its limitations, providing a richer framework for fermionic scotogenic dark-matter which we review in detail.
In Sec.~\ref{sec:scoto-seesaw} we discuss the scoto-seesaw picture, combining the seesaw and scotogenic paradigms for generating the atmospheric mass scale from seesaw and the solar scale from a dark loop. 

An intriguing variant involves generating neutrino masses via a low-scale seesaw mechanism, in which lepton number violation is driven by interactions in a dark sector. This setup naturally connects the origin of neutrino masses with the dynamics of dark matter, opening new ways for theoretical and experimental exploration, see 
 Sec.~\ref{sec:dark-low-scale-seesaw}.
 A simple realization is based on the inverse seesaw mechanism, in which a calculable dark loop triggers neutrino mass generation through the seesaw, as discussed in Sec.~\ref{sec:dark-inverse-seesaw}. 
An alternative way to relate neutrino masses and dark matter physics employs the linear seesaw mechanism, as presented in Section~\ref{sec:linear-dark-seesaw}.
In both cases we discuss how dark-matter phenomena can be probed experimentally.

In Sec.~\ref{sec:dark-matter-scotogenic} we review the case for scalar scotogenic WIMP dark matter and its phenomenological implications, stressing the similarities and differences with IHDM predictions.
 Interesting synergies with charged lepton flavor violation (cLFV) physics are described in Sec.~\ref{sec:cLFV}, and their promising collider prospects are discussed in Chapter~\ref{sec:collider}.
Cosmological implications of the scotogenic approach for baryogenesis via leptogenesis are briefly discussed in Sec.~\ref{sec:cosmology}, with emphasis on the possibility of lowering the Davidson-Ibarra bound. Overall conclusions are given in Sec.~\ref{sec:conclusion}.

\section{Preliminaries}
\label{sec:dark-preliminaries}

\subsection{Evidence for dark matter}

One of the most compelling open questions in contemporary physics is the nature of dark matter. Simple gravitational arguments, supported by a wide range of astrophysical and cosmological observations, suggest that most of the mass in the Universe—about 85\%—is composed of a non-luminous, non-baryonic component. Dark matter is not only responsible for more than a quarter of the total energy density of the Universe, but also plays a key role in cosmic structure formation, acting as the gravitational skeleton for the large-scale structure known as the cosmic web~\cite{Bertone:2016nfn}.
Given the growing body of observational evidence, the problems of dark matter and dark energy have become central to modern cosmology~\cite{Albrecht:2006um}, motivating major upcoming scientific missions aimed at their investigation~\cite{LSST:2008ijt,DESI:2025zgx,DESI:2024mwx,Euclid:2024few}. What once appeared to be a minor anomaly has now evolved into a deep challenge for fundamental physics, with the nature of dark matter standing as one of the most pressing issues in our understanding of the Universe.
There is overwhelming empirical evidence for the existence of dark matter and increasingly strong constraints on its properties. This evidence arises from a wide range of astronomical observations and cosmological measurements across different length scales, from individual galaxies to the largest observable structures in the Universe. One of the most compelling indications comes from galactic rotation curves. According to Newtonian dynamics the rotational velocity of stars and gas orbiting the galactic center at a radius $r$ should scale as
$v(r)\propto \sqrt{\frac{M}{r}}$ 
for scales far exceeding that of the luminous matter, where $M$ is the mass concentrated in the Galactic Center. 
However, observations of the rotational velocities of stars and gas clouds within galaxies reveal that these velocities do not decrease with increasing distance from the galactic center, as would be expected based just on the distribution of visible matter. Instead, the rotation curves remain approximately flat even out to the outermost observable stars~\cite{Zwicky:1933gu,Rubin:1978kmz}. This striking discrepancy strongly indicates the presence of an additional, non-luminous component—commonly referred to as dark matter—which provides the extra gravitational pull necessary to account for the observed dynamics. 
For example, this puzzle can be addressed by postulating the existence of an invisible, spherically symmetric dark matter halo surrounding the galaxy, with a mass density profile scaling as
$\rho(r)\propto \frac{1}{r^2}\Rightarrow M(r)\propto r$. \\[-.4cm]
\par Gravitational lensing, the bending of light by massive objects~\cite{Einstein:1936llh}, offers another compelling line of evidence for the existence of dark matter. It enables the mapping of the total mass distribution—including both luminous and dark components—by observing the distortion of background light from distant sources, such as galaxies or quasars. 
Observations of gravitational lensing in galaxy clusters reveal the presence of significantly more mass than can be accounted for by luminous matter alone, implying the existence of dark matter. One of the earliest insights came from studies of the motions of galaxies within the Coma Cluster~\cite{Zwicky:1937zza}. Zwicky compared the cluster's gravitational mass—inferred from the galaxies' velocities through the virial theorem—with its luminous mass, estimated from visible light. He found a large discrepancy, concluding that most of the cluster's mass must be dark. 
A more compelling illustration is provided by the Bullet Cluster~\cite{Clowe:2006eq}. In this system, two galaxy clusters collide, separating the hot gas (which interacts via electromagnetism and slows down) from the dark matter (which interacts weakly and passes through). Using gravitational lensing, astronomers mapped the total mass distribution, finding that it aligns not with the visible hot gas, but rather with the locations of the galaxies—implying that most of the mass is non-luminous and does not interact electromagnetically. This provides direct evidence for the existence of dark matter, as the lensing mass does not follow the distribution of ordinary matter detected via electromagnetic radiation.
\par Measurements of the CMB~\cite{Penzias:1965wn, Smoot:2007zz}, the relic radiation from the early universe~\cite{Dicke:1965zz}, offer important insights into the large-scale distribution of matter in the cosmos. 
The tiny fluctuations in temperature observed across the CMB are particularly sensitive to the gravitational influence of dark matter. These anisotropies encode information about the density perturbations in the early universe, which evolved under the influence of both visible and dark matter. As such, the pattern of anisotropies provides compelling indirect evidence for the presence of dark matter and helps constrain its properties and abundance. 
For example, primary CMB anisotropies trace the primordial density fluctuations present in the early universe. The presence and distribution of dark matter play a critical role in shaping the evolution of baryon acoustic oscillations (BAOs)—pressure waves in the photon-baryon plasma prior to recombination. These imprint characteristic patterns on the CMB temperature anisotropy spectrum, particularly in the form of acoustic peaks at specific angular scales, providing further evidence of dark matter’s gravitational influence on the early universe's dynamics~\cite{Planck:2018vyg}. 
\par Finally, the large-scale distribution of galaxies and galaxy clusters reveals a remarkable web-like structure, characterized by vast cosmic voids separated by filaments and walls of galaxies.
Numerical simulations of cosmic structure formation indicate that dark matter plays a key role in this process, acting as the scaffolding around which baryonic matter accumulates. These simulations successfully reproduce the observed large-scale structure when dark matter is present, highlighting its essential contribution to the formation and evolution of visible matter in the cosmos~\cite{Navarro:1995iw}.
Altogether, CMB and large-scale structure observations serve as powerful and complementary probes of the nature and distribution of dark matter, constraining cosmological models and giving insight into dark matter's basic properties.

There are alternatives to the dark matter hypothesis, the most prominent being Modified Newtonian Dynamics (MOND) and broader modifications of General Relativity~\cite{Milgrom:1983ca,Clifton:2011jh,CANTATA:2021ktz}.
While these aim to explain galactic rotation curves without invoking dark matter, none has succeeded in accounting for the full range of observational evidence that supports the dark matter paradigm.
A key example is the Bullet Cluster, which remains as one of the strongest challenges to modified gravity theories, providing a robust evidence for non-luminous, collisionless dark matter~\cite{Clowe:2006eq}.
\par The discovery of the accelerated expansion of the Universe~\cite{SupernovaSearchTeam:1998fmf} implies that, in addition to dark matter, there must also exist a distinct component known as dark energy. Together with dark matter and ordinary (baryonic) matter, dark energy accounts for the total energy-matter content of the Universe in the framework of the $\Lambda$CDM model.
\subsection{Dark matter candidates}
\label{sec:dark-matt-cand}
One of the main open questions in modern cosmology and particle physics is the fundamental nature of dark matter—what it is actually made of.
Cosmological observations indicate that dark matter accounts for roughly 26\% of the Universe’s total energy density and nearly 85\% of its total matter content.
Despite decades of searches, its identity remains unknown. 
In particular, the extent to which dark matter interacts non-gravitationally with known particles remains a mystery. 
Unvealing the true nature of dark matter will require not only continued astronomical observations, but also a broad set of experimental search strategies—including deep underground detectors, high-energy particle accelerators, and indirect detection efforts searching for possible DM annihilation or decay products.
If dark matter is composed of a new particle species~\cite{Bertone:2004pz,Feng:2010gw,Bergstrom:2012fi}, it must satisfy several astrophysical, cosmological, and experimental requirements~\cite{Dodelson:2003ft}.  
\begin{itemize}
\item \textbf{Non-baryonic DM: }  
This essential property of dark matter is strongly supported by observations of \textit{Big Bang Nucleosynthesis} (BBN) and the cosmic microwave background (CMB). These constrain the total amount of baryonic matter in the Universe, showing that dark matter must be non-baryonic.
Table~\ref{table:cosmology} summarizes the cosmological density parameters for baryons, dark matter, visible matter, and dark energy as inferred from the latest \textit{Planck} data~\cite{Planck:2018vyg}. \\[-.6cm]
 \begin{table}[ht] 
 \begin{tabular}{| c | c |}
  \hline
  Cosmological quantity &Measured density \\
  \hline
 $\Omega_b h^2$ & $ 0.02242 \pm 0.00014$ \\
  \hline
 $\Omega_{\rm DM} h^2$ & $0.11933 \pm 0.00091$ \\
  \hline
 $\Omega_{\rm matter} h^2$ & $0.3111 \pm 0.00056$ \\
  \hline
 $\Omega_\Lambda h^2$ & $0.6889 \pm 0.00056$ \\
  \hline
 \end{tabular}
 \caption{ Values for different cosmological densities measured by the Planck satellite~\cite{Planck:2018vyg}.}
 \label{table:cosmology}
 \end{table}  \\[-.6cm]
 
   The discrepancy between $\Omega_b h^2$ and $\Omega_{\rm matter} h^2$ shows that non-baryonic dark matter dominates the mass density.  
\item \textbf{Non-relativistic DM: } 
There are two types of particle dark matter, i.e. \textit{cold} or \textit{hot} dark matter (CDM or HDM).
The distinction between CDM and HDM hinges on the thermal velocities of their constituent particles during the early universe's structure formation epoch. These velocities determine how dark matter clumps under gravity, shaping the cosmic web we observe today. 
CDM consists of slow-moving particles (e.g., WIMPs) with negligible thermal velocities relative to the universe's expansion. This allows them to cluster efficiently, seeding small-scale density fluctuations that grow into galaxies and galaxy clusters. 
HDM (e.g. massive neutrinos) involves particles moving near relativistic speeds. Their high thermal velocities lead to free streaming, hence particles escape gravitational potential wells, smoothing out density fluctuations on scales smaller than ~1 Mpc. This suppresses the formation of small structures like dwarf galaxies~\cite{Davis:1985rj,Blumenthal:1984bp}. 
Simulations show that HDM-dominated universes produce a top-down structure formation sequence (with large superclusters fragmenting into smaller objects), in conflict with the observed bottom-up growth pattern in which small structures merge into larger ones. CDM simulations align with the observed galaxy clustering. 
On the other hand high-resolution surveys reveal numerous dwarf galaxies and dark matter sub-halos, consistent with CDM predictions and not with HDM scenarios. 
Finally, the matter power spectrum derived from cosmological surveys shows a cutoff at small scales in HDM models (due to free-streaming), whereas CDM matches observations nicely. 
\item \textbf{Long-lived DM: } 
The longevity of dark matter particles is crucial for maintaining their cosmological abundance, and in most of this review we assume the existence of an unbroken stabilizing symmetry (usually $\mathbb{Z}_2$) so as to make the DM candidate absolutely stable. 
However, dark matter decay channels could exist while still preserving the dominant role of DM in structure formation. 
The KeV majoron—a hypothetical pseudo-Nambu-Goldstone boson arising from spontaneous lepton number symmetry breaking—illustrates this scenario, balancing stability with potential decay signatures~\cite{Berezinsky:1993fm}. 
The majoron decay rate is naturally suppressed by the small neutrino masses and the large lepton-number-violation symmetry-breaking scale~\cite{Lattanzi:2007ux}. 
Its mass lies typically in the KeV range, making it a candidate for warm or cold dark matter depending on production mechanisms. 
\item \textbf{Collisionless DM: } 
The collisionless nature of dark matter remains a cornerstone of the standard $\Lambda$CDM cosmological model. 
Cosmological observations, including the CMB, BAOs, and Lyman-alpha forest, provide constraints on the interactions between dark matter and ordinary matter~\cite{Buen-Abad:2021mvc}.
These complement other probes, such as direct DM detection experiments.
 However, dark matter self-interactions could exist at some level, while remaining consistent with current constraints~\cite{Bondarenko:2017rfu,Buckley:2009in}.
\end{itemize}  
\par Within the Standard Model, the only potential dark matter candidates would be neutrinos. However, cosmological and astrophysical considerations show that SM neutrinos cannot account for the observed DM abundance nor its required properties. First of all, current upper bounds on neutrino masses imply that their total contribution to the present-day matter density is much smaller than observed. 
Indeed, the thermal relic neutrino density is~\cite{Lesgourgues:2006nd,Lattanzi:2017ubx} $$\Omega_\nu h^2= \frac{\sum_i m_{i}}{93.14\text{ eV}},$$ which leads to a tiny value in view of the current upper bounds on neutrino masses discussed below in Sec.~\ref{sec:neutrino-parameters}. 
Even if neutrinos could make up for a sizeable contribution, since they are relativistic at decoupling, they would be HDM, strongly disfavored by large-scale structure observations.  
Therefore, it follows that particle dark matter must come from new physics, beyond the Standard Model.
\par
\begin{figure}[h]
\centering
\includegraphics[height=2cm,width=0.8\textwidth]{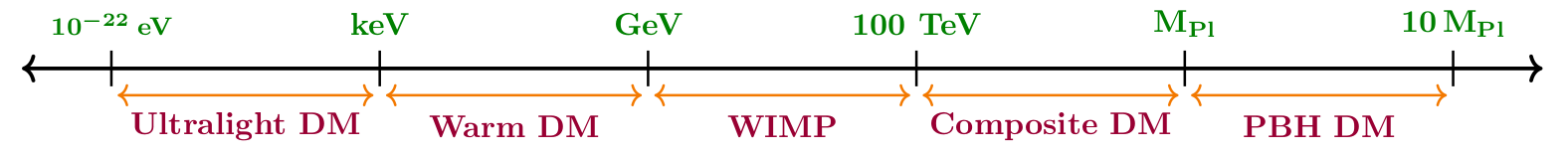}
\vskip .3cm
\caption{ Landscape of possible dark matter candidates, covering an incredibly broad range of masses.} 
\label{fig:dm-landscape}
\end{figure}

 There is a vast landscape of DM candidates that satisfy the core requirements—being long-lived, non-baryonic, and having negligible interactions with ordinary matter—spanning an enormous range of masses, from extremely light~\cite{Zimmermann:2024xvd} to extremely heavy DM, such as primordial black holes~\cite{Carr:2016drx,Green:2020jor,Zhu:2017plg}, as illustrated in Fig.~\ref{fig:dm-landscape}.
\par WIMPs are compelling dark matter candidates~\cite{Arcadi:2017kky,Feng:2022rxt,Bozorgnia:2024pwk,Boveia:2022adi} due to their weak-scale interactions, natural production in the early universe, and strong theoretical motivation from extensions of the Standard Model, such as supersymmetry with conserved R-parity~\cite{PhysRevLett.50.1419,Ellis:1983ew,Jungman:1995df}. 
WIMP signals include scattering with ordinary particles in the
laboratory, visible annihilation in the cosmos, and direct production at colliders. 
While WIMPs remain a leading dark matter candidate, the lack of definitive detection has spurred the exploration of a broader range of models. Here are some examples.
\par First of all the axion is one of the most prominent dark-matter candidates and a prime target for experimental searches.
Axions~\cite{Weinberg:1977ma,Wilczek:1977pj,Crewther:1979pi} arise in connection with the Peccei–Quinn solution to the strong CP-problem~\cite{Peccei:1977hh}, within the sub-eV mass regime. 
The axion is a pseudo-Nambu–Goldstone boson of an approximate anomalous U(1) Peccei–Quinn symmetry and can arise within~\cite{Dine:1981rt,Kim:1979if,Batra:2023erw} or beyond the minimal \SM gauge structure~\cite{Dias:2020kbj,Karan:2025pud}. 
Axion-like fields also arise in string compactifications, the so-called string axiverse~\cite{Arvanitaki:2009fg}.
Axions provide a cold, collisionless dark matter candidate produced through coherent oscillations in the early universe, primarily via the misalignment mechanism. This process explains their potential role as CDM candidates, despite their small mass
~\cite{Preskill:1982cy,Abbott:1982af,Dine:1982ah}, as they fulfill all necessary requirements discussed above; for recent reviews see~\cite{Irastorza:2018dyq,DiLuzio:2020wdo}. 
There are many phenomena associated to axions and also axion-like-particle~(ALP) generalizations, and fierce experimental efforts have been devoted to their detection see, for example~\cite{Ringwald:2012hr,HAYSTAC:2024jch,Candon:2024eah,Ning:2025tit,Angloher:2025fzw,axionpdg,axionlim}. 

\par Ultralight scalar bosons, also known as fuzzy dark matter, have been considered as a potential alternative to cold dark matter. It has been proposed that dark matter is composed of ultralight scalar particles which exhibit wave-like behavior on small scales, suppressing structure formation on these scales~\cite{Hui:2016ltb}.  \\[-.4cm]
\par
Dark-matter candidates can also be classified in terms of their velocity dispersion, or the associated free-streaming length. 
While typical TeV-scale WIMP dark matter candidates are considered cold, characterized by non-relativistic velocities at the time of structure formation, SM neutrinos have relativistic velocities and are taken  as a prominent \textit{hot} dark-matter candidate, severely disfavored by large scale structure formation studies.  
Between the cold and hot DM extremes lies an intermediate regime, referred to as warm dark matter (WDM). 
DM candidates in the keV range possess larger free-streaming lengths, which suppress the formation of small-scale structures and lead to observable differences in the matter power spectrum.
In this case dark matter particles have semi-relativistic velocities in the early universe and a moderate free-streaming length, resulting in partial suppression of small-scale structure, potentially addressing issues like the missing satellites and cusp-core problems~\cite{Penarrubia_2012}.

Prominent WDM candidates include the gravitino, the keV-scale majoron, and sterile (right-handed) neutrinos.
For example, the KeV majoron~\cite{Berezinsky:1993fm} is present in models where neutrino masses arise from the spontaneous breaking of a global lepton number symmetry, such as the type I seesaw mechanism~\cite{Chikashige:1980ui,Schechter:1981cv}. The associated Nambu-Goldstone boson can acquire mass from higher-order and/or gravitational effects~\cite{PhysRevLett.29.1698,hawking1975particle,Kallosh:1995hi,Antusch:2025ovm} and be
produced by mechanisms such as freeze-in in the early universe~\cite{
King:2024idj}.
Its decay rate is naturally suppressed by the small neutrino mass and also by the large symmetry-breaking scale, leading to detectable effects on the CMB~\cite{Lattanzi:2007ux}. 
Sub-leading electromagnetic decays lead to a plethora of associated X-ray signals~\cite{Bazzocchi:2008fh,Esteves:2010sh,Lattanzi:2013uza,Brune:2018sab}. 
Using N-body simulations one can show~\cite{Kuo:2018fgw} that the WDM decaying majoron picture can lead to a viable alternative to the $\Lambda$CDM scenario, with predictions that can differ substantially on small scales. Note that for the case of very light majorons ($\ll 100$~eV) non-thermal production from topological defects~\cite{Lazarides:2018aev} can also provide an efficient production mechanism~\cite{Reig:2019sok}. \par 

Sterile neutrinos with masses in the KeV range~\cite{Dodelson:1993je,Shi:1998km,Drewes:2016upu,Boyarsky:2018tvu} provide a simple candidate for non-baryonic dark matter.
They can be produced in the early universe via active–sterile neutrino oscillations, as originally proposed by Dodelson and Widrow~\cite{Dodelson:1993je}, or through resonant production in the presence of a lepton asymmetry, as described by Shi and Fuller~\cite{Shi:1998km}. 
Depending on the production mechanism, sterile neutrinos can behave as warm dark matter, with implications for structure formation, X-ray observations, and neutrino physics.
 Concerning their motivation, sterile neutrino dark-matter may fit into a compelling overall picture~\cite{Shaposhnikov:2024nhc}. One of its features is the existence of an X-ray signal similar to that of the above-mentioned KeV majoron dark-matter picture. \par 

Feebly Interacting Massive Particles~(FIMPs)~\cite{Hall:2009bx,Bernal:2017kxu,Borah:2025ema} represent a compelling alternative to conventional WIMP dark-matter. 
 The main difference with respect to WIMPs is that FIMPs interact so weakly that they never reach thermal equilibrium in the early universe, hence the relic density is not the result of a conventional \textit{freeze-out} mechanism, though there may be viable variants~\cite{henrich2025ultra}. 
 In \textit{freeze-in} scenarios, dark-matter is very slowly produced through decays or scattering of particles in the thermal plasma, and its abundance increases slowly but steadily as the Universe cools down to the so-called freeze-in temperature, after which it remains constant. 
 Given their extremely weak interactions with normal matter, the direct detection of FIMPs presents a great challenge. 
 A consequence of such feeble interaction is the absence of direct or indirect detection signals. 
 Hence, a positive signal in these experiments would potentially exclude the FIMP DM scenario.
Collider searches for missing-energy events, new detection techniques and/or combining indirect detection (like cosmic rays) with direct detection experiments using ultrasensitive detectors could provide ways to probe for FIMPs.
For discussions of FIMPs, constraints and experimental searches see Sec.~\ref{sec:FIMP-DM} and Sec.~\ref{subsec:fimp_col}. \par

Another compelling dark matter candidate is the dark photon, which arises from an additional U(1) gauge symmetry acting on a \textit{dark sector}~\cite{Holdom:1985ag,Essig:2013lka}.
In this framework, there can be a small  kinetic mixing between the dark photon and the SM photon leading to novel interaction terms with SM particles.
This kinetic mixing  opens various potential production mechanisms in the early universe, such as freeze-in, misalignment, or thermal production.
Dark photon abundance, mass range, and lifetime must obey cosmological and astrophysical constraints, including limits from the CMB, stellar cooling, Big Bang Nucleosynthesis, and direct detection experiments.
A detailed discussion of the associated phenomenology can be found in~\cite{Fabbrichesi:2020wbt}.\par
Last, but not least, we have the intriguing possibility that primordial black holes (PBHs) constitute some or all of the dark-matter~\cite{Carr:2016drx,Green:2020jor,Zhu:2017plg}. 
These black holes could have been formed during the radiation-dominated epoch, thereby avoiding the well-known BBN constraints.  Production mechanisms as well as cosmological and astrophysical constraints on their abundance and mass range, as well as potential observational signatures have been extensively discussed. 
With this we close our brief discussion of the broad landscape of conceivable candidates for dark-matter and their possible masses, partly captured in Fig.~\ref{fig:dm-landscape}.\par 

\subsection{WIMP dark-matter paradigm}
\label{sec:WIMP dark-matter}

Over the past few decades, WIMPs have been the leading candidates for particle dark matter, driving much of the experimental and theoretical research efforts. 
There is no precise definition of a ``WIMP'', but for our purposes we assume that WIMPs interact mainly via the weak interactions present in the Standard Model and lie within the 10 GeV and 10 TeV mass range~\footnote{
We focus mainly on WIMP dark matter above 1 GeV, as this fits most naturally within the context of models with radiative neutrino masses arising from a dark sector. However, sub-GeV WIMP-like dark matter as light as the KeV scale may be conceivable under some circumstances. A new round of experiments is now designed to probe such low-mass dark matter~\cite{SENSEI:2023zdf,DAMIC-M:2023gxo,SuperCDMS:2022kse,aguilararevalo2022oscuraexperiment,DarkSide:2018bpj}.} 
Although this is a relatively narrow mass range in Fig.~\ref{fig:dm-landscape}, WIMPs have attracted great interest in the community due to their strong theoretical motivation. 
\par
\begin{figure}[h]
\centering
\includegraphics[height=4.5cm,scale=0.3]{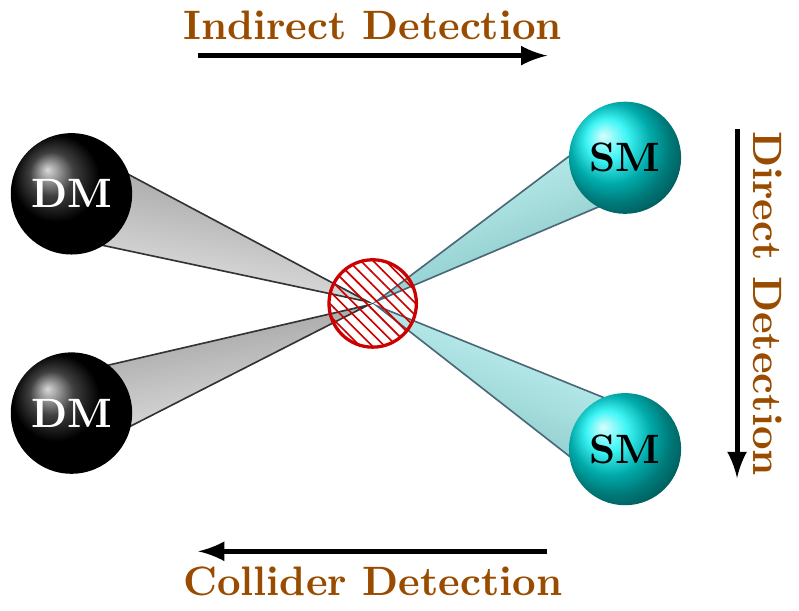}
\caption{
  Complementarity of WIMP dark-matter detection techniques~\cite{Arcadi:2017kky,Feng:2022rxt,Bozorgnia:2024pwk,Boveia:2022adi}.
  The horizontal forward direction indicates DM annihilation to SM particles, the upward direction indicates DM scattering off-nuclei, whereas the horizontal backward direction denotes DM production via SM particle collisions.} 
\label{fig:dm-collider}
\end{figure}

  WIMPs are generally argued to be a good DM candidate because they interact weakly with the SM particles, providing a non-relativistic CDM candidate compatible with all the observations. 
The \textit{freeze-out} scenario is the most common mechanism for WIMP DM production in the early Universe. The WIMP DM annihilation rate exceeds the Hubble expansion rate, so it is in thermal equilibrium with the primordial thermal bath. 
As the Universe continues to expand and cool, the WIMP number density drops to the point where they can no longer efficiently annihilate with each other, nor can they be replenished through the inverse process. This leads to their thermal freeze-out, leaving a relic abundance that persists to the present day.
WIMPs with mass near the electroweak scale, i.e. $\mathcal{O}(100\,\,\text{GeV})$ 
yield the right relic abundance through the thermal production via freeze-out. This fact is sometimes called the \textit{WIMP miracle}. 
 The thermal WIMP freeze-out in the early Universe may be thought of as due to the generic effective 4-point interaction depicted in
Fig.~\ref{fig:dm-collider}. 
For convenience, we adopt a color code for Feynman diagrams throughout the text, e.g. Figs.~\ref{fig:annihilation-diagram} and~\ref{fig:DD_scoto}: the SM particles are shown in cyan, the dark sector ($\mathbb Z_2$-odd) particles are presented in black, while the $\mathbb Z_2$-even beyond-the-Standard-Model (BSM) particles depicted in magenta.

This interaction implies a broad range of complementary dark-matter detection techniques that can be employed within a variety of search experiments~\cite{Feng:2022rxt,Bozorgnia:2024pwk} falling into three different classes. 
We now outline these:
\begin{itemize}
 \item  direct detection  method~e.g.~the nuclear recoil~\cite{Goodman:1984dc}, where one measures the 
 scattering of dark-matter particles off nuclei~\cite{Aprile:2018dbl,XENON:2020gfr,XENON:2020kmp,XENON:2023cxc,XENON:2024hup,PandaX-4T:2021bab,PandaX-II:2016vec,PandaX:2018wtu,PandaX:2022aac,PandaX:2024muv,LUX:2016ggv,LZ:2018qzl,LZ:2022lsv,PandaX:2024qfu,LZ:2024zvo};
 \item  indirect detection method, using dark-matter annihilation (or decay) to SM particles~\cite{PhysRevLett.55.257}. This method involves searching for secondaries such as gamma rays (which typically propagate interstellar or intergalactic space unaffected), positrons, or neutrinos~\cite{HESS:2016mib,ANTARES:2015vis,HESS:2016mib,Fermi-LAT:2016afa,Cuoco:2016eej,Heisig:2020nse,VERITAS:2017tif,Fermi-LAT:2017opo,AMS:2019rhg,Arguelles:2019ouk,KM3NeT:2024xca,IceCube:2022clp}; 
 \item  collider method, to produce DM through SM particle annihilation at accelerators, such as the LHC~\cite{Kahlhoefer:2017dnp,Buchmueller:2017qhf,Boveia:2018yeb}.
 \end{itemize}
 \par
 These alternative methods for WIMP dark matter search have so far only produced limits, interpreted in terms of the relevant theoretical framework. 
 However, the WIMP  hypothesis will be tested with higher sensitivities in the coming years. This will lead either to a far-reaching discovery, or to more stringent rejection~\cite{Bertone:2018krk,Albuquerque:2025xqe}.
In this review we will be concerned with a specific type of WIMP
dark-matter candidate, namely those that act as the source of neutrino mass generation, either through a \textit{visible} or via a \textit{hidden} dark sector, providing a deep interconnection between neutrino phenomena and dark matter physics.

 \subsection{Neutrino physics recap}
\label{sec:neutrino-parameters}

The propagation of massive neutrinos gives rise to the quantum-mechanical phenomenon of neutrino oscillations, as confirmed experimentally.
We now give a brief summary on the current status of neutrino oscillation parameters, as derived from global neutrino fits~\cite{deSalas:2020pgw,10.5281/zenodo.4726908,Esteban:2020cvm,Capozzi:2021fjo}. We also discuss complementary absolute neutrino mass limits derived from tritium beta decay and neutrinoless double-beta decay experiments, as well as from cosmological studies.\\

\begin{center}
   {\bf \small Neutrino oscillations}
\end{center}

The discovery of neutrino oscillations~\cite{Super-Kamiokande:1998kpq,SNO:2002tuh} marks a turning point in particle physics,  revealing fundamental differences between lepton and quark mixing that challenge our understanding of the Standard Model at a fundamental level~\cite{Ding:2024ozt}. 
Within the three-neutrino paradigm
(hints for a fourth--sterile--neutrino have been mainly disfavored~\cite{MicroBooNE:2025nll}), assuming unitarity, the neutrino mixing matrix is characterized by six physical parameters, three angles and three phases and can be conveniently described in symmetrical form~\cite{Rodejohann:2011vc} (the most general neutrino mixing matrix was systematically characterized in~\cite{Schechter:1980gr}).
The phases include, besides the Dirac phase, analogous to that of the Cabibbo-Kobayashi-Maskawa (CKM) matrix, also 
two physical Majorana phases. Altogether, they characterize leptonic CP violation~\cite{Nunokawa:2007qh,Branco:2011zb} in the simplest neutrino mixing scheme. 
Although they are physical and very important, Majorana phases do not affect conventional oscillations~\cite{Schechter:1980gk,Bilenky:1980cx,Doi:1980yb}, 
hence the mixing matrix can be taken as
\begin{eqnarray}
\label{eq:Ulep}
U_{\rm lep} = \left(\begin{array}{ccc}
c_{12} c_{13} & s_{12} c_{13} & s_{13} e^{-i \delta_{\mathrm{CP}}} \\
-s_{12} c_{23}-c_{12} s_{23} s_{13} e^{i \delta_{\mathrm{CP}}} & c_{12} c_{23}-s_{12} s_{23} s_{13} e^{i \delta_{\mathrm{CP}}} & s_{23} c_{13} \\
s_{12} s_{23}-c_{12} c_{23} s_{13} e^{i \delta_{\mathrm{CP}}} & -c_{12} s_{23}-s_{12} c_{23} s_{13} e^{i \delta_{\mathrm{CP}}} & c_{23} c_{13}
\end{array}\right) 
\end{eqnarray}
\par Without loss of generality the angles $\theta_{ij}$ can be taken to lie in the first quadrant $\theta_{ij}\in [0,\pi/2]$, and the phase $\delta_{\rm CP} \in [0,2\pi]$.  
The neutrino oscillation probabilities are described by six parameters, $\Delta m_{21}^2$, $|\Delta m_{31}^2|$, $\theta_{12}$, $\theta_{13}$, $\theta_{23}$ and $\delta_{\rm CP}$. Depending on the sign of $\Delta m_{31}^2$, there are two possible mass orderings,
$\Delta m_{31}^2>0$ is referred to as the Normal Ordering~(\textbf{NO}), while $\Delta m_{31}^2 < 0$ is referred as the 
Inverted Ordering~(\textbf{IO}). 
Different oscillation parameters are measured by different experiments, such as solar and atmospheric experiments, 
and also reactor experiments, such as KamLAND, and long-baseline accelerator experiments, e.g. NOvA and T2K. 
Results from the latest global fit of the neutrino masses and mixing parameters from the Valencia group are given in Table.~\ref{tab:sum-2020}.
\renewcommand{\arraystretch}{1.4}
\begin{table}[h!]\centering 
\catcode`?=\active \def?{\hphantom{0}}
\begin{tabular}{|c|ccc|}
\hline
parameter & best fit $\pm$ $1\sigma$ & \hphantom{x} 2$\sigma$ range \hphantom{x} & \hphantom{x} 3$\sigma$ range \hphantom{x}
\\ \hline
$\Delta m^2_{21} [10^{-5}$eV$^2$]  &  $7.50^{+0.22}_{-0.20}$  &  7.12--7.93  &  6.94--8.14  \\[4mm]
$|\Delta m^2_{31}| [10^{-3}$eV$^2$] (NO)  &  $2.55^{+0.02}_{-0.03}$  &  2.49--2.60  &  2.47--2.63  \\
$|\Delta m^2_{31}| [10^{-3}$eV$^2$] (IO)  &  $2.45^{+0.02}_{-0.03}$  &  2.39--2.50  &  2.37--2.53  \\[4mm]
$\sin^2\theta_{12} / 10^{-1}$         &  $3.18\pm0.16$  &  2.86--3.52  &  2.71--3.69  \\
$\theta_{12} /\degree$                &  $34.3\pm1.0$  &  32.3--36.4  &  31.4--37.4
\\[4mm]
$\sin^2\theta_{23} / 10^{-1}$       (NO)  &  $5.74\pm0.14$  &  5.41--5.99  &  4.34--6.10  \\
$\theta_{23} /\degree$              (NO)  &  $49.26\pm0.79$  &  47.37--50.71  &  41.20--51.33  \\
$\sin^2\theta_{23} / 10^{-1}$       (IO)  &  $5.78^{+0.10}_{-0.17}$  &  5.41--5.98  &  4.33--6.08  \\
$\theta_{23} /\degree$              (IO)  &  $49.46^{+0.60}_{-0.97}$  &  47.35--50.67  &  41.16--51.25  \\[4mm]
$\sin^2\theta_{13} / 10^{-2}$       (NO)  &  $2.200^{+0.069}_{-0.062}$  &  2.069--2.337  &  2.000--2.405  \\
$\theta_{13} /\degree$              (NO)  &  $8.53^{+0.13}_{-0.12}$  &  8.27--8.79  &  8.13--8.92  \\
$\sin^2\theta_{13} / 10^{-2}$       (IO)  &  $2.225^{+0.064}_{-0.070}$  &  2.086--2.356  &  2.018--2.424  \\
$\theta_{13} /\degree$              (IO)  &  $8.58^{+0.12}_{-0.14}$  &  8.30--8.83  &  8.17--8.96  \\[4mm]
$\delta/\pi$                        (NO)  &  $1.08^{+0.13}_{-0.12}$  &  0.84--1.42  &  0.71--1.99  \\
$\delta/\degree$                    (NO)  &  $194^{+24}_{-22}$  &  152--255  &  128--359  \\
$\delta/\pi$                        (IO)  &  $1.58^{+0.15}_{-0.16}$  &  1.26--1.85  &  1.11--1.96  \\
$\delta/\degree$                    (IO)  &  $284^{+26}_{-28}$  &  226--332  &  200--353  \\[3mm]
\hline
\end{tabular}
\caption{
Neutrino oscillation parameters summary determined from the global analysis in~\cite{deSalas:2020pgw,10.5281/zenodo.4726908}. 
The ranges for inverted ordering refer to the local minimum for this neutrino mass ordering.}
\label{tab:sum-2020}
\end{table}

\par Besides the individual oscillation parameter determination~\cite{deSalas:2020pgw}, in good agreement with those of the NuFit and Bari groups~\cite{Esteban:2020cvm,Capozzi:2021fjo}, oscillation data also provide all pairwise oscillation parameter correlations, given explicitly in~\cite{10.5281/zenodo.4726908}.
The global analysis provides a precision determination of solar oscillation parameters, as well as the reactor mixing angle, 
the latter driven mainly by Daya-Bay results.
However, three ambiguities remain. 
The first concerns the neutrino mass ordering.
Currently normal mass ordering, \textbf{NO}, is preferred over inverted ordering, \textbf{IO}, with $2.5\sigma$ statistical significance.
The second ambiguity concerns the atmospheric angle $\theta_{23}$, whose current best fit value lies in the second octant, though the first octant solution remains allowed at $3\sigma$.
Finally, we have a poor determination of $\delta_{\rm CP}$, with the preferred value given as $1.08\pi$~($1.58\pi$) for \textbf{NO}~(\textbf{IO}) cases respectively, with the errors indicated in Table~\ref{tab:sum-2020}~\footnote{ Notice that the latest results presented at the Neutrino 2024 conference do not change the global picture in any important way, and have no visible impact on the dark-matter studies presented here. }.

Altogether, we have ambiguous determinations of the neutrino mass ordering, the octant of the atmospheric mixing, and the value of the CP phase.
Improving upon these will constitute the target of the next round of experiments, such as JUNO~\cite{JUNO:2015zny,abusleme2025first}, DUNE~\cite{DUNE:2015lol} and
Hyper-Kamiokande~\cite{Hyper-Kamiokande:2018ofw}.
Besides resolving these issues, many others remain, such as probing for the robustness of the neutrino oscillation
interpretation with respect to the presence of non-standard neutrino interactions~\cite{Miranda:2004nb,Coloma:2017egw}. Likewise, other non-standard neutrino properties, such as neutrino magnetic moments should be probed, e.g. through coherent elastic neutrino-nucleus scattering~\cite{Miranda:2019wdy} or measurements of anti-neutrinos from the Sun~\cite{Miranda:2003yh,Miranda:2004nz}.
Finally, probing the unitarity of the lepton mixing matrix~\cite{Antusch:2006vwa,Escrihuela:2015wra} may be essential for shedding light on the value of the CP phase in experiments like DUNE~\cite{Miranda:2016wdr} and indirectly probing the scale of neutrino mass generation and the nature of the underlying physics.\\ 

 \begin{center}
   {\bf \small Absolute neutrino masses}
\end{center}

Neutrino oscillation experiments do not provide information on the absolute neutrino mass scale.
There are three complementary experimental approaches for probing the absolute masses of neutrinos, including beta decay studies~\cite{KATRIN:2021uub,Katrin:2024cdt}, experiments searching for neutrinoless double-beta decay \cite{Dolinski:2019nrj,GERDA:2019ivs,EXO-200:2025obe,Adams:2022jwx,Cirigliano:2022oqy}, as well as cosmological studies~\cite{Planck:2018vyg}. \par

\par First of all we mention that one can constrain the sum of neutrino masses from cosmological data, 
such as those from the CMB, baryon acoustic oscillations,  and lensing measurements, especially from the Planck satellite.
These limits hold irrespective of whether neutrinos are Dirac or Majorana fermions. 
 By combining  Planck 2018 with baryon acoustic oscillations (BAO) data-sets one obtains
\begin{align}
\sum_{i} m_i < 0.12 \text{  eV},
\end{align} 
 as the stringent and relatively robust limit on the sum of neutrino masses at $95\%$ C.L.~\cite{Planck:2018vyg}.
{ Combined results from DESI and CMB studies place the most stringent bound on the sum of neutrino massest $95\%$ C.L.~\cite{DESI:2025zgx,DESI:2024mwx} as:
\begin{align}\sum_{i} m_i < 0.064 \text{  eV},\end{align}
which, however, is quite sensitive to the underlying models, data-sets and priors assumed.}
{ Therefore, }cosmological bounds, however intriguing, are no substitute for dedicated neutrino mass experiments.\\ \par

Indeed, using the kinematics of $\beta$ decay one can obtain information on the absolute scale of neutrino masses from
beta decay studies, which are sensitive to the following effective electron neutrino mass parameter combination $m_\beta$,
\begin{align}
m_\beta^2=\sum_{j=1}^3 |U_{ej}|^2 m_j^2.
\end{align}
Currently the most stringent upper limits on $m_\beta$ are set by the KATRIN experiment~\cite{Katrin:2024cdt}, which obtained 
$$m_\beta< 0.45~\mathrm{eV~at~90\%~C.L.} $$ \\[-1cm]

On the other hand, the lepton number violating neutrinoless double-beta decay process~\cite{Dolinski:2019nrj,GERDA:2019ivs,EXO-200:2025obe,Adams:2022jwx,Cirigliano:2022oqy} (or \znbb for short) is expected to take place if neutrinos are Majorana particles. 
In what follows we employ the preferred form of the lepton mixing matrix~\cite{Schechter:1980gr}, since it provides a neater description of the \znbb decay amplitude, 
whose relevant effective mass parameter is given as~\cite{Rodejohann:2011vc}
\begin{equation}
\label{eq:mbb}
|m_{\beta\beta}|  = \left|\sum_{j=1}^3 U_{ej}^2 m_j\right|=\left|c^{2}_{12}c^{2}_{13} m_1 + s^{2}
_{12}c^{2}_{13} m_2 e^{-2i\phi_{12} }+ s^{2}_{13} m_3 e^{-2i\phi_{13}}\right|~.
\end{equation}
We now consider the restrictions on the attainable values of $|m_{\beta\beta}|$ that follow from the existing oscillation experiments in some interesting situations.

\vskip 0.5cm
\begin{center}
   {\bf \small Neutrinoless double-beta decay for the general neutrino mass spectrum} 
\end{center}

\begin{figure}[!h]
    \centering 
\includegraphics[height=6cm,width=0.5\textwidth]{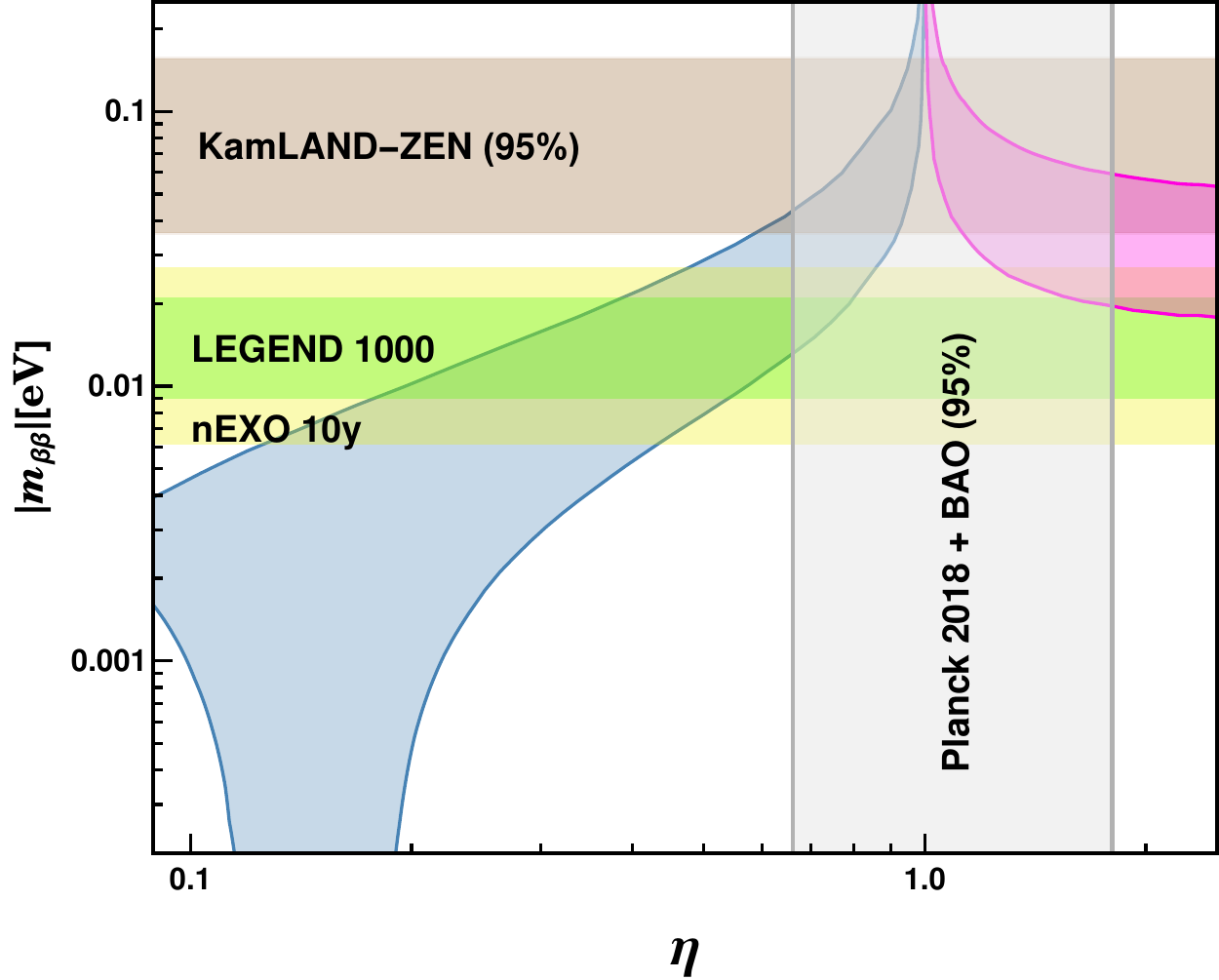}~~~~~%
\caption{
The \znbb decay amplitude in a general three-neutrino picture versus the degeneracy parameter $\eta$, taken from~\cite{Lattanzi:2020iik}. The blue and magenta branches are the regions allowed by neutrino oscillations~\cite{deSalas:2020pgw,10.5281/zenodo.4726908} for normal and inverted neutrino mass ordering.
The current experimental bound $|m_{\beta\beta}|<(36-156)\,$meV at $90\%$ C.L. from KamLAND-Zen~\cite{KamLAND-Zen:2022tow} and the sensitivities on  $|m_{\beta\beta}|<(9.0-21)\,$ meV expected at LEGEND-1000~\cite{LEGEND:2021bnm} and $|m_{\beta\beta}|<(6.1-27)\,$ meV at nEXO~\cite{nEXO:2021ujk}, as indicated by the horizontal bands in light brown,  yellow and green, respectively. The vertical grey band is excluded by the $95\%$ C.L. limit $\Sigma_{i}m_{i}<0.120\,\text{eV}$ from Planck~\cite{Planck:2018vyg,Gerbino:2022nvz}.}
\label{fig:dbd1}
\end{figure}

Fig.~\ref{fig:dbd1} shows the neutrinoless double-beta decay amplitude $|m_{\beta\beta}|$ for the general neutrino mass spectrum, 
plotted versus the degeneracy parameter $\eta$ which is related with the sum of light neutrino masses as~\cite{Lattanzi:2020iik} 
$$\sum m_\nu=\Big(\frac{\Delta m_{31}^2}{1-\eta^2}\Big)^{1/2}(1+2\eta).$$
The amplitude is very large when neutrinos are nearly degenerate in mass, or $\eta\approx 1$.

In order to generate oscillations, neutrino masses must deviate from exact degeneracy, the two curved branches to the left and to the right 
correspond to the two possible neutrino mass orderings: 
the normal-ordering (\textbf{NO}) branch is the left one in blue,
while the inverted ordering branch (\textbf{IO}) in the upper-right corner is depicted in magenta color. 
One sees that current data strongly disfavor the nearly degenerate neutrino spectrum.
Thanks to the presence of the Majorana phases in Eq.~(\ref{eq:mbb}), the \znbb decay amplitude can vanish due to possible destructive interference amongst the three individual neutrino exchange amplitudes, this happens only for normal mass ordering, see left branch.
The upper horizontal band is the current KamLAND-Zen limit~\cite{KamLAND-Zen:2022tow}, while the vertical band is excluded~\cite{Lattanzi:2020iik} as a result of the CMB and BAO observations~\cite{Gerbino:2022nvz} by the Planck collaboration~\cite{Planck:2018vyg}. 
The lower horizontal bands denote the projected sensitivities of upcoming \znbb experiments LEGEND and nEXO, as indicated in the figure. \\

\begin{center}
   {\bf \small Neutrinoless double-beta decay when one neutrino is nearly massless } 
\end{center}

The cancellation in the neutrinoless double-beta decay amplitude, Eq.~\ref{eq:mbb}, can be avoided, irrespective of the charged current form, for a lightest neutrino mass below $10^{-4}$ eV or so.
This situation happens naturally within the \textit{missing partner}
seesaw mechanism with three active neutrinos but just two singlet right-handed neutrinos, 
called (3,2) scheme~\cite{Schechter:1980gr}, for short.
This provides the minimal viable type-I seesaw that can describe the observed neutrino oscillations~\cite{King:1999mb,Frampton:2002qc,Raidal:2002xf,Barreiros:2018bju}.

Besides the tree-level \textit{missing partner} seesaw, this feature also holds in radiative models such as the singlet-triplet (or \textit{revamped}) scotogenic model that also incorporates the existence of WIMP dark-matter~\cite{Hirsch:2013ola,Merle:2016scw,Diaz:2016udz,Rocha-Moran:2016enp,Restrepo:2019ilz,Avila:2019hhv,Karan:2023adm} (as will be extensively discussed in Section \ref{sec:singlet-triplet-scoto}), and also many other incomplete multiplet schemes. 
The radiative bound-state dark-matter model proposed reference in~\cite{Reig:2018ztc} also displays this feature. \\[-.3cm]

Moreover, the existence of a \znbb lower bound is also a characteristic feature of schemes incorporating dark-matter according to the the scoto-seesaw prescription, the simplest of which employs the (3,1) \textit{missing partner} seesaw template~\cite{Rojas:2018wym,Mandal:2021yph}, combining tree level seesaw with radiative corrections, as discussed in Section~\ref{sec:scoto-seesaw}. 
\begin{figure}[!h]
\includegraphics[height=6cm,width=0.6\textwidth]{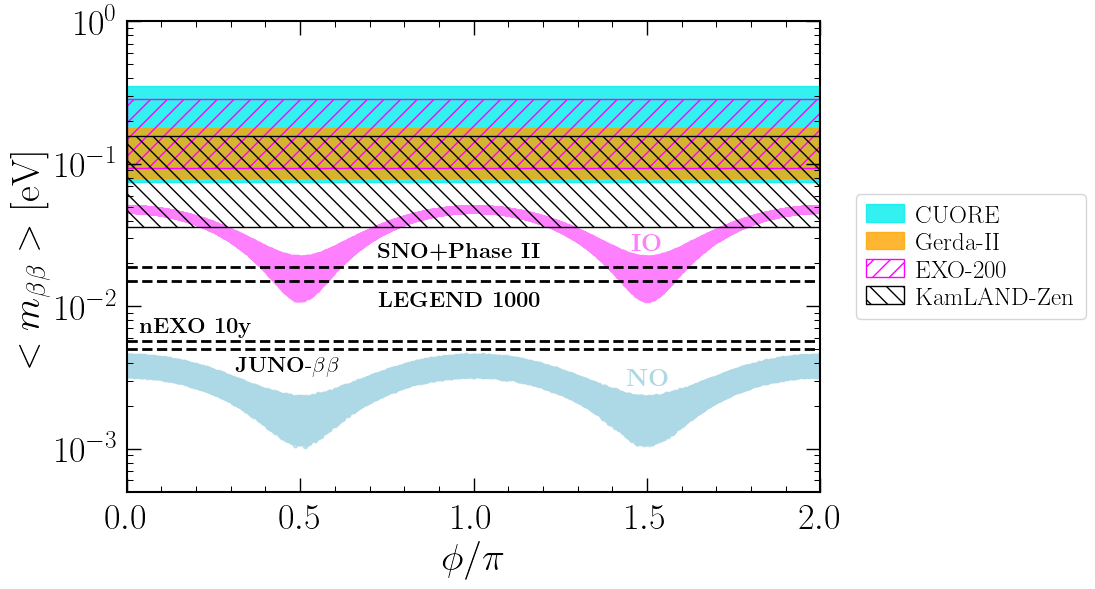}
\caption{
Allowed \znbb decay amplitude when one neutrino is massless (see text).
The periodic bands in blue and magenta are the current 3$\sigma$ C.L. regions for normal and inverted mass-ordering, respectively. The horizontal bands give the constraints from current experiments: CUORE~(cyan, $\braket{m_{\beta\beta}}<0.075-0.350$~eV)~\cite{CUORE:2020ymk},  EXO-200~(magenta, $\braket{m_{\beta\beta}}< 0.093-0.286$~eV)~\cite{EXO-200:2019rkq}, Gerda-II~(orange, $\braket{m_{\beta\beta}}<0.079-0.180$~eV)~\cite{GERDA:2020xhi} and KamLAND-Zen~(gray, $\braket{m_{\beta\beta}}<0.036-0.156$~eV)~\cite{KamLAND-Zen:2016pfg}. The black horizontal dashed lines are the projected sensitivities of SNO+ Phase-II (0.019 eV)~\cite{SNO:2015wyx}, LEGEND-1000 (0.015 eV)~\cite{LEGEND:2017cdu}, {JUNO~\cite{Zhao:2016brs}} and nEXO - 10yr (0.0057 eV)~\cite{nEXO:2017nam}. }
\label{fig:dbd2}
 \end{figure}

In the absence of cancellation in the neutrinoless double-beta decay amplitude, one has a lower bound even for normal neutrino mass ordering.
In Fig.~\ref{fig:dbd2} we give the allowed \znbb regions for all such shemes. 
They correspond to the colored bands seen in the figure, which are periodic in the relative Majorana CP phase $\varphi$, the only free parameter available. 
The colored/hatched horizontal bands give the current experimental limits, with the width of the bands reflecting nuclear matrix element uncertainties~\cite{Dolinski:2019nrj} relevant for the computation of the decay rates. The dashed horizontal lines indicate the projected experimental sensitivities.

Unfortunately the predicted theoretical lower bound on $|m_{\beta\beta}|$ lies below the sensitivities expected at the upcoming round of \znbb experiments.
However, for the case of inverse mass-ordering one would expect a guaranteed signal in these experiments~\cite{Agostini:2022zub,Gomez-Cadenas:2011oep,LEGEND:2025jwu,AMoRE:2024loj}.  %
This would also potentially enable one to underpin, up to periodicity, the value of the relative Majorana phase.\par

\begin{center}
   {\bf \small Neutrinoless double-beta decay in the presence of family symmetries} 
\end{center}

\begin{figure}[h]
\centering
\includegraphics[height=5cm,scale=0.95]{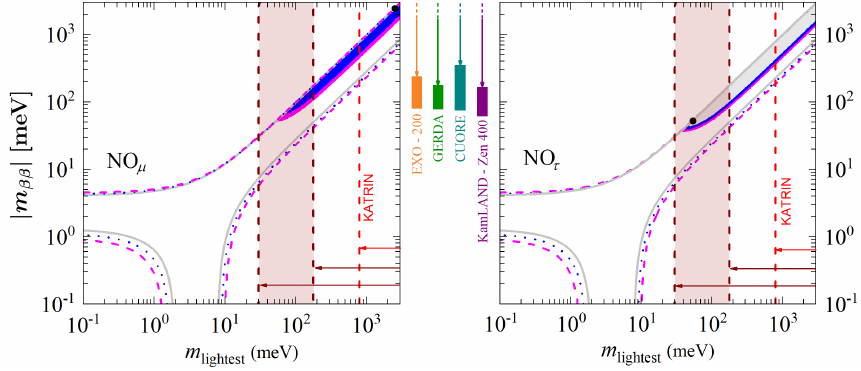}
\caption{ The solid (dotted) [dashed] lines delimit the $1\sigma$ ($2\sigma$) [$3\sigma$] $m_{\beta\beta}$ regions allowed by oscillations, and also the predictions of two flavor schemes taken from~\cite{Barreiros:2020gxu}. Vertical bars in mid-panels indicate 95\%CL $|m_{\beta\beta}|$ upper bounds from KamLAND-Zen~400~\cite{KamLAND-Zen:2016pfg}, GERDA~\cite{GERDA:2020xhi}, CUORE~\cite{CUORE:2020ymk} and EXO-200~\cite{EXO-200:2019rkq}. }
\label{fig:dbd3}
\end{figure}

We now turn to the general three-massive-neutrino case. 
Although in general there is no minimum value for the effective mass parameter $|m_{\beta\beta}|$ expected in this case, there can still be a lower-bound for the process, 
even for normal mass-ordering. 
This happens when the cancellation amongst the light-neutrino amplitudes is prevented by the structure of the leptonic weak interaction charged current vertex~\cite{King:2013hj,Hirsch:2005mc,Hirsch:2007kh,Dorame:2011eb,Dorame:2012zv,deAnda:2019jxw,deAnda:2020pti}.
This can result from the imposition of a family symmetry to account for the mixing pattern seen in oscillation experiments
\footnote{ 
Understanding the pattern of quark and lepton masses and mixing from basic principles constitutes one of the biggest challenges of present-day particle physics, see~\cite{King:2017guk,Feruglio:2019ybq,Xing:2020ijf,Chauhan:2023faf,Ding:2024ozt,Li:2025bsr} for recent review articles. }.
As an illustrative example, Fig.~\ref{fig:dbd3} shows the predictions of one such family symmetry scheme.\par
The heights of the bars shown between the two panels in Fig.~\ref{fig:dbd3} indicate the  nuclear matrix element uncertainties.
The hollow solid (dotted) [dashed] lines delimit the $1\sigma$ ($2\sigma$) [$3\sigma$] $|m_{\beta\beta}|$ allowed regions, while the colored sub-regions shown in gray, blue and magenta, respectively, correspond to the flavor-model predictions. 
The black dots are the best-fit points.  
One sees that the preferred predicted amplitude value in the right panel sits right inside the cosmologically interesting band, and close to the current \znbb limits, as indicated in between the panels.
Similar predictions for \znbb amplitudes occur in other family-symmetry models.

The interplay of the three approaches to probe the absolute neutrino mass scale can also be appreciated from the figure. 
Besides the \znbb regions, the vertical dashed red line gives the KATRIN tritium beta decay $m_{\beta}$ upper-limit~\cite{KATRIN:2021uub}, 
while the vertical shaded band indicates the current sensitivity of the cosmological data~\cite{Planck:2018vyg}. \par

In short, if neutrinos are Majorana fermions, the mixing pattern inferred from oscillation studies has an important imprint on the expectations for neutrinoless double-beta decay searches. \\

\begin{center}
   {\bf \small The significance of neutrinoless double-beta decay} 
\end{center}

Whether or not neutrinos are Dirac or Majorana fermions remains a profound open issue in particle physics.
Unless there are special symmetries that enforce the Dirac nature of neutrinos they are, in fact, 
generally expected to be Majorana-type~\cite{Valle:2015pba} so that \znbb should be expected to take place~\cite{Doi:1985dx}.
Figs.~\ref{fig:dbd2} and \ref{fig:dbd3} illustrate that there is a reasonable chance that neutrinoless double-beta decay could be detectable within the next round of experiments. This would constitute a major breakthrough in our understanding of particle physics. \par
\begin{figure}[h]
\centering
\includegraphics[height=4cm,width=0.45\textwidth]{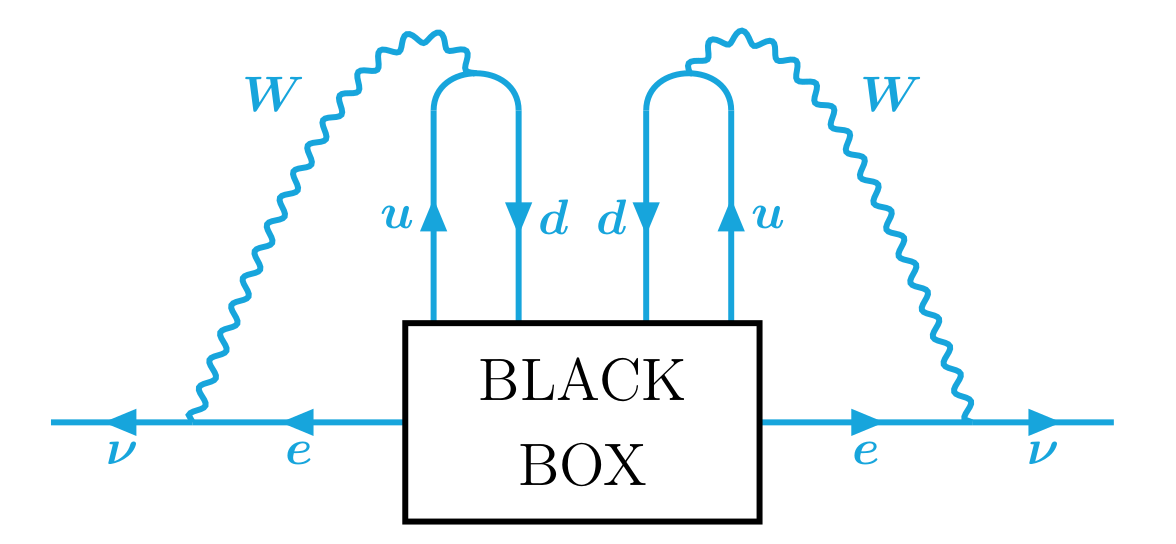}
\caption{
The observation of neutrinoless double-beta decay implies that neutrinos are Majorana fermions~\cite{Schechter:1981bd}.}
\label{fig:dbd-bb}
\end{figure}

Indeed, as a consequence of the black box theorem~\cite{Schechter:1981bd}, a positive \znbb decay detection would imply that neutrinos have Majorana nature. 
The argument is illustrated in figure~\ref{fig:dbd-bb}. 

Note that the black-box argument holds irrespective of the underlying physics responsible for generating the process~\cite{Schechter:1981bd,Duerr:2011zd,Graf:2022lhj}.
If one neutrino is (nearly) massless, as in figure~\ref{fig:dbd2}, the discovery of \znbb decay would also help underpinning the value of the relevant Majorana phase.
However, a null \znbb signal would not imply that neutrinos are Dirac-type, as the amplitude can be suppressed even for Majorana-type neutrinos, due to cancellation effects associated to the Majorana phases. 
\par Black-box-type arguments have also been suggested for the case of Dirac neutrinos~\cite{Hirsch:2017col}, but the situation is not conclusive in this case. A negative \znbb signal would also require a positive \zn4b quadruple beta decay signal~\cite{Heeck:2013rpa,NEMO-3:2017gmi} in order to ensure 
the Dirac nature of neutrinos. For the possibility of quasi-Dirac neutrinos see~\cite{Valle:1982yw}.

\begin{center}
   {\bf \small Charged lepton flavor violation} 
\end{center}

The discovery of lepton flavor non-conservation in the propagation of neutrinos was a remarkable indication of new physics in nature~\cite{McDonald:2016ixn,Kajita:2016cak}. 
The fact that charged leptons sit together with neutrinos as part of $\mathrm{SU(2)_L}$ doublets makes one expect that, at some level, charged lepton flavor violation
(cLFV) should also manisfest itself through the existence of rare processes such as $\mu \to e+\gamma$, $\mu\to 3e$ or muon-to-electron conversions in nuclei.

A summary of existing constraints and projected sensitivities for muon-related cLFV processes is provided in Table \ref{tab:LFV}.
These processes provide important restrictions for new theoretical models such as our those to be described in this review, relating dark matter to neutrino masses. 
It is worth noting that, by combining the full MEG dataset with data collected in 2021, MEG II has recently improved their limit on $\text{Br}(\mu\to e \gamma)$ to $3.1\times 10^{-13}$~\cite{MEGII:2023ltw}. 
This limit has been further tightened to $1.5\times10^{-13}$~\cite{MEGII:2025gzr} based on the 2022 dataset and an updated analysis of the 2021 data. 
\renewcommand{\arraystretch}{1.5}
\begin{table}[h!] 
    \begin{tabular}{| c | c | c |}
		\hline
		\;	cLFV Process \; & \; Current Limit \; & \; Future Sensitivity \;\\
		\hline
		$\mathcal B (\mu\to e \gamma)$& $1.5\times10^{-13}$ MEG-II~\cite{MEGII:2025gzr} & $6.0 \times 10^{-14}$  MEG-II~\cite{MEGII:2018kmf}\\
		\hline
		$\mathcal B (\mu\to 3e)$ & $1.0 \times 10^{-12}$ SINDRUM~\cite{SINDRUM:1987nra} & $\sim 10^{-16}$ Mu3e~ \cite{Blondel:2013ia,Mu3e:2020gyw}\\
		\hline
		$\mathcal C (\mu, Au \to e, Au)$ & $7.0\times 10^{-13}$ SINDRUMII~\cite{SINDRUMII:2006dvw} & --- \\
		\hline
		$\mathcal C (\mu, Pb \to e, Pb)$ & $4.6\times 10^{-11}$ SINDRUMII~\cite{SINDRUMII:1996fti} &  --- \\
		\hline
		$\mathcal C (\mu, Ti \to e, Ti)$ & $4.3\times 10^{-12}$ SINDRUMII~\cite{SINDRUMII:1993gxf} &  $\sim 10^{-18}$ PRISM~\cite{prism}\\
		\hline
		$\mathcal C (\mu, Al \to e, Al)$ & --- & $\sim 10^{-17}$ Mu2e~\cite{Mu2e:2014fns}, COMET~\cite{COMET:2018auw}\\
		\hline
	\end{tabular}
	\caption{
    Current experimental limits and future sensitivities on cLFV processes involving muons.}
	\label{tab:LFV}	
\end{table}
Severe cLFV restrictions also come from $\mu\to 3e$ decay searches and muon-to-electron conversion searches in the electromagnetic field of a gold nucleus~\cite{Lindner:2016bgg}.
Improved sensitivities are expected in the future, as seen in the Table.\\ \par

We notice that the current sensitivities on cLFV phenomena involving $\tau$ decays is relatively weak, $\mathcal O (10^{-8})$ \cite{Banerjee:2022xuw}. While cLFV in $\tau$ decays is not too constraining on theoretical models, it is expected to improve by one or two orders of magnitude by Belle II~\cite{Belle-II:2022cgf}.

\section{Inert Higgs Doublet Model }
\label{sec:Inert}

The Inert Higgs Doublet Model (IHDM) for dark matter is one of the simplest WIMP extensions of the Standard Model, which may be taken as 
a theory prototype
~\cite{Barbieri:2006dq,Branco:2011iw,Belyaev:2016lok,LopezHonorez:2006gr}.
Besides the conventional \sm Higgs doublet $\Phi$, assumed to be even under a ``dark'' $\mathbb{Z}_2$ symmetry, 
the model introduces a second complex scalar doublet $\eta$ which is odd under $\mathbb{Z}_2$. 
As a consequence, this doublet has no Yukawa couplings to SM fermions and does not acquire a nonzero vacuum expectation value (VEV).
 The most general scalar potential respecting the \SM $\otimes ~\mathbb{Z}_2$ symmetry is written as~\cite{Branco:2011iw,Belyaev:2016lok} (all parameters are assumed to be real in
order to keep CP-invariance) 
\begin{align}
 V (\Phi, \eta)=m_{\Phi}^{2}\Phi^{\dagger}\Phi + m_{\eta}^{2}\eta^{\dagger}\eta + \lambda_{1} (\Phi^{\dagger}\Phi)^{2}+\lambda_{\eta}(\eta^{\dagger}\eta)^{2}
 +\lambda_{3}(\Phi^{\dagger}\Phi)(\eta^{\dagger}\eta)+\lambda_{4}(\Phi^{\dagger}\eta)(\eta^{\dagger}\Phi) +\frac{\lambda_{5}}{2}\left((\Phi^{\dagger}\eta)^{2}+\text{h.c.}\right).
 \label{eq:scoto-pot}
\end{align}
Here $\Phi$ and $\eta$ both have hypercharge $Y=+1$, and can be written as 
\begin{equation}
\label{Higgs}
\Phi 	=
	\begin{pmatrix}
		G^+\\ \displaystyle\frac{v +h +i G^0}{\sqrt{2}} 
	\end{pmatrix}, ~~~~\ \  
\eta = \begin{pmatrix} \eta^+\\  \displaystyle\frac{\eta_{R} + i \eta_{I}}{\sqrt 2} \end{pmatrix},
\end{equation}
 where $h$ is the SM Higgs boson discovered at CERN~\cite{ATLAS:2012yve,CMS:2012qbp}, $v$ is the electroweak VEV, with $G^{+}, G^{0}$ denoting the charged and neutral unphysical Goldstone bosons, respectively. 
 The inert doublet $\eta$ contains the charged physical scalar $\eta^\pm$, along with the CP-even and CP-odd neutral scalars, $\eta_{\rm R}$ and $\eta_{\rm I}$ respectively. 
Note that, since $v_{\eta}= 0$, the $\mathbb{Z}_2$ symmetry is kept exact, stabilizing the WIMP dark-matter candidate. 

Since all the SM fermions are even under $\mathbb{Z}_2$, the scalar doublet $\eta$ does not couple to the SM fermions and therefore, the new scalars do not have fermionic interactions. The scalar-gauge boson interactions originate from the doublet kinetic terms, 
\begin{equation}
\label{HiggsKin}
{\cal L}_{kin} = (D_{\mu}\Phi)^{\dagger}  (D^{\mu} \Phi) + (D_{\mu}\eta)^{\dagger}  (D^{\mu} \eta),
\end{equation}
where $D_{\mu}$ denotes the gauge-covariant derivative.
After electroweak symmetry-breaking the tree-level masses of the physical scalars can be written as 
\begin{align}
& m_h^2 = 2\lambda_1 v^2, \nonumber\\
 & m_{\eta^{\pm}}^2 = m_{\eta}^2 + \frac{1}{2}\lambda_3 v^2 , \nonumber\\
 & m_{\eta_{R}}^2 = m_{\eta}^2 + \frac{1}{2}(\lambda_3+\lambda_4+\lambda_5)v^2 , \nonumber\\
 &m_{\eta_{I}}^2 = m_{\eta}^2 + \frac{1}{2}(\lambda_3+\lambda_4-\lambda_5)v^2. 
\label{mass_relation}
\end{align}
 Here $m_h$ is the SM-like Higgs boson mass, and $m_{\eta_{R(I)}}$ are the masses of the CP-even (odd) scalars from the inert doublet, while $m_{\eta^{\pm}}$ is the charged dark scalar mass.
 Within this framework either of the two inert neutral scalars can be the WIMP dark-matter candidate, since observations do not probe the CP-property.
 We define $\lambda_3+\lambda_4+\lambda_5 = \lambda_L$, which can be either positive or negative. 

\subsection{Relevant Constraints}
\label{Constraints0}

In order to address the phenomenology of dark-matter  one needs to implement a number of theory requirements and experimental constraints. 
We now perform the numerical analysis of our simplest theory benchmark setup, i.e. the IHDM dark-matter model. To do so we must impose the following restrictions:
\begin{itemize}
\item We ensure that the scalar potential is bounded from below, by imposing the following vacuum stability constraint,
\begin{align}
\lambda_{1},\,\, \lambda_\eta\geq 0, \,\, \lambda_3  > -2 \sqrt{\lambda_1\lambda_\eta}, \,\, \lambda_3 +\lambda_4 - |\lambda_5| > -2 \sqrt{\lambda_1\lambda_\eta},
\label{eq:boundary-condition}
\end{align}
so that the potential does not become negative for large field values.
\item We also enforce perturbativity of scalar quartic couplings, by taking them to be less than $\mathcal{O}(1)$. For such scalar quartic coupling choices, perturbative unitarity constraints outlined in Ref.~\cite{Bhardwaj:2019mts} will be satisfied.
\item The Inert Doublet Model also allows for observable signatures at current and future colliders. 
Data from the LEP collider do impose strong constraints, and we ensure that all of them are implemented in the analysis~\cite{Belyaev:2016lok}. For instance, the $W$ and $Z$ boson decay widths are precisely measured at LEP, and this gives 
\begin{align}
m_{\eta_R}+m_{\eta_I} > m_Z,\,\, 2m_{\eta^\pm} > m_Z,\,\, m_{\eta_{R,I}} + m_{\eta^\pm} > m_W.
\end{align}
For an almost degenerate mass spectrum with $\lambda_5 \to 0$ i.e. $m_{\eta_{R}} \sim m_{\eta_{I}}$, the first bound will push DM masses to $m_{\eta_{R}} > 45~ \text{GeV}$.
 The $\eta^{\pm}$ can be produced at LEP via the Drell-Yan mechanism, $e^{+}~e^{-} \to \eta^{+} \eta^{-}$, and this constrains the mass of $\eta^{\pm}$.
Using the OPAL collaboration results~\cite{OPAL:2003wxm,OPAL:2003nhx}, one obtains $m_{\eta^{\pm}} > 70~\text{GeV}$~\cite{Pierce:2007ut}. 
This is obtained by adapting and translating limits on charginos.  However, recent global fits suggest $m_{\eta^\pm}>90$ GeV~\cite{Coutinho:2024zyp}.
Likewise Ref.~\cite{Lundstrom:2008ai} used the DELPHI Collaboration study of neutralino pair production in order to constrain the masses of inert scalars. As a result, the region described by the intersection of the following conditions is ruled out,
\begin{equation}
m_{\eta_{R}} < 80~ \text{GeV},~~~ m_{\eta_{I}} < 100 ~\text{GeV},~~~ m_{\eta_{I}}- m_{\eta_{R}} > 8~ \text{GeV}.
\end{equation}

This constraint should be read as follows: if $m_{\eta_{R}}<80$ GeV and $m_{\eta_{I}}<100$ GeV then the mass splitting ($m_{\eta_{I}}-m_{\eta_{R}}$) should be smaller than 8 GeV.   
Hence $m_{\eta_{R}}<80$ GeV is allowed if $m_{\eta_{I}}<100$ GeV with mass splitting less than 8 GeV, or
$m_{\eta_{I}}>100$ GeV with an arbitrary mass splitting.
\item The production of dark states $\eta_{R,I}$ can modify the Higgs invisible decay branching ratio. In particular, if $\eta_R$ and $\eta_I$ are light enough, the following channel will contribute to the invisible decay~\cite{Mandal:2021yph} 
\begin{equation}
\begin{split}
\Gamma_{h\to \eta_{R} \eta_{R}} =& \frac{v_{\Phi}^2 \lambda_{345}^{2}}{32 \pi m_{h}} \sqrt{1-\frac{4 m_{\eta_{R}}^{2}}{m_{h}^{2}}},\\
\Gamma_{h\to \eta_{I} \eta_{I}} =& \frac{(m_{\eta_{I}}^{2} - m_{\eta_{R}}^{2} + \frac{\lambda_{345}}{2}v_{\Phi}^{2}  )^{2}}{8 \pi v_{\Phi}^{2} m_{h}} \sqrt{1-\frac{4 m_{\eta_{I}}^{2}}{m_{h}^{2}}}.
\end{split}
\end{equation}
The current best upper limit on the branching ratio to invisible modes comes from the ATLAS experiment~\cite{ATLAS:2023tkt}, $\text{BR}^{\text{inv}}(h)<0.107$ at $95\%$ C.L.
The SM-like Higgs boson $h$ also couples to $\eta^\pm$, contributing to the di-photon decay channel $h\to\gamma\gamma$~(charged scalar contributions to $h\to\gamma\gamma$ are generic features of inert doublet~\cite{Barbieri:2006dq,Cao:2007rm} as well as scotogenic schemes~\cite{Ma:2006km,Tao:1996vb,Bonilla:2016diq,Hirsch:2013ola,Merle:2016scw,Diaz:2016udz,Rocha-Moran:2016enp,Restrepo:2019ilz,Avila:2019hhv,Karan:2023adm}). In order to quantify the deviation from the SM prediction, we define the following parameter, 
\begin{align}
R_{\gamma\gamma}=\frac{\text{BR}(h\to\gamma\gamma)}{\text{BR}(h\to\gamma\gamma)^{\text{SM}}}.
\end{align}
 The ATLAS and CMS collaborations have studied this decay mode and their combined analysis with 7 and 8 TeV data gives $R_{\gamma\gamma}^{\text{exp}}=1.14_{-0.18}^{+0.19}$~\cite{ATLAS:2016neq}. 
 For the 13 TeV Run-2, there is no final combined data analysis so far, and the available data is separated by production processes~\cite{ATLAS:2019nkf}. 
 The ATLAS and CMS collaborations have presented their measurements as: $R_{\gamma\gamma}^{\text{exp}}=1.04_{-0.09}^{+0.10}$~\cite{ATLAS:2022tnm} and $R_{\gamma\gamma}^{\text{exp}}=1.13\pm0.09$~\cite{CMS:2022dwd} respectively. 
 However, using data from 7, 8, and 13 TeV collisions, the PDG (2024) gives the global average of this quantity as: $R_{\gamma\gamma}^{\text{exp}}=1.10\pm0.06$~\cite{ParticleDataGroup:2024cfk}.
\end{itemize}
Some of the above constraints, such as vacuum stability, are quite generic, and should be imposed to any dark-matter scheme, 
such as the inert doublet or the scotogenic setups to be discussed in this review.

\subsection{Dark-matter profile}
\begin{figure}[ht!]
\centering
\includegraphics[scale=0.25]{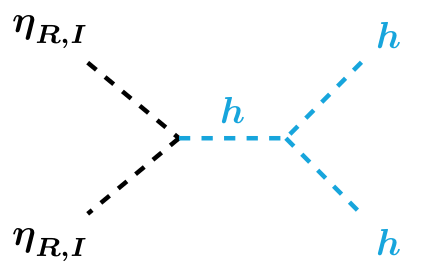}\hfil
		\includegraphics[scale=0.25]{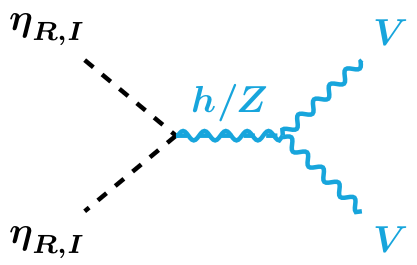}\hfil
		\includegraphics[scale=0.25]{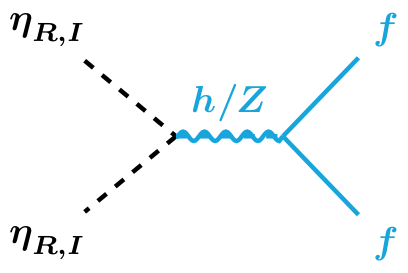}\hfil
		\includegraphics[scale=0.25]{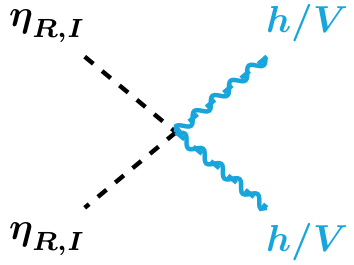}\hfil
		\includegraphics[scale=0.25]{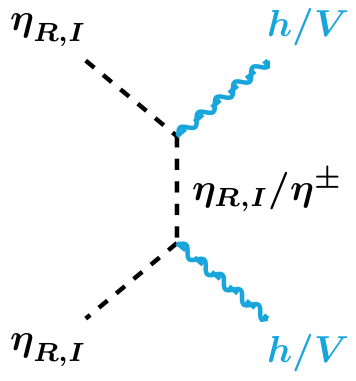}
\vspace{5mm}
\caption{
Annihilation diagrams contributing to the $\eta_R$ relic abundance, where $V=(W,Z)$ and $f$ denotes SM fermions}
\label{fig:annihilation-diagram}
\end{figure}

In this section we collect the main results of our analysis of IHDM dark-matter phenomenology.
This will be useful to understand subsequent scalar scotogenic DM results in Sec.~\ref{sec:dark-matter-scotogenic}. 
Note that the $\mathbb{Z}_2$ symmetry in the dark sector ensures the stability of the LSP.
For definiteness, we also impose the condition $\lambda_5<0$ on the quartic coupling $\lambda_5$. 
In this case $\eta_R$ will be our dark-matter candidate 
(the opposite scenario with $\lambda_5>0$ would have $\eta_I$ as the dark-matter particle).
We now summarize the relic dark-matter density calculation, determine the spin-independent direct dark-matter detection cross section, and briefly discuss the relevant experimental prospects.
In order to calculate vertices, mass matrices, tadpole equations, etc. the model is implemented in the SARAH package~\cite{Staub:2015kfa}. 
On the other hand, the thermal component of the dark-matter relic abundance, as well as the dark-matter-nucleon scattering cross section, are determined using micrOMEGAS-5.0.8~\cite{Belanger:2018ccd}.\\
\begin{center}
   {\bf \small Relic density} 
\end{center}
 There are several dark-matter annihilation/co-annihilation diagrams in the Inert Higgs Doublet Model that play a role in determining the relic abundance of cosmological dark matter. They involve annihilation to quarks and leptons, SM gauge bosons, as well as the SM Higgs boson, as depicted schematically in Fig.~\ref{fig:annihilation-diagram}.

Altogether, they determine the relic abundance of our assumed LSP, namely, $\eta_R$. 
Our numerical scan is performed varying the input parameters\footnote{ In general, in the IHDM scenario $\lambda_5$ can be larger than in the scotogenic one. However, since we will be mainly concerned with the latter,  where $\lambda_5$ must be very small to generate the tiny neutrino masses, we take the $\lambda_5$ in a similar range, given in Table~\ref{tab:scan_scoto}.} as given in Table~\ref{tab:scan_scoto}. 
\begin{table*}[ht]
\centering  
\begin{tabular}{|c|c|}
\hline
Parameters & Range \\
\hline
$m_\eta^2$ & $[100^2,5000^2]\,(\text{GeV}^2)$ \\
$\lambda_3$ & $[10^{-5},1]$ \\
$\lambda_4$ & $[10^{-5},1]$ \\
$|\lambda_5|$ & $[10^{-5},10^{-3}]$ \\
\hline
\end{tabular}
\caption{ Ranges of variation of the input parameters used in the IHDM numerical scan.}
 \label{tab:scan_scoto}
\end{table*} 

\begin{figure}[h]
\centering
\includegraphics[height=6cm,width=0.5\textwidth]{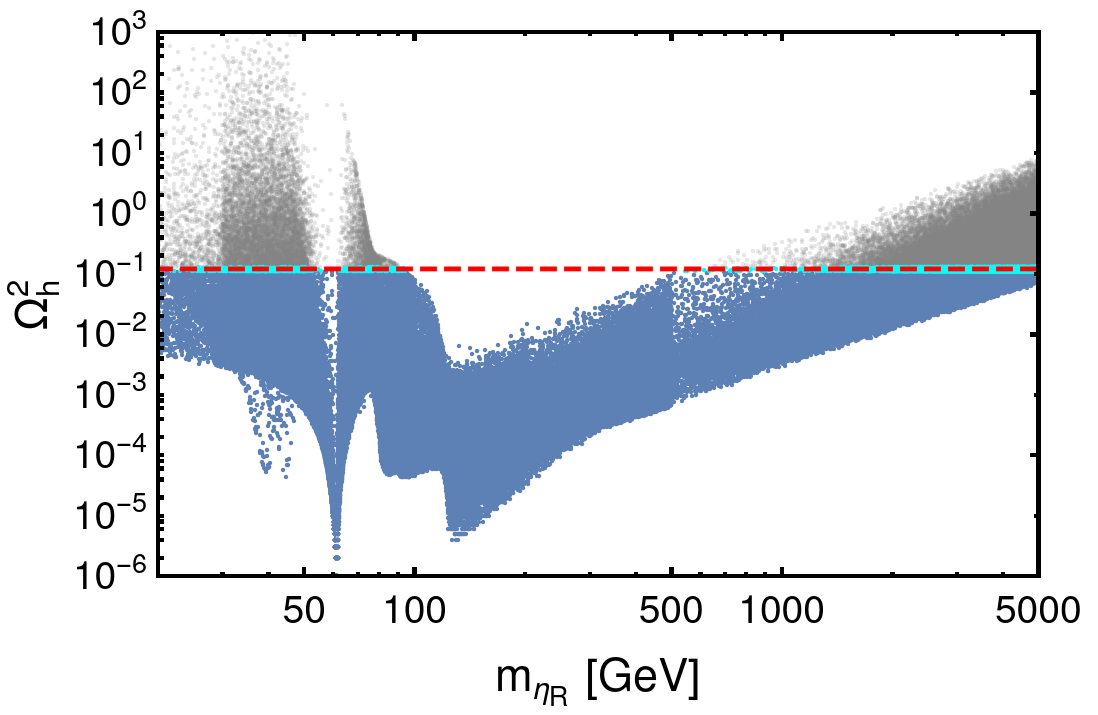}
\caption{\footnotesize{  Relic abundance versus the dark-matter mass $m_{\eta_R}$.
Cyan points on the red line fall within the measured $3\sigma$ CDM relic density range given by Planck data, Eq.~\eqref{eq:Pl}, while 
    blue/gray points outside the narrow band give DM under/over abundance, respectively. }}
\label{fig:Relic_scoto_seesaw}
\end{figure}

\begin{figure}[h]
\centering
\includegraphics[scale=0.26]{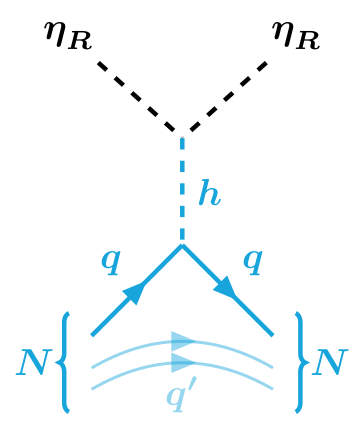}~~~~~~
\includegraphics[scale=0.26]{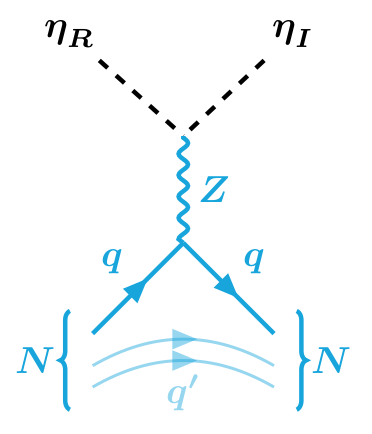}
\caption{ 
Higgs and $Z$-mediated tree-level Feynman diagrams contributing to the elastic scattering of $\eta_R$ off nuclei.}
\label{fig:DD_scoto}
\end{figure}

In Fig.~\ref{fig:Relic_scoto_seesaw} we show the relic density as a function of the mass of the scalar dark-matter candidate $\eta_R$.
The narrow horizontal band is the $3\sigma$ range for cold dark-matter derived from the Planck satellite data~\cite{Planck:2018vyg}:
\begin{align}
  \label{eq:Pl}
0.117 \leq \Omega_{\eta_R} h^2 \leq 0.122.
\end{align}
The totality of the dark-matter can be explained by $\eta_R$ only for solutions falling exactly within this band.  
The relic density for the cyan points in Fig.~\ref{fig:Relic_scoto_seesaw} lies within the above $3\sigma$ range, whereas the relic density for gray and blue points is above and below the $3\sigma$ range. 
One sees from Fig.~\ref{fig:Relic_scoto_seesaw} that the correct relic density can be obtained in three mass ranges: $m_{\eta_R}<50$~GeV, $70\,\text{GeV}< m_{\eta_R}< 100$~GeV and $m_{\eta_R}>550$~GeV.
The reasons for these mass gaps can be understood by looking in detail at the $\eta_R$ annihilation channels. 
The first dip occurs at $m_{\eta_R}\sim M_Z/2$ and corresponds to annihilation via s-channel $Z$-exchange.
The second depletion of the relic density occurs around $m_{\eta_R}\sim m_h/2$ and corresponds to annihilations via s-channel Higgs boson exchange.
This becomes very efficient when the SM-like Higgs $h$ is on-shell, precluding us from obtaining a relic density matching Planck observations.  
Notice that the second dip is more efficient than the first one, as the Z-mediated dip is momentum suppressed. For heavier $\eta_R$ masses, quartic interactions with gauge bosons become effective.
For $m_{\eta_R} \gsim 80$ GeV, annihilations of $\eta_R$ into $W^+W^-$ and $ZZ$ via quartic couplings are particularly important, thus explaining the third drop in the relic abundance. 
In the mass range $m_{\eta_R}\geq 120$ GeV, $\eta_R$ can annihilate also into Higgs boson pairs, $hh$. For $m_{\eta_R}\geq m_t$, a new channel $\eta_R\eta_R\to t\bar{t}$ opens up.  
These annihilation channels make DM annihilation very efficient decreasing the relic density. 
For larger $m_{\eta_R}$ values the relic density increases due to the suppressed annihilation cross section, which drops as $\propto 1/m_{\eta_R}^2$.
%
The gradual increase of the relic density in the intermediate and heavy $m_{\eta_R}$ region seen in the figure is controlled by the mass splitting within the scalar dark sector through $W^+W^-$ and $ZZ$ annihilation channels~\cite{Bhardwaj:2019mts}.
Finally, co-annihilation channels with $\eta_I$ and $\eta^\pm$ may occur in all parameter regions, 
lowering the relic dark-matter density. 
\begin{center}
   {\bf \small Direct detection} 
\end{center}
Let us now study the direct detection prospects of our dark-matter particle $\eta_R$. 
The tree-level spin-independent $\eta_R$-nucleon cross section is mediated by the Higgs and the $Z$ portals, see Fig.~\ref{fig:DD_scoto}. 

\begin{figure}[h]
\centering
\includegraphics[height=6cm,width=12cm]{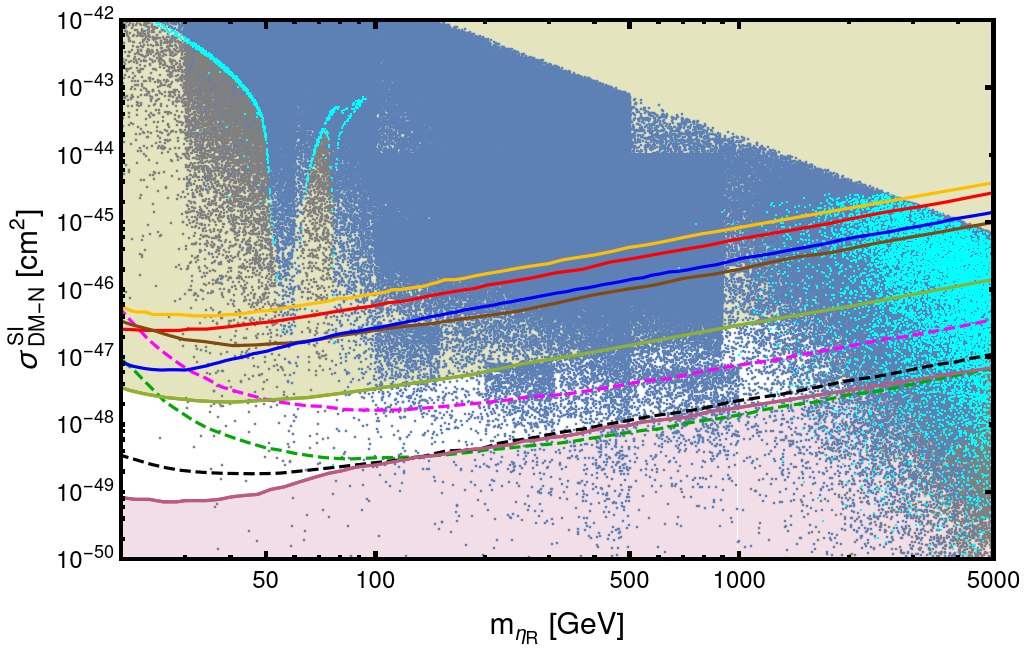}

\hspace*{0.6cm}\includegraphics[height=1.5cm,width=0.6\textwidth]{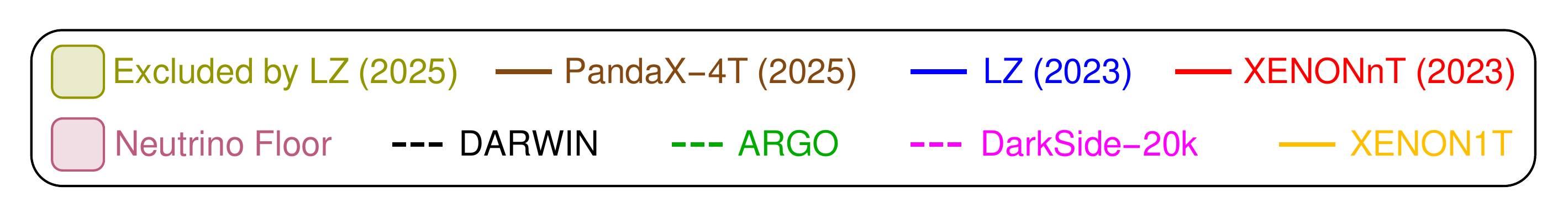}
\caption{\footnotesize{
Spin-independent WIMP-nucleon elastic scattering cross section $\sigma_\text{DM-N}^\text{SI}$ versus the dark-matter mass $m_{\eta_R}$. The cyan, blue and gray color code is as in Fig.~\ref{fig:Relic_scoto_seesaw}.
The solid yellow line denotes the upper bound from the XENON1T experiment~\cite{Aprile:2018dbl}, while the solid red and brown curves indicate the same from XENONnT with 20 ton-yr exposure~\cite{XENON:2020kmp} and PandaX-4T~\cite{PandaX:2024qfu}. The solid blue curve denotes the upper limit from LZ (2023)~\cite{LZ:2022lsv}.
The greenish shaded area is disallowed by the most stringent bound on $\sigma_\text{DM-N}^\text{SI}$ { from the LZ (2025)~\cite{LZ:2024zvo} experiment (solid green).}
Future projections from  DarkSide-20k~\cite{DS_ESPP} (magenta), DARWIN~\cite{DARWIN:2016hyl} (black) and ARGO~\cite{Billard:2021uyg} (dark green) are indicated by the dashed lines.
The lower purple shaded portion corresponds to the \textit{neutrino floor} emerging from coherent elastic neutrino scattering~\cite{Billard:2013qya}. 
The region between the greenish and purple areas includes viable nodel points that may be  experimentally detectable in the coming future. The upper triangular region violates perturbativity. }}
\label{fig:DD-constraints}
\end{figure}

%
%
%
%
For nonzero $\lambda_5$, the $\eta_R$-nucleon interaction via the Higgs portal will be the dominant one.
The coupling between $\eta_R$ and the Higgs boson depends on $\lambda_{345}=\lambda_3+\lambda_4+\lambda_5$ so that the $\eta_R$-nucleon cross section is given by
\footnote{ Note that, as the $\eta$ doublet has non-zero hypercharge, the $\eta_R$-nucleon spin-independent (SI) cross-section mediated by the Z-boson generally exceeds current limits from direct detection experiments.
However, this term is avoided by having non-zero $\lambda_5$ that causes a small mass splitting between $\eta_I$ and $\eta_R$, so the interaction through the Z-boson is kinematically forbidden or leads to inelastic scattering. }
\begin{align}
\sigma^{\text{SI}}=\frac{\lambda_{345}^2}{4\pi m_{h}^4}\frac{m_N^4 f_N^2}{(m_{\eta_R}+m_N)^2},
\end{align} 
where $m_h$ is the SM Higgs boson mass and $m_N$ is the nucleon mass, i.e. the average of the proton and neutron masses. 
Here $f_N$ is the form factor, which includes hadronic matrix elements.  
In Fig.~\ref{fig:DD-constraints} we show the spin-independent $\eta_R$-nucleon cross section as a function of the $\eta_R$ mass, for the range of parameters covered by our scan given in Table~\ref{tab:scan_scoto}.
The color code in Fig.~\ref{fig:DD-constraints} is the same as in Fig.~\ref{fig:Relic_scoto_seesaw}.

The yellow line denotes the latest upper bound from the XENON1T collaboration~\cite{Aprile:2018dbl}, while the red and brown lines indicate the bounds for XENONnT~\cite{XENON:2023cxc} and PandaX-4T~\cite{PandaX:2024qfu} collaborations. 
There are also constraints from other experiments like  LUX~\cite{Akerib:2016vxi} and PandaX-II~\cite{PandaX-II:2016vec}, which are not shown explicitly.
The strongest bound on dark-matter direct detection comes from the {LUX-ZEPLIN(LZ) collaboration in 2025~\cite{LZ:2024zvo}}. It rules out the entire upper greenish region in Fig.~\ref{fig:DD-constraints}. The blue line depicts the bound from the same collaboration in 2023.
We also show the projected sensitivities for the DarkSide-20k~\cite{DS_ESPP},
DARWIN~\cite{DARWIN:2016hyl} and ARGO~\cite{Billard:2021uyg} experiments by the magenta, black  and dark green dashed curves.

The lower limit corresponding to the \textit{neutrino floor} arising from coherent elastic neutrino-nucleus scattering (CEvNS) is also indicated. This gives an idea of the level at which neutrino backgrounds, particularly from solar neutrinos, mimic the WIMP dark-matter signal, making it difficult to distinguish between them.
 
Recent DM searches gave positive indications for the “neutrino fog” arising from coherent elastic neutrino-nucleus scattering of ${}^{8}\mathrm{B}$ solar neutrinos~\cite{PandaX:2024muv,XENON:2024hup}; for other related papers see \cite{Boehm:2018sux,OHare:2021utq,Blanco-Mas:2024ale}. 
One sees from Fig.~\ref{fig:DD-constraints} that there are perturbative low-mass solutions with the correct dark-matter relic density.
However, a big chunk of this region is ruled out by the LZ direct detection cross section limits~\cite{LZ:2022lsv}.

\begin{figure}[h]
\centering
\includegraphics[height=5cm,width=0.4\textwidth]{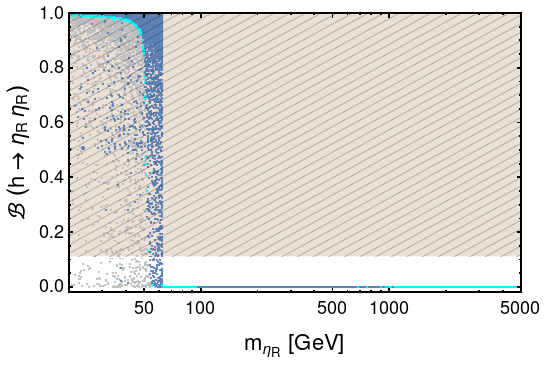}
\includegraphics[height=4.9cm,width=0.4\textwidth]{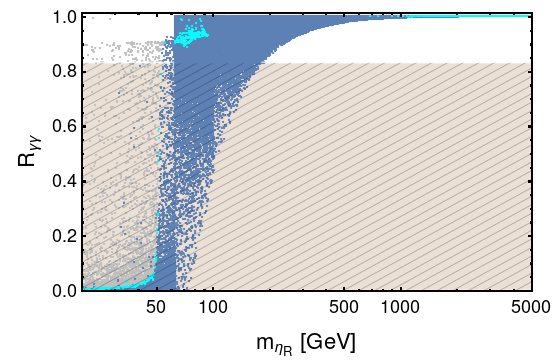}
\caption{ 
Invisible Higgs branching ratio~(left panel) and $R_{\gamma\gamma}$~(right panel) versus the dark-matter mass $m_{\eta_R}$. The color code is the same as in Fig.~\ref{fig:Relic_scoto_seesaw}.
  The shaded region in the left panel is excluded by the LHC constraint on the invisible Higgs decay~\cite{ATLAS:2023tkt}, while the same in the right panel is excluded by ATLAS measurements of $R_{\gamma\gamma}$~\cite{ATLAS:2018hxb}.     
    }
\label{fig:higgs-decay-scoto}
\end{figure}
Moreover, there are also tight constraints on low mass dark-matter from collider searches.
In the left panel of Fig.~\ref{fig:higgs-decay-scoto} we show the invisible Higgs branching ratio $\text{BR}(h\to\eta_R\eta_R)$ as a function of the dark-matter mass $m_{\eta_R}$.
In the right panel we give the expected $R_{\gamma\gamma}$ values for the same random scan of parameters. Cyan points give the relic density indicated by Planck observations~\cite{Planck:2018vyg}. 
The lower horizontal band in the left panel and the upper one in the right panel are the ones allowed by experiment. 
For the $\lambda_5$ values taken in our scan $m_{\eta_R} \approx m_{\eta_I}$ and the Higgs invisible width limit constrains $\text{BR}(h\to\eta_R\eta_R) + \text{BR}(h\to\eta_I \eta_I)$.  
For low dark-matter masses $m_{\eta_R}< 60$~GeV the invisible decay mode $h\to\eta_R\eta_R$ is open and violates the LHC limit $\text{BR}(h\to\text{Inv})\leq 0.107$~\cite{ATLAS:2023tkt}.
Likewise, the $R_{\gamma\gamma}$ measurement rules out the lower dark-matter mass band.
For intermediate dark-matter masses in the range $70\,\text{GeV}\leq m_{\eta_R}\leq 100\,\text{GeV}$, there are acceptable solutions with $R_{\gamma\gamma}\approx 1$.
In the large mass region $m_{\eta_R}>550$~GeV, the charged Higgs $\eta^{\pm}$ contribution to the di-photon decay mode $h\to\gamma\gamma$ is negligible, in such a way that $R_{\gamma\gamma}$ is close to unity, so that heavy dark-matter is  allowed by LHC constraints.
However, from this discussion, one can see that low-mass dark-matter with $m_{\eta_R}<60$~GeV is ruled out by LHC constraints.  
 Moreover, they cannot rule out the intermediate DM mass region $70\,\text{GeV}\leq\,m_{\eta_R}\leq \,100\,\text{GeV}$. 
Intermediate dark-matter masses between 100 and 550 GeV could also be possible in the presence of another (dominant) dark-matter component.

 \section{Simplest Scotogenic Model}
 \label{sec:simpl-scot-setup}

Radiative models of neutrino mass generation have a very long history, for reviews see~\cite{Boucenna:2014zba,Cai:2017jrq}. 
 Scotogenic schemes form a subclass of radiative neutrino mass models in which the calculability of neutrino masses goes hand-in-hand with the stability of dark-matter, identified as the lightest scotogenic particle, or LSP.
 The idea was first proposed in~\cite{Ma:2006km,Tao:1996vb}, revamped in~\cite{Hirsch:2013ola} and now constitutes one of the paradigms in neutrino mass generation,
 with a very extensive literature, see for example, Refs.~\cite{Merle:2015gea,Merle:2016scw,Merle:2016scw,Rocha-Moran:2016enp,Diaz:2016udz,  Kubo:2006yx,Ma:2008cu,Lozano:2025tst,Karan:2023adm,Leite:2019grf,VanDong:2023xmf,
Bonilla:2023egs,Chao:2012sz,Toma:2013zsa,Fraser:2014yha,Merle:2015ica,Lindner:2016kqk,Ma:2016mwh,Borah:2018rca,Hugle:2018qbw,Singh:2023eye,Babu:2019mfe,Nomura:2019lnr,Ahriche:2016cio,Escribano:2020iqq,Alvarado:2021fbw,Kumar:2024zfb,Garnica:2024wur,Leite:2023gzl,Lavoura:2012cv,CentellesChulia:2016rms,Farzan:2012sa,CentellesChulia:2016rms,Bonilla:2016diq,Wang:2017mcy,Bonilla:2018ynb,CentellesChulia:2019xky,Li:2022chc,Borah:2024gql,Restrepo:2019ilz,Avila:2019hhv,Kang:2019sab,CarcamoHernandez:2020ehn,Hernandez:2021zje,CentellesChulia:2024iom,Jobu:2025tto,Darricau:2025vcs}. 
  
 The terminology ``scotogenic" comes from the Greek word ``scotos," meaning darkness, and indicates that neutrino masses are mediated by dark-matter particles. 
 The simplest model contains, in addition to \sm particles 
 and a dark scalar doublet $\eta$ with hypercharge 1/2, three sequential singlet fermions $F_i$ (actually, two dark fermions are  enough, as our current evidence comes from neutrino oscillations and these involve only two mass scales: the \textit{solar} and the \textit{atmospheric} scale).
 The new ``dark-sector" particles and their quantum numbers under the \SM gauge group are given in Table.~\ref{tab:scoto}, where the family index $i$ runs from 1 to 3.  
 The additional $\mathbb{Z}_2$ symmetry is the ``dark parity" responsible for the stability of the dark-matter candidate. 
 All the Standard Model particles are $\mathbb{Z}_2$-even, while those of the dark sector, consisting of the fermions $F_i$ and the scalar doublet $\eta$, are odd. 
\par Various possible origins for such dark parity have been discussed.
 It could be a remnant parity arising from an extended gauge group~\cite{Kang:2019sab,Leite:2019grf,CarcamoHernandez:2020ehn,Hernandez:2021zje,Garnica:2024wur,VanDong:2023xmf,Leite:2023gzl} or from a global symmetry~\cite{Bonilla:2023egs}. Dark matter stability could also be enforced by other residual protecting $\mathbb{Z}_n$ symmetries~\cite{Bonilla:2018ynb} or arise accidentally from family symmetries~\cite{Lavoura:2012cv}. 

 \par
\begin{table}[!h] 
\centering
\phantom{~}\noindent
\includegraphics[scale=0.3]{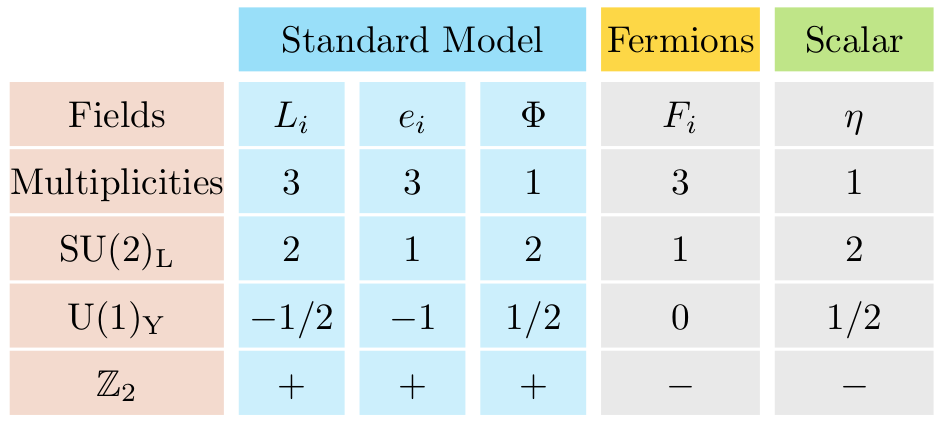}
\caption{ 
Quantum numbers of the original scotogenic model: SM and dark states differ in their $\mathbb{Z}_2$ assignment.}
\label{tab:scoto}
\end{table}

In contrast to the inert Higgs doublet scenario, discussed in Sec.~\ref{sec:Inert}, here the dark scalar doublet $\eta$ also couples to leptons, as it takes part in neutrino mass generation.
The relevant Yukawa Lagrangian is
\begin{align}
-\mathcal{L}=Y_e^{ij} \bar{L}_i \Phi e_j + Y_F^{ij} \bar{L}_{i}\tilde{\eta} F_j + \frac{1}{2} M_F^{ij} \overline{F^c_i} F_j + \text{h.c.} 
\end{align}
where $\tilde{\eta}=i\sigma_2 \eta^{*}$, $L=(\nu_{L}\,\, \ell_{L})^T$ and $e_i$ is right-handed. Notice that it
also includes an explicit Majorana mass term for the dark fermions. The \SM gauge invariant scalar potential is given by
\begin{align}
 V=m_{\Phi}^{2}\Phi^{\dagger}\Phi + m_{\eta}^{2}\eta^{\dagger}\eta + \lambda_{1} (\Phi^{\dagger}\Phi)^{2}+\lambda_{\eta}(\eta^{\dagger}\eta)^{2}
 +\lambda_{3}(\Phi^{\dagger}\Phi)(\eta^{\dagger}\eta)+\lambda_{4}(\Phi^{\dagger}\eta)(\eta^{\dagger}\Phi) +\frac{\lambda_{5}}{2}\left((\Phi^{\dagger}\eta)^{2}+\text{h.c.}\right).
 \label{eq:scoto-pot}
\end{align}
In order to ensure dark-matter stability the $\mathbb{Z}_2$ symmetry must remain unbroken. This implies that the $\mathbb{Z}_2$-odd scalar $\eta$ must not acquire a nonzero VEV. The components of $\eta$ have the following masses 
\begin{align}
 m_{\eta_{R}}^{2}&=m_{\eta}^{2}+\frac{1}{2}\left(\lambda_{3}+\lambda_{4}+\lambda_{5}\right)v_\Phi^{2},\\
  m_{\eta_{I}}^{2}&=m_{\eta}^{2}+\frac{1}{2}\left(\lambda_{3}+\lambda_{4}-\lambda_{5}\right)v_\Phi^{2},\\
 m_{\eta^{+}}^{2}&=m_{\eta}^{2}+\frac{1}{2}\lambda_{3}v_\Phi^{2}.
\end{align}
\subsection{Neutrino masses}  
\begin{figure}[h]
\centering
\includegraphics[height=3.5cm,scale=0.3]{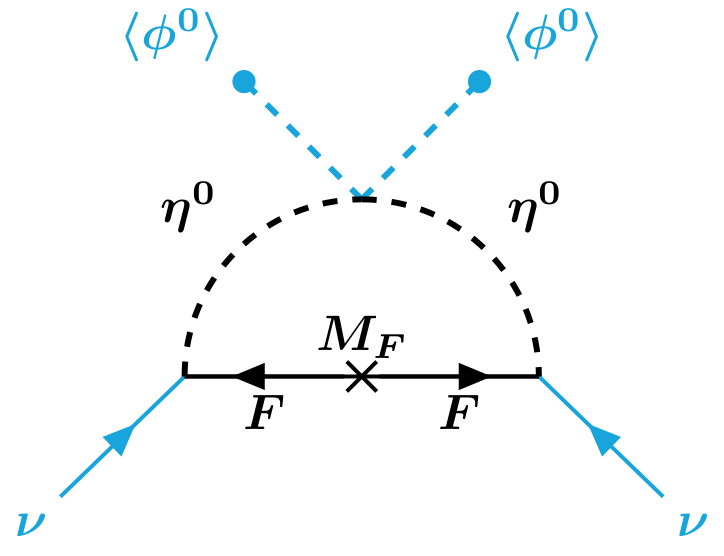}
\caption{ 
One-loop neutrino mass in the scotogenic model, where $F$ is a dark fermion and $\eta^0=(\eta_R,\eta_I)$ a dark scalar.} 
\label{fig:neutrino-loop-scotogenic}
\end{figure}

The $\mathbb{Z}_2$ symmetry forbids the tree-level contribution to neutrino masses. However, calculable masses result from the one-loop exchange of ``dark'' fermions and scalars~\cite{Ma:2006km,Tao:1996vb}, as seen in Fig.~\ref{fig:neutrino-loop-scotogenic}. 

The following expression gives the one-loop-generated neutrino mass~\cite{Mandal:2021yph}, 
\begin{align}
\label{eq:numass}
(m_\nu)_{\alpha\beta} &
= \sum_{k=1}^3 \frac{Y_F^{\alpha k} \, Y_F^{\beta k} M_{F_k}}{32 \pi^2}\left[ \frac{m_{\eta_R}^2}{m_{\eta_R}^2-M_{F_k}^2} \log \frac{m_{\eta_R}^2}{M_{F_k}^2} -  \frac{m_{\eta_I}^2}{m_{\eta_I}^2 - M_{F_k}^2} \log \frac{m_{\eta_I}^2}{M_{F_k}^2} \right],\\
& \equiv (Y_F \Lambda Y_{F}^T)_{\alpha\beta},
\end{align}
where $M_{F_k}$ are the dark fermion masses, and the $\Lambda$ matrix is defined as $\Lambda=\text{diag}(\Lambda_1,\Lambda_2,\Lambda_3)$ with 
\begin{align}
\label{eq:Lambda}
\Lambda_k
=  \frac{M_{F_k}}{32 \pi^2}\left[ \frac{m_{\eta_R}^2}{m_{\eta_R}^2-M_{F_k}^2} \log \frac{m_{\eta_R}^2}{M_{F_k}^2} -  \frac{m_{\eta_I}^2}{m_{\eta_I}^2 - M_{F_k}^2} \log \frac{m_{\eta_I}^2}{M_{F_k}^2} \right].
\end{align}
Note that in the lepton-number-symmetric limit where $\lambda_5\to 0$ one has $m_{\eta_R}^2=m_{\eta_I}^2$, implying an exact cancellation between the $\eta_R$ and $\eta_I$ loops and, as a consequence, vanishing neutrino masses. Therefore, neutrino masses are symmetry-protected.
In the small $\lambda_5$ limit one can simplify the above expression as,
\begin{align}
\label{eq:numass2}
(m_\nu)_{\alpha\beta} 
\approx \frac{\lambda_5 v_\Phi^2}{32\pi^2}\sum_{k=1}^3 \frac{Y_F^{\alpha k} \, Y_F^{\beta k} }{M_{F_k}}\left[ \frac{M_{F_k}^2}{m_{0}^2-M_{F_k}^2}  +  \frac{M_{F_k}^4}{(m_{0}^2 - M_{F_k}^2)^2} \log \frac{M_{F_k}^2}{m_0^2} \right],
\end{align}
where we assumed $m_{\eta_R}^2\approx m_{\eta_I}^2\equiv m_0^2$.  
One sees that there are many ways to accommodate the observed atmospheric and solar neutrino mass-squared differences from Eq.~\eqref{eq:numass2}. 
For instance, one can choose a sizeable Yukawa coupling $Y_F$ even with TeV-scale masses for $F_k$ and $\eta$. One can obtain the $Y_F$ using a Casas-Ibarra-like form~\cite{Casas:2001sr}, 
\begin{align}
Y_F=U_{\rm lep}^{*}\sqrt{\hat{m}_\nu} R \sqrt{\Lambda}^{-1},
\label{eq:Ynu}
\end{align}
where $R$ is a $3\times 3$ complex orthogonal matrix and $U_{\rm lep}$ is the leptonic mixing matrix 
determined through global analyses of the neutrino oscillation data~\cite{deSalas:2020pgw,10.5281/zenodo.4726908,Esteban:2020cvm,Capozzi:2021fjo}. 

\subsection{Relevant Constraints}  
\label{Constraints1}
According to the scotogenic paradigm, by construction, the \textit{lightest scotogenic particle} in the model should be the WIMP DM candidate.
The model can easily harbor either scalar or fermionic LSP.  
We will consider both alternatives. 
For scalar DM we have two choices: either $\eta_R$ or $\eta_I$.
We assume that $\eta_R$ is lighter than $\eta_I$, so it will play the role of scalar DM. 
For the fermionic case we will assume that $F_1$ is the lightest, hence the dark-matter particle.

As we did in Sec.~\ref{sec:Inert}, in our present analysis we take into account the relevant theoretical restrictions, such as perturbative unitarity and consistency of the symmetry-breaking. 
These are quite generic requirements and relatively simple to fulfill in this case, so we will not repeat the details. 
In contrast, observational constraints coming from neutrino oscillations, as well as from charged lepton flavor violation (cLFV) searches are characteristic features of scotogenic setups. They can be enforced in the following manner: 
\begin{itemize}
\item Reproducing the small neutrino masses in a natural way is one of the main goals of the scotogenic picture. This can be ensured throughout the parameter scan  by requiring compatibility with the neutrino oscillation parameters obtained in the global fit of~\cite{deSalas:2020pgw,10.5281/zenodo.4726908}, using Eq.~\eqref{eq:Ynu}. 
  For simplicity, we assumed the preferred neutrino mass ordering, \textbf{NO},  and set the poorly determined CP phases to zero.

\item For the case of scalar dark-matter $\eta_R$ we take $\lambda_4 +\lambda_5 <0$ and $\lambda_5 <0$ to make sure that $m_{\eta_R} < m_{\eta_I},\,m_{\eta^{\pm}}$. We also ensure that $m_{\eta_R} < m_{F_{i}}$. 
\end{itemize}

\subsection{Fermionic Dark-matter Relic Density}%
\begin{figure}[!h]
\centering
\includegraphics[height=4cm,scale=0.2]{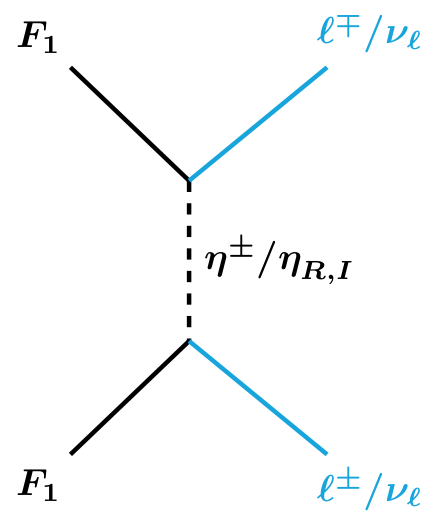}
\caption{
 Feynman diagram relevant for fermionic dark-matter annihilation in the scotogenic model.} 
\label{fig:relic-pure-singlet}
\end{figure}

\label{sec:dm-fermion-scoto}
Fermionic WIMP dark-matter is a leading dark-matter candidate naturally realized within the scotogenic approach. 
Indeed, for  $m_{F_1} < m_{\eta_R}, \,m_{\eta_I},\,m_{\eta^{\pm}}$, $F_1$ is the LSP and dark-matter particle.  
 The main annihilation channels  determining the $F_1$ relic abundance are $F_1 F_1\to\ell_i\ell_j, \nu_i\nu_j$,
 which proceed via the Yukawa coupling $Y_F$, see Fig.~\ref{fig:relic-pure-singlet}, involving the t-channel propagation of $\eta^\pm,\eta_{R,I}$ (for the case of a compressed mass spectrum with $M_{F_1}\approx m_{\eta^\pm}$, the $F_1-\eta$ co-annihilation channels are also important~\cite{Vicente:2014wga,Batra:2022pej}). 

The analytical expression for thermal averaged annihilation cross section can be written as~\cite{Kubo:2006yx,Ma:2008cu,Chao:2012sz,Duerr:2015vna} 
\begin{align}
\langle \sigma v \rangle_{F_{1}F_{1}\to \ell^{+}\ell^{-},\nu\nu}=\frac{6r_{1}^{2}(1-2r_{1}+2r_{1}^{2})\sum_{i,j}|Y_{F}^{i1} Y_{F}^{*j1}|^{2} }{24\pi M_{F_{1}}^{2}x_{f}},\hspace{0.51cm} r_1= M_{F_{1}}^{2}/(M_{F_{1}}^{2}+m_{\eta}^{2})
\label{eq:PAn1s}
\end{align}
 Assuming that the scalars are nearly degenerate $m_{\eta^\pm}\sim m_{\eta_R}\sim m_{\eta_I}$ and neglecting lepton masses one finds that the relic density is given as 
\begin{equation}
\begin{aligned}
\Omega_{F_{1}}h^{2}=1.756\times 10^{-11}\left(\frac{\langle \sigma v \rangle_{F_{1}F_{1}\to \ell^{+}\ell^{-},\nu\nu}}{x_{f}} \right)^{-1},
\label{eq:vanillascot}
\end{aligned}
\end{equation}
where $x_f$ is the ratio $M_{F_1}/T_{f}$ at the freeze-out temperature, 
\begin{align}
x_{f}=\ln \frac{0.0764 c(2+c) M_{\rm pl} M_{F_{1}}\langle \sigma v \rangle_{F_{1}F_{1}\to \ell^{+}\ell^{-},\nu\nu}}{\sqrt{g_{*} x_{f}}}.
\label{eq:Ftemp}
\end{align}
Here $c\approx 1/2$, $M_{\rm pl}=1.22\times 10^{19}$ GeV and and $g_{*}\approx 100$ is the number of relativistic degrees of freedom. 
\par Since the dark-matter annihilation cross section is determined by the Yukawa couplings $Y_F$,
these should be adequately chosen so as to produce the correct relic abundance. In order to explore the parameter space, a comprehensive  analysis should take into account all constraints, such as those coming from neutrino oscillations, lepton flavor violation and dark-matter relic density, since they all involve the same Yukawa interactions. 
\par The dark-matter analysis is performed varying the model parameters in the following ranges
\small
\begin{align}
\lambda_{3}, \lambda_{4} \in [10^{-5} , 1],\,\, \lambda_5 \in [10^{-12},10^{-9}],\,\, M_{F_1}\in [1,10^4] \text{ GeV}, \,\, M_{F_{2,3}}\in [M_{F_1},10 M_{F_1}] \text{ GeV},\,\, m_{\eta^\pm}\in [1.2 M_{F_1},10 M_{F_1}] \text{ GeV},
\end{align}
\normalsize
and implementing all relevant experimental and theoretical restrictions mentioned above. 
We also impose a perturbativity limit on the Yukawa and scalar couplings, $Y_{F}^{ij}\leq \sqrt{4\pi}$ and $|\lambda_i|\leq 1$. The latter are further required to satisfy the vacuum stability conditions in Eq.~\eqref{eq:boundary-condition}. For definitiness all scalars are taken in the range from 100 GeV to 100 TeV.  
Since existing collider data do not constrain the masses of the singlet fermions $M_{F_i}$, we allow them to vary from a small minimum value of 1 GeV. 
The minimum value of $m_{\eta^\pm}$ is set at 1.2 times $M_{F_1}$, so as to neglect $F_1-\eta$  co-annihilation effects. Notice that the ratio between the singlet fermion and  dark scalar doublet  masses is strongly restricted, to ensure an adequate annihilation cross section for $F_{1}F_{1}\to \ell^{+}\ell^{-},\nu\nu$. \\[-.3cm]
 \par The relic density, particle spectrum and decay rate determination, as well as the relevant parameter constraint implementation can be performed using the packages \textbf{SARAH}~\cite{Staub:2013tta}, \textbf{micrOmegas 5.0.2}~\cite{Belanger:2014vza}, \textbf{Spheno 4.0.3}~\cite{Porod:2011nf} and \textbf{flavorKit}~\cite{Porod:2014xia}. 
 As already mentioned, the Yukawa couplings $Y_F^{\alpha 1}$ govern the dark-matter annihilation cross section. Hence, in order to ensure that the dark-matter density is consistent with observation, at least one of these couplings needs to be sizeable. 
\begin{figure}[h]
\centering
\includegraphics[height=0.25\textwidth,width=0.42\textwidth]{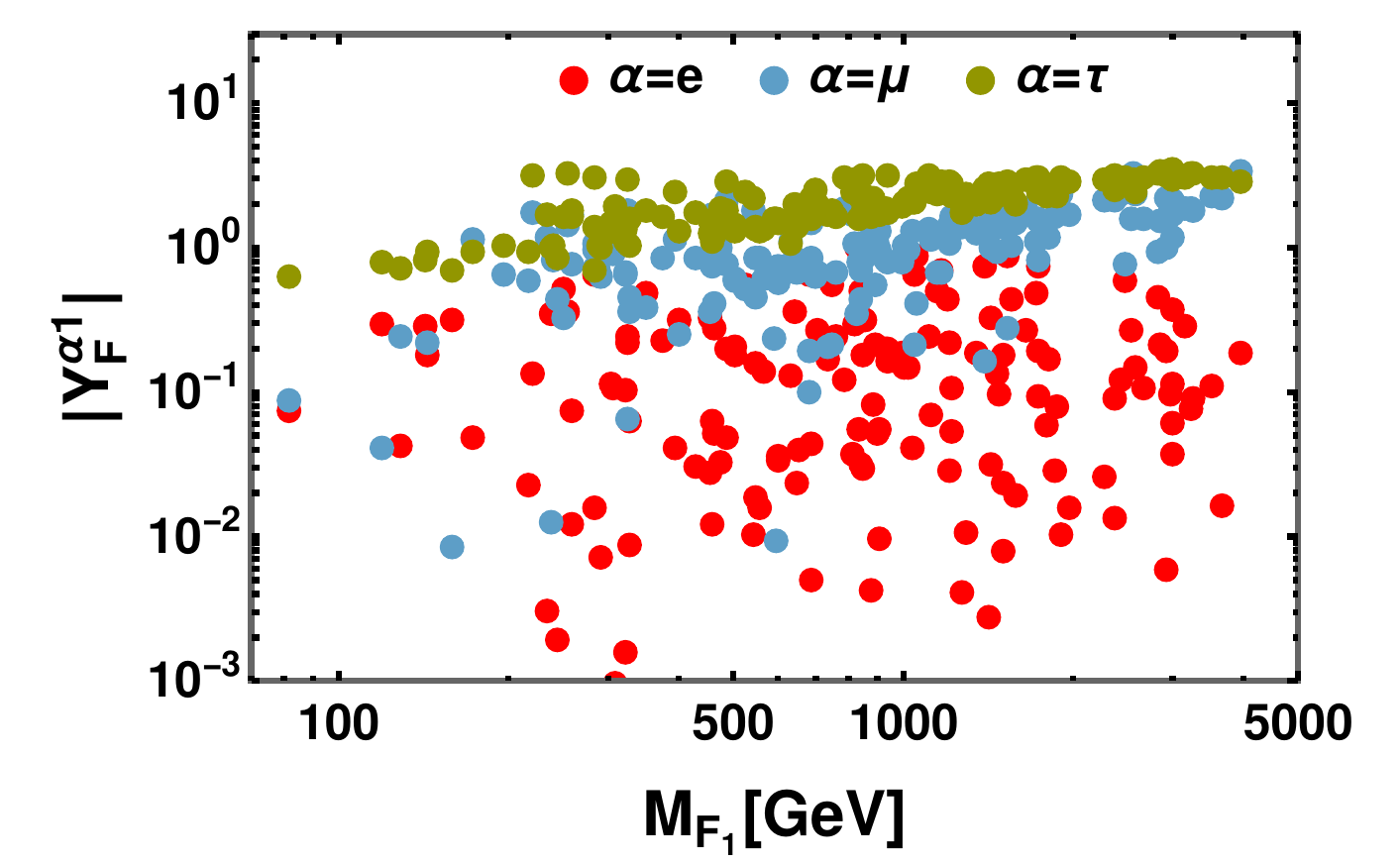}
\caption{
Region in the $M_{F_1}-|Y_F^{\alpha 1}|$ plane allowed by cLFV, dark-matter relic density and neutrino constraints. Colors correspond to $\alpha =e$ (red), $\mu$ (blue), and $\tau$ (green).} 
\label{fig:Y-lam5}
\end{figure}

Fig.~\ref{fig:Y-lam5} illustrates various parameter choices
in the plane $M_{F_1}-|Y_{F}^{\alpha 1}|$ that lead to viable scenarios.
The associated DM Yukawa couplings $|Y_{F}^{\alpha 1}|$ are in the range between $10^{-3}$ and $\sqrt{4\pi}$. 
In order to fulfill restrictions coming from cLFV processes such as $\mu\to e\gamma$ decay rates, one finds that these Yukawa couplings should obey $|Y_{F}^{e1}|\ll |Y_F^{\mu 1}|<|Y_F^{\tau 1}|$. 
It follows that dark-matter annihilates predominantly into third-generation leptons: $\tau^+\tau^-$ and $\nu_\tau\overline{\nu_\tau}$. 
One also sees from Fig.~\ref{fig:Y-lam5} that the range of DM masses is sensitive to the imposed Yukawa perturbativity limit. 

The values of $\lambda_5$ that reproduce the required neutrino masses are naturally very small, a characteristic feature of scotogenic models. 
Indeed, we find that the allowed $\lambda_5$ values consistent with neutrino mass, cLFV and relic density restrictions should lie around $\lambda_5 \sim \mathcal O (10^{-10})$.
 One sees from Eq.~\eqref{eq:numass2} that larger values of $\lambda_5$ require smaller Yukawa couplings, and this may lead to overabundant dark-matter.  
On the other hand, too small values of $\lambda_5$ would require larger Yukawa couplings, leading to potential conflict with cLFV experiments. 
Taking into account projected sensitivities of upcoming searches for
$\mu\to e\gamma$, $\mu\to 3e$, and $\mu -e$ conversion in nuclei, the parameter space for fermionic dark-matter ($F_1$) becomes restrictive~\cite{Vicente:2014wga}. 
However, there are ways to handle this issue. If the mass-splitting between $F_1$ and $\eta^\pm$ is small enough, then relic density will be determined via $F_1-\eta$ co-annihilations~\cite{Vicente:2014wga,Batra:2022pej}. 
 In this case freeze-out is delayed due to co-annihilation so that dark-matter can remain in thermal equilibrium for longer, leading to a much smaller relic abundance. 
 As a result, the Yukawa couplings can be small enough that the neutrino constraints can be satisfied. 
Concerning colliders, such scenario would yield soft final-state-leptons difficult to reconstruct using di-lepton signatures~at hadron colliders (see Sec.~\ref{subsec:collider-scotogenic}), although in lepton colliders one may still probe such regime, see Ref.~\cite{Baumholzer:2019twf}. 
The overproduction problem can also be avoided when the fermionic DM relic density arises via the freeze-in mechanism,
 as in the FIMP scotogenic scenario~\cite{Molinaro:2014lfa,Hessler:2016kwm}. We will discuss this issue later.


\subsection{Fermionic Dark-matter Direct Detection }
\label{subsec:dd-minimal-scoto}
\begin{figure}[ht!]
\begin{center}
    \includegraphics[scale=0.22]{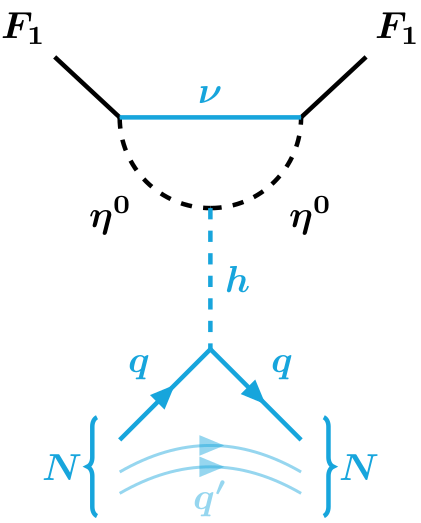}
    \includegraphics[scale=0.22]{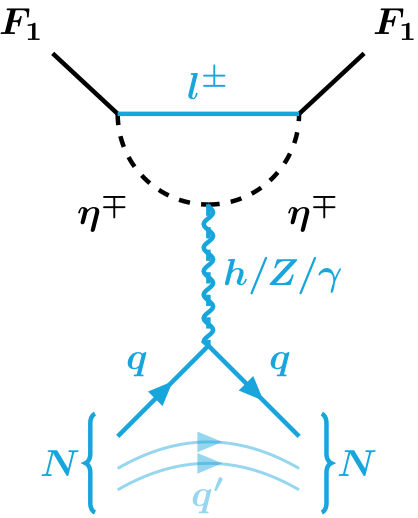}
    \includegraphics[scale=0.22]{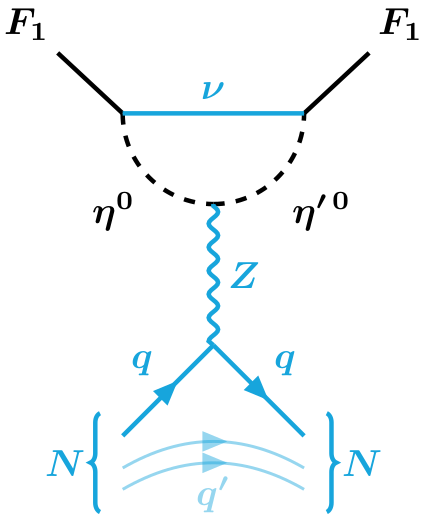}
    \includegraphics[scale=0.22]{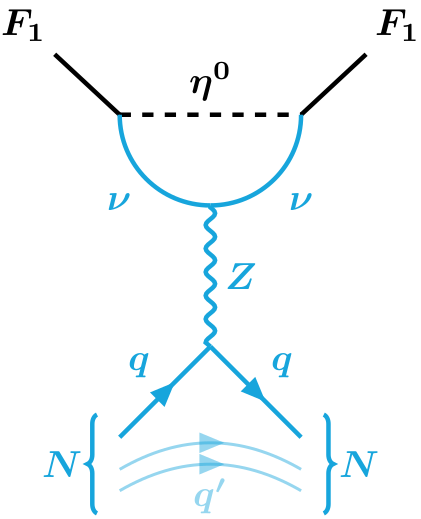}
    \includegraphics[scale=0.22]{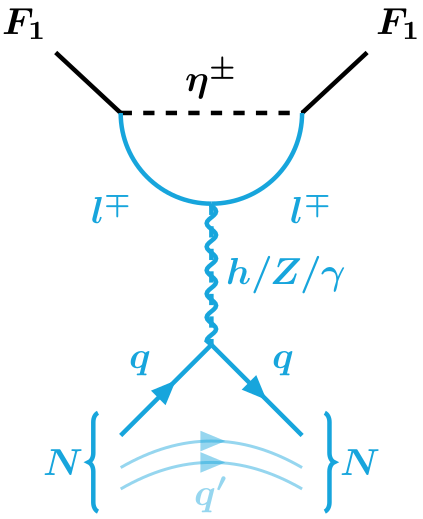}
\end{center}
    \caption{ 
    Feynman diagrams for fermionic DM-nucleon scattering cross section in the simplest scotogenic setup. Here $\eta$ is the dark scalar doublet.}    \label{fig:feyn_loop_DD}
\end{figure}

We now give a brief summary of the DM-nucleon scattering cross section in the minimal scotogenic model. As already mentioned this scattering process appears only at the one-loop level~\cite{Schmidt:2012yg,Ibarra:2016dlb}, as seen in Fig.~\ref{fig:feyn_loop_DD},
which shows the relevant Feynman diagrams for DM-matter-nucleon-scattering in the simplest scotogenic setup. 
\begin{figure}[ht!]
\begin{center}
    \includegraphics[height=6cm,width=10cm]{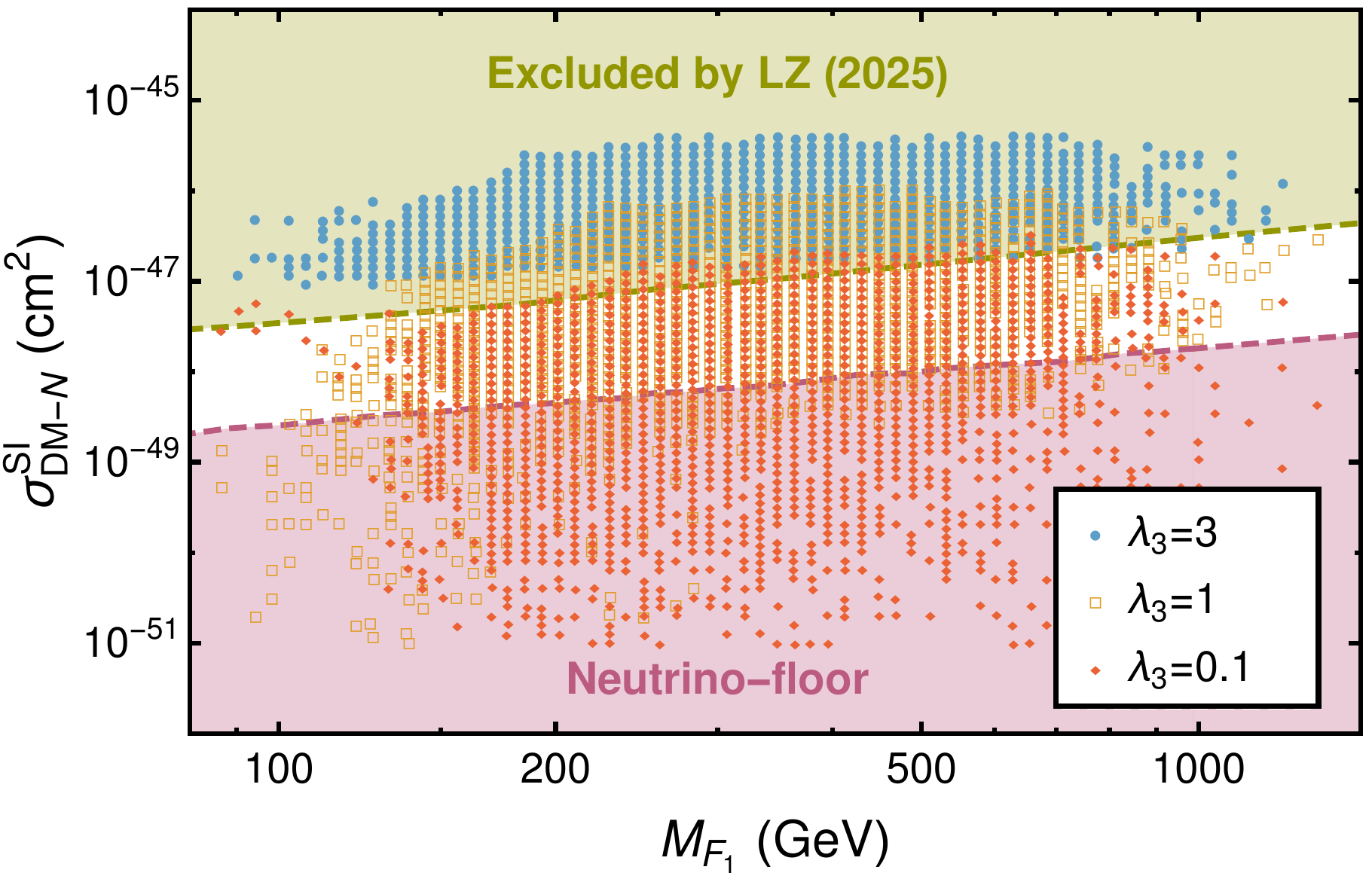}
\end{center}
    \caption{ 
    Fermionic DM-nucleon spin-independent scattering cross-section at one-loop in the simplest scotogenic scenario \cite{Ibarra:2016dlb}. The blue (circle), yellow (square) and red (diamond) points give the cross-section for different $\lambda_3$ values. The greenish region is { excluded by LZ (2025)~\cite{LZ:2024zvo}}, whereas the purple region is the neutrino-floor~\cite{Billard:2013qya}.}
    \label{fig:loop_DD}
\end{figure}

For tiny values of $\lambda_{3,4}$ the trilinear couplings between the Higgs boson and the dark scalars becomes negligible. 
In this case DM-nucleon coupling is dominated by gauge boson exchange. 
The contribution of the $Z$-boson leads to the spin-dependent scattering cross-section per nucleon similar to supersymmetric dark matter~\cite{Jungman:1995df}, i.e.
\begin{equation}
    \sigma_{DM-N}^{SD}= \frac{16}{\pi}\frac{M_{F_1}^2 m_N^2}{(M_{F_1}+ m_N)^2} J_N(J_N+1)\zeta_N
\end{equation}
where, $m_N$ and $J_N$ are the mass and total angular momentum of the nucleon. The parameter $\zeta_N$ is expressed as: $\zeta_N=\sum_{q=u,d,s} \Delta_q^N \zeta_q$ with $\Delta_u^N=0.842$, $\Delta_d^N=-0.427$ and $\Delta_s^N=-0.085$~\cite{HERMES:2006jyl} where $\zeta_q$ reads~\cite{Ibarra:2016dlb}
\begin{equation}
    \zeta_q=\frac{a_q \sum_\alpha |Y_F^{\alpha 1}|^2}{32\pi^2 M_Z^2}\Bigg[(v_l+a_l)\,\mathcal{G}_2\Big(\frac{M_{F_1}^2}{m_{\eta^\pm}^2}\Big)+(v_\nu+a_\nu)\,\mathcal{G}_2\Big(\frac{M_{F_1}^2}{m_{0}^2}\Big)\bigg].
\end{equation}
Here $m_{\eta_R}\approx m_{\eta_I} =m_{0}$, and $v_f$ and $a_f$ ($f=l,\nu,q$) are the vector and axial vector couplings, where the fermion to $Z$-boson interaction is written as: $-i \bar f \gamma^\mu (v_f-a_f\gamma^5)fZ_\mu$, and the loop function $\mathcal{G}_2(x)=1+\frac{2}{x^2}[x+(1-x)\ln(1-x)]$. 
However, in the simplest scotogenic scenario, the spin-dependent cross-section $\sigma_{DM-N}^{SD}$ remains well below the current strongest bound~\cite{Ibarra:2016dlb}, which is provided by \footnote{ Although the LZ (2025) collaboration \cite{LZ:2024zvo} provides the most stringent constraint on $\sigma_{\text{DM-N}}^{\text{SD}}$, this suffers from a huge uncertainty due to the nuclear structure function.} the  PICO-60~collaboration~\cite{PICO:2019vsc}.
Note that the diagrams involving Higgs exchange become dominant when the quartic couplings $\lambda_{3,4}$ are sizable. 

Concerning the spin-independent cross-section, one finds that it can lie within the detectable range. Indeed, this cross-section per nucleon reads
\begin{equation}
    \sigma_{DM-N}^{SI}=\frac{4}{\pi} \frac{M_{F_1}^2 m_N^2}{(M_{F_1}+ m_N)^2} \bigg(\frac{\Lambda_q^2}{m_q^2}\bigg)f_N^2~,
\end{equation}
where $m_q$ is the quark mass, $f_N\approx 0.3$ is the scalar form-factor for the nucleons. 
Here the effective coupling constant for the scalar interaction between dark-matter and quarks is given by: 
\begin{equation}
    \Lambda_q=-\frac{\sum_\alpha|Y_F^{\alpha 1}|^2}{16\pi^2 m_h^2 M_{F_1}}\bigg[\lambda_3\, \mathcal G_1\Big(\frac{M_{F_1}^2}{m_{\eta^\pm}^2}\Big)+\frac{1}{2}(\lambda_3+\lambda_4)\,\mathcal G_1 \Big(\frac{M_{F_1}^2}{m_{\eta_0}^2}\Big)\bigg] \quad \text{with} \quad \mathcal G_1(x)= 1-\Big(1-\frac{1}{x}\Big) \ln(1-x)~.
\end{equation}
Figure \ref{fig:loop_DD} compares the spin-independent DM-nucleon scattering cross-section for different values of $\lambda_3$ to the strongest current bounds coming from LZ (2025) experiment~\cite{LZ:2024zvo}. 
One sees that the loop-induced scattering cross-section enables viable fermionic dark-matter detection within the mass range of 100~GeV to 1~TeV.
\par Concerning indirect detection, fermionic scotogenic dark-matter models are similar to supersymmetric
neutralino dark-matter~\cite{PhysRevLett.50.1419,Ellis:1983ew,Jungman:1995df}. Due to the Majorana nature of $F_1$, the fermion dark-matter annihilation rate is p-wave suppressed~\cite{Duerr:2015vna}, due to the velocity dependence of the annihilation cross-section. 
This feature leads to small indirect dark-matter detection signals, given the low velocities typical of the present-day universe.
Moreover, in addition to the p-wave suppression, the annihilation rate may be further suppressed due to the small Yukawa coupling which is required in order to satisfy the neutrino and cLFV constraints.

\subsection{High energy behavior of dark parity}
\label{subsec:limitations}
 
Dark parity conservation is a key feature of scotogenic models, ensuring both the stability of dark-matter as well as the radiative origin of the neutrino masses.
It is crucial for the scotogenic approach to be consistent.
The simplest scotogenic model suffers from a theoretical limitation concerning the high energy behavior of the dark parity symmetry.
Indeed, within the simplest scotogenic model it was shown~\cite{Merle:2015gea} that renormalisation group evolution~(RGE) can alter the scalar potential at high energies, leading to $\mathbb{Z}_2$ breaking. 
In order to understand the source of this phenomenon, one should analyze the evolution of the $m_{\eta}^2$ parameter with the renormalisation scale $\mu$.

\begin{figure}[h]
\centering
\includegraphics[height=4.5cm,width=0.42\textwidth]{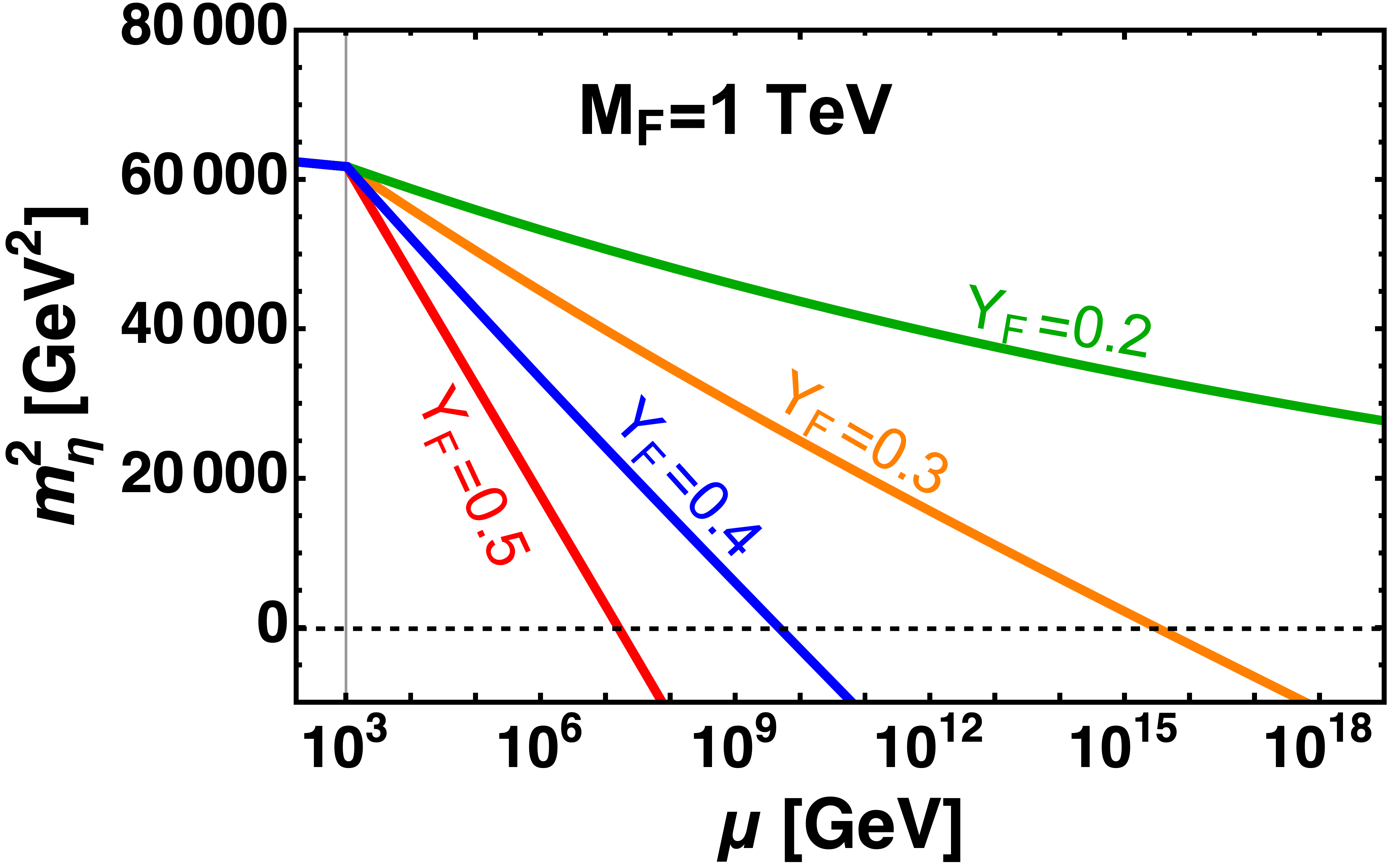}~~~
\includegraphics[height=4.5cm,width=0.42\textwidth]{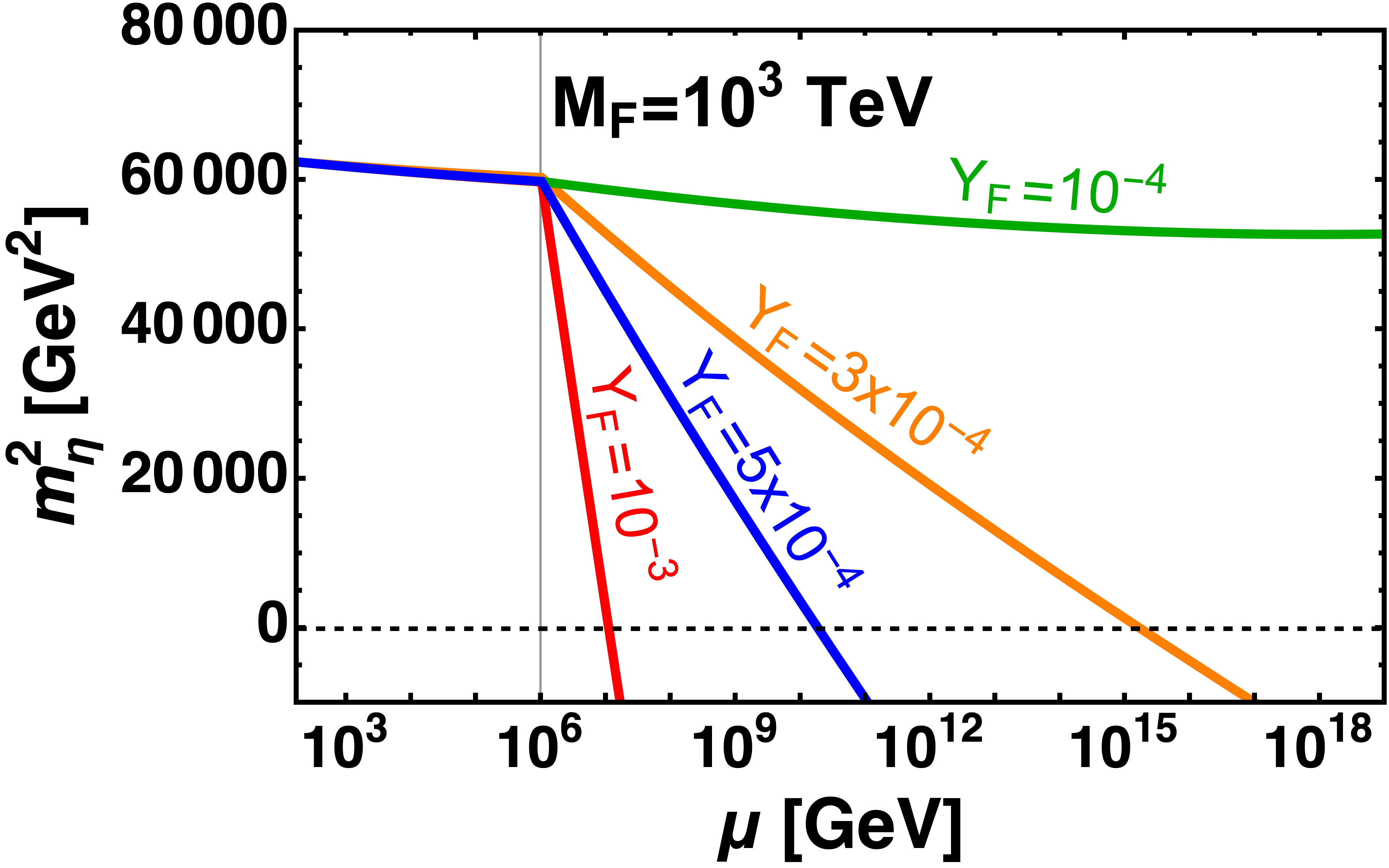}
\caption{ RGE evolution of $m_\eta^2$ as a function of the renormalization scale $\mu$. In both panels we fix the input values as $m_{\eta}^2=250^2\text{ GeV}^2$ and $\lambda_{\eta,3,4}(\mu=m_{\eta})=0.1$.  The lines in each panel correspond to different values of $Y_F(\mu=M_F)$. Comparing the panels we see that for larger $M_F$, the allowed range for the $Y_F$ Yukawa coupling becomes smaller.} 
\label{fig:z2parity-scotogenic}
\end{figure}
The beta function can be computed at one-loop as follows~\cite{Merle:2015ica} 
\begin{align}
\beta_{m_{\eta}^2}^{(1)}=12\lambda_\eta m_{\eta}^2 + (4\lambda_3+2\lambda_4)m_{\Phi}^2 + m_{\eta}^2 \Big[2\text{Tr}(Y_F^\dagger Y_F)-\frac{3}{2}(g_1^2+3 g_2^2)\Big] -4 \sum_i M_{F_i}^2 (Y_F Y_F^\dagger)_{ii} .
\label{eq:mbeta-running}
\end{align}
We need to focus on the terms which contribute negatively to the evolution of $m_{\eta}^2$. 
These involve the Yukawa coupling $Y_F$. One sees that, if  is the latter is large enough and $M_F^2 > m_{\eta}^2$, the last term dominates the running of $m_{\eta}^2$. 
In such case, $m_{\eta}^2$ can become negative very quickly and induce a minimum of the scalar potential with $\braket{\eta}\neq 0$. 
Note, however, that Eq.~\eqref{eq:mbeta-running} also contains terms that can counter this negative effect. For instance, terms involving the quartic couplings $\lambda_{3,4}$ may  contribute positively, if the adequate sign choice is made. 
\par Fig.~\ref{fig:z2parity-scotogenic} shows the evolution of the scalar mass $m_{\eta}^2$ as a function of the renormalisation scale $\mu$, for different choices of the model parameters. 
The left and right panels correspond to different $M_F$ values, $M_F=10^3$ GeV and $M_F=10^6$ GeV, respectively. 
For both panels we fix the input values as $m_{\eta}^2=250^2~\text{GeV}^2$, $\lambda_{\eta,3,4}(\mu=m_{\eta})=0.1$.  
For simplicity we assume the Yukawa coupling matrix to be diagonal, with diagonal entries denoted as $Y_F$.
The red, blue, orange and green lines in the left~(right) panel correspond to different values of $Y_F$. 
One sees from Fig.~\ref{fig:z2parity-scotogenic} that, in order to preserve the $\mathbb{Z}_2$ symmetry all the way up to the Planck scale, for larger values of $M_F$ the allowed value of the Yukawa coupling should be smaller. 
For example, one finds that the allowed Yukawa coupling is $Y_F\leq 0.2$ for $M_F=10^3$ GeV, whereas for a larger mass $M_F=10^6$ GeV, it should be smaller, $Y_F\leq 10^{-4}$.  
Choosing different input values for the quartic couplings $\lambda_{3,4}$, will alter the details, but not the generic picture.
One should also be careful with the choices of input values for the quartic couplings, since relatively large input values may quickly make them non-perturbative. 
Moreover, as discussed in Sec.~\ref{sec:darkmatterscoto}, for the scalar dark-matter case, $M_F^2 > m_{\eta_{R,I}}^2$, dark parity can only be preserved up to the Planck scale for restricted values of the $Y_F$ Yukawa coupling, suppressing the rates for cLFV processes.\\[-.8cm]

\subsection{Scotogenic FIMPs}
\label{sec:FIMP-DM}
We now inquire under what circumstances the scotogenic picture may provide a natural framework for FIMP dark matter. The basic condition for the FIMP mechanism is that the dark matter particle must remain out of thermal equilibrium in the early Universe. Within the simplest scotogenic model, only the dark fermions $F_i$ are viable FIMP candidates. To prevent these particles from reaching equilibrium, their Yukawa couplings must be very small.
\par On the other hand, the basic idea of the scotogenic picture is that neutrino masses arise radiatively from dark sector exchange, as in Eq.~\eqref{eq:numass2}. For WIMP dark matter this typically requires much larger dark Yukawa couplings than indicated, making it unviable to have a minimal dark-matter loop-mediation of the neutrino mass, with a single dark fermion responsible for both the relic DM density and the required neutrino mass scale.
However, two of the three dark fermions can mediate the required oscillation mass-squared splittings, while the other one can provide the critical DM density. 

In contrast to WIMPs, which are produced by freeze-out, FIMPs will most naturally be produced by the freeze-in mechanism. The production of the DM fermion $F_1$ will be driven by the decays of the dark scalars~($\eta_{R,I}$, $\eta^\pm$), while they are still in equilibrium with the thermal bath. The decay rate for the production of $F_1$ is approximately given by $\Gamma(\eta\to F_1 L)\approx m_{\eta} y_1^2/8\pi$, where $y_1^2=\sum_\alpha |Y_F^{\alpha 1}|^2$. Then, the out-of-equilibrium condition for this decay
reads,
\begin{align}
\Gamma(\eta\to F_1 L) \lesssim H(T\sim m_{\eta}), \end{align}
which for $m_\eta\sim \mathcal{O}(100\text{ GeV})$ implies $y_1\lesssim 10^{-8}$. For larger $y_1$ values, there would be sufficient production of $F_1$ to achieve thermal equilibrium~\cite{Molinaro:2014lfa}~\footnote{ 
Note that, if kinematically allowed, the heavier dark fermions, $F_{2,3}$, can also be produced via the same scalar decays, or via the inverse decay $F_{2,3}+L\to \eta$ if they are heavier than scalars. Since adequate neutrino masses require $y_{2,3}\gtrsim 10^{-6}$, $F_{2,3}$ necessarily reach thermal equilibrium in the early Universe. Hence, the scotogenic model admits only one FIMP candidate, $F_1$.}. Such small Yukawa coupling implies that $F_1$ gives a negligible contribution to neutrino masses. There can be an additional contribution to the $F_1$ production from the decays of the heavier dark fermions $F_{2,3}\to F_1 \bar{\ell}_{\alpha}\ell_\beta$, but this gives a negligible contribution to DM production, as the rate is suppressed by the fourth power of the Yukawa coupling. Therefore, the abundance of $F_1$, denoted as $\tilde{Y}_{F_1}$, is determined by the Yukawa couplings $Y_F^{\alpha 1}$ and the mass spectrum of the dark scalar particles $m_{\eta^\pm}, m_{\eta^{R,I}}$. 

\begin{figure}[h]
\centering
\includegraphics[height=5.5cm,width=0.45\textwidth]{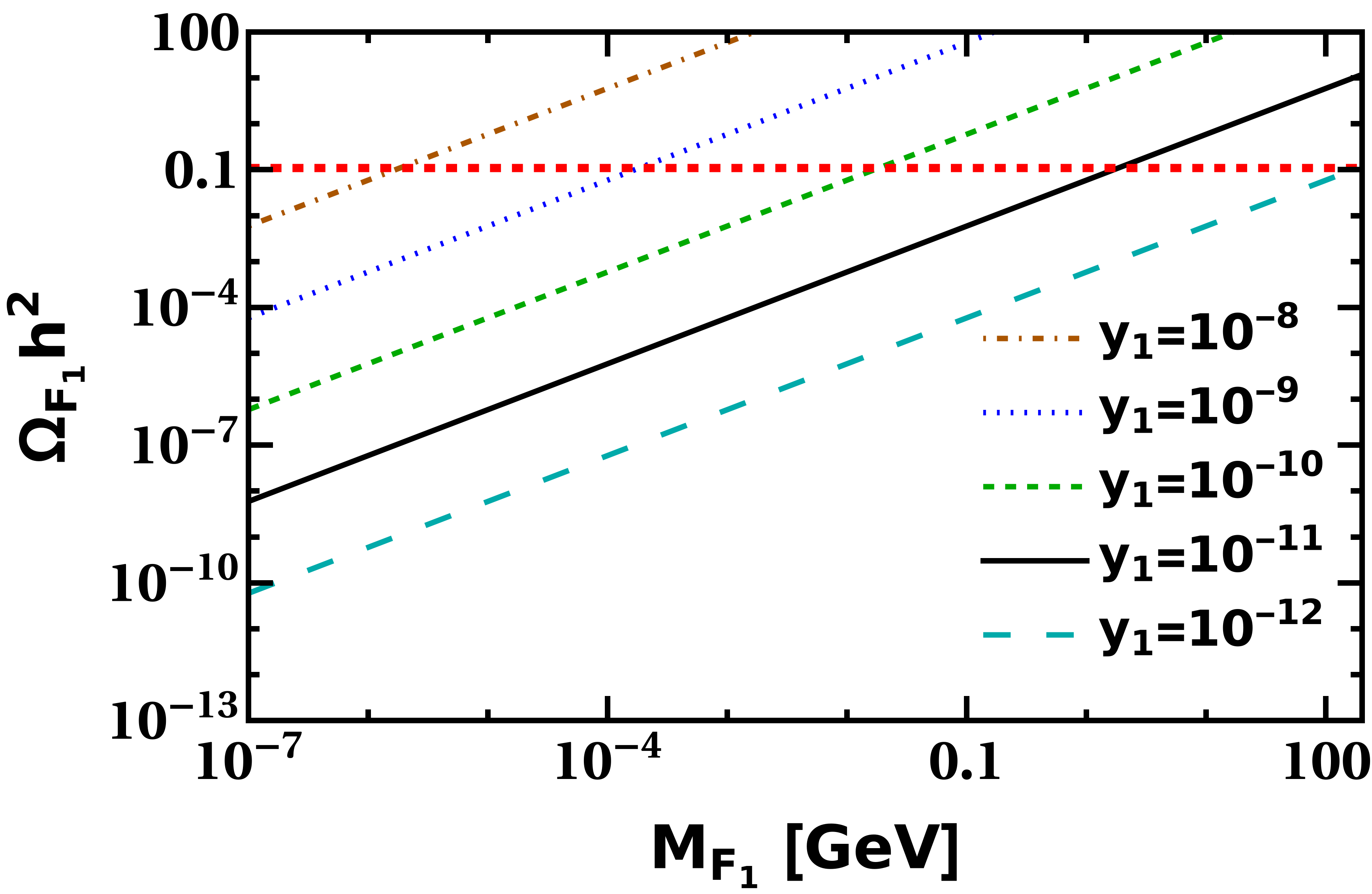}~~~
\includegraphics[height=5.5cm,width=0.45\textwidth]{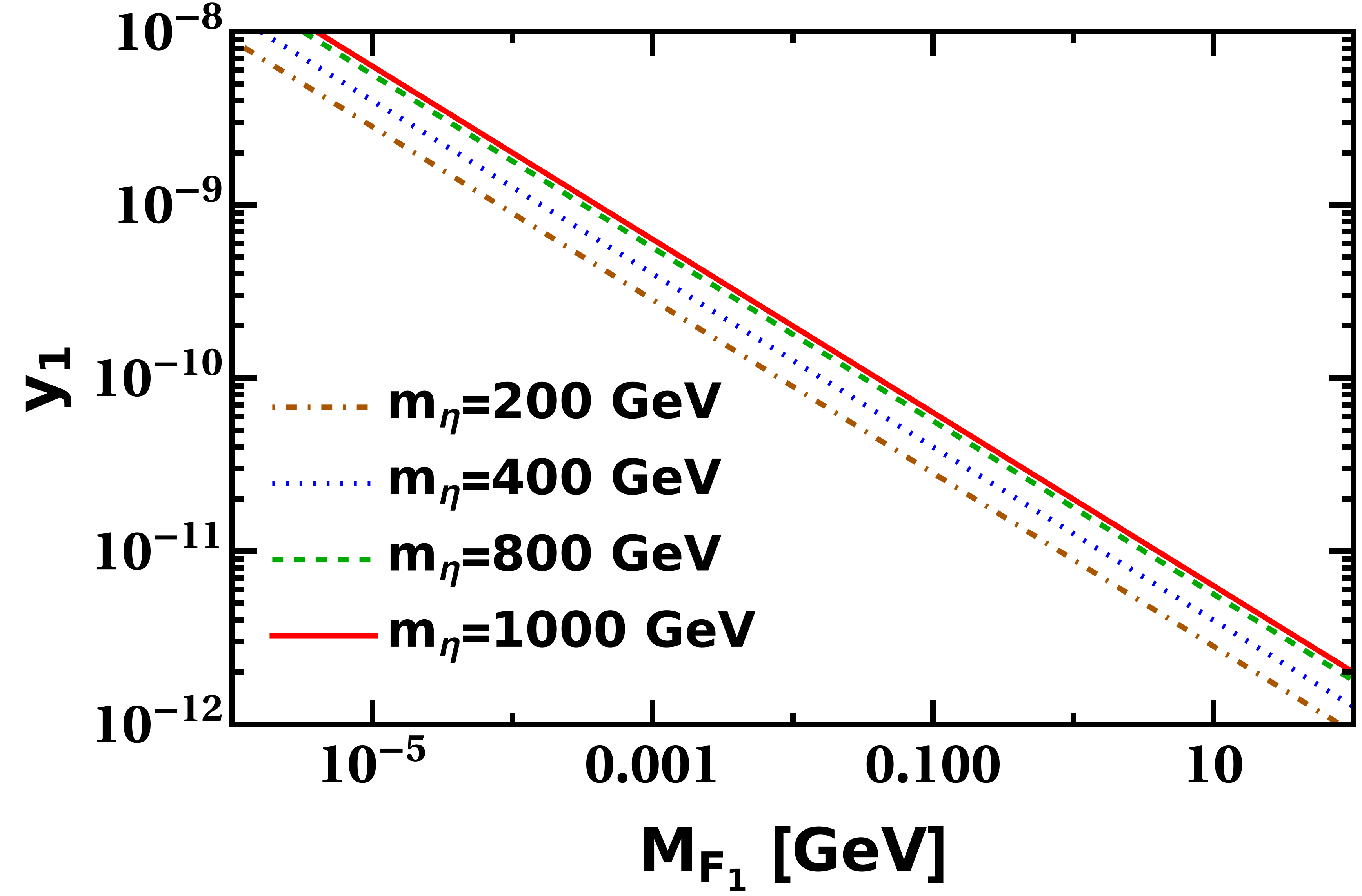}
\caption{ 
Left panel: Freeze-in relic density versus dark matter mass for different FIMP Yukawa couplings $y_1$. We have fixed the dark particle mass as $m_{\eta}=500$~GeV. The horizontal dashed red band is the region  consistent with Planck observations. Right panel: Correct relic contour in the $M_{F_1}-y_1$ plane for different scalar mass values. } 
\label{fig:FIMP_12}
\end{figure}

For simplicity,  we consider all dark scalars to be degenerate and denote their common mass by $m_\eta$. In this case we can calculate the $F_1$ yield, $\tilde{Y}_{F_{1}}(T)=n_{F_1}(T)/s(T)$ by solving the following Boltzmann equation \cite{Hall:2009bx}
\begin{equation}
	s\, T\,\frac{d \tilde{Y}_{F_{1}}}{dT} \; = \; -\frac{\gamma_{F_{1}}(T)}{H(T)}\,,\label{eq:BE}
\end{equation}
where $H(T)$ is the expansion rate of the Universe at temperature $T$, $s$ is the entropy density and $\gamma_{F_1}(T)$ is the thermal-averaged $F_1$ production rate. The relic  dark-matter density, $\Omega_{F_{1}}h^{2}$, is related to the asymptotic value of $\tilde{Y}_{F_{1}}$ at low temperatures by
\begin{equation}
	\Omega_{F_{1}}\,h^{2}\;=\; 2.744\times 10^{8}\, \frac{M_{F_1}}{\text{GeV}}\,\tilde{Y}_{F_{1}}(T_{0})\,,\label{eq:OmegaF1}
\end{equation}
where $T_0=2.752$~K is the present-day CMB temperature.
Solving Eq.~\eqref{eq:BE}, we get the following value for the present-day abundance,
\begin{equation}
	\tilde{Y}_{F_{1}}\left(T_0 \right)\;\approx\;  10^{-4}\,\left(\frac{1~\text{TeV}}{m_{\eta}}\right)\,\left(\frac{y_{1}}{10^{-8}}\right)^{2}\,.\label{Yapprox}
\end{equation}
The dark-matter relic density is then approximately given as,
\begin{align}  
\Omega_{F_1} h^2\approx 0.12 \left(\frac{M_{F_1}}{10\,\text{KeV}}\right)\left(\frac{100\,\text{GeV}}{m_\eta}\right)  \left(\frac{y_1}{2\times 10^{-9}}\right)^2, \quad \text{with }\quad  y_1=\sqrt{\sum_\alpha |Y_F^{\alpha 1}|^2}.
\label{eq:FIMP-relic}
\end{align} 
 From Eq.~\eqref{eq:FIMP-relic}, the size of the Yukawa couplings leading to the correct relic abundance can be approximated as
\begin{align}
y_1\approx 2\times 10^{-9} \left(\frac{10\,\text{KeV}}{M_{F_1}}\right)^{\frac{1}{2}} \left(\frac{m_\eta}{100\,\text{GeV}}\right)^{\frac{1}{2}} . 
\label{eq:FIMPy1}
\end{align}
One sees that the freeze-in mechanism can account for the observed relic density over a wide range of DM masses, all the way from the keV to the TeV scale. This is illustrated in Fig.~\ref{fig:FIMP_12}, where we show the relic density as a function of dark matter mass $M_{F_1}$~(left panel) for fixed dark scalar mass $m_\eta=500$~GeV, whereas in the right panel we show the required Yukawa coupling to get the correct relic density as a function of dark matter mass $M_{F_1}$ for various dark scalar masses. For $M_{F_1}$ we considered a minimum value of 1 keV which can play the role of warm dark matter~\cite{Boyarsky:2008ju,Gorbunov:2008ka}.  

We see that larger dark matter masses require a smaller Yukawa $y_1$ to be consistent. Hence, the FIMP mechanism requires tiny Yukawa couplings to be consistent with the relic density.
All in all one sees that, in this weaker sense, the option of scotogenic FIMPs is a perfectly viable one. However, in this review we will be mainly concerned with the most conventional WIMP scotogenic dark matter scenario.

\section{Revamped scotogenic model}
\label{sec:singlet-triplet-scoto}
The simplest scotogenic Model can be revamped by including triplets~\cite{Hirsch:2013ola}. Such singlet-triplet construction provides a variant of the original model with improved theoretical features~\cite{Merle:2016scw} and with a richer phenomenology~\cite{Diaz:2016udz,Rocha-Moran:2016enp,Diaz:2016udz,Restrepo:2019ilz,Avila:2019hhv,Karan:2023adm,Lozano:2025tst}.
In this Section we will briefly review the singlet-triplet scotogenic model. The new particles and their quantum numbers are given in Table~\ref{tab:revampedModel}.
Besides the \sm particle content there are very few new fields: one singlet fermion $F$ (as opposed to three in the original scotogenic model) and one hyperchargeless $\mathrm{SU(2)_L}$ triplet fermion $\Sigma$.
These dark fermions are odd under $\mathbb{Z}_2$. Moreover, we include two new scalars: besides the dark $\mathrm{SU(2)_L}$ doublet $\eta$ of the original scotogenic model, one includes a $\mathbb{Z}_2$-even $\mathrm{SU(2)_L}$  triplet scalar $\Omega$. 
\begin{table}[!h]
\centering
\includegraphics[scale=0.3]{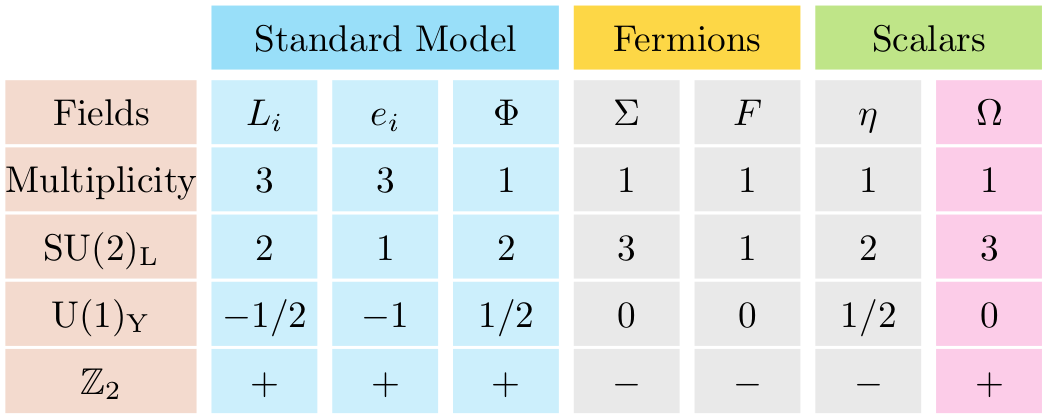}
\caption{ Revamped scotogenic model quantum numbers: 
dark states have opposite $\mathbb{Z}_2$ parity as SM states and $\Omega$}
\label{tab:revampedModel}
\end{table}

In contrast to the dark doublet $\eta$, which does not acquire a vacuum expectation value (VEV), the neutral component of the triplet scalar field $\Omega$ acquires a nonzero VEV, as it is even under $\mathbb{Z}_2$. 
The presence of the triplets $\Sigma$ and $\Omega$ brings theoretical and phenomenological advantages with respect to the original scotogenic model. 
The most general \SM $\otimes \mathbb{~Z}_2$ invariant Yukawa Lagrangian with the particle content and charge assignment given in Table~\ref{tab:revampedModel} can be written as 
\begin{equation}
- \mathcal{L}_Y = Y_e^{ij}\,\overline{L}_{i} \, \Phi \, e_{j} + Y_{F}^i \, \overline{L}_{i} \, \tilde{\eta} \, F + Y_{\Sigma}^i \, \overline{L}_{i} \,C\Sigma^{\dagger}\, \tilde{\eta} \, + Y_{\Omega} \, \text{Tr}\big[\overline{\Sigma} \Omega\big] \, F + \text{h.c} \, . 
\label{eq:yukawa}
\end{equation}
where $i,j=1,2,3$ are the flavor indices and $\tilde{\eta}=i\sigma_2\eta^*$. The Majorana mass term of $\Sigma$ and $F$ has the following form,  
\begin{equation}
- \mathcal{L}_M = 
\frac{1}{2} \, M_\Sigma \, \text{Tr}\big[\overline{\Sigma}^{c} \Sigma\big]
+ \frac{1}{2} \, M_F \, \overline{F}^{c} F
+ \text{h.c} \, . 
\label{eq:mass}
\end{equation}
\\[-.2cm]
The scalar potential invariant under the $\mathrm{SU(2)_L\otimes U(1)_Y~\otimes}~\mathbb{Z}_2$ symmetry is \\[-.2cm]
\begin{align}
\mathcal V &= -m_{\Phi}^2 \Phi^\dagger \Phi + m_{\eta}^2 \eta^\dagger \eta + \frac{\lambda_1}{2} \left( \Phi^\dagger \Phi \right)^2 + \frac{\lambda_2}{2} \left( \eta^\dagger \eta \right)^2 + \lambda_3 \left( \Phi^\dagger \Phi \right)\left( \eta^\dagger \eta \right) 
  + \lambda_4 \left( \Phi^\dagger \eta \right)\left( \eta^\dagger \Phi \right) \nonumber \\
 &+ \frac{\lambda_5}{2} \left[ \left(\Phi^\dagger \eta \right)^2 + \text{h.c} \right] - \frac{m_\Omega^2}{2} \, \text{Tr}(\Omega^\dagger \Omega)  
  + \frac{\lambda^{\Omega}_1}{2} \left( \Phi^\dagger \Phi \right) \text{Tr}\left( \Omega^\dagger \Omega\right) + \frac{\lambda^{\Omega}_2}{4} \, \text{Tr}(\Omega^\dagger \Omega )^2 \nonumber \\
&+ \frac{\lambda^{\eta}}{2} \left( \eta^\dagger \eta \right) \text{Tr}\left( \Omega^\dagger \Omega\right) 
+ \mu_1 \, \Phi^\dagger \, \Omega \, \Phi + \mu_2 \, \eta^\dagger \, \Omega \, \eta \, . \label{eq:scpot}
\end{align} 

\par 
We now turn to a discussion of the particle spectrum in all sectors of the revamped scotogenic theory,
standard and dark, bosons and fermions. We will follow the same notation given in~\cite{Avila:2019hhv}.
 As mentioned, besides the gauge singlet $F$, the model contains a dark fermion triplet,
\begin{equation}
\Sigma = \left( \begin{array}{cc}
\frac{\Sigma^0}{\sqrt{2}} & \Sigma^+ \\
\Sigma^- & -\frac{\Sigma^0}{\sqrt{2}}
\end{array} \right).
\end{equation}

Concerning scalars one has,
besides the \sm Higgs doublet $\Phi$ and dark doublet $\eta$, also a real triplet scalar multiplet $\Omega$,
\begin{equation}
\eta=\left(
\begin{array}{c}
\eta^+\\
\frac{1}{\sqrt{2}} (\eta_R+i\eta_I)
\end{array}\right),\,\,
\Phi=\left(
\begin{array}{c}
\phi^+\\
\frac{1}{\sqrt{2}} (v_\phi+\phi^0+i\psi)
\end{array}\right),\,\, \Omega = \left(\begin{array}{cc}
(\Omega^{0}+v_\Omega)/\sqrt{2} &\Omega^{+}  \\
 \Omega^{-} &- (\Omega^{0}+v_\Omega)/\sqrt{2} 
\end{array}\right)~,  \label{eq:triplets}
\end{equation}
 The scalar doublet $\eta$ in the dark sector is the same as present in the minimal scotogenic scenario discussed in Sec.\ref{sec:simpl-scot-setup}.
In order to ensure dark-matter stability the $\mathbb{Z}_2$ symmetry must remain unbroken. This implies that the $\mathbb{Z}_2$-odd
scalar $\eta$ should not acquire a nonzero VEV, so that spontaneous electroweak symmetry-breaking will be associated only to the VEVs of the neutral components of $\Phi$ and $\Omega$.

After symmetry-breaking the scalar spectrum contains four physical neutral scalars. 
Besides the neutral dark scalars, $\eta_R,~\eta_I$, these include two physical $CP$-even neutral 
Higgs bosons $(h,H)$, coming from the mixing of $\phi^0$ and $\Omega^0$.
The mass matrix of $CP$-even neutral Higgs scalars in the basis $(\phi^0, \Omega^0)$ reads as 
\begin{eqnarray}
	\label{eq:scalar_mass}
\mathcal{M}_S^2 &=& \left(\begin{array}{cc}
\lambda_1 v_\phi^2
& \lambda_1^\Omega v_\Omega v_\phi - \mu_1 \frac{v_\phi}{\sqrt{2}} \\
~~\lambda_1^\Omega v_\Omega v_\phi - \mu_1 \frac{v_\phi}{\sqrt{2}}
& ~~2 \lambda_2^\Omega v_\Omega^2 + \frac{\mu_1}{2\sqrt{2}} \frac{v_\phi^2}{v_\Omega}
\end{array}\right) \equiv \left(\begin{array}{cc}
A  &  B \\
B  &  C \\
\end{array}\right),
\end{eqnarray}
with the mass eigenvalues given by
\begin{align}
m_{h,H}^2=\frac{1}{2}(A+C \mp \sqrt{(A-C)^2 + 4 B^2}),
\end{align}
where by convention $m_h^2\leq m_H^2$, with $h$ identified as \sm Higgs boson discovered at LHC~\cite{ATLAS:2012yve,CMS:2012qbp}. \\[-.2cm]

On the other hand, the physical charged sector contains, besides the dark scalar $\eta^\pm$, a physical charged scalar~($H^\pm$).
The latter is obtained from the charged-scalar mass matrix (in the basis $(\phi^{\pm},\Omega^{\pm})$), given as 
\begin{eqnarray}
\mathcal{M}_{\pm}^2 &=& \left(\begin{array}{cc}
\sqrt{2}\mu_1 v_\Omega
& \mu_1 \frac{v_\phi}{\sqrt{2}} \\
\mu_1 \frac{v_\phi}{\sqrt{2}}
& \mu_1 \frac{v_\phi^2}{2\sqrt{2}v_\Omega}
\end{array}\right) \, . \nonumber 
\end{eqnarray}
The above mass matrix has one zero eigenvalue, corresponding to the would-be Goldstone boson absorbed by the charged gauge boson $W^{\pm}$.
The massive one is a physical charged Higgs boson, with mass
\begin{align}
m_{H^\pm}^2=\frac{\mu_1 (v_\phi^2+4 v_\Omega^2)}{2\sqrt{2}v_\Omega}.
\end{align}
Note that the VEV of $\Omega$ contributes to the $W$ boson mass, 
\begin{eqnarray}
m_W^2 &=& \frac{1}{4} \, g^2 \left( v_\phi^2 + 4 \, v_\Omega^2 \right), 
\label{eq:mW} 
\end{eqnarray}
Electroweak precision observables place a limit on the triplet VEV $v_\Omega$~\cite{Schechter:1980gr}, i.e. $v_\Omega \lesssim 4$ GeV~\cite{Gunion:1989ci,Gunion:1989we,ParticleDataGroup:2024cfk}.\\

The conservation of the $\mathbb{Z}_2$ symmetry implies that the 
$\mathbb{Z}_2$-odd scalars $\eta_{R,I}$ and $\eta^{\pm}$ do not mix with other scalars. These dark scalar boson masses are given by  
\begin{eqnarray}
m_{\eta_R}^2 &=& m_{\eta}^2 + \frac{1}{2}\left(\lambda_3 + \lambda_4 + \lambda_5 \right) v_\phi^2 + \frac{1}{2}\lambda^\eta v_\Omega^2 - \frac{1}{\sqrt{2}} \, v_\Omega \, \mu_2, \, \label{eq:etrSTSM}\\
m_{\eta_I}^2 &=& m_{\eta}^2 + \frac{1}{2}\left(\lambda_3 + \lambda_4 - \lambda_5 \right) v_\phi^2 + \frac{1}{2}\lambda^\eta v_\Omega^2 - \frac{1}{\sqrt{2}} \, v_\Omega \, \mu_2, \, \label{eq:etiSTSM} \\
m_{\eta^{\pm}}^2 &=& m_{\eta}^2 + \frac{1}{2}\lambda_3 v_\phi^2 + \frac{1}{2}\lambda^\eta v_\Omega^2 + \frac{1}{\sqrt{2}} \, v_\Omega \, \mu_2 \, .
\end{eqnarray}
Notice that the mass difference $m_{\eta_R}^2-m_{\eta_I}^2$ depends only on the parameter $\lambda_5$, which is also responsible for smallness of neutrino masses. The limit $\lambda_5\to 0$ restores lepton number conservation, a key feature of the scotogenic picture. 
\par The triplet scalar $\Omega$ induces a mixing between the singlet and triplet dark fermion fields $F$ and $\Sigma$ through the term proportional to Yukawa coupling $Y_\Omega$, given in Eq.~\eqref{eq:yukawa}. The fermion mass matrix in the basis $(\Sigma, F)$ is given as  
\begin{equation}
	\label{eq:mass_fermion}
\mathcal{M}_\chi = \left(\begin{array}{cc} M_\Sigma & Y_\Omega v_\Omega \\ 
Y_\Omega v_\Omega & M_F \end{array}\right), 
\end{equation}
leading to the following tree-level dark fermion masses
\begin{align}
m_{\chi^\pm}&=M_\Sigma,\\
m_{\chi_1}&=\frac{1}{2}\left((M_\Sigma+M_F) - \sqrt{(M_\Sigma-M_F)^2+4 Y_\Omega^2 v_\Omega^2}\right),\\
m_{\chi_2}&=\frac{1}{2}\left((M_\Sigma+M_F)+\sqrt{(M_\Sigma-M_F)^2+4 Y_\Omega^2 v_\Omega^2}\right).
\end{align}
The mass-eigenstates $\chi_{1,2}$ are determined by the $2\times 2$ orthogonal matrix $V(\alpha)$,
\begin{equation}
\left(\begin{array}{c}\chi_1\\ \chi_2\end{array}\right) = \left( \begin{array}{cc}
\cos \alpha & \sin \alpha \\
-\sin \alpha & \cos \alpha
\end{array} \right) \, \left(\begin{array}{c} \Sigma^0\\ F\end{array}\right) = V(\alpha)\left(\begin{array}{c} \Sigma^0\\ F\end{array}\right)\,\,\,\text{with}\,\,\, \tan(2\alpha) = \frac{2 \, Y_\Omega v_\Omega}{M_\Sigma - M_F} .
\end{equation}
Note that due to the $\mathbb{Z}_2$ symmetry, the lightest of the neutral fermion eigenstate, $\chi_1$ or $\chi_2$ can play the role of the fermionic scotogenic dark-matter, see Sec.~\ref{subsec:dmSTFM}.\\[-1.2cm]

\subsection{Neutrino masses}   

\begin{figure}[h!]
\centering
\includegraphics[height=4.cm,scale=0.35]{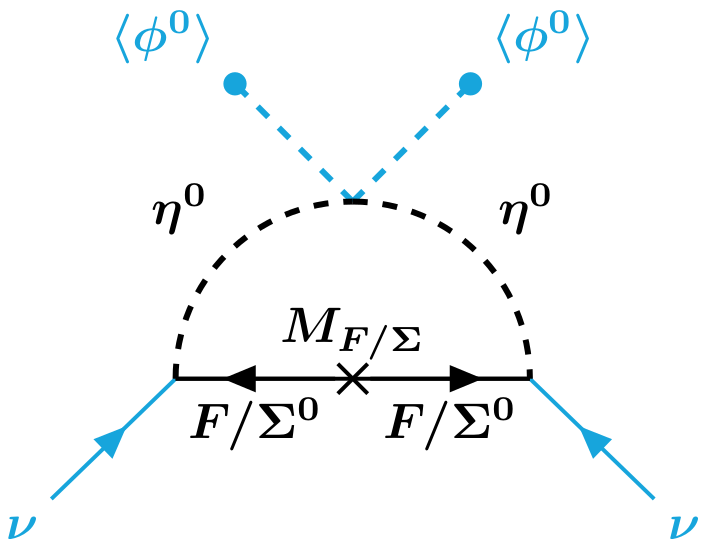}\hspace{10mm}\raisebox{-1.1mm}[0pt][0pt]{\includegraphics[height=4cm,scale=0.3]{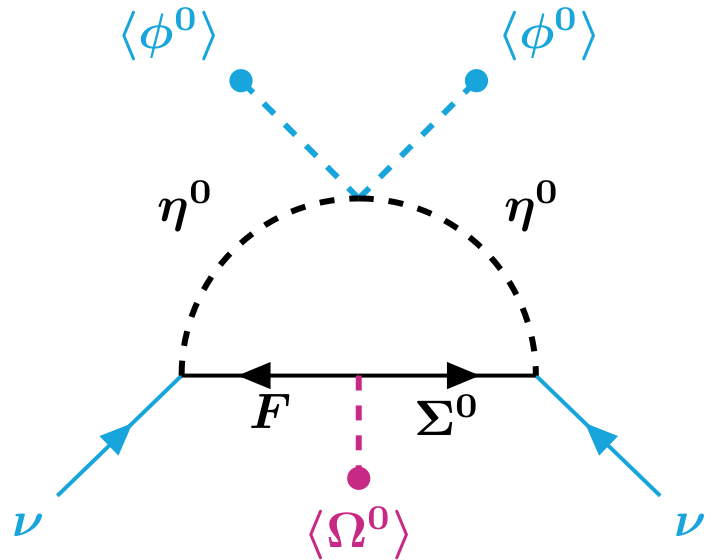}}
\vspace*{5mm}
\caption{ 
One-loop neutrino mass in the revamped (singlet-triplet) scotogenic model. Here $\eta^0=(\eta_R,\eta_I)$. } 
\label{fig:neutrino-loop-triplet}
\end{figure}

In this scheme neutrinos acquire a radiatively induced mass through the exchange of the dark fermions and scalars, as illustrated in Fig.~\ref{fig:neutrino-loop-triplet}. 
Eqs.~\eqref{eq:yukawa}, \eqref{eq:mass} and \eqref{eq:scpot} provide the relevant interactions for neutrino mass generation. 

The resulting neutrino mass matrix can be expressed as~\cite{Hirsch:2013ola,Merle:2016scw},
\begin{eqnarray}
(\mathcal{M}_\nu)_{ij}&=&\sum_{\sigma=1}^2\left(\frac{ih_{i \sigma}}{\sqrt{2}}\right)\left(\frac{-ih_{j \sigma}}{\sqrt{2}}\right)\left[I(m_{\chi_\sigma}^2,m_{\eta_R}^2)-I(m_{\chi_\sigma}^2,m_{\eta_I}^2)\right] \nonumber \\
&=&\sum_{\sigma=1}^2 \frac{h_{i \sigma} \, h_{j \sigma} \, m_{\chi_\sigma}}{2 \, (4\pi)^2} \left[\frac{m_{\eta_R}^2\ln\left(\frac{m_{\chi_\sigma}^2}{m_{\eta_R}^2}\right)}{m_{\chi_\sigma}^2-m_{\eta_R}^2}
-\frac{m_{\eta_I}^2\ln\left(\frac{m_{\chi_\sigma}^2}{m_{\eta_I}^2}\right)}{m_{\chi_\sigma}^2-m_{\eta_I}^2}\right] \, , \label{eq:mnu}
\end{eqnarray}
where $i,j$ are family indices, $m_{\chi_\sigma}$ are the masses of the $\chi_{1,2}$ and $h$ is the new Yukawa matrix, 
\begin{equation}
h=\left(\begin{array}{cc} 
\frac{Y_\Sigma^1}{\sqrt{2}} & ~~Y_F^1 \\
\frac{Y_\Sigma^2}{\sqrt{2}} & ~~Y_F^2 \\
\frac{Y_\Sigma^3}{\sqrt{2}} & ~~Y_F^3
\end{array}\right) \cdot V^T(\alpha). \, 
\end{equation}
The matrix $V(\alpha)$ is a $2\times 2$ orthogonal matrix that diagonalizes the fermionic mass matrix $\mathcal{M}_\chi$ given in Eq.~\eqref{eq:mass_fermion}.  Note that in the limit $\lambda_5\to 0$, $m_{\eta_R}\approx m_{\eta_I}$, hence neutrino masses vanish and the lepton number symmetry is restored. Thus the choice of $\lambda_5\ll 1$ is natural in the sense of t'Hooft~\cite{tHooft:1979rat}. As there is only one $\Sigma$ and one $F$, the lightest neutrino is massless, this model providing an example of radiative missing partner seesaw, see discussion in Sec.~\ref{sec:neutrino-parameters}. \\[-.3cm] 

It proves convenient to write the neutrino mass matrix in Eq.~\eqref{eq:mnu} as
\begin{equation} \label{eq:mnumat}
\mathcal{M}_\nu = h \, \Lambda \, h^T ~~~~~ \mathrm{where} 
\end{equation}
\begin{equation}
\Lambda = \left( \begin{array}{cc}
\Lambda_1 & 0 \\
0 & \Lambda_2
\end{array} \right) \, , \quad \Lambda_\sigma = \frac{m_{\chi_\sigma}}{2 \, (4\pi)^2} \left[\frac{m_{\eta_R}^2\ln\left(\frac{m_{\chi_\sigma}^2}{m_{\eta_R}^2}\right)}{m_{\chi_\sigma}^2-m_{\eta_R}^2}
-\frac{m_{\eta_I}^2\ln\left(\frac{m_{\chi_\sigma}^2}{m_{\eta_I}^2}\right)}{m_{\chi_\sigma}^2-m_{\eta_I}^2}\right] \, .
\end{equation}
Similar to Ref.~\cite{Casas:2001sr}, the Yukawa matrix $h$ can be extracted in terms of measured quantities, so that it automatically satisfies the neutrino oscillation restrictions as follows:
\begin{equation} \label{eq:CI}
h = U_{\rm lep}^\ast \, \sqrt{\widehat{\mathcal{M}}_\nu} \, R \, \sqrt{\Lambda}^{-1} \, \text{ 
  with   }\, \, U_{\rm lep}^{T} \, \mathcal{M}_\nu \, U_{\rm lep}=\widehat{\mathcal{M}}_\nu=\text{diag}(m_1,m_2,m_3) .
\end{equation}

The complex matrix $R$ for normal and inverted neutrino mass-ordering is given through a complex angle $\gamma$ as follows
\begin{eqnarray}
R &=& \left( \begin{array}{cc} 0 & 0 \\ \cos \gamma & \sin \gamma
  \\ -\sin \gamma & \cos \gamma
\end{array} \right) \quad \text{\bf for NO} \, (m_1 = 0) \,\,  \text{~~and~~}\,\,
R = \left( \begin{array}{cc}
\cos \gamma & \sin \gamma \\
-\sin \gamma & \cos \gamma \\
0 & 0
\end{array} \right) \quad \text{\bf for IO} \, (m_3 = 0) \, .
\end{eqnarray}

\subsection{Relevant Constraints}  
\label{sec:fdm-Constraints}

In order to quantify the detection prostects 
of fermionic dark matter in the revamped scotogenic scheme in a realistic manner one must take into account all the relevant constraints on the model parameters. 
Apart from neutrino oscillation data, mentioned in Sec.~\ref{sec:neutrino-parameters}, there are several other experimental constraints and theoretical consistency requirements restricting the parameters of our scotogenic dark-matter model.

$\bullet$ \textbf{Theoretical constraints:}  To keep the scalar potential bounded from below, the quartic couplings should be restricted~\cite{Kannike:2012pe}, so we impose the following conditions on these model parameters~\cite{Merle:2016scw}: 
\begin{align}
	\label{eq:th_cons}
	&\hspace*{-0.2cm}i)\;\lambda_1\geq 0, \qquad ii)\;\lambda_2\geq 0, \qquad iii)\;\lambda_2^\Omega\geq 0,\qquad iv)\;\lambda_3+\sqrt{\lambda_1 \lambda_2}\geq 0,\nonumber\\
	&\hspace*{-0.2cm}v)\;\lambda_3+\lambda_4-|\lambda_5|+\sqrt{\lambda_1 \lambda_2}\geq 0,\qquad vi)\;\lambda_1^\Omega+\sqrt{2\lambda_1 \lambda_2^\Omega} \geq 0, \qquad vii)\;\lambda^\eta+\sqrt{2\lambda_2 \lambda_2^\Omega} \geq 0,\nonumber\\
	&\hspace*{-0.2cm}viii)\;\sqrt {2\lambda_1\lambda_2\lambda_2^\Omega}+\lambda_3\sqrt{2\lambda_2^\Omega}+\lambda_1^\Omega\sqrt\lambda_2+\lambda^\eta\sqrt \lambda_1+\sqrt{\Big(\lambda_3+\sqrt{\lambda_1 \lambda_2}\Big)\Big(\lambda_1^\Omega+\sqrt{2\lambda_1\lambda_2^\Omega}\Big)\Big(\lambda^\eta+\sqrt{2\lambda_2\lambda_2^\Omega}\Big)}\geq 0,
\end{align}
assuming $\lambda_4+|\lambda_5|\geq 0$.
For $\lambda_4+|\lambda_5|< 0$, inequality (viii) should be modified by replacing the $\lambda_3$ term by
$(\lambda_3+\lambda_4-|\lambda_5|)$.

In order to ensure perturbative unitarity, all the Yukawa couplings ($Y^{\alpha\beta}$, $Y_F^\alpha$, $Y_\Sigma^\alpha$, $Y_\Omega$) and the scalar quartic couplings ($\lambda_i$) must be smaller than $\sqrt{4\pi}$~\cite{Allwicher:2021rtd} and $4\pi$~\cite{Durand:1993vn,AkeroydArhribNaimi2000} respectively, at any particular energy scale. 
Moreover, in order to avoid the breaking of the dark parity symmetry at some higher scale due to the running of different parameters, one should choose $\mu_\Omega^\eta \lesssim \mathcal{O} (1 \text{ TeV})$ \cite{Merle:2016scw}.\\ 

$\bullet$ \textbf{Electroweak Precision Observables (EWPO):}
The scalar triplet $\Omega$ increases the mass of $W$-boson at tree level, keeping the mass of the $Z$-boson unchanged. The $\rho$-parameter in the singlet-triplet model at tree-level is:
$\rho=1+4\,v_\Omega^2/v_\phi^2$. The $3\sigma$ range of the current global fit, i.e. $\rho=1.00031\pm 0.00019$ \cite{ParticleDataGroup:2024cfk}, constrains the triplet VEV as: $v_\Omega\lsim 4$ GeV.
Moreover, the S, T and U parameters obtained from the current electroweak global fits, i.e. $S=-0.04 \pm 0.10,$ $T=0.01\pm 0.12$ and $U= -0.01\pm 0.09$ \cite{ParticleDataGroup:2024cfk} can be used to restrict the splitting between the masses of charged and neutral scalars. \\  

$\bullet$ \textbf{Collider constraints:} LEP and LHC experiments lead to limits on the masses of additional scalars or fermions. 
The lowest mass for fermiophobic neutral scalars allowed by the L3 collaboration is 107 GeV \cite{L3:2003ieq}, while the bound arising from ALEPH by using the invisible decays is 114 GeV \cite{ALEPH:2001roc}. 
The combination of these results with the searches at ATLAS \cite{ATLAS:2012yxc} and CMS \cite{CMS:2013zma} for neutral scalar di-photon decay modes pushes the mass of $H$ above 150 GeV.
In order to comply with the EWPO constraints, the mass of $H^+$ also gets pushed above a similar limit. 
Although the constraints from LEP \cite{ALEPH:2013htx} and LHC \cite{ATLAS:2014otc,CMS:2015lsf} do not  strictly apply to our $H^+$, 
they are indicative of charged scalar exclusion up to 150 GeV of mass. 
However, these constraints do not apply to the dark scalars ($\eta^\pm, \eta_{I},\eta_R$).
\\[-.3cm] 

The lower limit on the mass of long-lived or stable charged lepton is set by LEP as 102 GeV \cite{OPAL:2003zpa,L3:2001xsz}. 
This bound restricts the mass of $\Sigma^+$, i.e. $M_\Sigma$, requiring also the triplet-like neutral fermion to lie above a similar limit~\cite{L3:2001xsz}.
However, that limit assumes the neutral fermion to decay into a SM lepton and the $W$ boson, and hence does not apply to our $\mathbb{Z}_2$-odd neutral fermion. \\[-.3cm]

If kinematically allowed, all new particles present in the model can contribute to the widths of the Higgs, Z and W bosons, which are precisely measured in experiments.
Therefore, the model parameters must obey the updated constraint:
$Br\,(h\to \text{BSM}\, +\text{BSM})\lsim 10.7\%$ ~\cite{ATLAS:2023tkt},
$\Gamma\,(Z\to \text{BSM}\, +\text{BSM})\lsim 5$ MeV ($\sim2\sigma$) \cite{ParticleDataGroup:2024cfk},
$\Gamma\,(W\to \text{BSM}\, +\text{BSM})\lsim 90$ MeV ($\sim2\sigma$) \cite{ParticleDataGroup:2024cfk}, where BSM generically denotes the new particles present.\\

$\bullet$ \textbf{Dark-matter  constraints:} 
The first constraint is set by the precise relic density measurement by the Planck collaboration~\cite{Planck:2018vyg}  : $\Omega h^2 = 0.120\pm 0.001.$
In addition, many searches have been made for nuclear recoils caused by DM scattering.  
They lead to limits on the spin-independent dark-matter-nucleon scattering cross-section from various experiments and different dark-matter mass ranges. 
The most stringent constraints below 2.5 GeV and in the window from 2.5 GeV to 5 GeV come from DarkSide-50 \cite{DarkSide-50:2022qzh} and the recent PandaX-4T results \cite{PandaX:2025rrz}, respectively. 
In the window from 5 GeV to 9 GeV we have XENONnT \cite{XENON:2024hup} and LZ \cite{LZ:2025igz}, while
 LZ \cite{LZ:2024zvo} again comes in for the range from 9 GeV to 10 TeV. Note that nuclear recoil arising from ``coherent neutrino-nucleus scattering'' acts as background for the direct detection experiments, the so-called neutrino floor~\cite{Billard:2013qya}.

\subsection{Fermionic dark-matter relic density}

It is instructive to compare the main features of fermionic dark-matter in the present revamped scenario with those of the pure singet or pure triplet scotogenic setups.  To do so we consider two limits, singlet-like versus triplet-like, depending on the dark-matter composition.
These limits draw an immediate analogy with the \textit{vanilla} case of supersymmetric dark-matter~\cite{PhysRevLett.50.1419,Ellis:1983ew,Jungman:1995df}. 
Indeed, our scenario mimics the main features of neutralino dark-matter within the Minimal Supersymmetric Standard Model (MSSM) with conserved R-parity. The latter is analogous to our dark $\mathbb{Z}_2$ symmetry. 

\begin{figure}[h]
\centering
\includegraphics[height=2.8cm,width=0.225\textwidth]{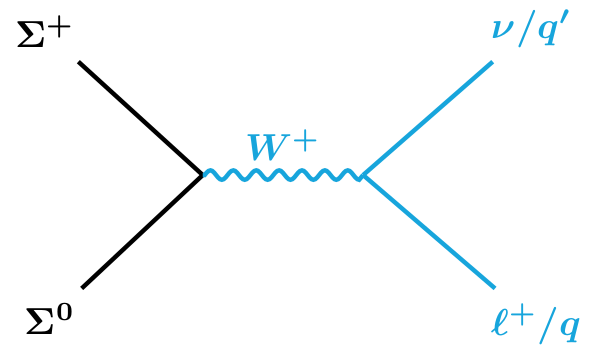}
\includegraphics[height=2.8cm,width=0.225\textwidth]{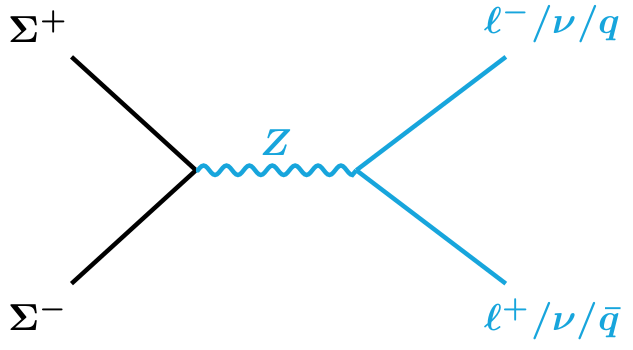}
\includegraphics[height=2.8cm,width=0.225\textwidth]{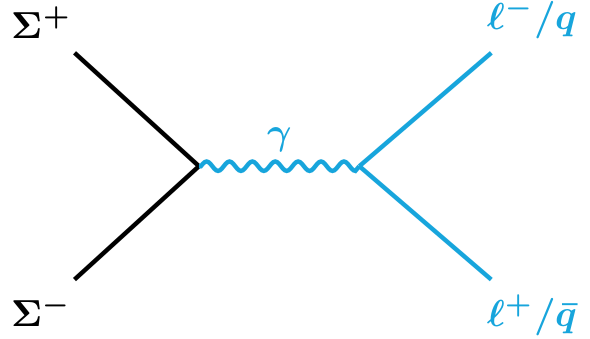}
\includegraphics[height=2.8cm,width=0.145\textwidth]{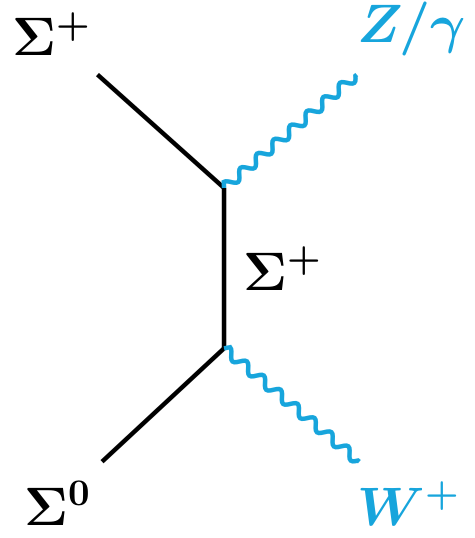}
\includegraphics[height=2.8cm,width=0.145\textwidth]{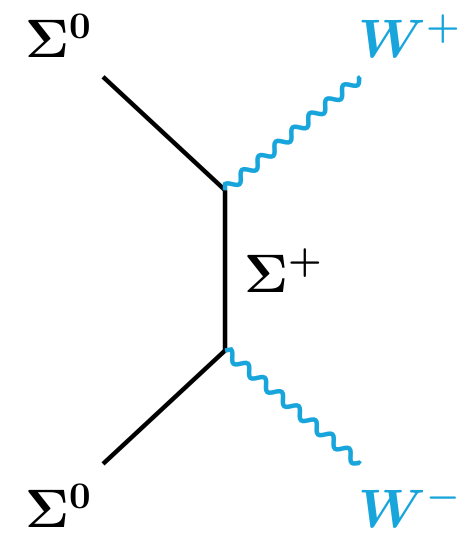} 
\vskip .2cm
\caption{
Fermion annihilation and co-annihilation diagrams for pure-triplet scotogenic dark-matter.} 
\label{fig:ann-pure-triplet}
\end{figure}

Let us first start with the case of pure-triplet~($\Sigma^0$) fermionic dark-matter, which occurs by taking the limit $M_F\to\infty$. 
The relic dark-matter abundance is determined by the annihiliation and co-annihiliation of $\Sigma^0$ and $\Sigma^{\pm}$, as shown in Fig.~\ref{fig:ann-pure-triplet}.  
Note that co-annihilations between $\Sigma^0$ and $\Sigma^{\pm}$ are also efficient, due to their small mass splitting~\footnote{
At tree level $M_{\Sigma^{\pm}}=M_{\Sigma^0}$ but,
for large $M_{\Sigma}$, ${\Sigma^{\pm}}$ becomes $\mathcal{O}(170\,\text{MeV})$ heavier than ${\Sigma^0}$ due to one-loop radiative corrections~\cite{Cirelli:2005uq}.}. \\[-.2cm]
\begin{figure}[h]
\centering
\includegraphics[width=0.4\textwidth]{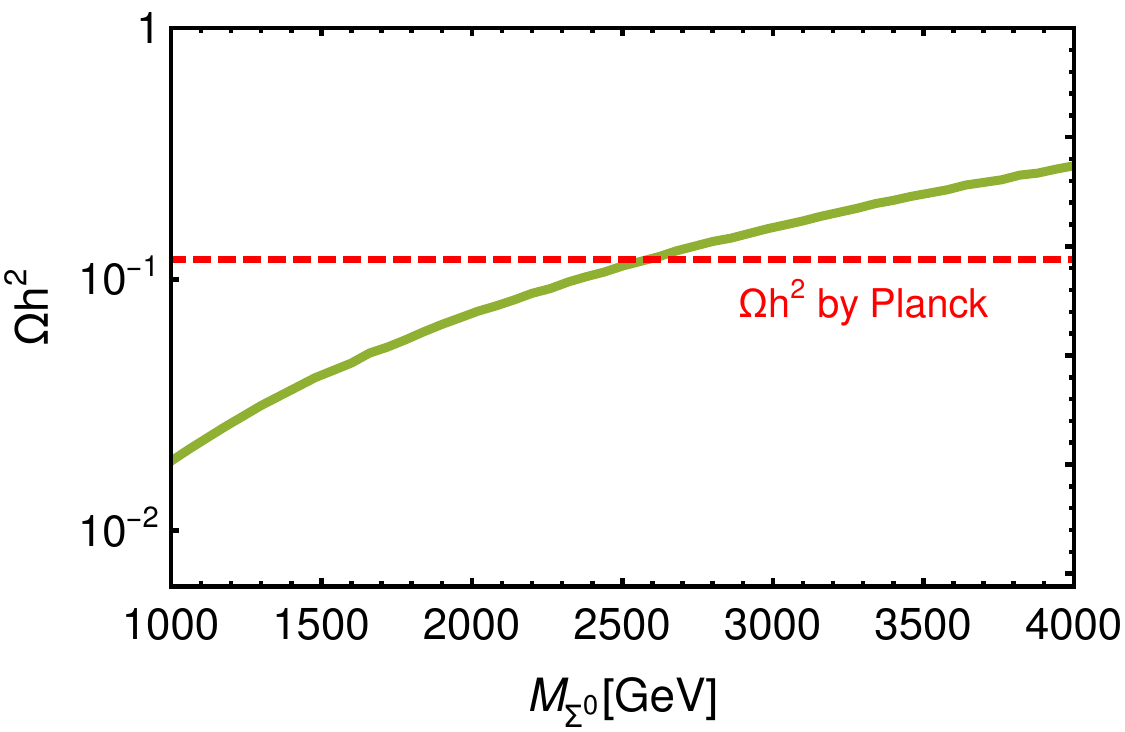}\hfil
\includegraphics[width=0.38\textwidth]{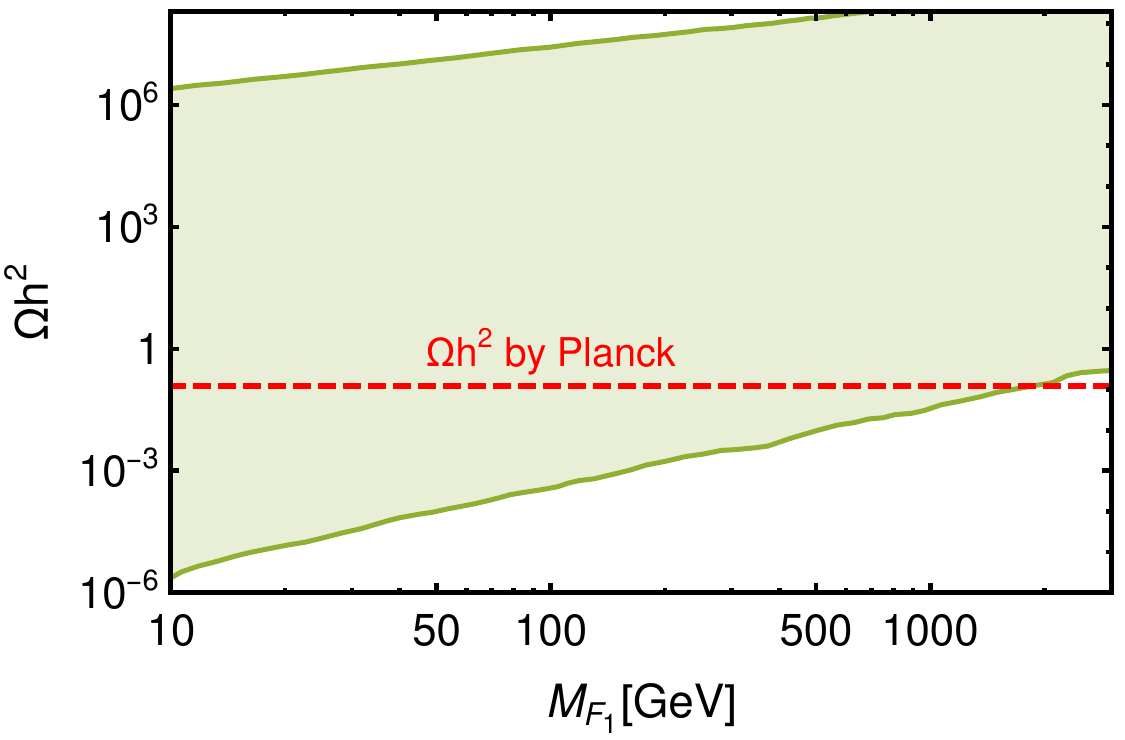}
\caption{
 Fermionic relic dark-matter density for pure-triplet (left) and pure-singlet scotogenic dark-matter (right).} 
\label{fig:relic-pure-triplet}
\end{figure}

These co-annihilations force the triplet DM mass to lie in the narrow range from $2.3$~TeV to 2.4 TeV, see left panel of Fig.~\ref{fig:relic-pure-triplet}. 
The production cross-section for such a heavy DM particle at the LHC with 13 TeV will be very small. 
Likewise, its direct detection by nuclear recoil, as it this occurs only at the loop level.
On the other hand, for the pure-singlet DM fermion $F$, as discussed in Sec.~\ref{sec:dm-fermion-scoto}, the main annihilation channels which determine the DM relic abundance are $F F\to\ell_i\ell_j, \nu_i\nu_j$ via the Yukawa coupling $Y_F$, see Fig.~\ref{fig:relic-pure-singlet}. 
Depending on the value of the $Y_F$ Yukawa coupling, the allowed DM mass can have a broad range of possible values, compared with the pure-triplet case, as seen in the right panel of Fig.~\ref{fig:relic-pure-triplet}.  
However, direct detection proceeds only at the loop-level in both cases.
\begin{figure}[h!]
		\includegraphics[scale=0.25]{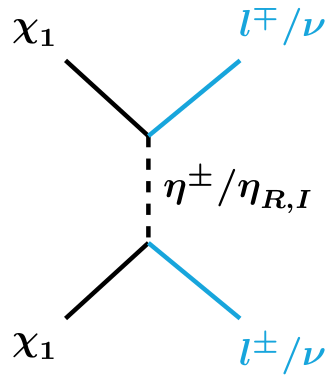}\hfil
		\includegraphics[scale=0.25]{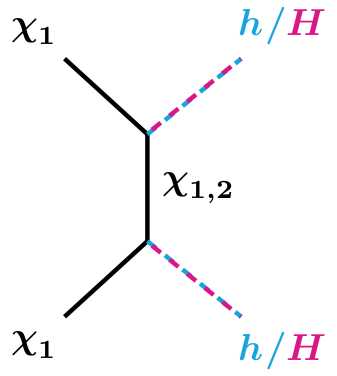}\hfil
		\includegraphics[scale=0.25]{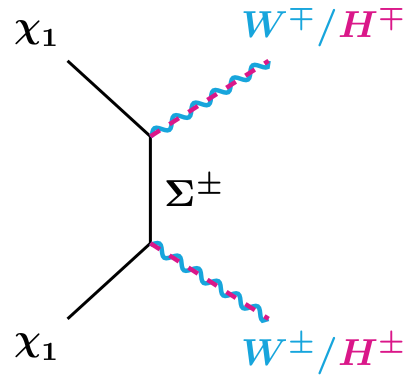}\hfil

\vspace{5mm}
  
		\includegraphics[scale=0.25]{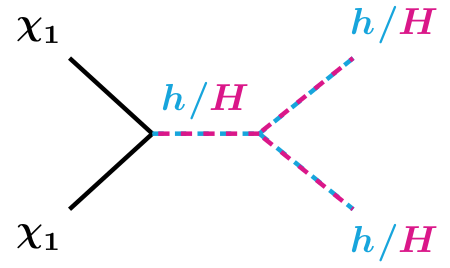}\hfil
		\includegraphics[scale=0.25]{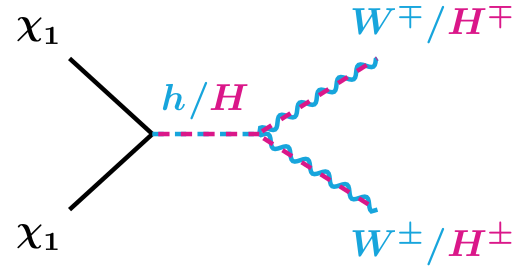}\hfil
		\includegraphics[scale=0.25]{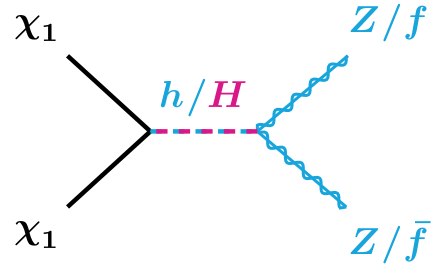}
	\caption{
    Fermionic DM-DM annihilation. While the first three diagrams show $\eta$, $\chi_{1,2}^{}$ and $\Sigma^{\pm}$--mediated t-channel processes, those below correspond to s-channel resonant annihilation through the neutral Higgs scalars $h^{}\text { and } H^{}$.}
	\label{fig:ann-DM}
\end{figure}
\begin{figure}[h]
\centering
\includegraphics[height=6cm,scale=0.25]{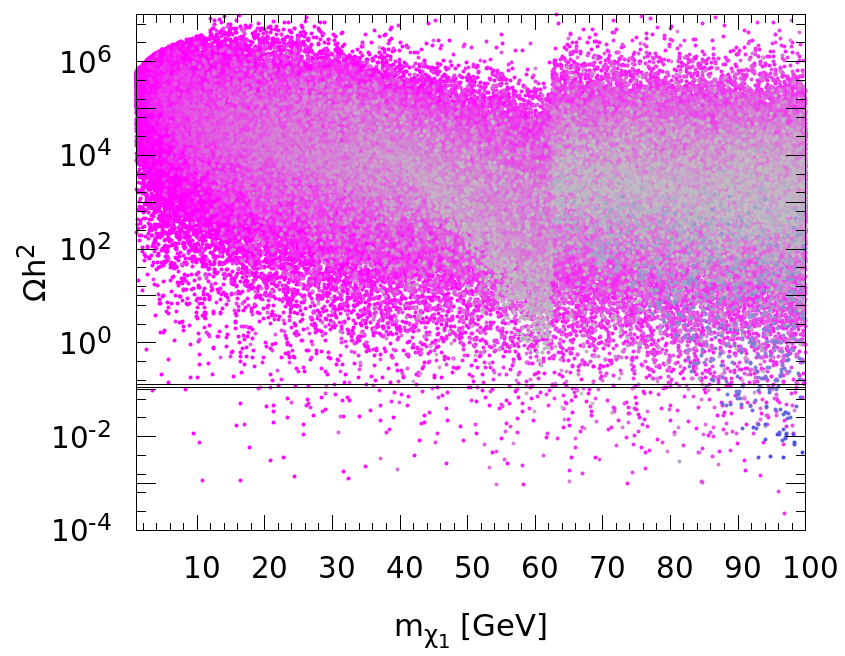}
\includegraphics[height=6cm,scale=0.25]{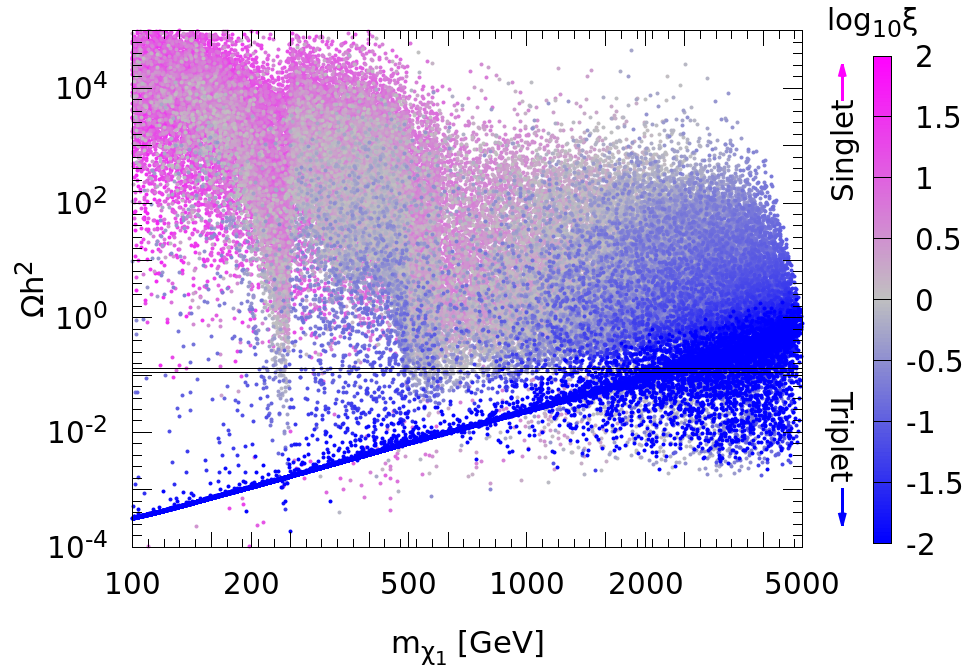}
\caption{ The fermionic dark-matter relic density in the revamped scotogenic model without imposing theoretical or experimental constraints. The presence of singlet-triplet mixing allows for light fermionic dark-matter.} 
\label{fig:relic-mixed}
\end{figure}
\vskip .2cm We now turn to the full-fledged revamped scotogenic scenario.
In this case the mixing between $\Sigma_0$ and $F$ implies two mechanisms for fermionic dark-matter annihilation:
\begin{enumerate} 
\item Through t/u-channel process mediated by the $\mathbb{Z}_2$-odd scalars $\eta^\pm,\,\eta_{I,R}^{}$ and fermions $\Sigma^\pm,\,\chi_{1,2}^{}$, as depicted by the first three diagrams in Fig. \ref{fig:ann-DM}. As a result, the Yukawa couplings $Y_{F,\,\Sigma}$ will control the relic density of the singlet-like dark-matter fermion. 

	\item Through Higgs $(h,H)$ mediated s-channel processes, shown by the last three diagrams in Fig.~\ref{fig:ann-DM}, and thus the heavy Higgs mass also plays a crucial role in the relic density. 
\end{enumerate}
However, if the dark-matter mass becomes very close to the mass of another dark particle, the second particle also takes part in the annihilation process, which is termed as ``co-annihilation''~\cite{Griest:1990kh}. 
Since the thermal-averaged co-annihilation cross-section gets an exponential suppression of the form $e^{-(m_i-m_{DM})/{T}}$, it remains negligible for larger mass differences between the dark-matter candidate and the other dark sector particles. In Fig.~\ref{fig:relic-mixed} we show the relic density as a function of the DM mass $m_{\chi_{1}}$ in the presence of mixing between $\Sigma_0$ and $F$. 
In order to quantify the composition of the fermionic dark-matter , we define the quantity
\begin{align}
\xi=\frac{|M_\Sigma-m_{\chi_{1}}|}{m_{\chi_{1}}} .
\label{eq:dmness}
\end{align}
Low values of $\xi$ correspond to triplet DM, while high values indicate singlet dark-matter.
Note that this is a naive estimate. A more thorough and realistic analysis will be presented later, including theoretical and experimental constraints. In any case one sees that the revamped (singlet-triplet) scotogenic model can potentially account for the critical dark-matter density for a broad range of DM masses.  \\

\begin{center}
   {\bf \small Fermionic dark-matter detection} 
\end{center}
\par We now turn to the direct DM detection prospects in the revamped scotogenic model.
While we already saw that loop-induced direct detection leads to reasonable rates, tree-level cross-sections could substantially enhance direct detection prospects and open new possibilities.
That is exactly what happens in the revamped scotogenic scenario.
In the presence of a nonzero VEV of the scalar triplet $\Omega$, there is a mixing between the fermions $\Sigma_0$ and $F$. 
For a large $F$-$\Sigma$-$\Omega$ Yukawa coupling the fermionic dark-matter candidate $\chi_i$ will be a substantially mixed state of $F$ and $\Sigma_0$. 
This implies the existence of a tree-level interaction term leading to direct dark-matter detection~\cite{Hirsch:2013ola}. This improves the phenomenological prospects of the revamped scotogenic scenario with respect to either the pure-singlet~\cite{Ma:2006km,Tao:1996vb} or pure-triplet expectations~\cite{Kubo:2006yx,Ma:2008cu,Chao:2012sz}.

\begin{figure}[h!]
	\centering
	\includegraphics[scale=0.22]{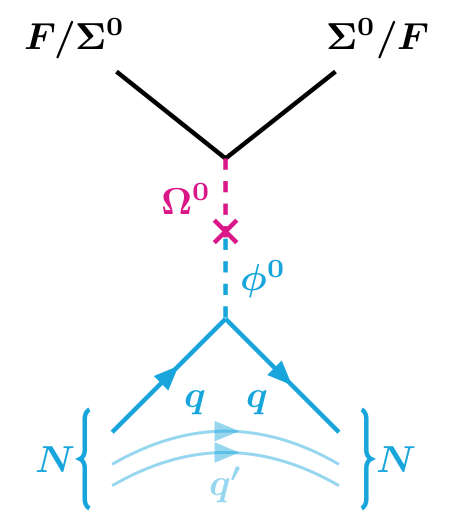}
	\caption{ Feynman diagram for direct fermionic dark-matter detection in the revamped scotogenic model}
	\label{fig:DD}
\end{figure}

Dark-matter nucleon scattering will proceed through a scalar mediated t-channel diagram, shown in Fig. \ref{fig:DD}, involving the mixing of scalars as well as the mixing of fermions. 
For illustration we give the the tree-level approximation for the spin-independent scattering cross-section per nucleon, expressed as 
\begin{align}
	\label{eq:DD_cr}
	\sigma^{\text{SI}}_{\rm DM-N}\approx \frac{\mu_{\text{red}}^2}{\pi}\Big[\frac{Y_\Omega f_N m_N^{}}{2v} \sin2\alpha\, \sin2\beta\,\Big(\frac{1}{m_{h}^2}-\frac{1}{m_{H}^2}\Big)\Big]^2,
\end{align}
with the nucleon form factor $f_N \approx 0.3$.  Here $m_N^{}$ is the nucleon mass and $\mu_{\text{red}}=m_{\chi_1^0}\, m_N^{}/(m_{\chi_1^0}+ m_N^{})$ is the reduced mass of the $N-\chi_1^{}$ system, while $\beta$ is the scalar mixing angle, given as
$\tan2\beta=\frac{4v_\Omega v_\phi(\mu_1-\sqrt 2\lambda_1^\Omega v_\Omega)}{4\sqrt 2\lambda_2^\Omega v_\Omega^3-2\sqrt 2\lambda_1 v_\Omega v_\phi^2+ \mu_1 v_\phi^2}.$ 

The direct detection cross-section, given by Eq.~\eqref{eq:DD_cr}, depends on five model parameters: $ \lambda_1, \lambda_{\Omega}^\phi, v_\Omega, \Delta m_{\Sigma F}, Y_\Omega$,  
 the first three of which regulate the mass of $H$ and the mixing angle $\beta$, while the last two determine the fermionic mixing angle $\alpha$. 
Notice that the direct detection cross-section gets a $Y_\Omega^2$ dependence from the $F\Sigma^0\Omega^0$ vertex.
However, it is not too sensitive to the dark-matter mass $m_{\chi_1}$, since $\mu_{\text{red}}$ reduces to $m_N$ for $m_{\chi_1}\gg1$ GeV. In order to estimate the attainable values for this cross section one must implement the relevant constraints.\\[-.2cm]

\subsection{Fermionic Dark-matter Direct Detection}
\label{subsec:dmSTFM}

We now discuss in detail the revamped scotogenic scenario. The model Lagrangian involves many parameters, fortunately not all of them independent, since several are measured, such as the values of the Higgs boson mass and electroweak VEV, the two light neutrino mass splittings, the three neutrino mixing angles, etc. 
In order to study co-annihilation effects, it is better to work with mass differences or squared differences, defined as follows:
\begin{equation}
	\Delta m_{\Sigma F}^{}=M_\Sigma-M_F, \qquad \Delta m_{\eta^+ F}=m_{\eta^+}-M_F \qquad\text{and} \qquad \Delta m^2_{\eta_I^0\eta^+}=m^2_{\eta^0_I}-m^2_{\eta^+}~.
\end{equation} 

\vskip 0.5cm
\begin{center}
   {\bf \small Reference benchmark} 
\end{center}

After using the oscillation data, the Yukawa couplings $Y_{F,\,\Sigma}$ are mainly regulated by $\lambda_5$, $\rm{Im} (\gamma)$ and $m_{\chi_1}$,
as can be seen from Eqs. \eqref{eq:mnumat}--\eqref{eq:CI}. 
Therefore, the cLFV processes, to be discussed in section~\ref{sec:cLFV} mainly depend on these three parameters. Again, a modification in these parameters will change the relic density through the Yukawa dependence of the t-channel diagrams in Fig. \ref{fig:ann-DM}.
Similarly, the mass of $H$ and the scalar mixing angle $\beta$ mostly depends on $\lambda_1$, $\lambda_1^\Omega$ and $v_\Omega$, see Eq. \eqref{eq:scalar_mass}.
Changing these three parameters therefore regulates the relic density through the three s-channel diagrams in Fig. \ref{fig:ann-DM}.
The parameters $v_\Omega$, $Y_\Omega$ and \dsf control the singlet-triplet mixing angle $\alpha$ (see Eq. \eqref{eq:mass_fermion}) that enters into various couplings of $\chi_1$ with other particles and thus also contribute to the relic density. Finally, small values of \dsf and \depf help reducing the relic density by enhancing the fermion-fermion and fermion-scalar co-annihilation.

\renewcommand{\arraystretch}{1.5}
\begin{table}[h!] 
	\scalebox{1.1}{
		\begin{tabular}
			{||c|c|c|c||c|c|c||c|c||c|c|c|c|c|c|c||}
			\hline
			$M_F$& $\Delta m_{\Sigma F}$& $\Delta m_{\eta^+ F}$& $\Delta m^2_{\eta_I^0\eta^+}$&$\mu_2$& $v_\Omega$& \multirow{2}{*}{$Y_\Omega$} &\multirow{2}{*}{Re$(\gamma)$}&\multirow{2}{*}{Im$(\gamma)$}&\multirow{2}{*}{$\lambda_1$}&\multirow{2}{*}{$\lambda_2$}&\multirow{2}{*}{$\lambda_3$}&\multirow{2}{*}{$\lambda_5$}&\multirow{2}{*}{$\lambda_1^\Omega$}&\multirow{2}{*}{$\lambda_2^\Omega$}&\multirow{2}{*}{$\lambda^\eta$}\\
			(GeV)&(GeV)&(GeV)&(GeV$^2$)&(GeV)&(GeV)&&&&&&&&&&\\
			\hline
			[1, 1000]&200&500&1000&400&4.0&2.0&$\pi/4$&$\pi/4$& 0.2626 &0.5&0.5&$10^{-8}$&0.5&0.5&0.5\\
			\hline
	\end{tabular}}
	\caption{ Specification of the benchmark scenario BP0~\cite{Karan:2023adm}.}
	\label{tab:BP0}
\end{table}

\begin{figure}[h!]
	\includegraphics[height=5cm,scale=0.5]{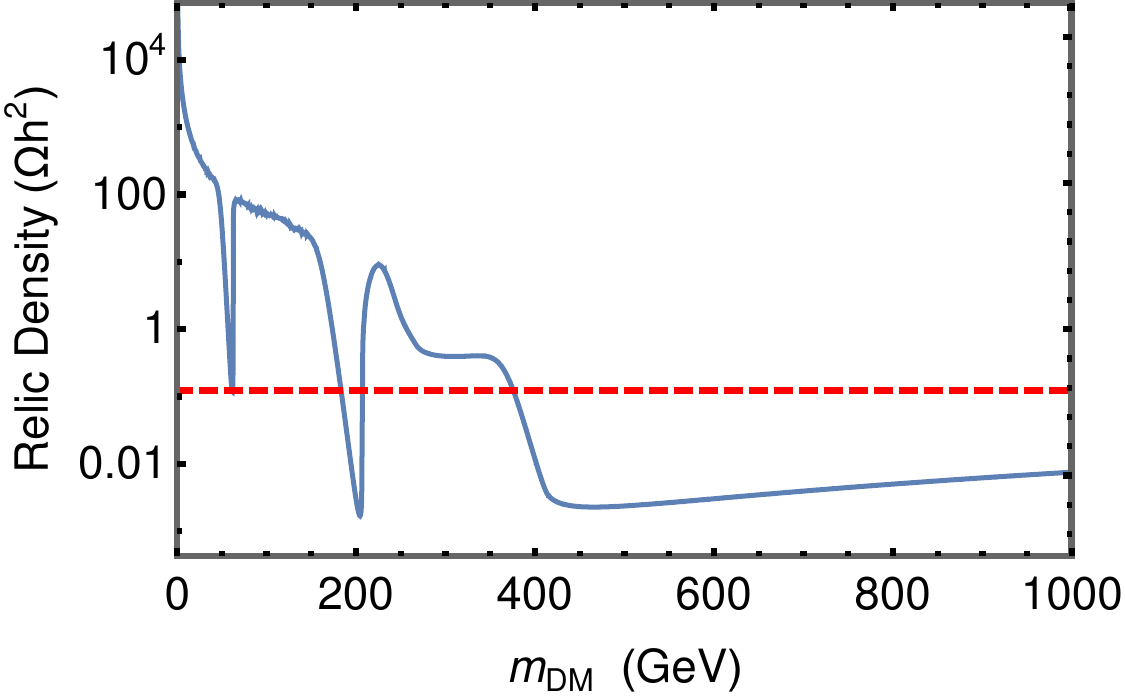}
	\caption{ 
    Relic density versus the fermionic dark-matter mass for benchmark point BP0~\cite{Karan:2023adm}. The observed Planck value is shown by the red dashed line. Notice the role of the second Higgs boson in explaining a low relic density.}
	\label{fig:rel}
\end{figure}

To study the variation of the relic density with the mass of the dark-matter, we choose a typical benchmark point BP0, shown in Tab. \ref{tab:BP0}.
The $H$ mass turns out to be roughly 400~GeV and the values of \dsf and \depf in BP0 are large enough to make the co-annihilation effects negligible.
Increasing the dark-matter mass, different annihilation channels open, as illustrated in Fig. \ref{fig:rel}.
Four dips in relic density at four different masses of the dark-matter are clearly visible in Fig. \ref{fig:rel}. The first two sharp dips at 62.5 GeV (i.e. $m_h/2$) and 200 GeV (i.e. $m_H/2$) correspond to the resonance production of $h/H$ in the s-channel annihilation processes $\chi_1 \chi_1 \to h/H \to SM~SM$.
Two more annihilation channels $\chi_1 \chi_1 \to W^\pm H^\pm$ and $\chi_1 \chi_1 \to h H$ open up at $m_{\rm DM}^{}=(m_{W^\pm}^{}+m_{H^\pm}^{})/2\approx242$ GeV and
$m_{\rm DM}^{}=(m_{h}^{}+m_{H}^{})/2\approx263$ GeV respectively which is indicated by the relic density drop near 250 GeV. Finally, for a DM mass of 400 GeV or so (i.e. $m_H\approx m_{H^+}$)
there is enough phase space for the annihilation processes  $\chi_1 \chi_1 \to H H$ and $\chi_1 \chi_1 \to H^+ H^-$ so one has another fall-off
in the relic density curve. Beyond 400 GeV, the relic density drops substantially. 

\vskip 0.5cm

To extract the salient features of this model concerning fermionic dark-matter, we scan the parameter-space.
We ensure the singlet-like fermion to be the dark-matter candidate by taking positive values for the  parameters \dsf, \depf and \deip.
The parameter $v_\Omega$ plays a key role in the mixing of both fermions and scalars. Hence, depending on the values of $v_\Omega$ we divide our analysis in two classes:
Scenario-I indicates higher value of $v_\Omega$, whereas Scenario-II assumes a lower value of $v_\Omega$, shown in Table~\ref{tab:BP12}.
\renewcommand{\arraystretch}{1.5}
\begin{table}[h!] 
	\scalebox{0.88}{
		\begin{tabular}
			{||c||c||c|c|c|c||c|c||c|c||c|c|c|c|c|c|c||}
			\hline
			\multirow{2}{*}{Cases}&$v_\Omega$&$M_F$& $\Delta m_{\Sigma F}$& $\Delta m_{\eta^+ F}$& $\Delta m^2_{\eta_I^0\eta^+}$&$\mu_2$& \multirow{2}{*}{$Y_\Omega$} &\multirow{2}{*}{Re$(\gamma)$}&\multirow{2}{*}{Im$(\gamma)$}&\multirow{2}{*}{$\lambda_1$}&\multirow{2}{*}{$\lambda_2$}&\multirow{2}{*}{$\lambda_3$}&\multirow{2}{*}{$\lambda_5$}&\multirow{2}{*}{$\lambda_1^\Omega$}&\multirow{2}{*}{$\lambda_2^\Omega$}&\multirow{2}{*}{$\lambda^\eta$}\\
			&(GeV)&(GeV)&(GeV)&(GeV)&(GeV$^2$)&(GeV)&&&&&&&&&&\\
			\hline\hline
			\multicolumn{17}{||l||}{\textbf{Scenario-I}}\\
			\hline
			BP$_1$&\multirow{3}{*}{\red\framebox{{4.0}}}&\multirow{3}{*}{[3, 10000]}&[100, 500]&[100, 500]&1000&\multirow{3}{*}{400}&\multirow{3}{*}{[0.1, 3.5]}&\multirow{3}{*}{$[-\pi, \pi]$}&\multirow{3}{*}{$[-2\pi, 2\pi]$}& \multirow{3}{*}{0.2626}&\multirow{3}{*}{0.5}&\multirow{3}{*}{0.5}&\multirow{3}{*}{$[10^{-9},0.5]$}&\multirow{3}{*}{0.5}&\multirow{3}{*}{0.5}&\multirow{3}{*}{0.5}\\
			BP$_1^{FF}$&&&[1, 50]&[100, 500]&1000&&&&& &&&&&&\\
			BP$_1^{FS}$&&&[100, 500]&[1, 30]&[1, 1000]&&&&& &&&&&&\\
			\hline\hline
			\multicolumn{17}{||l||}{\textbf{Scenario-II}}\\
			\hline
			BP$_2$&\multirow{3}{*}{\red\framebox{{1.5}}}&\multirow{3}{*}{[3, 10000]}&[100, 500]&[100, 500]&1000&\multirow{3}{*}{400}&\multirow{3}{*}{[0.1, 3.5]}&\multirow{3}{*}{$[-\pi, \pi]$}&\multirow{3}{*}{$[-2\pi, 2\pi]$}& \multirow{3}{*}{0.2626}&\multirow{3}{*}{0.5}&\multirow{3}{*}{0.5}&\multirow{3}{*}{$[10^{-9},0.5]$}&\multirow{3}{*}{0.5}&\multirow{3}{*}{0.5}&\multirow{3}{*}{0.5}\\
			BP$_2^{FF}$&&&[1, 50]&[100, 500]&1000&&&&& &&&&&&\\
			BP$_2^{FS}$&&&[100, 500]&[1, 30]&[1, 1000]&&&&& &&&&&&\\
			\hline
	\end{tabular}}
	\caption{ Parameter values in Scenario-I ($v_{\Omega}=4.0$~GeV) and Scenario-II ($v_{\Omega}=1.5$~GeV)~\cite{Karan:2023adm}. The values of $m_H$ in these two scenarios are about 400 GeV and 1100 GeV respectively.}
	\label{tab:BP12}
\end{table}\\

Moreover, each scenario in Table \ref{tab:BP12} is divided in three cases:
i) large values for both \depf and \dsf and no co-annihilation (no superscript),
ii) large \depf with small \dsf pointing towards  fermion-fermion co-annihilation (superscript `FF'),
iii)  small \depf with large \dsf suggesting fermion-scalar co-annihilation (superscript `FS').
The parameters that affect only the scalar sector of this model are kept fixed for simplicity.
Since $\lambda_5$ helps to regulate the Yukawa couplings $Y_F$ and $Y_\Sigma$ that obey the oscillation constraints we vary it in the scanning.

\begin{figure}[h!]
	\centering
	\includegraphics[scale=0.40]{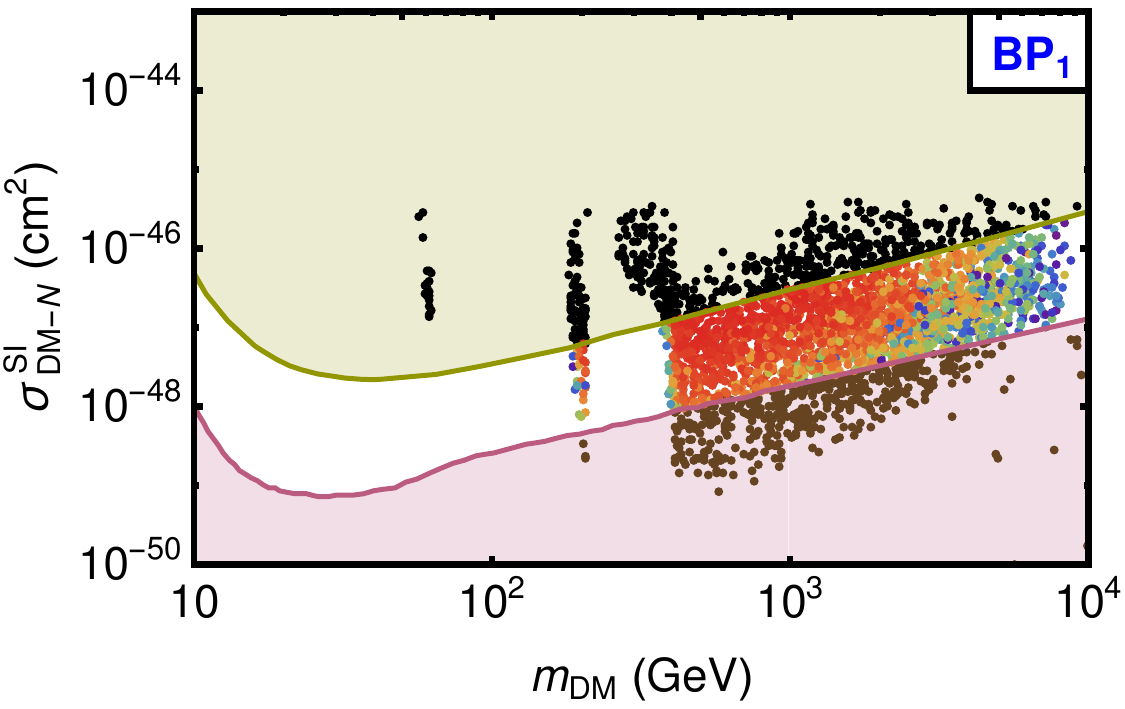}
	\hfil
	\includegraphics[scale=0.40]{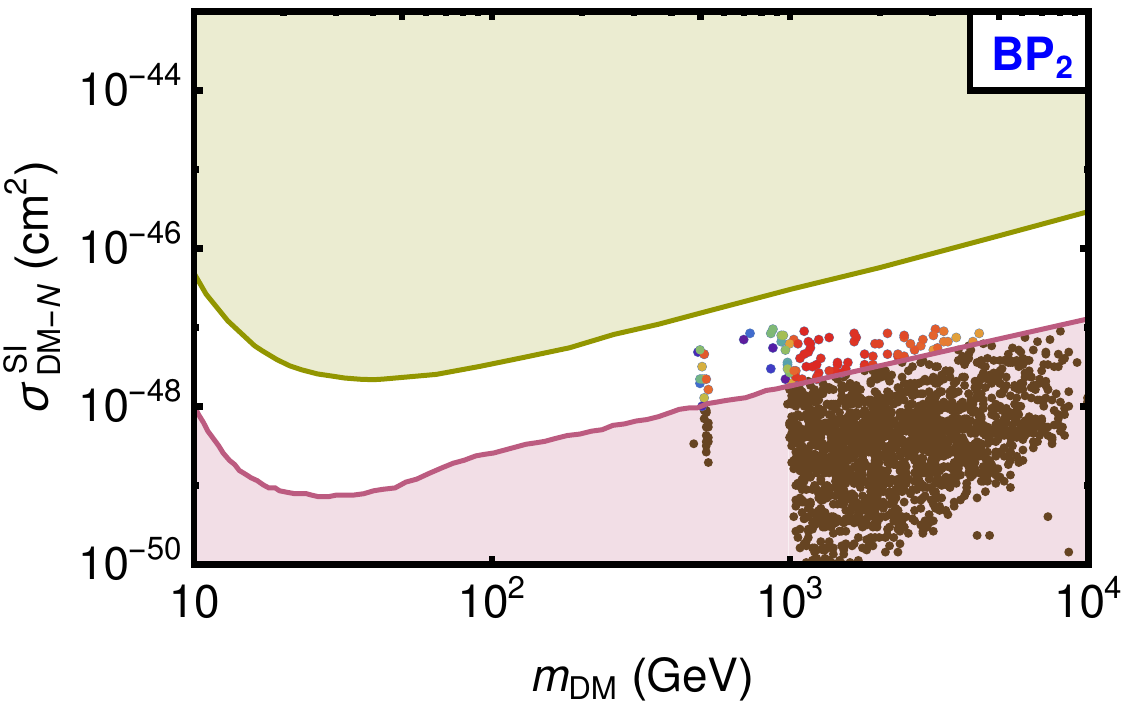}
	\text{\red \large \textbf{No co-annihilation}}
	
	\vspace*{0.5cm}
	\includegraphics[scale=0.40]{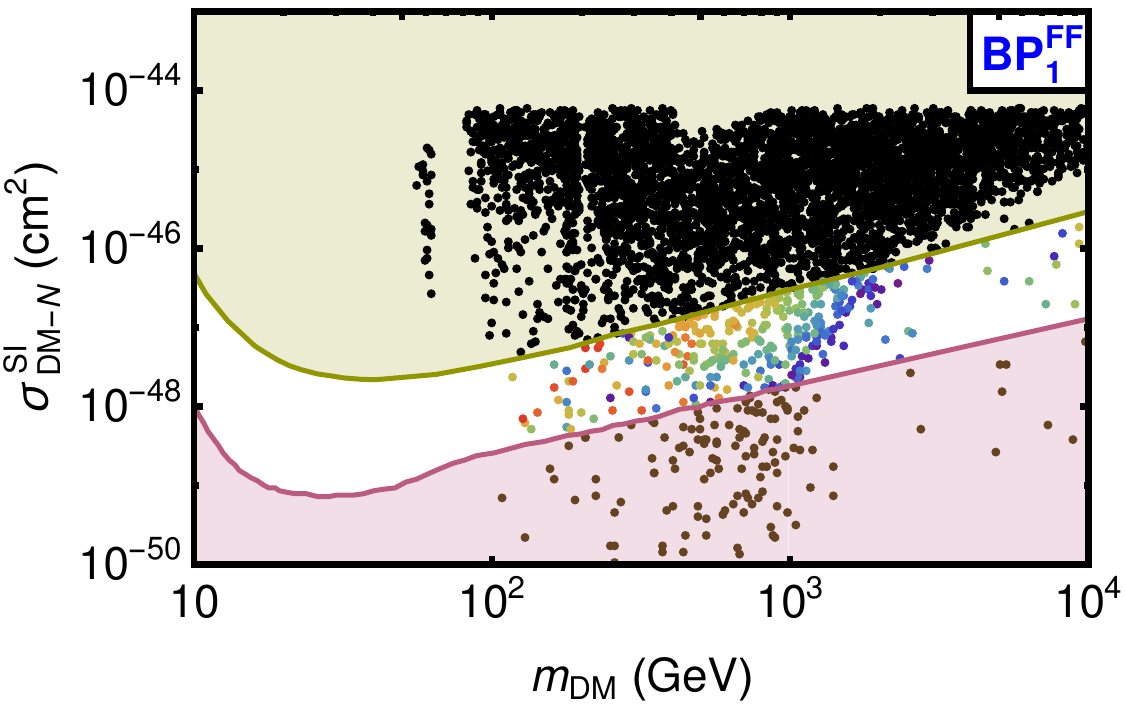}
	\hfil
	\includegraphics[scale=0.40]{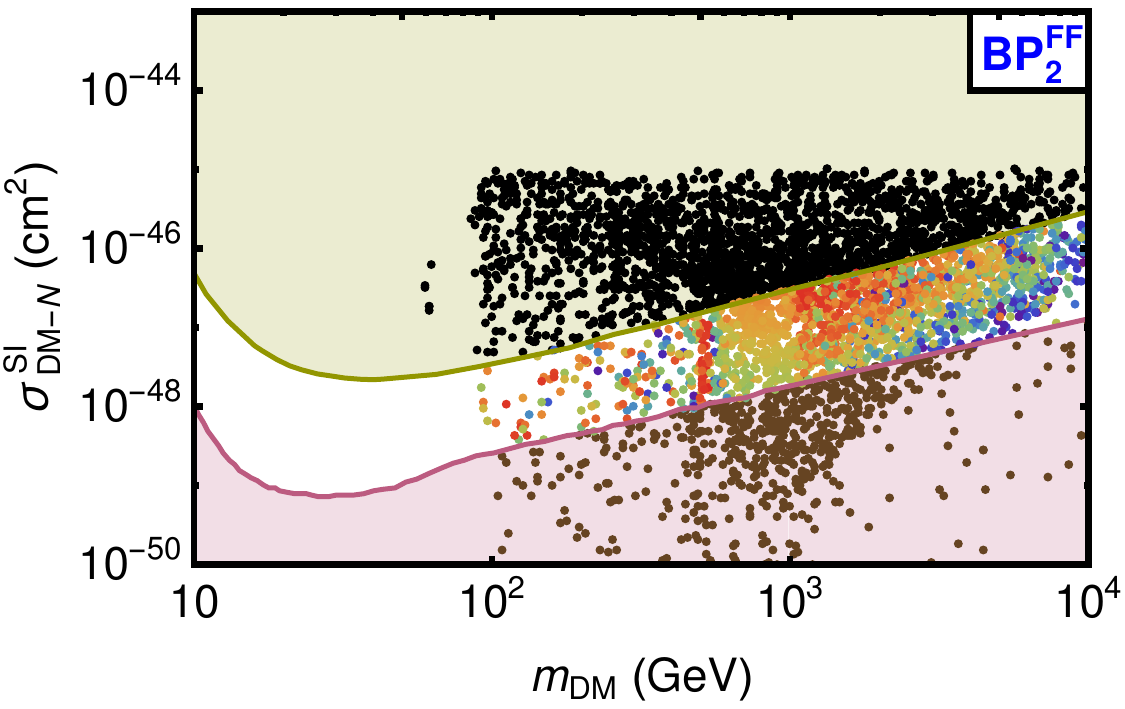}
	\text{\red \large \textbf{Fermion-Fermion co-annihilation}}
	
	\vspace*{0.5cm}
	\includegraphics[scale=0.40]{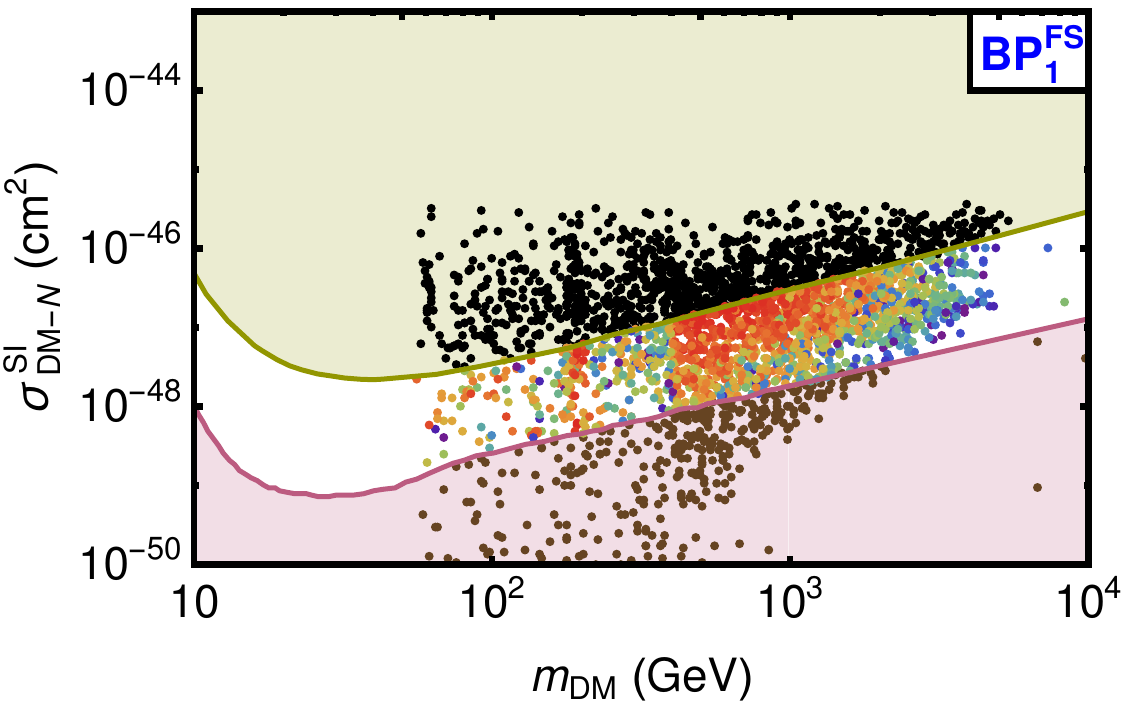}
	\hfil
	\includegraphics[scale=0.40]{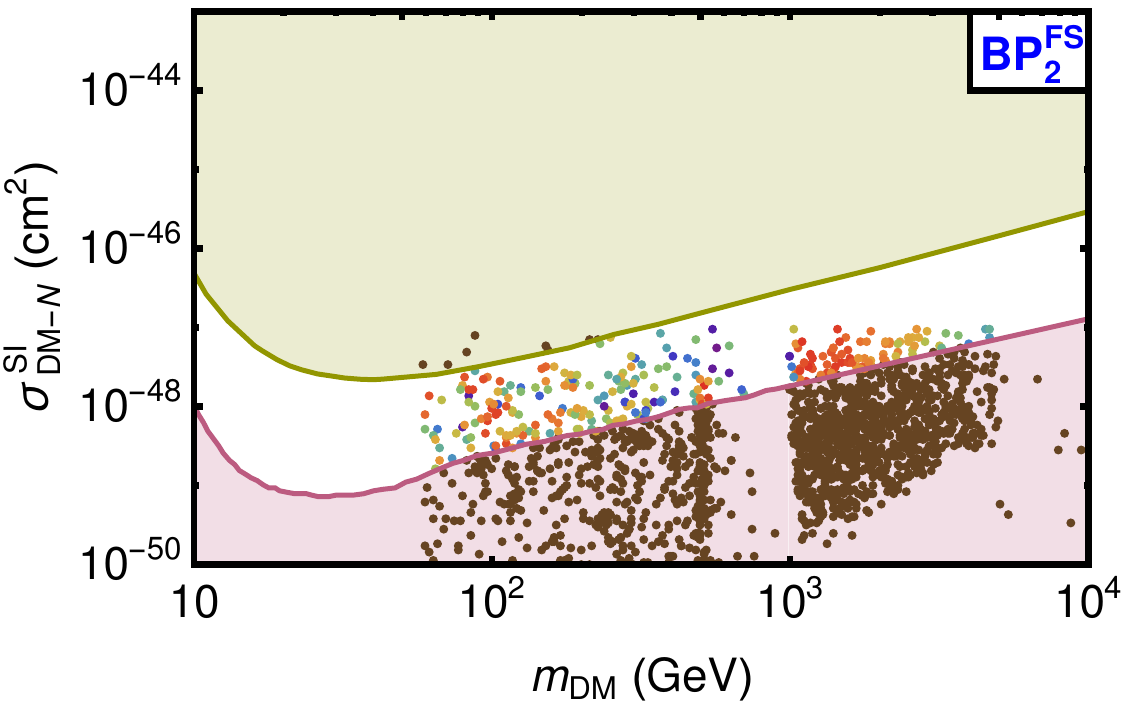}
	\text{\red \large \textbf{Fermion-Scalar co-annihilation}}
		
	\vspace*{0.2cm}
	\scalebox{1.1}{\hspace{-0.6cm}\includegraphics[height=1.4cm,width=4.5cm]{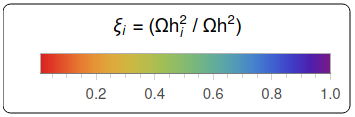}\hfill
		\raisebox{-0.2cm}{\includegraphics[height=1.8cm,width=13.2cm]{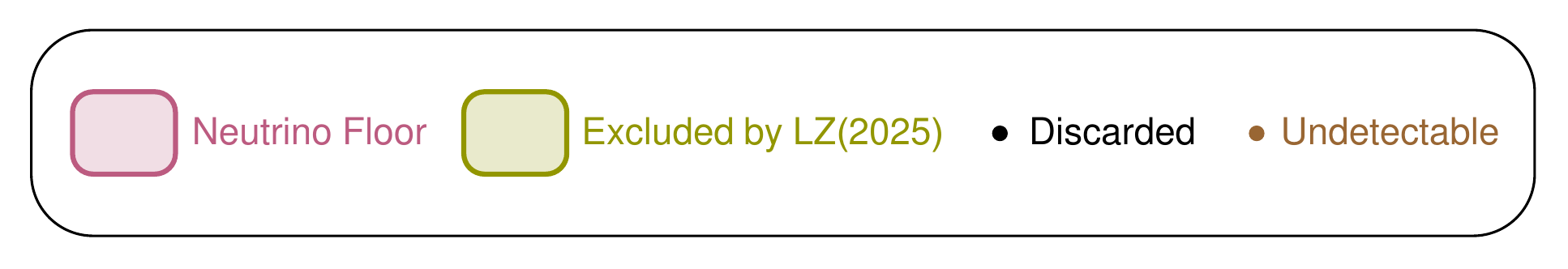}}}
	\caption{ 
    Dark-matter-nucleon spin-independent scattering cross section (in cm$^2$) for Scenario-I and Scenario-II~\cite{Karan:2023adm}. Different color shades correspond to the DM fraction $1\%<(\Omega h_i^2/\Omega h^2)<100\%$, as indicated in the lower-left box.    }
	\label{fig:DD_sc}
\end{figure}

First we generate parameter-points that satisfy neutrino-oscillation data as well as all the theoretical constraints for each case. Then restrictions from EWPO, collider studies, cLFV and observed relic density are implemented. Finally, we collect the points satisfying $1\%$ to $100\%$ of the measured relic density and study the prospects for their direct detection in the most constraining experiment LZ, as depicted in Fig. \ref{fig:DD_sc}. As a result, the allowed white corridors in Fig~\ref{fig:DD_sc} are narrower than presented in Ref.~\cite{Karan:2023adm}. 

Though we sample the parameter-points uniformly in the dark-matter mass range of 3 GeV to 10 TeV, one can notice the absence of points below 60 GeV in all plots in Fig. \ref{fig:DD_sc}, mainly due to constraints from colliders and cLFV.
While the purple area in Fig. \ref{fig:DD_sc} represents the neutrino floor emerging from the coherent scattering of neutrinos, the greenish region is excluded by LZ, the most constraining experiment in the DM mass range from 10 GeV to 10 TeV.
Parameter-points disallowed by LZ are shown in black and the points below the neutrino floor are shown in brown. 
The left and right panel depict different cases of Scenario-I and Scenario-II respectively. Since higher $v_\Omega$ increases the mixing of scalars and fermions ($ \alpha$ and $\beta$), the maximum $\sigma^{\rm{SI}}_{\rm{DM-N}}$ is higher in Scenario-I than II.

In the no co-annihilation case, there are allowed points for \mdm$\gsim m_H$ (400 GeV and 1100 GeV respectively) apart from the resonance near \mdm
$\sim m_H/2$.
Scenario-I is more promising in this case, as a large chunk of parameter-space in Scenario-II lies below the neutrino floor. 
For the case of fermion-fermion co-annihilation, though one gets viable parameter-space from 100 GeV onward, many points get ruled out by the LZ experiment.
However, due to higher value of $v_\Omega$ there are more points ruled out in Scenario-I than the same in Scenario-II. Thus scenario-II seems more favorable in this case. 
Finally, for the case of fermion-scalar co-annihilation, one can have dark-matter masses between $m_h/2$ and 5 TeV.  However, Scenario-I looks more encouraging here, as most of points in Scenario-II remain below the neutrino floor. 

Therefore, in order to determine viability of DM masses between 60 GeV and 100 GeV, fermion-scalar 
co-annihilation 
 effects are instrumental. Such fermion-scalar co-annihilations are also very important for scenarios with higher values of $v_\Omega$ (like Scenario-I) . Again, fermion-fermion co-annihilation is propitious for scenarios with lower values of $v_\Omega$ (like Scenario-II). Notice that the direct detection results within the mass range of 100 GeV to 1~TeV could be improved by including loop-corrections to the scattering cross-section, described in section \ref{subsec:dd-minimal-scoto}.

\subsection{Fermionic Dark-matter Indirect Detection}

Turning to indirect detection of fermionic dark-matter, one notices that, besides the present-day $p$-wave suppression of the thermal averaged dark-matter annihilation rates $\langle \sigma v\rangle$, there is a suppression due to the smallness of the Yukawa couplings.
The later is required in order to satisfy the neutrino data, as already seen in Sec.~\ref{sec:simpl-scot-setup}.
However, the presence of the triplet in the revamped scenario brings in new features reminiscent of supersymmetric neutralino dark-matter annihilation~\cite{PhysRevLett.50.1419,Ellis:1983ew,Jungman:1995df}.
As in the case supersymmetric case ~\cite{PhysRevLett.50.1419,Ellis:1983ew,Jungman:1995df}, one can consider two limits in the revamped scotogenic model, namely, the limit in which DM is mainly triplet or mainly singlet, depending on the magnitude of the singlet-triplet mixing strength. 
In contrast to supersymmetric dark-matter, however, DM annihilation rates are  constrained by the smallness of the Yukawa couplings required to satisfy the neutrino oscillation and cLFV constraints~\cite{Restrepo:2019ilz,Karan:2023adm,Avila:2019hhv}.
\begin{figure}[h!]
	\centering
\includegraphics[scale=0.3]{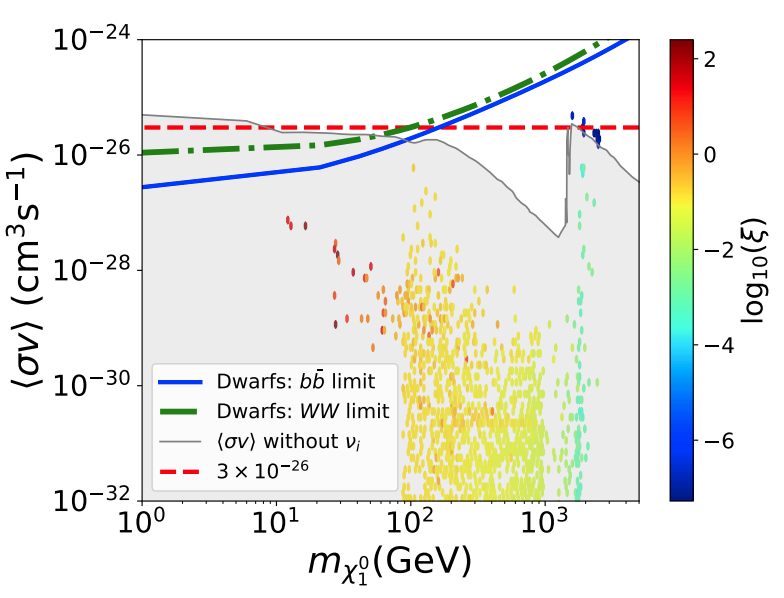} 
\caption{ Indirect detection of the gamma ray signal  from DM annihilation into $b\bar{b}$ (blue) and WW (green). The canonical thermal WIMP annihilation cross-section is shown by the red dashed line.  The region compatible with relic density but not the neutrino data are shown in gray~\cite{Restrepo:2019ilz}. The colors of the points indicate $\xi=|M_\Sigma-m_{\chi_1^0}|/m_{\chi_1^0}$. }	\label{fig:ID_st}
\end{figure}

Fig. \ref{fig:ID_st} depicts the velocity-averaged  fermionic DM annihilation cross-section in the revamped scotogenic scenario. The constraints on DM annihilating to $b \bar b$ and $WW$ channels, which emerge from the 95\% C.L. upper limit on gamma rays from Dwarf Spheroidal Galaxies (dSphs)~\cite{Fermi-LAT:2015att}, are shown in blue and green curves. 

 One finds that, before imposing neutrino constraints (the gray region), $\langle{\sigma} v\rangle$ is competitive with current dark-matter indirect detection upper bounds, such as the one placed by the Fermi-LAT observations~\cite{Kerszberg:2023cup,Fermi-LAT:2015att}. 
The imposition of restrictions 
arising, e.g., from neutrino oscillations and cLFV limits~
(Sec.~\ref{sec:neutrino-parameters}),
as well as collider constraints (Sec.~\ref{sec:collider})  substantially reduces the detection prospects, though gamma-ray telescopes, such as  HESS~\cite{HESS:2016mib}, may still partly cover the revamped scotogenic parameter space.

Notice that, despite the fact that vanilla supersymmetric dark-matter is unrelated to neutrinos (and hence unrestricted by neutrino data), the associated indirect detection sensitivities are not especially promising~\cite{Han:2014nba,Abdughani:2018bhj,Barman:2022jdg}. 
The revamped scotogenic scheme mimics the case of supersymmetric DM in a simpler manner and directly related to neutrino physics. Yet one may still have some chance of probing it at gamma-ray telescopes such as, for example, the upcoming Cherenkov
Telescope Array (CTA)~\cite{Acharya:2017ttl,CTA:2020qlo}, expected to significantly enhance our ability to search for dark-matter indirectly. \\[-.3cm]

Before closing this section, let us mention that fermionic scotogenic dark-matter can also exhibit neutrino annihilation channels.
Hence the well-motivated physics scotogenic picture may also be potentially probed through neutrino telescope searches at IceCube~\cite{Arguelles:2019ouk} as well as KM3NeT~\cite{KM3NeT:2024xca}, in addition to high-energy gamma-ray telescopes.


\subsection{Advandages of the revamped scenario}

In the revamped singlet-triplet scotogenic model, dark matter candidates arise from a combination of singlet and triplet fermions or scalars, stabilized by a dark symmetry.
Besides a clear prediction for neutrinoless double-beta decay, Fig.~\ref{fig:dbd2}, this brings several advantages to the revamped scotogenic scenario, absent in the original one, overcoming its limitations and widening the phenomenological scope. We now turn to these issues.  \\

\begin{center}
   {\bf \small Neutrinoless double-beta decay } 
\end{center} 

Here we note that the revamped scotogenic model has just one singlet and one triplet dark fermion mediating the one-loop diagram responsible for neutrino mass generation.
As a consequence, it provides an example of a \emph{missing partner} radiative seesaw mechanism, so that one of the three active neutrinos will remain massless.
As we saw in Sec.~\ref{sec:neutrino-parameters}, this leads to promising predictions for the \znbb decay processes, displayed in Fig.~\ref{fig:dbd2}, with promising expectations for a possible experimental detection.

\begin{center}
   {\bf \small High energy behaviour of dark parity} 
\end{center}

\begin{figure}[!h]
\centering
\includegraphics[height=5.5cm,width=0.6\textwidth]{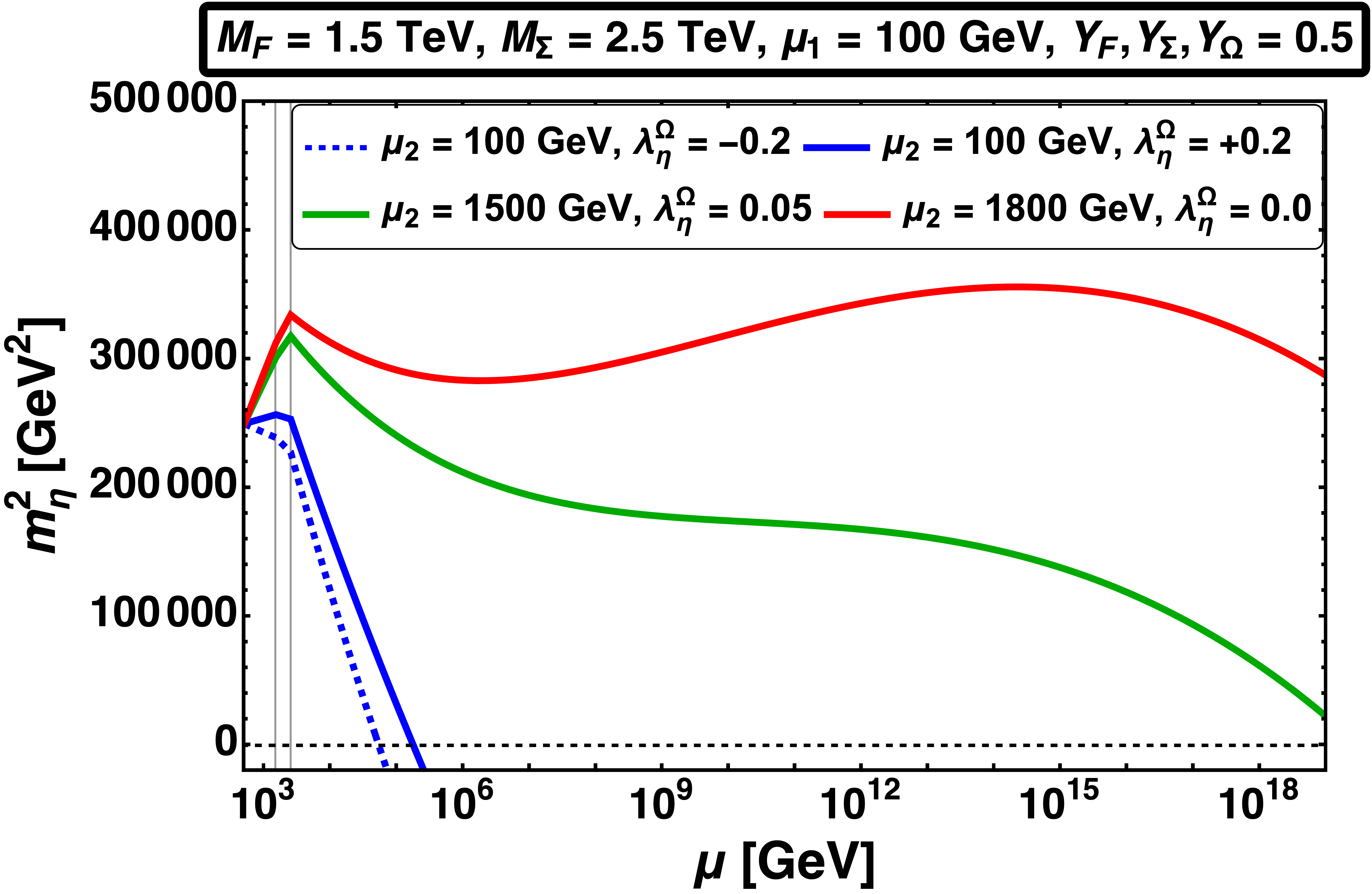}
\caption{
$m_\eta^2$ evolution of as a function of the renormalisation scale $\mu$. We fixed the parameters $M_F$, $M_\Sigma$, $Y_{\Sigma/F}$ and $\mu_1$ as indicated. 
The blue solid and dashed lines correspond to the indicated $\mu_2$ and $\lambda_{\eta}^{\Omega}$ values.
See text for details.}
\label{fig:z2parity-singlet-triplet}
\end{figure}

The original scotogenic model suffers from the problem associated to the unwanted breaking of the dark parity symmetry. This can be naturally circumvented in the revamped (singlet-triplet) version due to the presence of the scalar triplet $\Omega$. Let us focus on the fermionic dark-matter case with~$M_F^2 < m_{\eta_{R,I}}^2$.
In order to understand the source of $\mathbb{Z}_2$ symmetry-breaking, one should analyze the evolution of $m_{\eta}^2$ with the renormalisation scale $\mu$, described by
\begin{align}
\beta_{m_\eta^2} & =  
-m_\eta^2\Big(\frac{3}{2} g_{1}^{2} + \frac{9}{2} g_{2}^{2} - 6 \lambda_2 \Big) -2 m_\Phi^2\Big(2\lambda_3 + \lambda_4\Big) - 3 \lambda_\eta^\Omega m_\Omega^2 +3 \mu_{2}^{2} +2 \Big(m_\eta^2 -2 |M_F|^2 \Big) |Y_F|^2 \nonumber \\ 
 &+3 \Big( m_\eta^2 -2 |M_\Sigma|^2 \Big) |Y_\Sigma|^2.
\end{align}
One sees that the situation is similar to the original scotogenic model in the sense that, once the renormalisation scale $\mu\geq M_{\Sigma/F}$,
the Yukawa couplings $Y_{\Sigma/F}$ starts running and this can drive $m_{\eta}^2$ to negative values and induce the breaking of the dark $\mathbb{Z}_2$ symmetry. 
However, the presence of a scalar triplet $\Omega$ can counteract this effect, making an important difference when compared to the original scotogenic setup. 
Indeed, in addition to the negative fermionic contribution, there are other contributions such as $\beta_{m_{\eta}^2}\sim -3\lambda_{\eta}^\Omega m_{\Omega}^2$, where a negative $m_{\Omega}^2$ is required by virtue of the tadpole equation.
Hence, depending on the sign of $\lambda_{\eta}^\Omega$, this term can give a positive contribution, so the breaking of $\mathbb{Z}_2$ can be postponed to higher scales, or avoided altogether. 
This effect is limited if $\lambda_{\eta}^\Omega$ is restricted to lie within the perturbative regime. 
This is shown in Fig.~\ref{fig:z2parity-singlet-triplet}, adapted from~\cite{Merle:2016scw} where blue solid and blue dotted lines correspond to $\lambda_{\eta}^\Omega=0.2$ and $\lambda_{\eta}^\Omega=-0.2$, respectively. 

We also find that the dimensionful triplet scalar mass parameter $\mu_2$ yields potentially large and positive contribution through the term $\beta_{m_{\eta}^2}\sim +3\mu_2^2$ associated to the $\Omega$ scalar. 
For sufficiently large $\mu_2$ values, this positive contribution can exceed the negative ones from the fermions $\Sigma/F$, such that the dark $\mathbb{Z}_2$ symmetry is preserved up to the Planck scale, as shown by the red and green lines in Fig.~\ref{fig:z2parity-singlet-triplet}. 

It follows that in the revamped scenario, one can easily preserve the dark parity up to the Planck scale, even with large $Y_F$ values characterizing fermionic scotogenic dark-matter, as long as $\mu_2$ is sufficiently large.
In conclusion, one finds that in the singlet-triplet scotogenic model the dark $\mathbb{Z}_2$ symmetry can remain exact all the way up to the Planck scale, even for sizable Yukawa couplings.\\

\section{Scoto-seesaw picture: two paradigms in one}
\label{sec:scoto-seesaw}
Neutrino oscillations determine two mass parameters, namely the two mass squared splittings associated with the “atmospheric”  and “solar” neutrino oscillations~\cite{McDonald:2016ixn,Kajita:2016cak}. 
Precision measurements at reactor and accelerator-based experiments~\cite{DayaBay:2012fng,T2K:2017hed} further strengthen the three-neutrino oscillation picture. The corresponding ratio of squared solar-to-atmospheric mass splittings for normal and inverted mass hierarchy are found to be~\cite{deSalas:2020pgw,10.5281/zenodo.4726908} 
\begin{align}  
\textbf{NO:}\, \frac{\Delta m_{\text{SOL}}^{2}}{\Delta m_{\text{ATM}}^{2}}=0.0294^{+0.0027}_{-0.0023},\,\,\,\,\textbf{IO:}\,\frac{\Delta m_{\text{SOL}}^{2}}{\Delta m_{\text{ATM}}^{2}}=0.0306^{+0.0028}_{-0.0025}.
  \label{eq:sol-atm-obs}
\end{align} 
This ratio is required in order to describe the neutrino oscillation data successfully. The existence of two different neutrino mass scales-or oscillation lengths-could be an indication that, perhaps, two different mechanisms are responsible for generating these mass scales. 
The idea then is that the above ratio could arise from a loop suppression factor.
The \textit{scoto-seesaw} proposal combines the seesaw and the scotogenic mechanisms, so that the atmospheric scale arises from the seesaw mechanism, while the solar mass scale arises from the scotogenic radiative seesaw mechanism. For recent descriptions of the seesaw mechanism see, for example, Ref.~\cite{Ding:2024ozt}.

\subsection{The simplest \textit{scoto-seesaw} }
\label{sec:scotoseesaw}

As mentioned, in this framework neutrino masses have a hybrid origin, combining a type-I seesaw structure (including the possibility of inverse or linear seesaw) with a scotogenic (radiative seesaw) mechanism to naturally explain the smallness of neutrino masses, the mediator role of dark-matter, as well as ”explaining” Eq.~\eqref{eq:sol-atm-obs}.
The simplest \textit{scoto-seesaw} extension of the \sm can be constructed by minimally combining the scotogenic paradigm with the type-I seesaw mechanism~\cite{Rojas:2018wym,Mandal:2021yph}.
\begin{table}[!h]
\centering 
%
\includegraphics[scale=0.3]{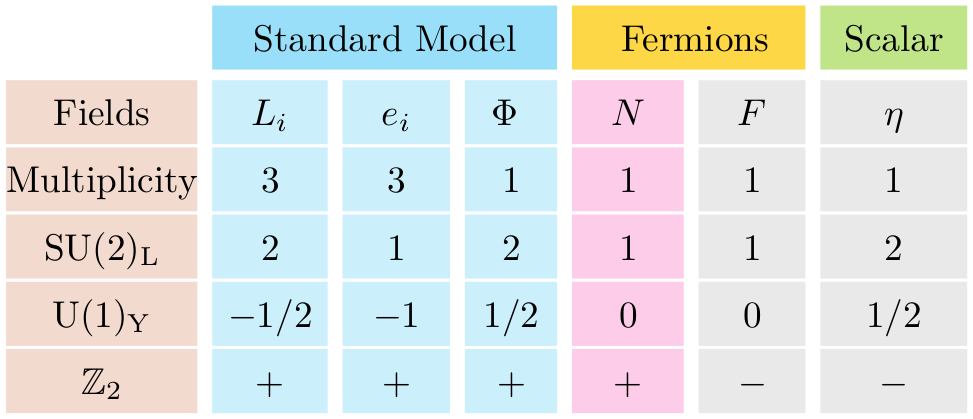}
\caption{ 
Assignments of the minimal scoto-seesaw model, dark states have opposite $\mathbb{Z}_2$ parity as SM states and $N$}
\label{tab:scoto-seesaw}
\end{table}
This is simply the (3,1) type-I seesaw mechanism already discussed in~\cite{Schechter:1980gr}, 
we call it~\textit{missing partner seesaw}.  
This is now \textit{cloned} with the minimal scotogenic model proposed in~\cite{Tao:1996vb,Ma:2006km}.
In such hybrid scenario the \textit{atmospheric} scale comes from the tree level seesaw, while the \textit{solar} scale is mediated by the radiative exchange of dark states.
The new particles and their charges are given in Table.~\ref{tab:scoto-seesaw}, where the family index $i$ runs from 1 to 3.
All the SM particles and $N$ are even under the dark parity $\mathbb{Z}_2$, while the dark sector, consisting of a fermion $F$ and a scalar $\eta$, is $\mathbb{Z}_2$-odd.
The full Yukawa Lagrangean can be split as
\begin{align}
 \mathcal{L}=\mathcal{L}_{\text{SM}}+\mathcal{L}_{\text{ATM}}+\mathcal{L}_{\text{DM,SOL}},
\end{align}
where $\mathcal{L}_{\text{SM}}$ is the SM Lagrangian, while
\begin{align}
 \mathcal{L}_{\text{ATM}}= -  Y_{N}^{i}\bar{L}^{i}  \tilde{\Phi} N  + 
 \frac{1}{2}  M_{N} \overline{N^{c}}N + \text{h.c}.
\label{eq:ATM-Lagrangian-scoto}
\end{align}
 After electroweak symmetry-breaking this generates the following type-I seesaw mass matrix 
\begin{align}
\mathcal{M}_{\nu}^{ij} = \frac{1}{\sqrt{2}}\begin{pmatrix} 
0 & 0 & 0 & Y^{1} _{N}  v_\Phi \\
0 & 0 & 0 & Y^{2} _{N}  v_\Phi\\
0 & 0 & 0 & Y^{3} _{N}  v_\Phi\\
Y^{1} _{N}  v_\Phi & Y^{2} _{N}  v_\Phi & Y^{3} _{N}  v_\Phi & \sqrt{2}\,M_N \\           \end{pmatrix}.
\end{align}
Perturbative diagonalization~\cite{Schechter:1981cv} leads to
\begin{align}
\mathcal{M}_{\nu\text{TREE}}^{ij}=-\frac{v_\Phi^2}{2M_N}Y_N^i Y_N^j,
\label{atmospheric scale}
\end{align} 
where $i,j=1,2,3$ are family indices of the lepton doublets. 
One sees that the $N$ pairs-off with one combination of the doublets in $L^i$ through their Dirac-like Yukawa couplings. 
As a result one has a projective structure for the tree-level neutrino mass matrix
Eq.~\eqref{atmospheric scale},
leading to one neutrino mass parameter, to be identified with the \textit{atmospheric} neutrino mass scale.
Taking $\mathcal{O}(1)$ Yukawa couplings $Y_N$ one can reproduce the required value of the atmospheric scale
if the heavy neutrinos lie at some large mass scale $M_N\sim\mathcal{O}(10^{14}\,\text{GeV})$. 
Smaller values of the Yukawa coupling $Y_N$ would require correspondingly lower seesaw scale $M_N$.     
\par The Lagrangian responsible for the generating the \textit{solar} neutrino mass scale is given by
\begin{align}
 \mathcal{L}_{\text{DM,SOL}} = Y_{F}^{i}\bar{L}^{i} \tilde{\eta}~F+\frac{1}{2}M_{F}\overline{F^{c}}F+ \text{h.c}.
\label{eq:SOL-Lagrangian-scoto}
\end{align}
The radiative exchange of the dark scalar and fermionic
mediators $\eta$ and $F$ induces a \textit{calculable} solar neutrino mass matrix~(see Fig.~\ref{fig:neutrino-loop-scotogenic}), 
\begin{equation}
 \mathcal{M}_{\nu \text{Loop}}^{ij}= \frac{Y_F^{i} \, Y_F^{j} M_{F}}{32 \pi^2}\left[ \frac{m_{\eta_R}^2}{m_{\eta_R}^2-M_{F}^2} \log \frac{m_{\eta_R}^2}{M_{F}^2} -  \frac{m_{\eta_I}^2}{m_{\eta_I}^2 - M_{F}^2} \log \frac{m_{\eta_I}^2}{M_{F}^2} \right].
 \label{eq:mnu-scoto-seesaw}
\end{equation}
This lifts the tree-level mass degeneracy. Note that, although these corrections are also projective, they involve \textit{different} Yukawa couplings, breaking the tree-level structure of the (3,1) type-I seesaw  mechanism, in a way analogous to neutrino mass generation in bilinear broken R-parity supersymmetry~\cite{Hirsch:2000ef,Diaz:2003as,Hirsch:2004he}. 
The total neutrino mass is 
\begin{align}
 \mathcal{M}_{\nu \text{TOT}}^{ij}=\mathcal{M}_{\nu \text{Tree}}^{ij}+\mathcal{M}_{\nu \text{Loop}}^{ij},
 \end{align}
where the first term is the tree-level seesaw part associated to the atmospheric mass scale and the second is the one-loop
scotogenic contribution
responsible for the solar mass scale. Note that one out of the three neutrinos remains massless, as both contributions have a projective nature. The atmospheric and solar square mass differences can be calculated using the eigenvalues of the neutrino mass matrix $\mathcal{M}_{\nu\text{TOT}}^{ij}$ as follows:
\begin{align}
 \Delta m_{\text{ATM}}^{2}= \left(\frac{v_\Phi^{2}}{2M_{N}}\mathbb{Y}_{N}^{2}\right)^{2},\,\,\, 
 \Delta m_{\text{SOL}}^{2}\approx \left(\frac{1}{32\pi^{2}}\right)^{2}\left(\frac{\lambda_{5}v_\Phi^{2}}{M_{F}^{2}-m_{\eta_{R}}^{2}}M_{F}\mathbb{Y}_{F}^{2}\right)^{2}.
 \label{eq:atm-sol-mass}
\end{align}
where we take $M_{F}^{2}$, $m_{\eta_{R}}^{2}$, $M_{F}^{2}-m_{\eta_{R}}^{2}\gg\lambda_{5}v_\Phi^{2}$ and $\mathbb{Y}_{\alpha}^{2}=(Y_{\alpha}^{e})^{2}+(Y_{\alpha}^{\mu})^{2}+(Y_{\alpha}^{\tau})^{2}$ for $\alpha=N,\,F$. As a result, the ratio of the squared solar and the atmospheric  mass differences is loop-suppressed and can be expressed as
\begin{align}
\frac{\Delta m_{\text{SOL}}^{2}}{\Delta m_{\text{ATM}}^{2}}\approx \left(\frac{1}{16\pi^2}\right)^2  \left(\lambda_5\frac{M_N M_F}{M_F^2-m_{\eta_R}^2}\right)^2 \left(\frac{\mathbb{Y}_F^2}{\mathbb{Y}_N^2}\right)^2 .
\label{eq:ratio}
\end{align}
A priori, the mass of the right-handed neutrino $M_N$ and that of the dark sector fermion $M_F$ can be very different.
As an example, the case of scalar WIMP dark-matter can be realized by choosing $M_N\, \sim\, 10^{12}\,\text{GeV}$, $M_F\sim 10^{4}\,\text{GeV}$, $m_{\eta}^{R}\, \sim\, 10^{3}\,\text{GeV}$, $\mathbb{Y}_{N}\, \sim\, 10^{-1}$, $\mathbb{Y}_{F} \sim 10^{-4}$, 
and can easily fit the “solar”  and “atmospheric”  scales in Eq.~(\ref{eq:sol-atm-obs}), as long as one takes an adequately small value for $\lambda_5$. 
The latter indicates symmetry protection since, as $\lambda_5 \to 0$, lepton number is recovered in the theory.
However, we note that the masses of $N$ and $F$ can also be similar to each other. For example, keeping other parameters as before one can have $M_N \approx M_F \approx 10^4$ GeV for $\mathbb{Y}_{(N)}\, \sim\, 10^{-5}$ and $\mathbb{Y}_{F} \sim 10^{-4}$. Another benchmark with intermediate mass values $M_N \approx M_F \approx 10^6$ GeV is obtained for $\mathbb{Y}_{N}\, \sim\, 10^{-4}$ and $\mathbb{Y}_{F} \sim 10^{-3}$. 
Clearly, many other choices for masses and couplings in Eq.~\eqref{eq:ratio} can be consistent with Eq.~(\ref{eq:sol-atm-obs}). \par

The phenomenology of fermionic dark matter in the context of this simplest scoto-seesaw scheme, i.e. relic dark matter abundance and nuclear recoil cross sections are generically very similar to the results found in the original scotogenic model of Sect.~\ref{sec:dm-fermion-scoto}.
Concerning collider signals, those involving scalars are similar to results found for the IHDM and for the simplest scotogenic scenarios.
Besides explaining the ratio in Eq.~\eqref{eq:ratio}, the present scoto-seesaw scenario yields detectable rates for cLFV processes~\cite{Rojas:2018wym,Mandal:2021yph}, in contrast to the IHDM, where these are simply absent.

\subsection{Dynamical \textit{scoto-seesaw} }

One can naturally combine the seesaw and scotogenic paradigms within a fully dynamical scoto-seesaw scenario with a loop-suppressed \textit{solar-to-atmospheric} scale ratio~\cite{Leite:2023gzl} . 
In Table \ref{dyn-SCS} we give the particle content and the symmetry properties of the model. 
Besides the SM leptons, we introduce new SM singlet fermions $f_{aR}$ ($a=1,2$) and $N_R$, that are charged under $U(1)_{B-L}$. 
The scalar sector is extended beyond the SM-like Higgs doublet $H$, by introducing two extra $\mathrm {SU(2)_L}$ doublets, $\Phi$ and $\eta$ and four singlets, $\varphi_{1,2,3}, \sigma$. 
Our goal here is to sketch the main features of this proposal. 
The extra fields play well-defined roles in neutrino mass
generation, dark matter, and their interconnection.
By construction the model is free of B-L anomalies, as required for a consistent local gauge symmetry.
\begin{table}[h!]
\begin{center} 
%
\includegraphics[scale=0.28]{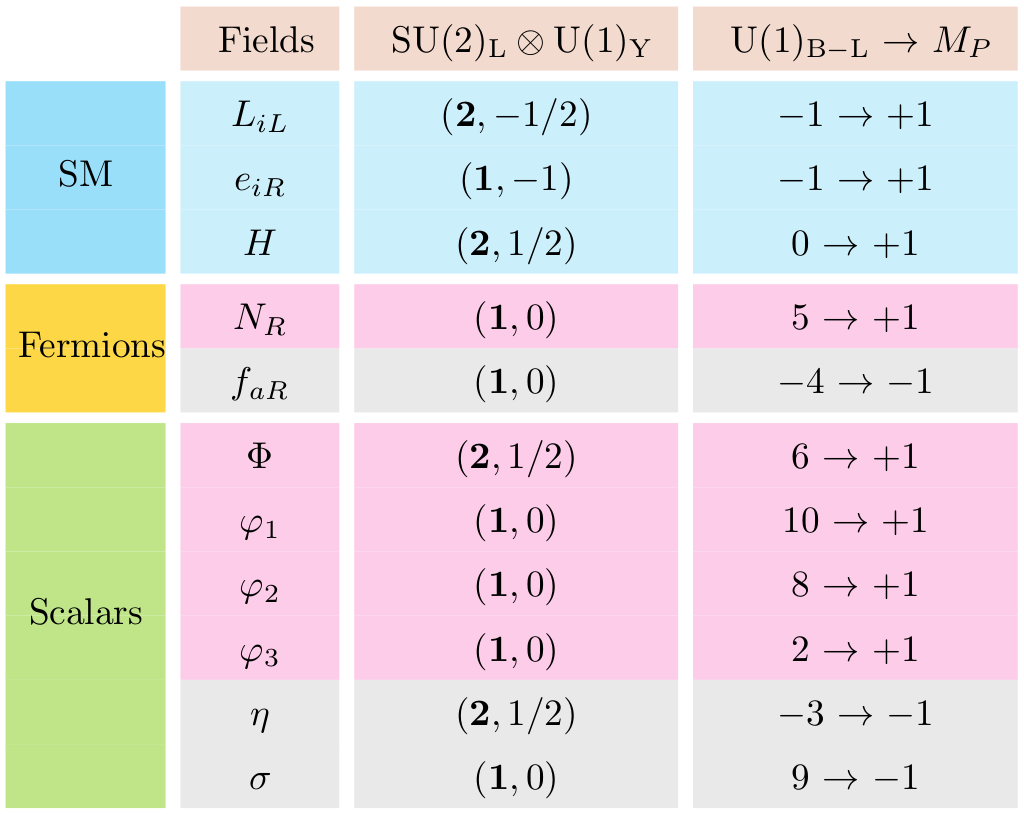}
\end{center}
\caption{
Quantum number assigments in the dynamical scoto-seesaw ($i=1,2,3$ and $a=1,2$), Ref.~\cite{Leite:2023gzl}. 
Dark fermions and scalars are odd under the residual matter parity symmetry, 
in contrast to SM and all other new states.}
 \label{dyn-SCS}
\end{table}

The most general renormalizable lepton Yukawa Lagrangian consistent with Table~\ref{dyn-SCS} is given as,
\begin{eqnarray}
\label{eq:Yuk}
-\mathcal{L}_Y &=& Y^H_{ij} \overline{L}_{iL}H e_{jR} + Y^{\eta} _{ia} \overline{L}_{iL}\tilde{\eta}f_{aR}+Y^{\Phi} _{i} \overline{L}_{iL}\tilde{\Phi}N_{R}+\frac{Y^{f}_{a}}{2} \varphi_2 \overline{(f_{aR})^c}f_{aR}+\frac{Y^{N}}{2}\varphi_1^* \overline{(N_{R})^c}N_{R}+h.c.,
\end{eqnarray}
Once the Higgs doublet $H$ acquires its vacuum expectation value the charged fermions obtain masses as in the SM.
In contrast, neutrino mass generation involves the new scalar bosons. 

 One assumes that only neutral scalars that have even B-L charges acquire VEVs.
As a result, a residual matter-parity symmetry defined as  
\begin{equation}
\label{eq:MP}
M_P = (-1)^{3(B-L)+2s}
\end{equation}
survives after spontaneous symmetry-breaking takes place. 
Only $f_{aR}$, $\eta$ and $\sigma$ are odd under $M_P$; these constitute the dark sector. The lightest electrically neutral  scotogenic particle (LSP) will be stable due to $M_P$ conservation, and hence a possible dark-matter candidate.

Neutrino mass generation involves the exchange of $M_P$-even and $M_P$-odd (or dark) fields at tree-level (seesaw) and loop-level (scotogenic), respectively. 
At tree-level, in the basis $(\nu_{iL}, (N_R)^c)$, one has the following mass matrix  
\begin{align}
M^{\nu,N} = \frac{1}{\sqrt{2}}\begin{pmatrix} 
0 & 0 & 0 & Y^{\Phi} _{1}  v_\Phi \\
0 & 0 & 0 & Y^{\Phi} _{2}  v_\Phi\\
0 & 0 & 0 & Y^{\Phi} _{3}  v_\Phi\\
Y^{\Phi} _{1}  v_\Phi & Y^{\Phi} _{2}  v_\Phi & Y^{\Phi} _{3}  v_\Phi & Y^N v_{\varphi_1}\\  \end{pmatrix},
\end{align}
very much analogous to the one discussed in section~\ref{sec:scotoseesaw}. 
Diagonalization in the limit $v_{\varphi_i}\gg v_H \gg v_\Phi$ leads to the seesaw-suppressed active neutrino mass matrix, 
\begin{equation}
\label{eq:numass0}
M^{\nu(0)}_{ij} \simeq -\frac{Y^{\Phi} _i Y^{\Phi} _j}{2} \frac{v_\Phi^2}{m_N},
\end{equation}
which has only one non-vanishing eigenvalue $\sim -\frac{v_\Phi^2}{m_N}\sum_i (Y^{\Phi} _{i})^2$, similar to Eq.~\eqref{atmospheric scale}.
Here $m_N \simeq v_{\varphi_1} Y^N/\sqrt{2}$ is the mass of $N_{R}$ and $v_\Phi\equiv v_H\epsilon$ is the tiny induced VEV of the leptophilic doublet $\Phi$, where $\epsilon$ is a small parameter (details given in Ref.~\cite{Leite:2023gzl}).
\begin{figure}[h]
\centering
\includegraphics[height=4.0cm,width=0.35\textwidth]{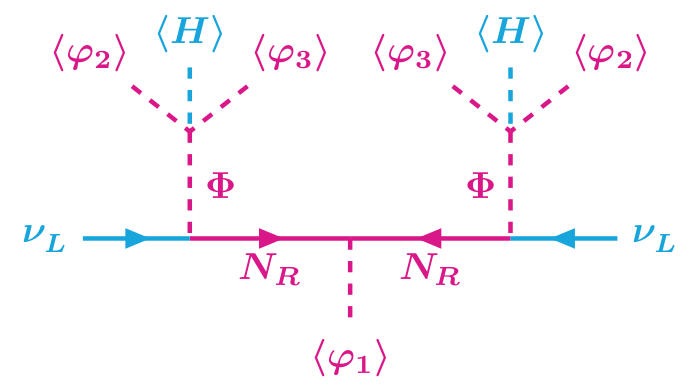} \hfil
\includegraphics[height=4.5cm,width=0.35\textwidth]{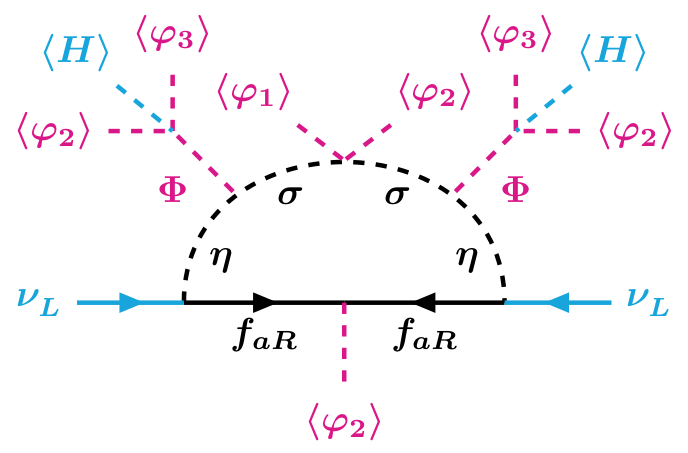}
\caption{
Tree-level seesaw diagram generating the atmospheric scale (left) and dark-mediated loop diagrams inducing the solar mass scale (right).}
\label{fig:scoto-seesaw}
\end{figure}

In contrast to the conventional type-I seesaw mechanism, in which neutrino mass suppression follows from the large size of $m_N$ with respect to the electroweak scale ($v_{EW}\sim v_H$), here it comes from the small induced VEV of $\Phi$, characterized by the tiny $\epsilon$~\cite{Leite:2023gzl}.
This allows for moderate values for $m_N$ accessible to collider experiments, where $N$ would be produced through a $Z^ \prime$-mediated Drell-Yan process.  
Concerning the other two active neutrinos, their masses are generated at one-loop through the scotogenic mechanism,  
\begin{eqnarray}\label{eq:numass1l}
M^{\nu(1)}_{ij} &=& \sum_{c=1}^2 Y^{\eta} _{ic} \mathcal{M}_c\, Y^{\eta} _{jc},\quad\quad\mbox{with}\\
\mathcal{M}_{c} &=& \frac{m_{f_c}}{16 \pi^2} 
\left[ \frac{\cos^2 \theta_s m^2_{s_1}}{m^2_{s_1} - m_{f_c}^2} \ln \frac{m^2_{s_1}}{m_{f_c}^2} 
- \frac{\cos^2 \theta_a m^2_{a_1}}{m^2_{a_1} - m_{f_c}^2} 
\ln \frac{m^2_{a_1}}{m_{f_c}^2} + \frac{\sin^2 \theta_s m^2_{s_2}}{m^2_{s_2} - m_{f_c}^2} \ln \frac{m^2_{s_2}}{m_{f_c}^2} - \frac{\sin^2 \theta_a m^2_{a_2}}{m^2_{a_2} - m_{f_c}^2} 
\ln \frac{m^2_{a_2}}{m_{f_c}^2} \right],\nonumber\label{scotomass}
\end{eqnarray}
where $m_{f_c}$ are the two dark fermion masses, $m_{s_i}$/ $m_{a_i}$ are CP-even/odd scalar masses.
One can check that, as before, the scotogenic loop has the required symmetry protection properties. The full neutrino mass generation mechanism is illustrated in Fig.~\ref{fig:scoto-seesaw}.

In short, the atmospheric neutrino mass scale is generated at tree level by a TeV-scale seesaw mediator, while two dark fermions carrying different B-L charges induce the solar scale
through the scotogenic loop,
reproducing the scoto-seesaw structure~\cite{Rojas:2018wym,Mandal:2021yph}. 
This framework favors the normal  neutrino mass ordering, preferred experimentally~\cite{deSalas:2020pgw,10.5281/zenodo.4726908,Esteban:2020cvm,Capozzi:2021fjo}, and provides an understanding of the origin of the smallness of the ratio $\Delta m^2_{sol}/\Delta m^2_{atm}$. \\[-.3cm]

We will not discuss here the details of the phenomenology of dark matter in this scheme. However, one can say that as before, it allows for a viable WIMP scotogenic dark matter scenario to be realized.
Moreover, it can have a rich collider and cLFV phenomenology. This arises from the gauged B-L symmetry, and from the existence of a Majoron-like Nambu-Goldstone boson~\cite{Leite:2023gzl}.
In particular, we will highlight in Sect.~\ref{sec:cLFV} the important role of the dark sector in mediating cLFV processes.
Before concluding this section let us mention also some recent attempts to combine neutrino mass generation, in particular within the scoto-seesaw picture, with family symmetries~\cite{Ding:2024ozt}; see Refs.~\cite{Barreiros:2020gxu,Barreiros:2022aqu,Ganguly:2022qxj,Bonilla:2023pna,Kumar:2023moh,PATHAK2025100055,Li:2025bsr,Nasri:2026nbf}.

\section{Dark low-scale seesaw mechanism}
\label{sec:dark-low-scale-seesaw}

 Neutrino masses may result from physics at the electroweak scale~\cite{Valle:2015pba} in a variety of ways~\cite{Boucenna:2014zba,Cai:2017jrq}.
 One can incorporate dark matter within a low-scale seesaw setup,
for example, within inverse seesaw schemes with explicitly~\cite{Mohapatra:1986bd} or spontaneously broken lepton number~\cite{Gonzalez-Garcia:1988okv}. 
 Alternatively, one can have linear seesaw models~\cite{Akhmedov:1995ip,Akhmedov:1995vm,Malinsky:2005bi} involving a leptophilic doublet Higgs boson~\cite{Fontes:2019uld,Batra:2022arl,Batra:2023mds,Batra:2023ssq}.
 Either way, one can realize the basic idea that a dark sector sources neutrino mass generation, that proceeds \emph{a la seesaw}~\cite{Mandal:2019oth,CarcamoHernandez:2023atk,Batra:2023bqj}.
 Here we sketch the relevant theory benchmarks.

\subsection{Dark inverse seesaw }
\label{sec:dark-inverse-seesaw}

We first consider the simpler case of explicit lepton number breaking in the inverse seesaw mechanism. One extends the standard \SM model with new singlet fermions $N^c$, $S$. In the minimal setup, dubbed (3,2,2), one adds two copies of $N^c$ and $S$ to account for both the \textit{atmospheric} and \textit{solar} mass splittings. However, here we take the more common, sequential choice of having three $N^c$, $S$ pairs, called $(3,3,3)$ scheme. 
The simplest way to introduce a dark sector to seed neutrino masses is through the presence of a complex dark scalar $\xi$ and dark fermion $F$.
The new particles and their charges are given in Table.~\ref{tab:dark-inverse-seesaw}. The $\mathbb{Z}_2$ symmetry is the dark parity responsible for the stability of the dark-matter candidate.
\begin{table}[!h] 
\includegraphics[scale=0.27]{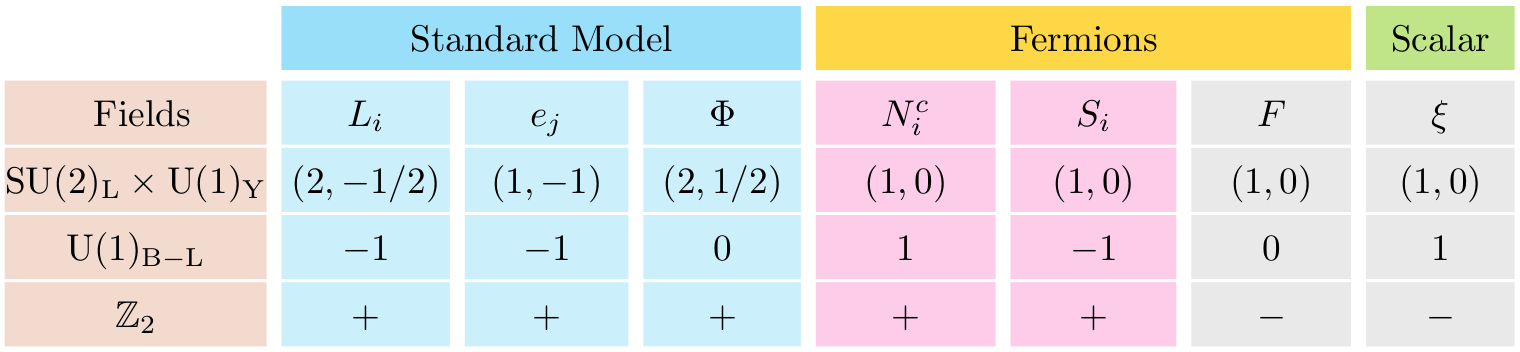}
\caption{\label{tab:dark-inverse-seesaw} 
Quantum numbers in the dark inverse seesaw model with explicit lepton number breaking.
Dark states have opposite $\mathbb{Z}_2$ parity as SM states and the seesaw mediators $N^c_i$ and $S_i$.
}
\end{table}
\renewcommand{\arraystretch}{1.2}

The relevant terms in the Yukawa Lagrangian responsible for dark matter and neutrino masses are the
following:
\begin{eqnarray}
\mathcal{L}^{\text{Yukawa}} &=& -\,Y_{N} {L} i\tau_2 \Phi N^c\, -\,Y_\xi \xi F S\,-\,M N^c S \,-\, \frac{1}{2}\mathcal{M}_F F F\, + \text{H.c.}
\label{eq:dis-Yukawa-Lagrangian} 
\end{eqnarray}
\ignore{\begin{eqnarray}
\mathcal{L}^{\text{Yukawa}} &=& -\,Y_{N} \overline{L} i\tau_2 \Phi^* N^c\, -\,Y_\xi \xi F S\,-\,M N^c S \,-\, \frac{1}{2}\mathcal{M}_F F F\, + \text{H.c.}
\label{eq:dis-Yukawa-Lagrangian} 
\end{eqnarray} }
%

The above Lagrangian is invariant under the \SM gauge group as well as under $\mathrm{U(1)_{B-L}\otimes\mathbb{Z}_2}$. The scalar potential is given by, 
\begin{eqnarray} 
\mathcal{V}_{(s)} & = & -m^2 \Phi^\dagger \Phi \,+ \, \frac{\lambda_\Phi}{2}\left( \Phi^\dagger \Phi\right)^2
\,- \, m_\xi^2 \xi^* \xi \,+\,\frac{\lambda_\xi}{2}\left( \xi^* \xi\right)^2 
\,+ \, \lambda_{\Phi\xi} \left(\Phi^\dagger \Phi \right)\left(\xi^* \xi \right) 
\,+\,\frac{\mu_\xi^2}{4}\left(\xi^2 \, + \, \text{H.c.}\right) 
\label{eq:scapot-explicit}
\end{eqnarray}
 
This scalar potential is $\mathbb{Z}_2$ symmetric, but the last term violates lepton number symmetry by two units, explicitly but softly~\cite{Ahriche:2016acx}.
This is required to generate neutrino masses.
The $\mathbb{Z}_2$ symmetry should remain unbroken so as to ensure dark-matter stability, therefore the $\mathbb{Z}_2$ odd scalar $\xi$ does not accquire any VEV.  The fields $\Phi$ and $\xi$ can be written as 
\begin{align}
\Phi=
\begin{pmatrix}
\phi^+\\
(v_\Phi+h+i\phi^0)/\sqrt{2}
\end{pmatrix},\,\,\,\,
~~~~~\xi=(\xi_R+i\xi_I)/\sqrt{2}
\end{align}

 The Higgs mass is exactly the same as in the SM, $m_h^2=\lambda_\Phi v_\Phi^2$. The real and imaginary components of $\xi$ have the following masses
\begin{align}
& m_{\xi_R}^2=m_\xi^2+\frac{1}{2}\left(\lambda_{\Phi\xi}v_\Phi^2+\mu_\xi^2\right),\,\,\,
m_{\xi_I}^2=m_\xi^2+\frac{1}{2}\left(\lambda_{\Phi\xi}v_\Phi^2-\mu_\xi^2\right)
\end{align}
The difference $m_{\xi_R}^2-m_{\xi_I}^2$ depends only on the parameter $\mu_\xi$ which, we see below, is also responsible for smallness of neutrino masses.  
Dark parity conservation also ensures the stability of the lightest of the two eigenstates $\xi_R$ or $\xi_I$.
As we will show below, this will be a viable dark-matter candidate.  \\[-.4cm] 

We now turn to the issue of  neutrino mass generation in the case of explicit breaking of lepton number. 

\begin{figure}[!h]
\centering 
\vglue -.2cm
\includegraphics[height=3.5cm,scale=0.3]{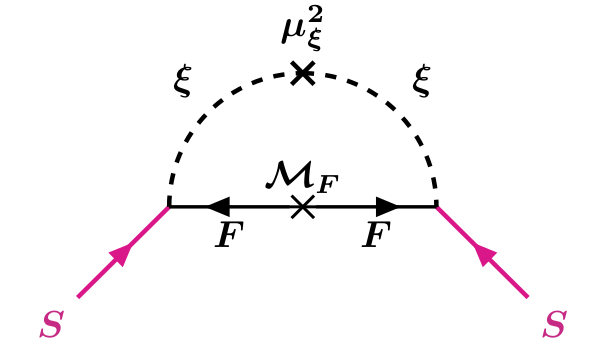}
\caption{ 
\textit{Dark} origin for the lepton number violating $\mu$ term in the simplest inverse seesaw mechanism.} 
\label{fig:dis-loop}
\end{figure}

The neutral lepton mass matrix has the following \textit{block} structure in the basis $(\nu, \nu^c, S)$ of our (3,3,3) scheme, 
\begin{eqnarray}
\mathcal{M}_{F^0} &=& \left[ \begin{array}{ccc}
                               0 & m_D & 0 \\
                               m_D & 0 & M \\
                               0 & M & \mu 
                               \end{array} \right] \,\,
                   \,\,=\,\, \left[ \begin{array}{ccc}
                               0 & Y_{\nu^c}^i \frac{v_\Phi}{\sqrt{2}} & 0 \\
                               Y_{\nu^c}^j \frac{v_\Phi}{\sqrt{2}} & 0 & M \\
                               0 & M & \mu
                               \end{array} \right]. 
\label{eq:treemass}
\end{eqnarray}
where the lepton number violating entry $\mu$ arises
through the one-loop diagram in Fig.~\ref{fig:dis-loop} namely, 
\begin{align}
\mu&=\frac{Y_\xi Y_\xi}{16\pi^2}\mathcal{M}_f\left(B_0(0,\mathcal{M}_f^2,m_{\xi_R}^2)-B_0(0,\mathcal{M}_f^2,m_{\xi_I}^2)\right)\nonumber \\
&=\frac{Y_\xi Y_\xi}{16\pi^2}\mathcal{M}_f\left( \frac{m_{\xi_R}^2}{m_{\xi_R}^2-\mathcal{M}_{f}^2}\log \left( \frac{m_{\xi_R}^2}{\mathcal{M}_{f}^2}\right) \,-\, \frac{m_{\xi_I}^2}{m_{\xi_I}^2-\mathcal{M}_{f}^2}\log \left( \frac{m_{\xi_I}^2}{\mathcal{M}_{f}^2} \right) \right)
\end{align}
where $B_0(0,\mathcal{M}_f^2,m_{\xi_{R/I}}^2)$ is a Passarino-Veltman function evaluated in the limit of zero external momentum. 
One sees that this term is induced by the soft-breaking-term $\frac{\mu_\xi^2}{4} (\xi^2+\text{h.c.})$ in the potential. 

The nonzero $SS$ mass-entry leads to neutrino masses, through 
\begin{align}
\mathcal{M}_\nu=m_D M^{-1}\mu M^{-1T}m_D^T.
  \label{eq:iss}
\end{align}
The smallness of neutrino masses is protected by lepton number symmetry, hence natural in t'Hooft's sense.
Since neutrino masses  result from a calculable radiative loop-correction indicated in Fig.~\ref{fig:dis-loop}, neutrinos are massless at the tree approximation, and one has an extra protection for their small masses coming from the loop factor. 

\begin{center}
   {\bf \small Spontaneous lepton number breaking} 
\end{center}

\begin{figure}[h!]
\centering
\includegraphics[scale=0.25]{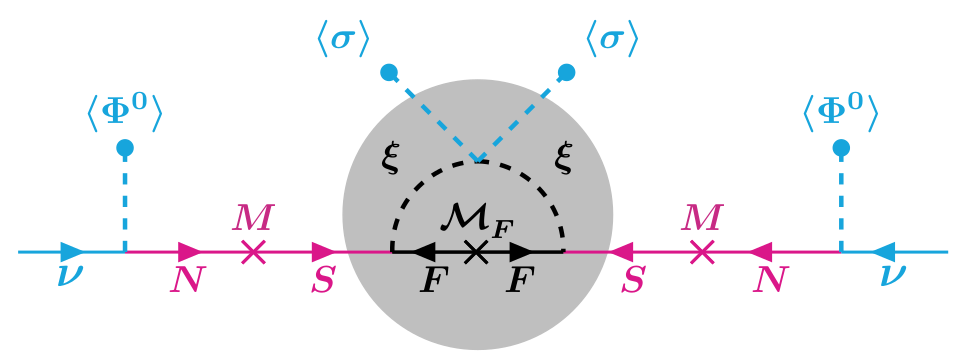}~~~
\caption{ 
Neutrino mass from \textit{dark} lepton number violating dynamical inverse seesaw mechanism.  }
\label{fig:scoto-loop}
\end{figure}

Let us now consider the dynamical generation of the $\mu$ parameter, which allows for a richer phenomenology. 
In such dynamical completion of the above scheme B-L is promoted to a spontaneously broken symmetry.
In order to do this, we add a complex scalar singlet $\sigma$, even under $\mathbb{Z}_2$ and with B-L=1 defined by the last term of the relevant scalar potential, given by  
\begin{eqnarray} 
\mathcal{V}_{(s)} & = & -m^2 \Phi^\dagger \Phi \,+ \, \frac{\lambda_\Phi}{2}\left( \Phi^\dagger \Phi\right)^2
\,- \, m_\xi^2 \xi^* \xi \,+\,\frac{\lambda_\xi}{2}\left( \xi^* \xi\right)^2 
\, - \, m_\sigma^2 \sigma^* \sigma \, + \,  \frac{\lambda_\sigma}{2}\left( \sigma^* \sigma \right)^2 \nonumber \\
& + &  \lambda_{\Phi\sigma} \left(\Phi^\dagger \Phi \right)\left(\sigma^* \sigma \right)
\, + \, \lambda_{\xi\sigma} \left(\xi^* \xi \right)\left(\sigma^* \sigma \right)
\,+ \, \lambda_{\Phi\xi} \left(\Phi^\dagger \Phi \right)\left(\xi^* \xi \right) 
\,+\,\frac{\lambda_5}{2} \left(\xi^* \sigma \right)^2 \, + \, h.c. 
\label{eq:scapot}
\end{eqnarray}
The lepton number symmetry is then broken spontaneously by the VEV of this complex singlet $\sigma$. 
We assume that this breaking happens at a relatively low scale~\cite{Gonzalez-Garcia:1988okv}.
The dark symmetry $\mathbb{Z}_2$ remains exactly conserved, enforcing the stability of the LSP.
In order to obtain the mass spectrum for the scalars after gauge and global symmetry-breaking, we expand the scalar fields as  
\begin{align}
\phi^0=\frac{1}{\sqrt{2}}(v_\Phi + R_1+i I_1),\,\, \sigma=\frac{1}{\sqrt{2}}(v_\sigma+R_2+i I_2),\,\,
\xi=\frac{1}{\sqrt{2}}(\xi_R+i \xi_I).
\end{align}
The scalar sector resulting from \eqref{eq:scapot} leads to two massive neutral CP-even scalars $h,H$ and a physical massless Goldstone boson, namely the majoron $J=\text{Im}\,\sigma$.
The CP-even Higgs scalar mass matrix in the basis $(R_1,R_2)$ is~\cite{Joshipura:1992hp} 
\begin{align} 
M_R^2=
\begin{bmatrix}
\lambda_\Phi v_\Phi^2  &  \lambda_{\Phi\sigma}v_\Phi v_\sigma \\
\lambda_{\Phi\sigma} v_\Phi v_\sigma & \lambda_\sigma v_\sigma^2 
\end{bmatrix}
\end{align}
with the mass eigenvalues given by 
\begin{align}
m_{h}^2 &=\frac{1}{2}\left(\lambda_\Phi v_\Phi^2 +\lambda_\sigma v_\sigma^2 - \sqrt{(\lambda_\sigma v_\sigma^2 - \lambda_\Phi v_\Phi^2)^2+(2\lambda_{\Phi \sigma}v_\Phi v_\sigma)^2}\right)\\
m_{H}^2 &=\frac{1}{2}\left(\lambda_\Phi v_\Phi^2 +\lambda_\sigma v_\sigma^2 + \sqrt{(\lambda_\sigma v_\sigma^2 - \lambda_\Phi v_\Phi^2)^2+(2\lambda_{\Phi \sigma}v_\Phi v_\sigma)^2}\right)
\end{align}
Taking $m_{h}^2\leq m_{H}^2$ the $h$ scalar must be identified with Standard Model Higgs boson~\cite{ATLAS:2012yve,CMS:2012qbp}. 
The two mass-eigenstates $h,H$ are related with the $R_1, R_2$ through the rotation matrix $O_R$ as,  
\begin{align}
\begin{bmatrix}
h\\
H\\
\end{bmatrix}
=O_R
\begin{bmatrix}
R_1\\
R_2\\
\end{bmatrix}
=
\begin{bmatrix}
\cos\theta & \sin\theta \\
-\sin\theta & \cos\theta \\
\end{bmatrix}
\begin{bmatrix}
R_1\\
R_2\\
\end{bmatrix},
\label{mixing relation}
\end{align}
where $\theta$ is the mixing angle in the CP-even Higgs sector.  We can also express $\lambda_\Phi$, $\lambda_\sigma$, $\lambda_{\Phi\sigma}$ in terms of mixing angle $\theta$ and the scalar masses $m_{h,H}$, i.e.
\begin{align}
\lambda_\Phi =\frac{m_{h}^2\cos^2\theta+m_{H}^2\sin^2\theta}{v_\Phi^2},\,\,
\lambda_\sigma =\frac{m_{h}^2\sin^2\theta+m_{H}^2\cos^2\theta}{v_\sigma^2},\,\, \text{~~and~~ }
\lambda_{\Phi\sigma} =\frac{\sin 2\theta (m_{h}^2-m_{H}^2)}{2 v_\Phi v_\sigma}.
\label{eq:lam123}
\end{align}
The masses of the real and imaginary components of the complex field $\xi$ are given by  
\begin{align}
m_{\xi_R}^2=m_\xi^2+\frac{\lambda_{\Phi\xi}}{2}v_\Phi^2+\frac{\lambda_{\xi\sigma}+\lambda_5}{2}v_\sigma^2,~~~~~~
  m_{\xi_I}^2=m_\xi^2+\frac{\lambda_{\Phi\xi}}{2}v_\Phi^2+\frac{\lambda_{\xi\sigma}-\lambda_5}{2}v_\sigma^2
\label{eq:limit}
\end{align}
Note that the masses in Eq.~(\ref{eq:limit}) become degenerate as $\lambda_5 v_\sigma^2 \to 0$, and this limit restores lepton number conservation. 
In such dynamical breaking scenario the parameter $\mu$ characteristic of the inverse seesaw is generated through the diagram shown in Fig.~\ref{fig:scoto-loop}, and is given as
\begin{align}
\mu&=\frac{1}{16\pi^2}Y_\xi\mathcal{M}_f\left( \frac{m_{\xi_R}^2}{m_{\xi_R}^2-\mathcal{M}_{f}^2}\log \left( \frac{m_{\xi_R}^2}{\mathcal{M}_{f}^2}\right) \,-\, \frac{m_{\xi_I}^2}{m_{\xi_I}^2-\mathcal{M}_{f}^2}\log \left( \frac{m_{\xi_I}^2}{\mathcal{M}_{f}^2} \right) \right)Y_\xi^T
\end{align}

Taking $Y_\xi$ and $\mathcal{M}_f$ diagonal for simplicity, we can factorize the above loop functions as follows: \vskip -.5cm 
\begin{equation}
\mu = \left( \begin{array}{ccc}
\mu_1 & 0 & 0 \\
0 & \mu_2 & 0 \\
0 & 0 & \mu_3 
\end{array} \right) \, , \quad \mu_i=\frac{Y_{\xi}^{(i)2}M_f^{(i)}}{16\pi^2}\left( \frac{m_{\xi_R}^2}{m_{\xi_R}^2-M_f^{(i)2}}\log \left( \frac{m_{\xi_R}^2}{M_f^{(i)2}}\right) \,-\, \frac{m_{\xi_I}^2}{m_{\xi_I}^2-M_f^{(i)2}}\log \left( \frac{m_{\xi_I}^2}{M_f^{(i)2}} \right) \right),
\end{equation}
Note that $m_{\xi_R}^2=m_{\xi_I}^2$ leads to an exact cancellation between the $\xi_R$ and $\xi_I$ loops, and vanishing neutrino masses.
The scalars in $\xi$ meet the requirements for playing the role of WIMP dark-matter candidates.
Moreover, under the approximation $M_f^{(i)2},m_{\xi_R}^2$ and $M_f^{(i)2}-m_{\xi_R}^2\gg \lambda_5 v_\sigma^2$, the parameter $\mu$ is given as,
\begin{align}
\mu_i\approx \frac{1}{16\pi^2}\frac{\lambda_5 v_\sigma^2}{M_f^{(i)2}-m_{\xi_R}^2}M_f^{(i)}Y_{\xi}^{(i)2}
\end{align}
Once $\mu$ is generated this induces the following active neutrino mass matrix
\begin{eqnarray}
\mathcal{M}_\nu &= m_D M^{-1}\mu M^{-1T}m_D^T\equiv m_D \mathcal{M}_R^{-1} m_D^{T}
\label{eq:neutrino-mass}
\end{eqnarray}
where we have defined $\mathcal{M}_R^{-1}=M^{-1}\mu M^{-1T}$. From the above equations it is clear that the smallness of neutrino mass will be symmetry- as well as loop-protected.
One can express the Yukawa coupling as~\cite{Deppisch:2004fa},  
\begin{align}
Y_{\nu^c}=\frac{\sqrt{2}}{v_\Phi}U^{\dagger}_{\text{lep}}\sqrt{\widehat{\mathcal{M}}_\nu} R \sqrt{\mathcal{M}_R}
\label{eq:Ynu-1}
\end{align}
where $R$ is a $3 \times 3$ complex matrix so that $R R^T = \mathbb{I}_{3}$, where $\mathbb{I}_{3}$ is the $3 \times 3$ unit matrix.  \\
\begin{center}
   {\bf \small Collider restrictions on the dynamical dark inverse seesaw scheme} 
\end{center}

Let us now discuss the constraints on the relevant parameters of the dark inverse seesaw. Since limits from neutrino physics and precision electroweak physics are common to most of the schemes already discussed, we focus here on limits which follow from collider searches performed at the LHC. 
First note that, due to the presence of the heavy Higgs boson $H$, the coupling of the SM Higgs boson gets modified according to  
\begin{align}
h_{\text{SM}}\to \cos\theta \,h - \sin\theta\, H
\label{eq:substitution}
\end{align}
Due to the spontaneous breaking of lepton number, there is a massless Nambu-Goldstone boson, called majoron $J$. 
Since the \lnv scale is relatively low, the SM Higgs boson $h$ can have potentially large invisible decays~\cite{Joshipura:1992hp}, $h\to JJ$.
\begin{figure}[h!]
\centering
\includegraphics[height=5cm,width=0.45\textwidth]{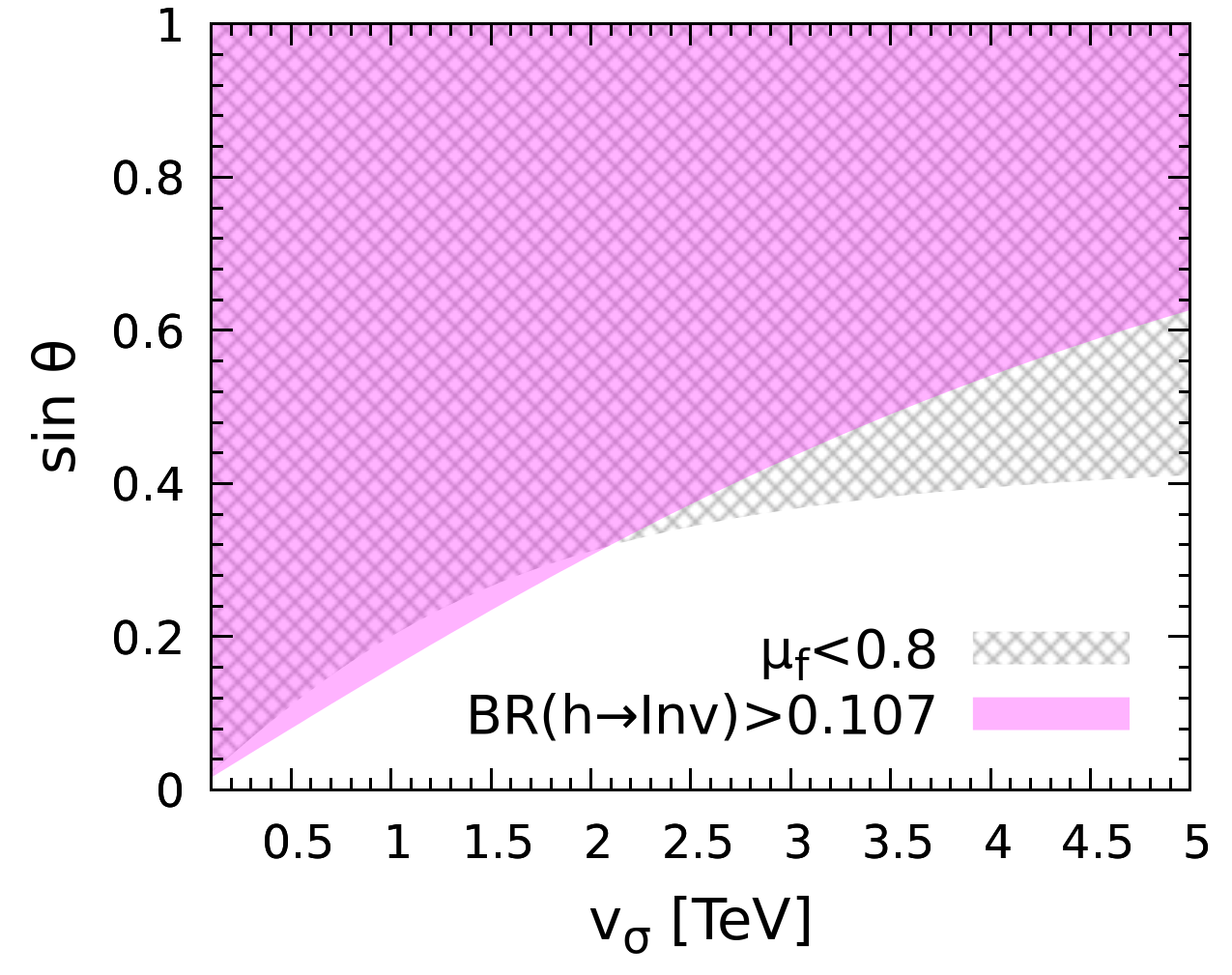}
\caption{ 
The shaded areas on $\sin\theta$ versus $v_\sigma$ are ruled out by current limit on the invisible Higgs decay, Eq.~(\ref{eq:invisible}), (magenta) and from the signal strength parameter limits in Eq.~(\ref{eq:muf})~(gray), assuming that $m_{\xi_{R,I}}>m_h/2$.}
\label{fig:limit-from-inv-mu}
\end{figure}
In the dynamical dark inverse seesaw, if either of $m_{\xi_R}$ or $m_{\xi_I}$ is smaller than half of the Higgs mass, then these two channels will also contribute to the invisible decay, so that the total invisible decay width is given as 
\begin{align}
\Gamma^{\text{inv}}(h)=\Gamma(h\to JJ)+\Gamma(h\to\xi_R\xi_R)+\Gamma(h\to\xi_I\xi_I)
\label{eq:inv-Higgs}
\end{align}
Therefore, the invisible branching ratio for $h$ becomes 
\begin{align}
\text{BR}^{\text{inv}}(h)=
\frac{\Gamma^{\text{inv}}(h)}{\cos^2\theta\Gamma^\text{SM}(h)+\Gamma^{\text{inv}}(h)}
\label{eq:inv-BR_Higgs}
\end{align}
Given the invisible Higgs decays and modified Higgs coupling, Eq.~(\ref{eq:substitution}) and Eq.~(\ref{eq:inv-Higgs}) one finds that the branching ratio to SM final states gets modified as 
\begin{align}
\text{BR}_f(h)=
\frac{\cos^2\theta \Gamma_f^\text{SM}(h)}{\cos^2\theta \Gamma^\text{SM}(h)+\Gamma^{\text{inv}}(h)}
\label{eq:BR-visible}
\end{align}
Since the coupling of $h$ and $H$ to SM fermions and gauge bosons are suppressed by $\cos\theta$ and $\sin\theta$, their production cross-sections are also modified as,  
\begin{align}
\sigma(pp\to h) &= \cos^2\theta\sigma^{\text{SM}}(pp\to h) \\
\sigma(pp\to H) &= \sin^2\theta\sigma^{\text{SM}}(pp\to H ) 
\end{align}
where $\sigma^{\text{SM}}(pp\to h)$ and $\sigma^{\text{SM}}(pp\to H)$ are the \sm  Higgs production cross-sections.  \\[-.4cm] 

The collider implications of invisibly decaying Higgs bosons have been extensively discussed~\cite{Romao:1992zx,Eboli:1994bm,DeCampos:1994fi,Romao:1992dc,deCampos:1995ten,deCampos:1996bg,Diaz:1998zg,Hirsch:2004rw,Hirsch:2005wd,Bonilla:2015uwa,Bonilla:2015jdf} with dedicated studies from the
 ATLAS/CMS  collaborations at the LHC ~\cite{CMS:2018yfx,ATLAS:2019cid}. 
Invisible Higgs decays will also be in the agenda of planned lepton collider experiments such as CEPC, FCC-ee, ILC and CLIC~\cite{CEPCStudyGroup:2018ghi,FCC:2018evy,Bambade:2019fyw,CLIC:2018fvx,Antusch:2025lpm,gao2025vision,wang2025new}. 
The upper limit given by the Particle Data Group on the Higgs decay branching ratio to invisible decay modes is~\cite{ParticleDataGroup:2024cfk}
\begin{align}
\text{BR}(h\to \text{Inv})\leq 0.107, 
\label{eq:invisible}
\end{align}
 If the dark-matter mass $m_{\xi}>m_h/2$, then only the majoron channel $\Gamma(h\to JJ)$ is open, so that the magenta region the $\sin\theta-v_\sigma$ plane shown in the Fig.~\ref{fig:limit-from-inv-mu} gets excluded by Eq.~\eqref{eq:invisible}. 
Notice that in the opposite case of $m_{\xi}<m_h/2$, the decay modes $\Gamma(h\to\xi_{R/I}\xi_{R/I})$ also contribute to the invisible Higgs decay, Eq.~(\ref{eq:inv-Higgs}). 
These depend on other quartic couplings such as $\lambda_{\Phi\xi}$, $\lambda_{\xi\sigma}$ and $\lambda_5$. 
In this case, neglecting $\sin\theta\sim 0$, the invisible Higgs decay constraint translates to an upper bound on the quartic coupling $\lambda_{\Phi\xi}$, 
\begin{align}
\lambda_{\Phi\xi}\left(1-\frac{4m_\xi^2}{m_h^2}\right)^{\frac{1}{4}}\leq 9.8\times 10^{-3}.
\end{align}

In addition we have constraints from the LHC measurements of various visible Higgs boson decay modes,  
given in terms of the so-called signal strength parameters, 
\begin{align}
\mu_f =\frac{\sigma^{\text{NP}}(pp\to h)}{\sigma^{\text{SM}}(pp\to h)} \frac{\text{BR}^{\text{NP}}(h\to f)}{\text{BR}^{\text{SM}}(h\to f)} ,
\end{align}
where $\sigma$ is the Higgs production cross-section, NP and SM stand for new physics and \sm respectively. 

The full combination of 7 TeV and 8~TeV results from both ATLAS and CMS is presented in Ref. \cite{ATLAS:2016neq}. Table \ref{tab:2} shows the updated Higgs signal strength results from ATLAS Run-2 data at 13~TeV~\cite{ATLAS:2022vkf}. Comparable updates have also been reported by CMS~\cite{CMS:2022dwd}. For a recent extensive review see~Ref.~\cite{ATLAS:2024itc}.

\renewcommand{\arraystretch}{2.0}
\begin{table}[h!] 
\begin{center}
\begin{tabular}{|c|c|c|c|c|c|c|} 
    \cline{1-7}
 \hspace{0.25cm} {\textbf{Production}} \hspace{0.25cm} &\multicolumn{6}{c|}{\textbf{Decay Channels}}\\
 \cline{2-7}
  {\textbf{Modes}} & \hspace{0.35cm} $\bm{bb}$ \hspace{0.35cm} & \hspace{0.35cm} $\bm{WW}$ \hspace{0.35cm} & \hspace{0.35cm} $\bm{\tau\tau}$ \hspace{0.35cm} & \hspace{0.35cm} $\bm{ZZ}$ \hspace{0.35cm} & \hspace{0.35cm} $\bm{\gamma\gamma}$ \hspace{0.35cm} & \hspace{0.35cm} $\bm{\mu\mu}$ \hspace{0.35cm}  \\
  \hline
    $\bm {th}$&\multirow{2}{*}{$0.348^{+0.341}_{-0.330}$}&\multirow{2}{*}{$1.639^{+0.650}_{-0.614}$}&\multirow{2}{*}{$1.373^{+0.859}_{-0.753}$}&\multirow{2}{*}{$1.683^{+1.682}_{-1.107}$}&$2.610^{+4.226}_{-3.378}$&\multirow{3}{*}{$0.537^{+0.891}_{-0.885}$}\\
    $\bm{tth}$&&&&&$0.901^{+0.328}_{-0.309}$&\\
    $\bm{ggF + bbh}$& \multirow{2}{*}{$0.980^{+0.376}_{-0.362}$} &$1.140^{+0.130}_{-0.126}$&$0.898^{+0.291}_{-0.256}$& $0.952^{+0.110}_{-0.105}$&$1.038^{+0.103}_{-0.099}$&\\ $\bm{VBF}$&&$1.128^{+0.195}_{-0.184}$&$0.998^{+0.207}_{-0.183}$&$1.331^{+0.519}_{-0.432}$&$1.358^{+0.297}_{-0.265}$&\multirow{3}{*}{$2.314^{+1.330}_{-1.258}$}\\ $\bm{WH}$&$1.058^{+0.284}_{-0.264}$&$2.264^{+1.212}_{-1.022}$&\multirow{2}{*}{$1.004^{+0.624}_{-0.595}$}&\multirow{2}{*}{$1.500^{+1.166}_{-0.939}$}&$1.528^{+0.560}_{-0.507}$&\\$\bm{ZH}$&$1.001^{+0.246}_{0.226}$&$2.861^{+1.836}_{-1.335}$&&&$-0.221^{+0.606}_{-0.541}$&\\
  \hline
\end{tabular}
\end{center}
\caption{ 
ATLAS 13 TeV 
 visible Higgs decay measurements normalised to SM, taken from Ref.~\cite{ATLAS:2022vkf}.}
\label{tab:2}
\end{table}

One sees that, although compatible at $1\sigma$, current limits still have quite large errors. For simplicity, we adopt the conservative range, 
\begin{align}
0.8\leq \mu_f\leq 1.2.
\label{eq:muf}
\end{align}
The gray shaded region in Fig.~\ref{fig:limit-from-inv-mu} is ruled out by Eq.~\eqref{eq:muf} assuming that  $h\to JJ$ is the only invisible mode open.
As seen from Fig.~\ref{fig:limit-from-inv-mu}, for low $v_\sigma$ values up to around 1~TeV, Eq.~\eqref{eq:invisible} and Eq.~\eqref{eq:muf} give similar limits on $\sin\theta$.  
However, for $v_\sigma > 1$~TeV the limit from Eq.~\eqref{eq:invisible} gets relaxed since, the larger the $v_\sigma$, the smaller the invisible decay mode $h\to JJ$.
As a result, for large $v_\sigma$ values the Higgs invisible decay gives a weaker exclusion limit on $\sin\theta$ than that coming from $\mu_f$. 
Note that for $m_{\xi}\leq m_h/2$, the exclusion region depends on the values of quartic couplings such as $\lambda_{\Phi\xi}$, $\lambda_{\xi\sigma}$.
Additional constraints will come from the direct search of the heavy CP-even Higgs boson $H$. 
For example, direct Higgs production $pp\to H$ with successive decay to \sm particles
e.g., $WW$, $ZZ$ and subsequent decays $WW\to 2\ell 2\nu$ and $ZZ\to 4\ell$~\cite{CMS:2017dua, ATLAS:2017uhp}.
Alternatively, it can also decay to a pair of SM Higgs bosons $h$ if kinematically allowed, as discussed in Ref.~\cite{Mandal:2021acg}. 
In short, we find that the signal strength constraints are the strongest ones, when $v_\sigma\gg v_{H}$.\\[-.5cm]

\subsection{Dark linear seesaw } 
\label{sec:linear-dark-seesaw}
As a variant of the previous approach one can also envisage that the linear seesaw mechanism~\cite{Akhmedov:1995ip,Akhmedov:1995vm,Malinsky:2005bi} can be  seeded by a dark sector. This was, in fact, developed explicitly in Refs.~\cite{Batra:2023bqj,CarcamoHernandez:2023atk}.
Here we sketch the main features of the model proposed in~\cite{CarcamoHernandez:2023atk} which can be regarded as a minimal extension of the inert doublet model where the linear seesaw mechanism produces the tiny neutrino masses at the one-loop level, seeded by the dark sector. 
The SM leptons include the neutral leptons characteristic of low-scale seesaw schemes, 
$N _{i}^{c}$ and $S_{i}$. We take three copies of all fermions, i.e. $i=1,2,3$, i.e. the sequential (3,3,3) setup.
The dark sector contains three copies of SM singlet two-component Majorana fermions $F_{i}$, plus a SM doublet dark scalar $\eta $, and a dark gauge singlet $\xi$. 
The \sm $\mathrm{SU(3)_c \otimes SU(2)_L \otimes U(1)_Y}$ gauge symmetry is supplemented by the inclusion of the global U(1) lepton number symmetry, which spontaneously breaks to a preserved $\mathbb{Z}_{2}$ symmetry.
 This remnant matter-parity symmetry $$(-1)^{3 \mathcal{B} + \mathcal{L} + 2s}$$ involves the baryon,
  lepton and spin quantum numbers respectively, ensuring the stability of the LSP (dark-matter candidate), as well as the radiative nature of neutrino mass generation through the linear seesaw mechanism. 

The scalar sector also requires Higgs bosons to drive spontaneous breaking of the gauge symmetry and the partial breaking of the global symmetry. 
Besides the SM doublet $\Phi$, we include a {complex scalar isotriplet $\Xi$} whose vacuum expectation value (VEV)
is restricted by precision electroweak measurements, i.e. the $\rho$ parameter~\cite{ParticleDataGroup:2024cfk}. 
 The leptons and scalars of the model and their transformation properties under the $\mathrm{SU(3)_c \otimes SU(2)_L \otimes U(1)_Y}$ gauge symmetry, the global lepton number symmetry, and the remnant $\mathbb{Z}_2$ symmetry are given in Table.~\ref{tab:MatterModel}. \\[-0.5cm]
\renewcommand{\arraystretch}{1.5}
\begin{table}[h]  
\includegraphics[scale=0.3]{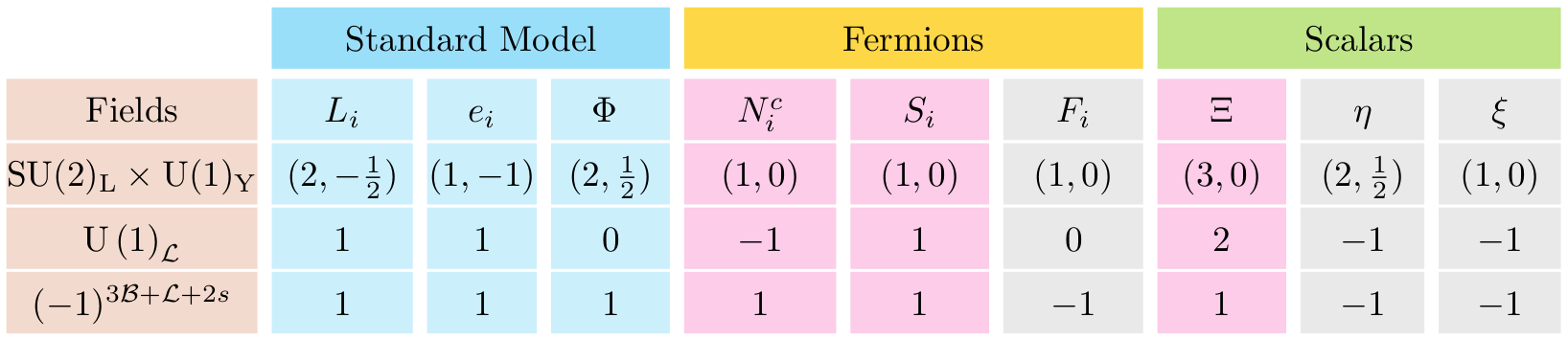}
 \caption{ 
 Dark linear seesaw model quantum numbers~\cite{CarcamoHernandez:2023atk}. Dark states have opposite $\mathbb{Z}_2$ matter-parity as SM states, the triplet scalar $\Xi$, and the seesaw mediators $N^c_i$,  $S_i$.}
\label{tab:MatterModel}
\end{table}
\renewcommand{\arraystretch}{1.2}

 \par
\begin{figure}[tbh]
\centering
\includegraphics[scale=0.25]{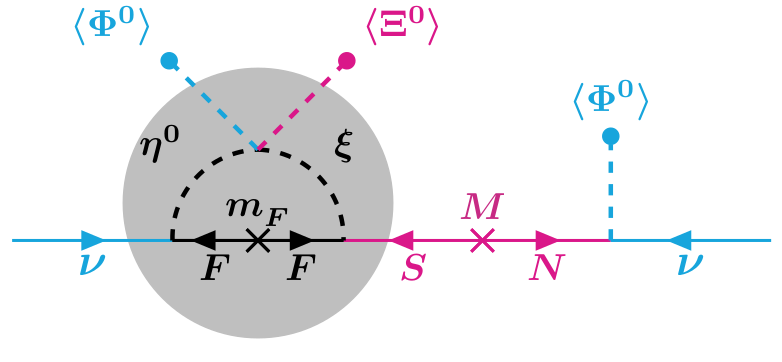}
\caption{ 
Feynman diagram for neutrino mass generation in dark linear seesaw mechanism (plus symmetrization).}
\label{fig:Neutrinoloopdiagram}
\end{figure}

 Notice that the leptons have the conventional lepton number assignment characteristic of low-scale seesaw schemes.
The dark scalars $\eta$ and $\xi$ and the dark fermions $F_{i}$ induce neutrino masses through the linear-seesaw, as seen in Fig~\ref{fig:Neutrinoloopdiagram}. 
Except for the \sm scalar doublet $\Phi$, all scalars carry non-zero lepton number.
The relevant neutrino Yukawa couplings and mass terms invariant under these symmetries are, 
\begin{align}
  \label{eq:Yuk}
-\mathcal{L}_{Y}^{\left( \nu \right)
}=& \sum_{i,j=1}^{3}Y_{ij}^{\left( \Phi \right)
}L_{i}^{T}CN_{j}^{c}\Phi +\sum_{i,j=1}^{3}Y_{ij}^{\left( \eta
\right) }L_{i}^{T}CF_{j}\eta +\sum_{i,j=1}^{3}Y_{ij}^{\left( \xi
\right) }S_{i}^T C F_{j}\xi \notag \\& +\sum_{i=1}^{3}\left( m_{F}\right)
_{i}F_{i}^T C F_{i}+\sum_{i,j=1}^{3}M_{ij}{N_{i}^{c}}^TCS_{j}\,{ + \sum_{i,j=1}^{3} {Y^\prime}^{(\xi)}_{ij}F_i^T C {N^c}_j \xi^*} +  H.c.
\end{align}
The scalar potential contains,   
\begin{align}
  \label{eq:V1}
\mathcal{V}_{\left( s\right) }=& -\mu _{\Phi }^{2}(\Phi ^{\dagger }\Phi
\,)-\mu _{\Xi }^{2}Tr(\Xi ^{\dagger }\Xi )  + \mu _{\eta }^{2}(\eta ^{\dagger }\eta )+\mu _{\xi }^{2}(\xi ^{\ast }\xi
) ~{~+A_{\Phi }(\Phi ^{\dagger }\Xi \Phi + \Phi ^{\dagger }\Xi^\dagger \Phi)} \notag  \\
& +\lambda _{1}(\Phi ^{\dagger }\Phi \,)^{2}+\lambda _{2}(\eta ^{\dagger
}\eta )^{2}+\lambda _{3}(\xi ^{\ast }\xi )^{2}+\lambda _{4}\left[ Tr(\Xi
^{\dagger }\Xi )\right] ^{2}+\lambda _{5}Tr\left[ (\Xi ^{\dagger }\Xi )^{2}%
\right]  \notag \\
& +\lambda _{6}(\Phi ^{\dagger }\Phi \,)(\eta ^{\dagger }\eta )+\lambda
_{7}(\Phi ^{\dagger }\eta )(\eta ^{\dagger }\Phi \,)+\lambda _{8}(\Phi
^{\dagger }\Phi \,)Tr(\Xi ^{\dagger }\Xi )+\lambda _{9}\Phi ^{\dagger }\Xi
\Xi ^{\dagger }\Phi  \notag \\
& +\lambda _{10}(\Phi ^{\dagger }\Phi \,)(\xi ^{\ast }\xi )+\lambda
_{11}(\eta ^{\dagger }\eta )Tr(\Xi ^{\dagger }\Xi )+\lambda _{12}\eta
^{\dagger }\Xi \Xi ^{\dagger }\eta +\lambda _{13}(\eta ^{\dagger }\eta )(\xi
^{\ast }\xi )  \notag \\
& +\lambda _{14}(\xi ^{\ast }\xi )Tr(\Xi ^{\dagger }\Xi )+ {\lambda
_{15}\left(\eta^{\dag }\Xi^{\ast }\Phi\xi ^{\ast }+h.c\right)} .
\end{align}

The U(1) lepton-number symmetry is broken by the VEV of the neutral part of $\Xi$.
The trilinear term $A_\Phi$ in Eq.~(\ref{eq:V1}) also breaks the global lepton number symmetry of Eq.~(\ref{eq:Yuk}), explicitly but softly. 
Dark-matter stability is ensured by the remnant unbroken $\mathcal{Z}_{2}$ symmetry that remains after the breaking of the U(1) symmetry. 
To ensure this we require that the $\mathbb{Z}_{2}$-odd scalars $\eta $ and $\xi$ not to acquire vacuum expectation values. 

The scalar fields $\Phi $, $\Xi $, $\eta $ and $\xi $ can be written as follows, 
\begin{eqnarray*}
\Phi =%
\begin{pmatrix}
\phi ^{+} \\ 
\frac{v_{\Phi }+\phi _{R}^{0}+i\phi _{I}^{0}}{\sqrt{2}}%
\end{pmatrix}%
,\qquad\eta =%
\begin{pmatrix}
\eta ^{+} \\ 
\frac{\eta _{R}^{0}+i\eta _{I}^{0}}{\sqrt{2}}%
\end{pmatrix}%
, \qquad \Xi =  {\left( 
\begin{array}{cc}
\frac{v_{\Xi } + \Xi ^{0}_R +  i\Xi ^{0}_I}{\sqrt{2}} & \Xi_1^{+} \\ 
\Xi_2^{-} & -\frac{v_{\Xi } + \Xi ^{0}_R +  i\Xi ^{0}_I}{\sqrt{2}}
\end{array}%
\right) } ,\qquad\xi =\frac{\xi _{R}+i\xi _{I}}{\sqrt{2}}.
\end{eqnarray*}

{There are two physical charged Higgs scalars $\Xi_1^{\pm}$ and $\Xi_2^{\pm}$ with mass-squared given as, 
\begin{equation}
m_{\Xi_{1,2}^{\pm       }}^2 = \frac{\sqrt{2} A_\Phi (v_\Phi^2 + 4 v_\Xi^2) \mp \sqrt{32 A_\Phi^2 v_\Xi^4 + v_\Phi^2 v_\Xi^2 (v_\Phi^2 + 8 v_\Xi^2)\lambda_9^2}}{4 v_\Xi}, 
 \end{equation} }
{Notice that the presence of the cubic term $A_\Phi$ allows the two charged components of the triplet scalar to have an adequate mass-squared term.}
The charged dark scalar $\eta^{\pm}$ has a mass-squared given as, 
\begin{equation}
m_{\eta^\pm}^2 =  \frac{1}{2} v_\Xi^2 (2 \lambda_{11} +\lambda_{12}) + 
\frac{1}{2} v_\Phi^2 \lambda_6  +
\mu_\eta^2.
\end{equation}

Electroweak symmetry-breaking is driven mainly by the VEV of $\Phi $.
The resulting mass squared matrices for the CP-even neutral Higgs scalars are given as, 
\begin{equation} 
M^2_{\phi_{R}^{0}~\Xi_R^{0}} = \left( 
\begin{array}{cc}
2 \lambda_1 v_\Phi^2 &  v_\Phi(-\sqrt{2}A_\Phi +  v_\Xi (2
\lambda_8 + \lambda_9)) \\ 
 v_\Phi(-\sqrt{2}A_\Phi +  v_\Xi (2
\lambda_8 + \lambda_9)) & 
\frac{A_\Phi v_\Phi^2}{\sqrt{2}v_\Xi} + 4 v_\Xi^2 (2 \lambda_4 + \lambda_5)%
\end{array}%
\right) ,
\end{equation}
while the corresponding neutral dark scalar mass squared matrices are given as, 
\begin{equation}
M^2_{\eta_{R}^{0}~\xi_{R}} = \left( 
\begin{array}{cc} \frac{v_\Xi^2}{2} ( 2 \lambda_{11} +
\lambda_{12}) + \frac{v_\Phi^2}{2}(\lambda_6 + \lambda_7) + \mu_\eta^2 & -%
\frac{1}{2}\lambda_{15} v_\Phi v_\Xi \\ 
-\frac{1}{2}\lambda_{15} v_\Phi v_\Xi & \frac{1}{2}\lambda_{10} v_\Phi^2  +
\lambda_{14} v_\Xi^2  + \mu_\xi^2 
\end{array}
\right) ,  \label{thetas}
\end{equation}
\begin{equation}
M^2_{\eta_{I}^{0}~\xi_{I}} = \left( 
\begin{array}{cc} \frac{v_\Xi^2}{2} ( 2 \lambda_{11} +
\lambda_{12}) + \frac{v_\Phi^2}{2}(\lambda_6 + \lambda_7) + \mu_\eta^2 & 
 \frac{1}{2}\lambda_{15} v_\Phi v_\Xi \\ 
 \frac{1}{2}\lambda_{15} v_\Phi v_\Xi & \frac{1}{2}\lambda_{10} v_\Phi^2  +
\lambda_{14} v_\Xi^2  + \mu_\xi^2 
\end{array}
\right) .  \label{thetaa}
\end{equation}
{There is also a CP-odd scalar coming from the imaginary part of the neutral component of $\Xi$, whose mass is, 
\begin{equation} 
m_{{\Xi^0}_I}^2 = \frac{A_\Phi v_\Phi^2}{\sqrt{2}v_\Xi}.
\end{equation}}
One sees that the physical neutral  scalar spectrum includes four CP-even scalars:
two neutral Higgs $H_1$ and $H_2$ arising from the mixing of $\Xi^{0}_R$ and $\phi_{R}^{0}$, and containing the 125~GeV SM Higgs boson~\cite{ATLAS:2012yve,CMS:2012qbp},
plus two dark neutral scalars $D_1$ and $D_2$ arising from the mixing of $\eta_{R}^{0}$ and $\xi_{R}$.
One has in addition two dark neutral CP-odd scalars, $D_{A_1}$ and $D_{A_2}$ arising from the mixing of $\eta_{I}^{0}$ and $\xi_{I}$ {and another CP-odd scalar associated with ${\Xi^0}_I$.}
The doublet-singlet mixing angles in these matrices are expected to be naturally small, thanks to the limit $v_\Xi \lsim 4$~GeV on the $\rho$ parameter. 

The presence of the cubic lepton number soft-breaking term $A_\Phi$ in Eq.~(\ref{eq:V1}) can make all physical scalars massive.
This avoids the existence of a Majoron~\cite{Schechter:1981cv,Chikashige:1980ui}, a physical Nambu-Goldstone boson associated to spontaneous lepton number
violation, which gets mass from the explicit $A_\Phi$-induced lepton-number-breaking term.
An alternative full-fledged Majoron scheme can also be implemented, along the lines of Ref.~\cite{Fontes:2019uld}.
However we do not pursue such extension here, as it is not essential. 
Radiative neutrino mass generation proceeds via the linear seesaw mechanism. 
Indeed, the lepton Yukawa interactions yield the following neutrino mass terms, 
\begin{equation}
-\mathcal{L}_{mass}^{\left( \nu \right) }=\frac{1}{2}\left( 
\begin{array}{ccc}
\nu^T & {N^{c}}^T & S^T%
\end{array}%
\right) M_{\nu }C\left( 
\begin{array}{c}
\nu \\ 
N^{c} \\ 
S%
\end{array}%
\right) +H.c., \quad \text{with} \quad M_{\nu }=\left( 
\begin{array}{ccc}
0_{3\times 3} & m_{D} & \varepsilon \\ 
m_{D}^{T} & 0_{3\times 3} & M \\ 
\varepsilon ^{T} & M^T & 0_{3\times 3}%
\end{array}%
\right) . 
\label{Lnu}
\end{equation}
Here the sub-matrix $M$ is a bare mass, and $m_{D}$ is generated at tree-level after electroweak symmetry-breaking,
\begin{equation}
  \label{eq:MD}
  m_{D} = Y^{\left( \Phi \right) }\frac{v_{\Phi }}{\sqrt{2}}.
\end{equation}
In contrast, the small entry $\varepsilon$ arises from calculable radiative corrections, mediated by the one-loop level exchange of the dark fermions and scalars. 
The one-loop level Feynman diagram in Fig.~\ref{fig:Neutrinoloopdiagram} yields the submatrix $\varepsilon$ as 
\begin{equation}
\varepsilon _{ij} = \sum_{k=1}^{3}\frac{Y_{ik}^{\left( \eta \right)
}Y_{jk}^{\left( \xi \right) }M_{F_k}}{16\pi ^{2}}\left\{ \left[ f\left(
m_{D_{1}}^{2},m_{F_{k}}^{2}\right) - f\left(
m_{D_{2}}^{2},m_{F_{k}}^{2}\right) \right] \sin 2\theta_D - \left[ f\left(
m_{D_{A_1}}^2,m_{F_{k}}^{2}\right) -f\left(
m_{D_{A_2}}^2,m_{F_{k}}^{2}\right) \right] \sin 2\theta_{D_A}\right\} ,
\end{equation}
where $f\left( m_{1},m_{2}\right) $ is the function defined as, 
\begin{equation}
f\left( m_{1},m_{2}\right) =\frac{m_{1}^{2}}{m_{1}^{2}-m_{2}^{2}}\ln \left( 
\frac{m_{1}^{2}}{m_{2}^{2}}\right) .
\end{equation}%
Here $m_{D_1}$ and $m_{D_2}$ are the masses of the physical CP-even dark scalars, whereas $m_{D_{A_1}}$ and $m_{D_{A_2}}$ are those that of the dark pseudoscalars.
Their mixing matrices are defined as, 
\begin{equation}
\left( 
\begin{array}{c}
D_{1} \\ 
D_{2}%
\end{array}%
\right) =\left( 
\begin{array}{cc}
\cos \theta _{D} & \sin \theta _{D} \\ 
-\sin \theta _{D} & \cos \theta _{D}%
\end{array}%
\right) \left( 
\begin{array}{c}
\eta _{R} \\ 
\xi _{R}%
\end{array}%
\right) ,\hspace{1cm}\left( 
\begin{array}{c}
D_{A_1} \\ 
D_{A_2}%
\end{array}%
\right) =\left( 
\begin{array}{cc}
\cos \theta _{D_A} & \sin \theta _{D_A} \\ 
-\sin \theta _{D_A} & \cos \theta _{D_A}%
\end{array}%
\right) \left( 
\begin{array}{c}
\eta _{I} \\ 
\xi _{I}%
\end{array}%
\right) .
\end{equation}
where the small doublet-singlet mixing angles $\theta _{D}$ and $\theta _{D_A}$ come from diagonalizing Eqs. (\ref{thetas}) and (\ref{thetaa}), respectively, {from which one sees $\theta_{D_A}= - \theta_A$.} 

The light active neutrino masses arise from the linear seesaw
mechanism~\cite{Akhmedov:1995ip,Akhmedov:1995vm,Malinsky:2005bi}, so that the resulting active-neutrino mass matrix has the form,
\begin{equation}
m_{\rm light}=-\left[ m_{D}(M^T)^{-1}\varepsilon ^{T}+\varepsilon M^{-1}m_{D}^{T}%
\right] .
\end{equation}
in terms of the submatrix $\varepsilon$. 
One sees that spontaneous lepton number violation through $v_\Xi$ provides a radiative seed for light neutrino mass generation that proceeds \textit{a la seesaw}.
The smallness of the light neutrino masses is ascribed to the smallness of loop-suppressed $\varepsilon$ as well as the $m_{D}/M$ ratio,
not necessarily negligible. 
The small neutrino masses are symmetry-protected, making the model natural in t'Hooft's sense. 
Finally, as in all low-scale seesaw models, the seesaw mediator sector consists of three pairs of quasi-Dirac~\cite{Anamiati:2016uxp,Arbelaez:2021chf} heavy neutrinos. 

\begin{center}
   {\bf \small Fermionic dark-matter } 
\end{center}

In this section we discuss the implications of our dark linear seesaw model for dark matter.
Due to the remnant $\mathbb{Z}_{2}$ symmetry which survives the spontaneous breaking of the global lepton number symmetry, our model will have a stable dark-matter candidate, the LSP. 
 As a warm up, we start by considering a simple scenario in which the LSP is fermionic, i.e. the lightest of the heavy Majorana fermions $F_{i}$ ($i=1,2,3$). 
It can annihilate into a pair of active neutrinos via the $t$-channel exchange of the
CP-even and CP-odd parts of the neutral component of the dark scalar doublet $\eta$, as shown in Fig.~\ref{fig:relic-pure-singlet}. 
In this case, the thermally-averaged annihilation cross section is given by~\cite{Bernal:2017xat}, 
\begin{equation}
  \label{eq:relic-fermion}
  \braket{\sigma v} \simeq \frac{9\left(Y^{\left(\eta\right)}_{11}\right)^{4}}{32\pi }\frac{m_{F_1}^{2}\left( 2m_{F_1}^{2}+m_{D_1}^{2}+m_{D _{A_1}}^{2}\right)^2 }{\left( m_{F_1}^{2}+m_{D
_{1}}^{2}\right) ^{2}\left( m_{F_1}^{2}+m_{D_{A_1}}^{2}\right) ^{2}},
\end{equation}
where we have assumed $F_1$ to be the lightest of the $F_i$. 
Here $Y^{\left(\eta\right)}_{11}$ is the Yukawa coupling with the dark scalar doublet $\eta $. 
From the previous relation, we find the following estimate for the DM relic abundance \cite{ParticleDataGroup:2024cfk},  
\begin{equation} 
\frac{\Omega _{DM}h^{2}}{0.12}=\frac{0.1pb}{0.12\braket{\sigma v}}=\frac{0.1pb}{0.12%
}\left[ \frac{9\left(Y^{\left(\eta\right)}_{11}\right)^{4}}{32\pi }\frac{m_{F_1}^{2}\left( 2m_{F_1}^{2}+m_{D_1}^{2}+m_{D _{A_1}}^{2}\right)^2 }{\left( m_{F_1}^{2}+m_{D
_{1}}^{2}\right) ^{2}\left( m_{F_1}^{2}+m_{D_{A_1}}^{2}\right) ^{2}}\right]
^{-1},
\end{equation}
which in turn can reproduce the observed DM relic abundance of~\cite{Planck:2018vyg}, see Table.~ \ref{table:cosmology}. 

\begin{figure}[tbh]
\centering
\includegraphics[width=0.5\textwidth,height=5cm]{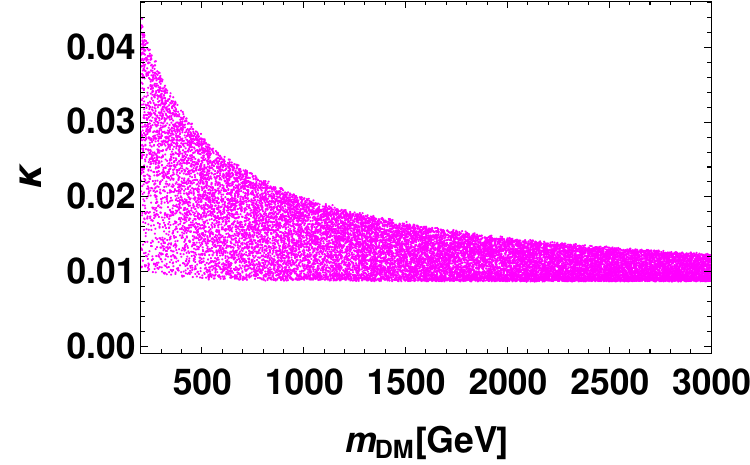}
\caption{
Allowed region in the $m_{DM} - \kappa $ plane that reproduces the correct fermionic dark matter relic density.
  Here $m_{DM}$ is the LSP mass $m_{F_1}$, and $\kappa=Y^{\left(\eta\right)}_{11}$ is the relevant Yukawa coupling. } 
\label{DMplotfermion}
\end{figure}

In Fig.~\ref{DMplotfermion} we present the allowed region in the $m_{F_1} - Y_{11}^{(\eta )} $ plane that reproduces the measured values of the dark-matter relic abundance.  
To generate this figure, the masses of the dark scalars $m_{D_1}$ and $m_{D_{A_1}}$ are both varied in the range 0.2-5 TeV.
One sees that the allowed range of the Yukawa couplings is broader for lower dark-matter masses. \\[-.2cm]

Concerning direct detection we note that, as already seen in Sec.~\ref{subsec:dd-minimal-scoto}, since the fermionic dark matter $F_1$ couples only to the dark scalars, this scenario can not be tested by the Higgs portal in leading order at nuclear recoil measurements.
However, the situation changes if the LSP is one of the neutral dark scalars. Scalar dark-matter candidates in various dark neutrino mass scenarios will be discussed next.


\section{Scalar scotogenic dark-matter}
\label{sec:dark-matter-scotogenic}

Scenarios with a scalar WIMP dark-matter candidate have been well studied within the framework of many extensions of the Standard Model of particle physics, such as
Higgs portal models~\cite{McDonald:1993ex,Lebedev:2021xey}.
 They can emerge naturally within the inert Higgs doublet~\cite{Barbieri:2006dq,Branco:2011iw,Belyaev:2016lok,LopezHonorez:2006gr} and also the scotogenic approach~\cite{Ma:2006km,Tao:1996vb,
 Hirsch:2013ola,
Merle:2015gea,Merle:2016scw,Merle:2016scw,Rocha-Moran:2016enp,Diaz:2016udz, Kubo:2006yx,Ma:2008cu,Lozano:2025tst,Karan:2023adm,Leite:2019grf,VanDong:2023xmf,
Bonilla:2023egs,Chao:2012sz,Toma:2013zsa,Fraser:2014yha,Merle:2015ica,Lindner:2016kqk,Ma:2016mwh,Borah:2018rca,Hugle:2018qbw,Singh:2023eye,Babu:2019mfe,Nomura:2019lnr,Escribano:2020iqq,Alvarado:2021fbw,Kumar:2024zfb,Garnica:2024wur,Leite:2023gzl,Lavoura:2012cv,CentellesChulia:2016rms,Farzan:2012sa,CentellesChulia:2016rms,Bonilla:2016diq,Wang:2017mcy,Bonilla:2018ynb,CentellesChulia:2019xky,Li:2022chc,Borah:2024gql,Restrepo:2019ilz,Avila:2019hhv,Kang:2019sab,CarcamoHernandez:2020ehn,Hernandez:2021zje,CentellesChulia:2024iom,Jobu:2025tto,Darricau:2025vcs} when the \textit{lightest scotogenic particle} or LSP is a scalar~\footnote{
Scalar dark matter can also emerge within extended Higgs
models \cite{Hernandez:2021iss}.}. 

Scotogenic scalar dark matter has a clear physics interpretation, namely that of mediator of neutrino mass generation.
  In this chapter we discuss some aspects of scalar dark matter in the simplest and revamped scotogenic models, as well as in the low-scale dark seesaw schemes discussed in Sections~\ref{sec:dark-inverse-seesaw} and \ref{sec:linear-dark-seesaw}.  

\subsection{ Simplest scotogenic}
\label{sec:darkmatterscoto}

  We now turn to scalar dark-matter phenomenology within the scotogenic scenario~\cite{Klasen:2013jpa,Avila:2021mwg}. 
  We follow the study given in~\cite{Avila:2021mwg}, performing the numerical analysis of the model parameters within the ranges specified in Table~\ref{tab:rangesscoto},
  using a Montecarlo analysis performed with Python. 
\begin{table}[!htb]
\centering 
\begin{tabular}{|c|c|}
\hline
{\bf Parameter} & {\bf Range}  \\ 
\hline  
\hline 
$m_{\eta}^2$ & [$10,\,  1\times 10^3 $]   $\rm (GeV^2)$     \\                          
$M_{F_1} $  & [$50,\, 5\times 10^3$]  \rm (GeV)    \\
$M_{F_2}  $ & [$5\times 10^3,\, 2\times10^6$]     \rm (GeV)     \\
$M_{F_3}  $ & [$5\times 10^3,\,3.5\times10^6$]     \rm (GeV)     \\
$|\lambda_{1}|$ &  $ [10^{-8} ,1$] \\
$|\lambda_{\eta}|$ &  $ [10^{-8} ,1$] \\
$|\lambda_{i}|$, $i=3...5$ & $\pm$[$10^{-8} ,1$] \\
\hline
\end{tabular}
\caption{ Parameter ranges used for the scalar dark-matter analysis in the simplest scotogenic model~\cite{Avila:2021mwg}.}
\label{tab:rangesscoto}
\end{table} 

\begin{center}
   {\bf \small  Scalar dark-matter relic density in simplest scotogenic} 
\end{center}

\begin{figure}[h]
\centering
\includegraphics[trim={0cm 0cm 3cm 0cm}, clip, height=5.8cm,width=0.44\textwidth]{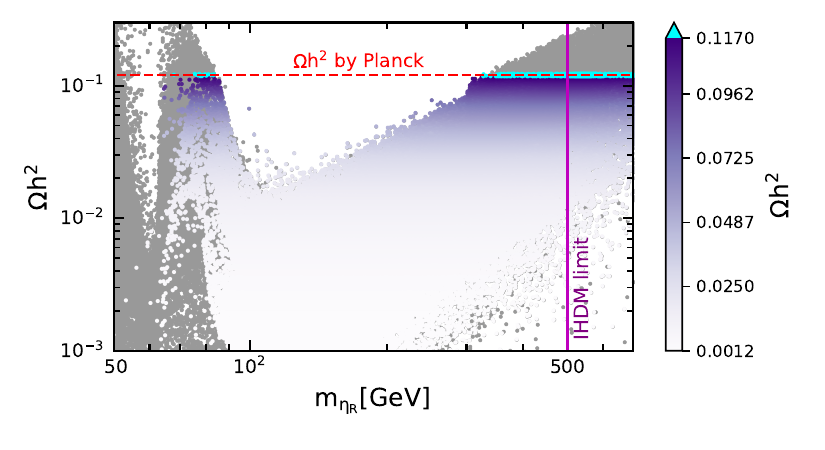}
\includegraphics[height=5.8cm,width=0.55\textwidth]{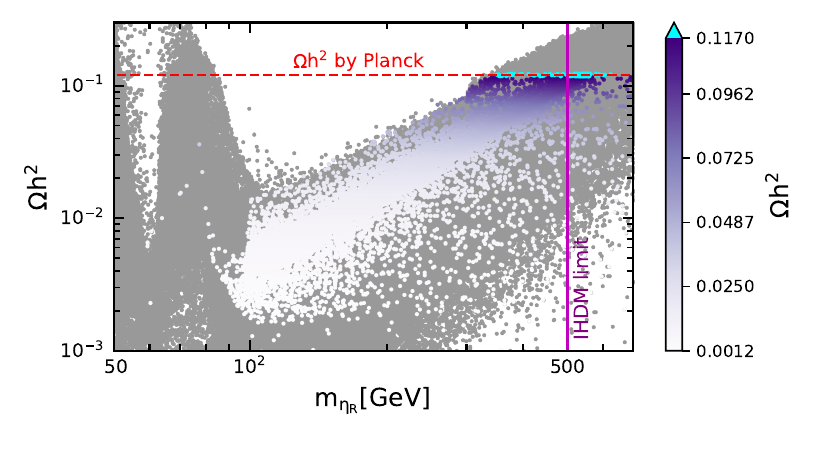}
\caption{
Relic density as a function of the scalar DM mass $m_{\eta_R}$ in the original scotogenic scheme~\cite{Avila:2021mwg}. The left panel does not include direct detection limits, while the right panel incorporates updated bounds from { LZ  (2025)~\cite{LZ:2024zvo}} as well as the neutrino floor restriction~\cite{Billard:2021uyg}. The thin horizontal band represents the $3\sigma$ C.L. Planck measurement for a freeze-out cold dark-matter scenario~\cite{Planck:2018vyg}. Masses below $m_{\eta_R}=500$ GeV (vertical line) are allowed by all relevant constraints in Sec.~\ref{sec:fdm-Constraints}, but would lead to overabundant dark-matter in the IHDM scenario.}
\label{fig:relicscoto}
\end{figure}

 Fig.~\ref{fig:relicscoto} shows the relic density versus the dark-matter mass $m_{\eta_R}$ in the original scotogenic scheme of Sec.~\ref{sec:simpl-scot-setup}. 
 The red horizontal line denotes the measurement by Planck~\cite{Planck:2018vyg}. 
  The vertical magenta line located at $m_{\eta_R} = 500$ GeV marks the \textit{desert region} in the inert Higgs doublet case. Indeed, as discussed in~\cite{Borah:2017dfn,Barman:2021ifu}, the IHDM is inconsistent with the total relic abundance below this line. 
  In contrast, the cyan points located around the Planck band in the range $m_{\eta_R}= 300 - 900$ GeV can account for the total scalar dark-matter abundance present in the Universe within the minimal scotogenic model. 
  A comparison of the left and right panels reveals that low-mass dark matter scenarios are excluded by current direct detection limits~\cite{LZ:2024zvo}~\footnote{
   Note that an 80~GeV dark matter mass is inconsistent with the required relic density. However, wider dark matter ranges may be allowed in more complicated multi-component dark matter scenarios, see for example, Ref.~\cite{Frank:2025jjt}.}.  
   Points indicated by a different intensity of purple color correspond to the case where $\eta_R$ is a subdominant DM candidate, requiring another contribution in order to account for the total observed relic abundance. 
  The color intensity decreases as the contribution to the relic abundance gets 
  lower, as indicated in the right scale. Grey points are excluded by the same constraints imposed in~\cite{Avila:2021mwg}.

\begin{figure}[h]
\centering
\includegraphics[height=5.5cm,scale=0.55]{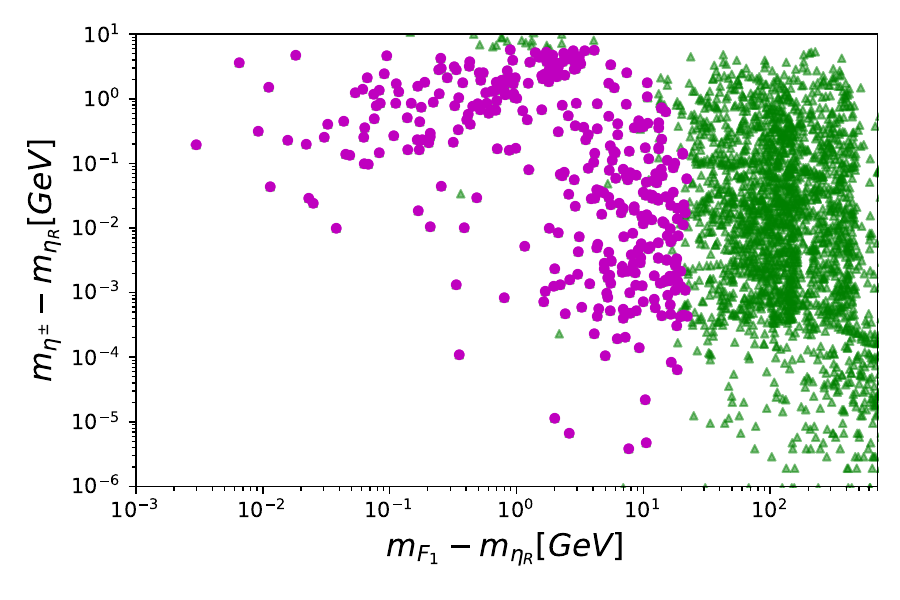}
\caption{ 
Mass differences in the dark sector, $\Delta m_{\eta^\pm}$ versus $\Delta m_{F_1}$~\cite{Avila:2021mwg}. Magenta points lead to $m_{\eta_R} < 500$ GeV, while green points correspond to large scalar DM masses, $m_{\eta_R} > 500$ GeV. All points fit the total relic abundance.} 
\label{fig:relicmassdiff}
\end{figure}

The reason why IHDM and scotogenic model results differ has to do with the role of co-annihilations.
When the mass splittings between the dark particles is small enough, co-annihilation of $\eta_R$ with other dark states may appear in some regions of parameters. 
Indeed, in contrast to IHDM results~\cite{Borah:2017dfn, Sarma:2020msa}, for the scotogenic model one finds that for $m_{\eta_R}$ in the $300-500$ GeV range one can get the correct relic density. This results from mainly from the possibility of small mass splitting between $\eta_R$ and the dark fermion, as seen in Fig.~\ref{fig:relicmassdiff}.
Indeed, the viability of dark-matter masses below 500~GeV in the scotogenic scenario follows from the small mass splitting between $\eta_{R,I}$ and $F_1$, highlighting the role of fermion-scalar co-annihilation.  Green triangles correspond to $\eta_R$ heavier than $500$ GeV, while magenta dots denote $m_{\eta_R} < 500$ GeV. 
Such compressed spectra in the scotogenic framework  translates into the possibility of long-lived particles (LLPs) at colliders~\cite{Blondel:2022qqo}, since small mass splittings can result in particles with macroscopic decay lengths.

\begin{center}
   {\bf \small Scalar dark-matter detection in simplest scotogenic} 
\end{center}

\par
\begin{figure}[h]
\centering
\scalebox{1}{
\includegraphics[trim={5mm 0 5mm 0},clip,scale=0.72]{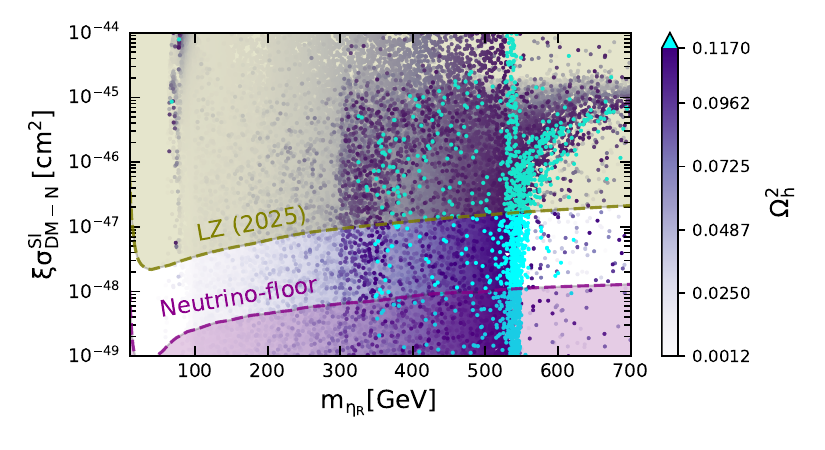} \hfill
\includegraphics[trim={3mm 0 0 0},clip,scale=0.70]{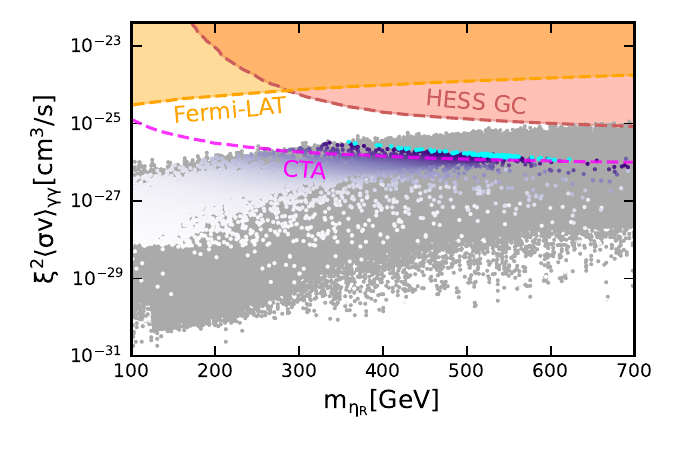}}
\caption{ 
Left: DM-nucleon spin-independent elastic scattering cross-section versus DM mass~\cite{Avila:2021mwg} in the original scotogenic model. The {LZ (2025) upper limit}~\cite{LZ:2024zvo} is shown by the dashed green curve, while the neutrino-floor~\cite{Billard:2021uyg} is seen in purple. Right: DM annihilation cross section versus the DM mass. Direct detection limits exclude the band of gray points. The gamma ray final states arise from $W^+W^-$ annihilation. Orange and pink regions are excluded by Fermi-LAT~\cite{Fermi-LAT:2016afa} and H.E.S.S~\cite{HESS:2016mib}. The magenta dashed curve is the projected CTA sensitivity~\cite{CTAConsortium:2017dvg}.}
\label{fig:ddscoto}
\end{figure}

 We now turn to scalar dark-matter detection through nuclear recoil. The left panel of Fig.~\ref{fig:ddscoto} shows the $\eta_R$-nucleon spin-independent elastic scattering cross-section versus the DM mass.  
The scattering is mediated by a neutral Higgs boson and a Z boson, where the Higgs boson channel gives the main contribution. 
Each result in the plot indicates the corresponding fraction $\xi=\Omega_{\eta_R}/\Omega_{Planck}$ of  the relic density, which is specified by the vertical bar. 
In the cyan region dark-matter makes up the total relic abundance present in the Universe, with the allowed points located mainly in a vertical band between $m_{\eta_R}=500-600$ GeV. 
The light green region is excluded by the LZ (2025) collaboration at 95\% C.L.~\cite{LZ:2024zvo}, while the purple one represents the neutrino floor~\cite{Billard:2013qya}. Comparing with the left panel of Fig. \ref{fig:relicscoto}, one sees that the cyan points near 62.5 GeV (i.e. $m_h/2$) are excluded by the LZ experiment.
\\[-.3cm] 

\par The results for indirect dark-matter detection are given in the right panel of Fig.~\ref{fig:ddscoto}. We plot the dark-matter annihilation cross section times the $\xi^2$ parameter--describing the fraction of the relic density--  versus the DM mass $m_{\eta_R}$. 
All constraints and limits from direct detection experiments have been taken into account.
Motivated by the results obtained in~\cite{Avila:2021mwg}, one can study the dark-matter annihilation into $W^+W^-$, with gamma rays as final states. 
The orange dashed line represents the $95\%$ C.L. upper limit set by Fermi-LAT from observations of dwarf spheroidal satellite galaxies in the Milky Way, assuming $W^+W^-$ annihilation. 
The pink dashed line shows the upper limit set by the H.E.S.S telescope at $95\%$ C.L. considering an Einasto DM density profile~\cite{einasto2010dark}. 
Cyan points reproduce the total observed relic abundance, while others (see vertical legend) are under-abundant, requiring an extra DM candidate to account for the total current relic density present in the universe. 
 Gray points are excluded by theoretical and experimental constraints~\cite{Avila:2021mwg}. The figure also includes the projected sensitivity of the future Cherenkov Telescope Array (CTA), indicated by the magenta dashed line, assuming dark-matter annihilation into $W^+W^-$ and an Einasto density profile. 
One sees that the points satisfying the total relic dark-matter density, lying between $m_{\eta_R}=300-600$ GeV, could be tested at CTA, which is a very interesting feature of the model.

\subsection{Revamped scotogenic }
\label{subsec:dmSTSM}
Details of the model parameter sampling procedure are similar to what we have considered previously. The ranges of variation of the input parameters are summarized  table~\ref{tab:rangesSTSM}.
In particular, in order to ensure that the LSP in the revamped scotogenic model of Sec.~\ref{sec:singlet-triplet-scoto} will be the $\eta_R$ scalar, we choose $\lambda_5 \leq 0$, see Eqs.~\eqref{eq:etrSTSM} and~\eqref{eq:etiSTSM}. 
\begin{table}[!htb]
\centering 
\begin{tabular}{|c|c|} 
\hline
{\bf Parameter} & {\bf Range}  \\ 
\hline  
\hline                             
$M_F $  & [$5\times 10^3,\,10^4$]  \rm (GeV)    \\
$M_{\Sigma} $ & [$5\times 10^3,\,10^4$]     \rm (GeV)     \\
$m_{\eta}^2$ & [$100,\,  5\times 10^3 $]   $\rm (GeV^2)$     \\
$\mu_{1,2} $ & [$10^{-8},\, 5\times 10^3$]   \rm (GeV)     \\
$v_\Omega $& [$10^{-5},\, 5$]  \rm (GeV)    \\
$|\lambda_{i}|$, $i=1...4$ & [$10^{-8} ,1$] \\
$|\lambda_{5}|$ &  $ [10^{-5} ,1$] \\
$|\lambda_{1,2}^{\Omega}|$ & [$10^{-8} ,1$] \\
$|\lambda_{\eta}^{\Omega}|$ & [$10^{-8} ,1$] \\
$|Y_{\Omega}|$  & [$10^{-8},1$] \\
\hline
\end{tabular}
\caption{
Ranges of variation of the input parameters used in the numerical scan~\cite{Avila:2019hhv}.}
\label{tab:rangesSTSM}
\end{table}  
\vskip .5cm
\begin{center}
   {\bf \small Scalar dark-matter relic density in revamped scotogenic} 
\end{center}

\begin{figure}[h]
\centering
\begin{minipage}{0.15\textwidth}
\hspace{\textwidth}
\end{minipage}
\begin{minipage}{0.45\textwidth}
    \includegraphics[width=\textwidth]{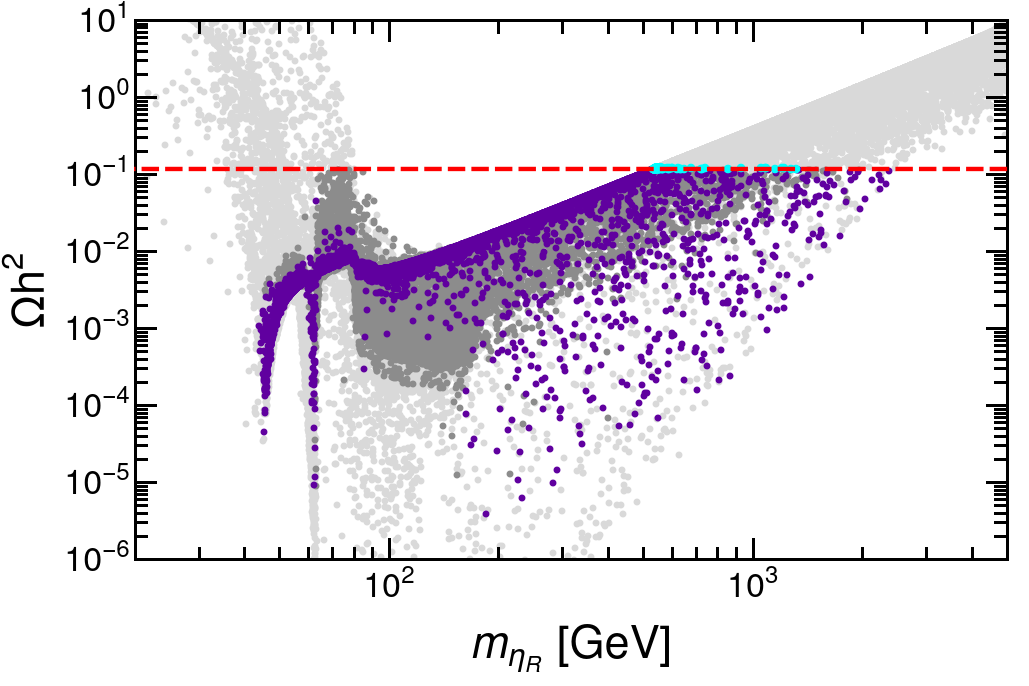}
\end{minipage}
\begin{minipage}{0.35\textwidth}   \includegraphics[width=\textwidth,height=3.5cm]{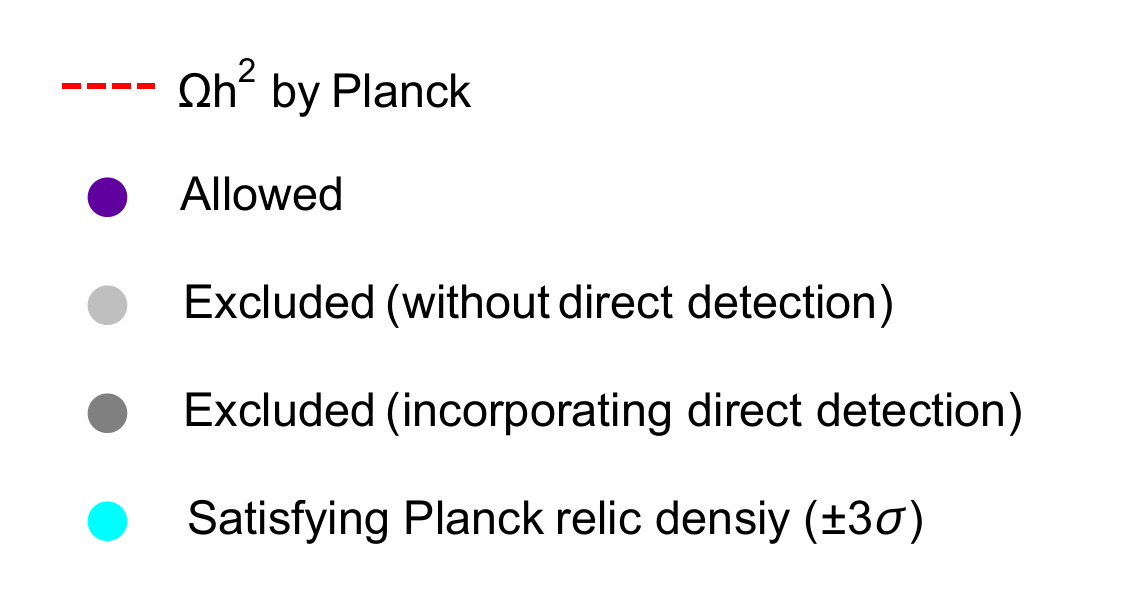}
\end{minipage}
\caption{
Dark-matter relic density $\Omega h^2$ versus its mass $m_{\eta}$~\cite{Avila:2019hhv} in the revamped scotogenic model.  Cyan points fall into the $3\sigma$ region of the Planck-measurement for the total relic dark matter, indicated by the red dashed line~\cite{Planck:2018vyg}.
The light gray points are excluded by other constraints, while the dark gray points are ruled out by the { LZ (2025) limit~\cite{LZ:2024zvo}} on the WIMP-nucleon spin-elastic scattering cross section~\cite{LZ:2022lsv}.} 
\label{fig:dmSTSM}
\end{figure}

 Fig.~\ref{fig:dmSTSM} shows the relic density versus the dark-matter mass $m_{\eta_R}$ in the revamped scotogenic model of Sec.~\ref{sec:singlet-triplet-scoto}. The horizontal red dashed line denotes the Planck $3\sigma$~\cite{Planck:2018vyg} measurement of the total relic abundance within the cold dark-matter revamped scotogenic freeze-out scenario.  
 The theoretical and experimental constraints described in~\cite{Avila:2019hhv} are imposed in the analysis. 
  The points in cyan color represent the total dark-matter abundance that survive all the constraints~\cite{Avila:2019hhv}. 
  One sees that within the ranges of variation of the input parameters given in table~\ref{tab:rangesSTSM}, the mass of $\eta_R$ should exceed $500$ GeV. 
  Note that the inclusion of scalar-fermion co-annihilation effects could lower the value of this mass limit, similar to the discussion of scalar DM phenomenology given in Sec.~\ref{sec:darkmatterscoto}~\cite{Klasen:2013jpa,Avila:2021mwg}.
  
\begin{figure}[!t]
\centering
\begin{minipage}{0.15\textwidth}
\hspace{\textwidth}
\end{minipage}
\begin{minipage}{0.45\textwidth}
    \includegraphics[width=\textwidth]{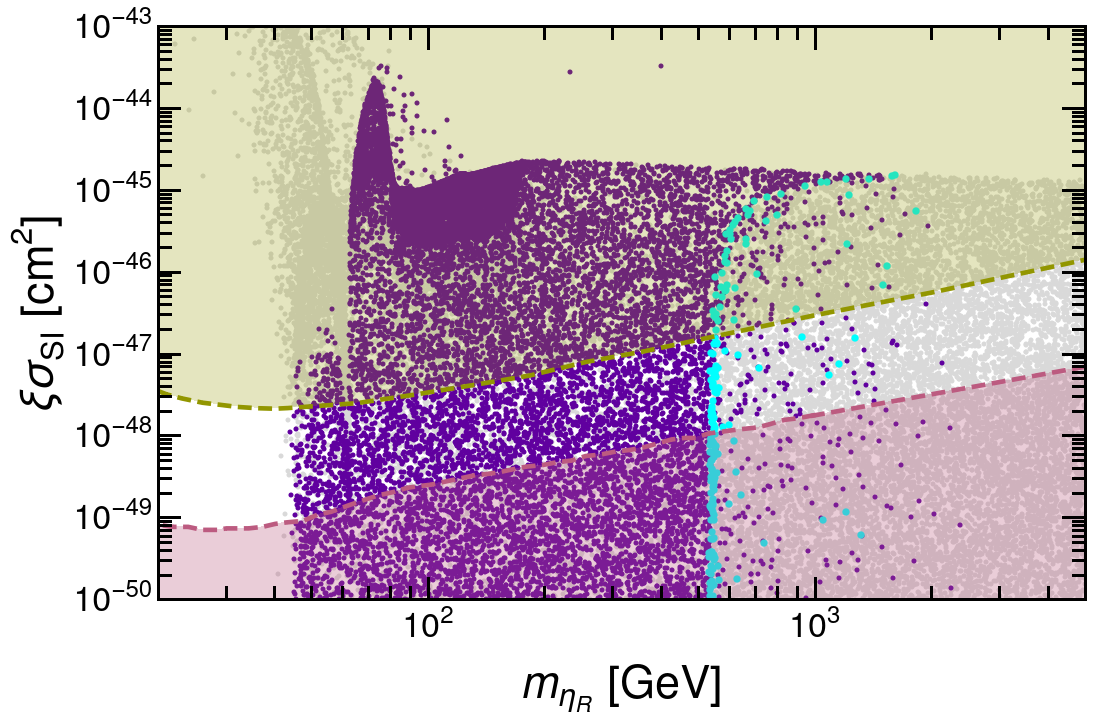}
\end{minipage}
\begin{minipage}{0.35\textwidth}
    \includegraphics[width=\textwidth,height=3.5cm]{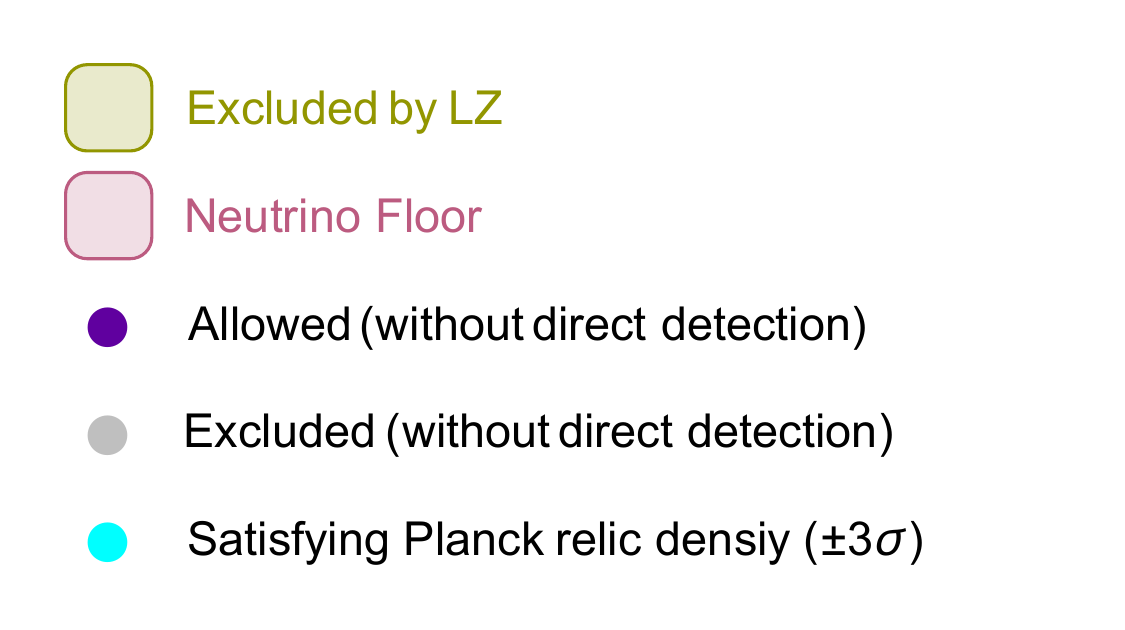}
\end{minipage}
\caption{
Spin-independent $\eta_R$-nucleon elastic scattering cross section versus $m_{\eta_R}$ in revamped scotogenic model~\cite{Avila:2019hhv}. The greenish region is excluded by { constraints from LZ (2025)~\cite{LZ:2024zvo}.}  The purple region is the neutrino floor from CE$\nu$NS~\cite{Billard:2021uyg,OHare:2021utq}. The cyan points account for the totality of dark matter indicated by the Planck measurements.}  
\label{fig:ddscalar}
\end{figure}

  Purple dots in Fig.~\ref{fig:dmSTSM} represent under-abundant  points where only a fraction of the dark-matter is obtained, so that another DM matter candidate is required. 
 Dark gray dots are excluded by current direct detection measurements.
  Light gray results are in conflict with some of the constraints described in~\cite{Avila:2019hhv}. 
  The first dip in Fig.~\ref{fig:dmSTSM} is located at $m_{\eta_{R}}\sim40-50$ GeV, i.e.  the Z-pole, where the relevant co-annihilation is via s-channel Z boson exchange.
  The second dip of the relic density lies around $m_{\eta_R}\sim 63$ GeV and corresponds to Higgs-boson-mediated s-channel annihilation. 
  The third drop of the relic density at $m_{\eta_R}\sim 80 $ GeV, is related to the annihilation of dark-matter into $W^{+}W^{-}$ via quartic couplings. 
  For DM masses larger than $120$ GeV the annihilation cross section drops as $1/m_{\eta_R}^2$ and the relic density increases. 
  As seen in~\cite{Avila:2019hhv}, for large DM masses, $\eta_R$ mainly annihilates into $W^{+}W^{-}$, $h^{0}h^{0}$ and $HH$. 
\begin{center}
   {\bf \small Scalar dark-matter detection in revamped scotogenic} 
\end{center}
The spin-independent ${\eta_R}-\text{nucleon}$ elastic scattering cross section results are shown in figure~\ref{fig:ddscalar}.
The color code for the relic density results is the same that is used in Fig.~\ref{fig:dmSTSM}, where the points giving the total dark-matter abundance are shown in cyan. Most of them cluster around a vertical band around $\eta_{R}\sim500-600$ GeV.
The gray points in~Fig.~\ref{fig:dmSTSM} falling in the green region in Fig.~\ref{fig:ddscalar} are excluded, as the dashed dark green line denotes the upper bound set by LZ (2025)~\cite{LZ:2024zvo}. The dashed purple line corresponds to the neutrino floor from coherent elastic neutrino-nucleus scattering (CE$\nu$NS)~\cite{Billard:2013qya}.
Notice that here the parameters are chosen so that co-annihilation effects can be neglected.
Their inclusion could lower this mass limit, as discussed earlier for the minimal scotogenic case.\\[-.3cm]

\begin{figure}[h]
\centering
\includegraphics[width=0.45\textwidth]{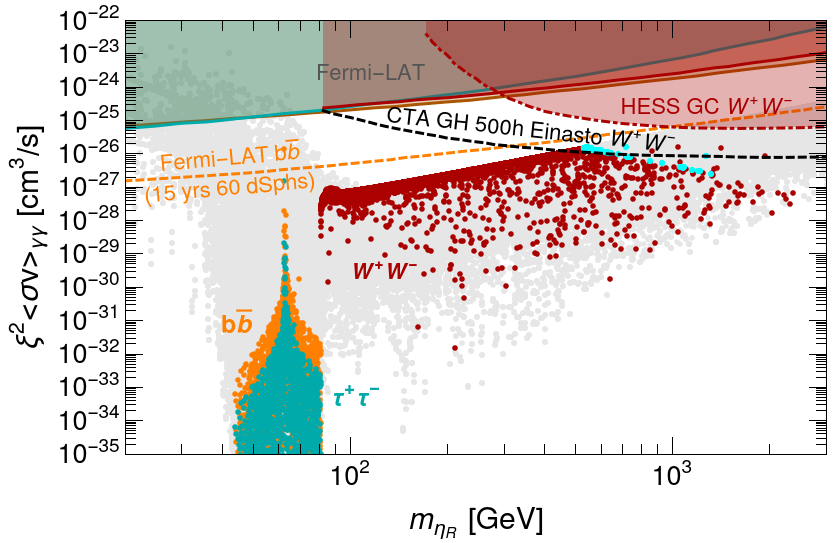}
\caption{ 
Scalar dark-matter revamped scotogenic annihilation cross section into $\gamma$ rays coming from 
$b\bar{b}$, 
$\tau^{+}\tau^{-}$, 
and $W^{+}W^{-}$ 
channels (indicated in different colors) versus experimental limits and sensitivities, from~\cite{Avila:2019hhv}.} 
\label{fig:idscalar}
\end{figure}

Turning to indirect detection of scalar dark matter, the preferred messengers are $\gamma$ rays, as they typically propagate interstellar or intergalactic space unaffected. In Fig.~\ref{fig:idscalar} we show the results for the dark-matter annihilation into $b\bar{b}$, $\tau^{+}\tau^{-}$ and $W^{+}W^{-}$ with $\gamma$ rays as final products. 
The results for the annihilation cross section are given in terms of the fraction $\xi=\frac{\Omega_{\eta_R}}{\Omega_{Planck}}$ and the corresponding branching ratio. 
The colored lines are the current 95\% C.L. upper limits set by Fermi-LAT~\cite{Fermi-LAT:2015att}. The results for different dark-matter annihilation mechanisms, $b\bar{b}$, $\tau^{+}\tau^{-}$,  $W^{+}W^{-}$ are indicated by different colors.

The region $m_{\eta_R} < 70$ GeV cannot be tested in current experiments, but future projections for Fermi-LAT~\cite{Fermi-LAT:2016afa} are promising. In this region we have dark-matter annihilation into $b\bar{b}$ and $\tau^{+}\tau^{-}$ (the bright orange is for annihilation into $b\bar{b}$, while the turquoise is for $\tau^{+}\tau^{-}$).
The dashed orange line represents the Fermi-LAT sensitivity projection from an analysis of 60 dwarf spheroidal satellite galaxies and 15 years of data into $b\bar{b}$.  \par
On the other hand, for $m_{\eta_R} >70$ GeV, the points in dark red correspond to dark-matter annihilation into $W^{+}W^{-}$. Notice that points in light cyan provide the total dark-matter abundance observed by Planck.
The red dot-dashed curve is the upper limit obtained by the H.E.S.S. telescope~\cite{HESS:2016mib}, using the $\gamma$ rays data accumulated in the galactic center (GC) over 10 years.  The black curve is the sensitivity projection for the Cherenkov Telescope Array (CTA), for the Milky Way galactic halo target annihilation into $W^{+}W^{-}$ and an Einasto dark-matter density profile. \par
\vskip .3cm
In summary, one sees that, although current measurements by H.E.S.S. cannot test our scalar dark matter within the scotogenic scenario, the projected sensitivity of upcoming CTA observations~\cite{CTAConsortium:2017dvg} will partly cover the relevant model parameter space. 

   \subsection{Dark inverse seesaw}

We now turn to the scalar dark-matter phenomenology within the dark inverse-seesaw mechanism of Sec.~\ref{sec:dark-inverse-seesaw}. As seen in Fig.~\ref{fig:scoto-loop} the dark loop involves a hidden DM sector containing dark singlet fermions and scalars, $F$ and $\xi$.
In addition to ensuring radiative generation of neutrino masses, the dark-parity symmetry is responsible for the stability of the \textit{lightest scotogenic dark particle} or LSP, for which there are three options to consider: the lightest of the real or imaginary parts of the $\xi$ scalar, $\xi_R$ or $\xi_I$, or the dark fermion $F$.
Here we focus on scalar DM and take, for definiteness,  $\lambda_5< 0$,~so that the dark-matter particle is $\xi_R$ (for the opposite sign the dark-matter particle is $\xi_I$).

In contrast to previous cases discussed in this section, instead of scanning the model parameter space, here we fix benchmarks in order to discuss the dark-matter phenomenology. Their choice is in turn motivated by the interactions determining the scalar DM annihilation properties, as discussed below.
\begin{center}
   {\bf \small 
   Scalar dark-matter relic abundance in dark inverse seesaw} 
\end{center}

There are several dark-matter annihilation and co-annihilation diagrams which contribute to the relic abundance of our assumed dark-matter particle, $\xi_R$. 
The relic $\xi_R$ dark-matter density is mainly determined by CP-even scalar-mediated s-channel annihilation to \sm
final states~($\ell^+\ell^-$, $q\bar{q}$, $W^+W^-$, $ZZ$, $\gamma\gamma$, $hh$) as well as to $HH$ and $JJ$ final states, where $J$ denotes the majoron associated to the spontaneous  spontaneous lepton number violation. 
A sub-dominant role is played by annihilation into $hh,HH$ and $JJ$ via the direct 4-scalar vertices $h^2\xi_R^2$, $H^2\xi_R^2$ and $J^2\xi_R^2$, respectively.
There could also be additional contributions from $\xi_{R/I}$ exchange in the t-channel.  \par

In Fig.~\ref{fig:annihilation-diagram} of Sec.~\ref{sec:Inert}, we collect relevant Feynman diagrams contributing to $\xi_R$ annihilations and co-annihilations. 
One sees that dark-matter annihilation is mainly determined by mixed quartic couplings, such as $\lambda_{\Phi\xi}$, $\lambda_{\xi\sigma}$, $\lambda_5$ and $\lambda_{\Phi\sigma}$. 
Notice that the effect of $\lambda_5$ is negligible, as it must be very small in order to generate light neutrino masses with potentially sizable charged lepton flavor violation. 
One also sees from Eq.~\eqref{eq:lam123} that the mixed quartic coupling $\lambda_{\Phi\sigma}$ is equivalent to the mixing angle $\sin\theta$ once we fix the free parameters $m_H$ and $v_\sigma$.
Hence non-zero $\sin\theta$ implies non-zero $\lambda_{\Phi\sigma}$. 
As we discussed before, the mixing angle $\sin\theta$ is tightly constrained from the LHC experiments.
In order to illustrate the main features of our chosen dark-matter candidate $\xi_R$ we fix four benchmark points: 
\begin{align}
&\text{\textbf{BP1:} }\sin\theta =0,  \lambda_{\Phi\xi}=0.01,~~~~
\text{\textbf{BP2:} } \sin\theta =0, \lambda_{\Phi\xi}=0.1. \\
&\text{\textbf{BP3:} } \sin\theta =0.1,  \lambda_{\Phi\xi}=0.01,\,\,\,
\text{\textbf{BP4:} }\sin\theta =0.1, \lambda_{\Phi\xi}=0.1.
\end{align}
while other parameters are taken as $m_H=1\,\text{TeV}, v_\sigma=3 \text{ TeV}, \lambda_\xi=0.1$ and $\lambda_{\xi\sigma}=0.1$. 
\begin{figure}[htb!]
\includegraphics[height=4.6cm,width=0.5\textwidth]{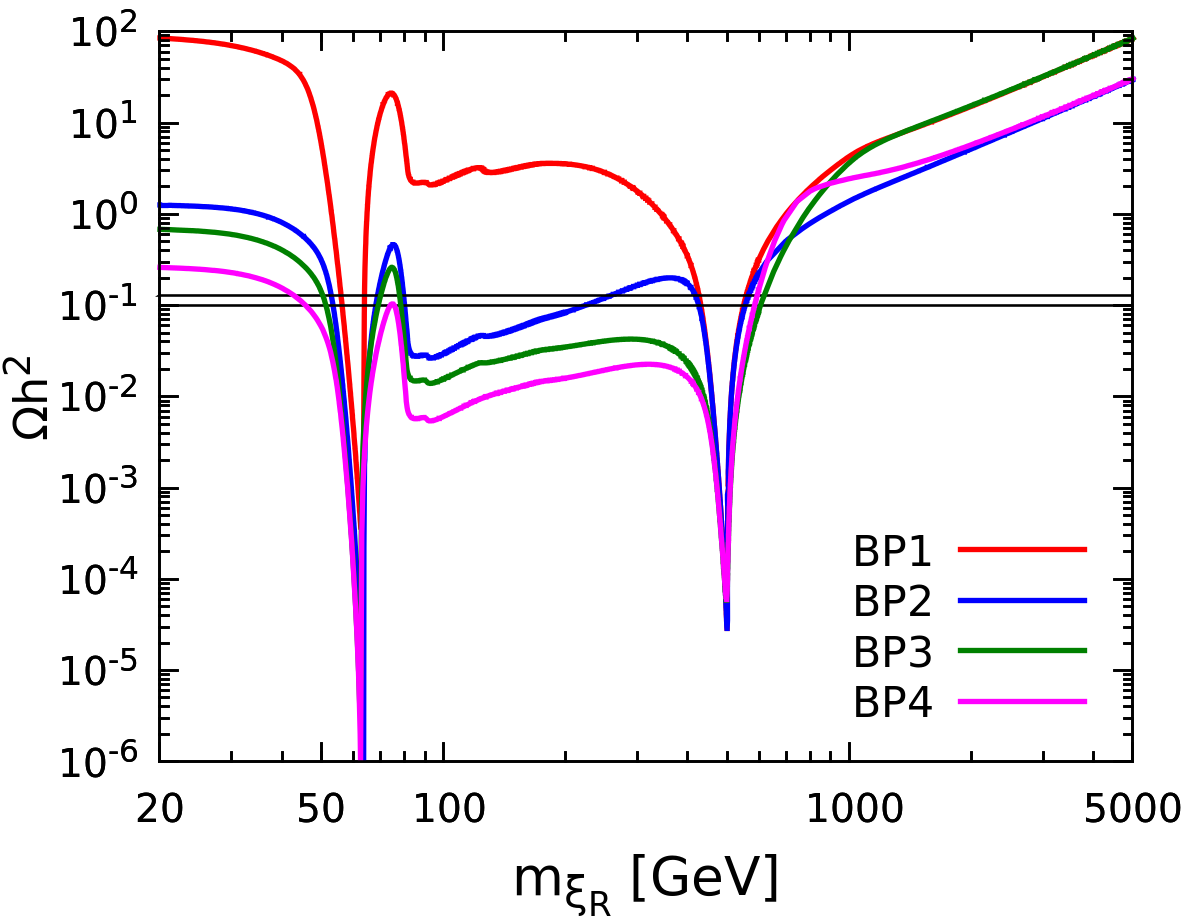}
\caption{
Relic abundance versus dark-matter mass in dark inverse seesaw.
  The horizontal black lines are the measured $3\sigma$ Planck range, Eq.~\eqref{eq:Pl}.}
\label{fig:relic}
\end{figure}
The relic $\xi_R$ density as a function of its mass is shown in Fig.~\ref{fig:relic}. 
The calculation corresponds to the four benchmarks \textbf{BP1}~(red line), \textbf{BP2}~(blue line), \textbf{BP3}~(green line) and \textbf{BP4}~(magenta line). 
The narrow horizontal band is the $3\sigma$ CDM range derived from the Planck satellite data~\cite{Planck:2018vyg}.
Only for solutions falling inside this band one can explain the total cosmological dark matter by $\xi_R$.  
Various features of the relic density in Fig.~\ref{fig:relic} can be understood by looking in detail into the $\xi_R$ annihilation channels
shown in Fig.~\ref{fig:annihilation-diagram}.
The two dips at $m_{\xi_R}\sim m_h/2$ and $m_{\xi_R}\sim m_H/2$ correspond to annihilation via s-channel $h$ and $H$ exchange. 
These become very efficient when $h$ or the heavy Higgs $H$ are on-shell, precluding us from obtaining a relic density matching Planck observations.
For $m_{\xi_R} \gsim 80$ GeV, the annihilation of $\xi_R$ into $W^+W^-$ and $ZZ$ become important, thus explaining the drop in the relic abundance at $m_{\xi_R}\sim 80$ GeV. 
For very heavy $m_{\xi_R}$ the relic density increases due to the suppressed annihilation cross section, which drops as $\sim\frac{1}{m_{\xi_R}^2}$.  
Note that for large $m_{\xi_R}$, the Higgs mediated annihiliation channel $\xi_R\xi_R\to W^+W^-, ZZ$ dominates, and involves the quartic coupling $\lambda_{\Phi\xi}$ for fixed
$\lambda_{\xi\sigma}$ and small $\sin\theta$.
Note that one can have significant annihilation to the $JJ$ final state, depending on the mixed quartic coupling strength. 
Notice also that, for small mass difference $m_{\xi_R}^2-m_{\xi_I}^2$, one must include the contributions of both $\xi_R,\xi_I$ species, as 
this can lower the relic dark-matter density. 


\begin{center}
   {\bf \small 
   Scalar dark-matter detection in dark inverse seesaw} 
\end{center}

Let us now study the direct detection prospects of our scalar dark-matter candidate $\xi_R$ in the dark inverse seesaw. 
Its elastic scattering with a nucleon proceeds via two t-channel diagrams mediated by $h$ and $H$, see Fig.~\ref{fig:DD_scoto}.\par 

\begin{figure}[htb!]
\includegraphics[height=5.6cm,width=0.5\textwidth]{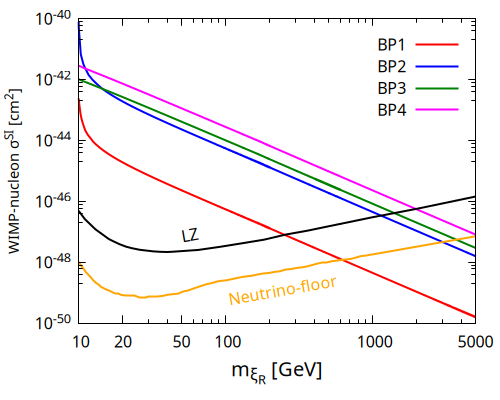}
\caption{ 
  Spin-independent $\xi_R$-nucleon cross-section versus the DM mass for different benchmarks in dark inverse seesaw. The black line denotes the upper bound from the {LZ experiment~\cite{LZ:2024zvo}} while the orange line corresponds to the neutrino floor~\cite{Billard:2021uyg}.}
\label{fig:direct-detection}
\end{figure}

The resulting spin-independent scattering cross section is given by~\cite{Basak:2021tnj}, 
\begin{align}
\sigma^{\text{SI}}=\frac{\mu_N^2 m_N^2 f_N^2}{4\pi m_{\xi_R}^2 v_\Phi^2}\Big(\frac{\lambda_{h\xi_R\xi_R}}{m_h^2}\cos\theta - \frac{\lambda_{H\xi_R\xi_R}}{m_H^2}\sin\theta\Big)^2,
\label{eq:SI}
\end{align}
where $\mu_N=\frac{m_N m_{\xi_R}}{m_N+m_{\xi_R}}$ is the reduced mass for nucleon-dark-matter system. Here $f_N$ is the form 
factor, which depends on hadronic matrix elements.
The trilinear couplings $\lambda_{h\xi_R\xi_R}$ and $\lambda_{H\xi_R\xi_R}$ are given as 
\begin{align}
\lambda_{h\xi_R\xi_R}&=\lambda_{\Phi\xi}v_\Phi\cos\theta+(\lambda_{\xi\sigma}+\lambda_5)v_\sigma\sin\theta \\
\lambda_{H\xi_R\xi_R}&=-\lambda_{\Phi\xi}v_\Phi\sin\theta+(\lambda_{\xi\sigma}+\lambda_5)v_\sigma\cos\theta 
\end{align}
where, as we have seen, $\lambda_5$ is tiny and can be neglected. 
Note that Eq.~\eqref{eq:SI} generalizes the expression given in~\cite{Cline:2013gha} for the case of complex singlet scalar dark matter. 
The relative sign between the $h$ and $H$ contributions comes from the substitution rule in Eq.~\eqref{eq:substitution}.
Depending on parameters, there can be a destructive interference between the two channels, so that direct detection becomes very small. 
In Fig.~\ref{fig:direct-detection} we show the spin-independent $\xi_R$-nucleon cross-section versus DM mass for the same benchmarks as in Fig.~\ref{fig:relic}. 
The black-solid line denotes the upper bound coming from the LZ collaboration~\cite{LZ:2024zvo}. 
Other experiments e.g. LUX~\cite{LUX:2016ggv} and PandaX-II~\cite{PandaX-II:2016vec} give weaker (undisplayed) constraints. 
We also indicate the neutrino floor associated to coherent elastic neutrino scattering from several astrophysical sources~\cite{Billard:2013qya}. 
Clearly there are high-mass solutions with the correct CDM relic density while,
for our chosen benchmarks, most of the low-mass DM region is ruled out by current direct detection upper limits. 
As discussed in Sect.~\ref{sec:cLFV} and Sect.~\ref{sec:collider} these solutions can be probed at upcoming cLFV and collider experiments. 
\begin{figure}[h!]
\centering
\includegraphics[height=5.6cm,width=0.5\textwidth]{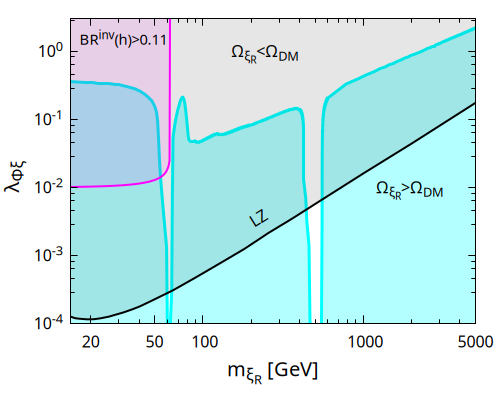}
\caption{
Allowed dark inverse seesaw quartic coupling versus the DM mass from relic density, direct detection and invisible Higgs decay.}
\label{fig:fitdm}
\end{figure}

In Fig.~\ref{fig:fitdm} we put together the constraints on $\lambda_{\Phi\xi}$ from the relic density, direct detection and invisible Higgs decay.
Here we fix the \textbf{BP1} benchmark, keeping now the quartic coupling $\lambda_{\Phi\xi}$ as free parameter.
 The magenta region is excluded by the invisible Higgs decay constraint~\cite{ParticleDataGroup:2024cfk}. The region above the black line is excluded by the LZ direct detection limit~\cite{LZ:2024zvo}. 
%
Along the cyan line the real scalar singlet ${\xi_R}$ gives the DM relic abundance within $3\sigma$, Eq.~\eqref{eq:Pl}. 
The region below this cyan line gives over-abundant dark matter.   
We see that, for this particular benchmark, our dark-matter candidate mass must lie close to the dips associated to the SM-like Higgs boson or in the heavier CP-even scalar boson mass region, consistent with all existing constraints. 

\subsection{Dark linear seesaw}

We now turn to the possibility of scalar dark matter realized within the linear seesaw mechanism. 
This can arise in the model discussed in Sec.~\ref{sec:linear-dark-seesaw} by assuming that the LSP is the lightest among the neutral scalar particles $D_{1}$, $D_{2}$, $D_{A_1}$, $D_{A_2}$,
and lighter than the heavy neutral Majorana fermions $F_i$. 
In the following discussion, we take small doublet-singlet mixing angles so that $D_{A_1}$, taken as the DM candidate,
  is mainly the imaginary part of the neutral component of the dark doublet $\eta$. 
Such scalar DM candidate would scatter-off a nuclear target through Higgs boson exchange in the $t$-channel, giving rise to a direct Higgs portal dark-matter detection mechanism. 
The main co-annihilation channels for the DM candidate lead to a pair of SM particles as well as the charged and neutral components of the scalar triplet.
In this benchmark scenario, we take all dark sector quartic couplings except $\lambda_{6,7,11}$ to be small.
The dominant dark-matter annihilation cross sections are then given as \cite{Hambye:2009pw,Bhattacharya:2016ysw},   
\begin{eqnarray}
v_{rel}\sigma \left( D_{A_{1}}D_{A_{1}}\to WW\right)  &=&\frac{1}{8\pi m_{DM}\sqrt{s}}\frac{g^4}{4}\Big(1+ \frac{m_{DM^4}}{m_W^4} \Big( \frac{\Delta m^2 + \kappa v_\Phi^2}{m_{DM}^2} \Big)^2\Big)\sqrt{1-\frac{4 m_W^2}{s}}, \\
v_{rel}\sigma \left( D_{A_{1}}D_{A_{1}}\to ZZ\right)  &=&\frac{1}{16\pi m_{DM}\sqrt{s}}\frac{g^2}{4 c_w^4}\Big(1+ \frac{m_{DM^4}}{m_Z^4} \Big( \frac{\Delta m^2 + \kappa v_\Phi^2}{m_{DM}^2} \Big)^2\Big)\sqrt{1-\frac{4 m_Z^2}{s}}, \\
v_{rel}\sigma \left( D_{A_{1}}D_{A_{1}}\to q\overline{q}\right)  &=&
\frac{N_{c}\kappa^{2}m_{q}^{2}}{16\pi }\frac{\sqrt{\left( 1-\frac{4m_{f}^{2}}{s}\right) ^{3}}}{\left( s-m_{H_1}^{2}\right) ^{2}+m_{H_1}^{2}\Gamma_{H_1}^{2}}, \\
v_{rel}\sigma \left( D_{A_{1}}D_{A_{1}}\to H_1 H_1\right)  &=&\frac{\kappa^{2}}{64\pi s}\left( 1+\frac{3m_{H_1}^{2}}{s-m_{H_1}^{2}}-\frac{2\lambda
v^{2}}{s-2m_{H_1}^{2}}\right) ^{2}\sqrt{1-\frac{4m_{H_1}^{2}}{s}},\\
v_{rel}\sigma \left( D_{A_{1}}D_{A_{1}}\to \Xi_i \Xi_i\right)  &=&\frac{\lambda_{11} ^{2}}{64\pi s}\sqrt{1-\frac{4m_{\Xi_i}^{2}}{s}},
\end{eqnarray}%
where $\sqrt{s}$ is the centre-of-mass energy, $N_{c}=3$ is the relevant color factor, $m_{H_1}=125.7$ GeV and $\Gamma _{H_1}$ is the total decay width of the SM Higgs boson, about 4.1 MeV. 
Moreover we have
$$\kappa = \frac{\lambda_6 + \lambda_7}{4},~~~m_{DM} = m_{D_1},~~~\Delta m^2 = m_{D_{A_1}}^2 - m_{D_1}^2,$$ 
the mass splitting between the CP even and the odd parts of the neutral component of $\eta$.
 The final state represented as $\Xi_i \Xi_i$ stands for $\Xi_1^+\Xi_1^-$, $\Xi_2^+\Xi_2^-$, $\Xi_I^0\Xi_I^0$, or $\Xi_R^0 \Xi_R^0$. 
The present DM relic abundance is estimated as follows (c.f.~\cite{ParticleDataGroup:2022pth,Edsjo:1997bg}), 
\begin{equation}
\Omega h^{2} = \frac{0.1~~\textrm{pb}}{\langle \sigma v \rangle },\,%
\hspace{1cm}\langle \sigma v \rangle =\frac{A}{n_{eq}^{2}},
\end{equation}%
where $\langle \sigma v \rangle $ is the thermally averaged
annihilation cross section, $A$ is the total annihilation rate per unit
volume at temperature $T$ and $n_{eq}$ is the equilibrium value of the
particle density, which are given as~\cite{Edsjo:1997bg},
\begin{eqnarray}
A &=&\frac{T}{32\pi ^{4}}\int_{4m_{D_{A_{1}}}^{2}}^{\infty}\sum_{p=W,Z,t,b,H_1,\Xi_i} g_{p}^{2}\frac{s\sqrt{s-4m_{D _{A_1}}^{2}}}{2}%
v_{rel}\sigma \left( D_{A_{1}}D_{A_{1}}\rightarrow SM SM\right)
K_{1}\left( \frac{\sqrt{s}}{T}\right) ds,  \notag \\
n_{eq} &=&\frac{T}{2\pi ^{2}}\sum_{p=W,Z,t,b,H_1,\Xi_i}g_{p}m_{D_{A_1}}^{2}K_{2}\left( \frac{m_{D_{A_1}}}{T}\right) 
\end{eqnarray}
with $K_{1}$ and $K_{2}$ being the modified Bessel functions of the second kind of order 1 and 2, respectively.  
For the relic density calculation, we take $T=m_{m_{A_1}}/20$ as in Ref.~\cite{Edsjo:1997bg}, which corresponds to a typical freeze-out temperature.  
The DM relic density thus determined should match the required value indicated by the Planck measurement. \\ 

\begin{figure}[tbh]
\centering
\includegraphics[width=0.5\textwidth,height=4.5cm]{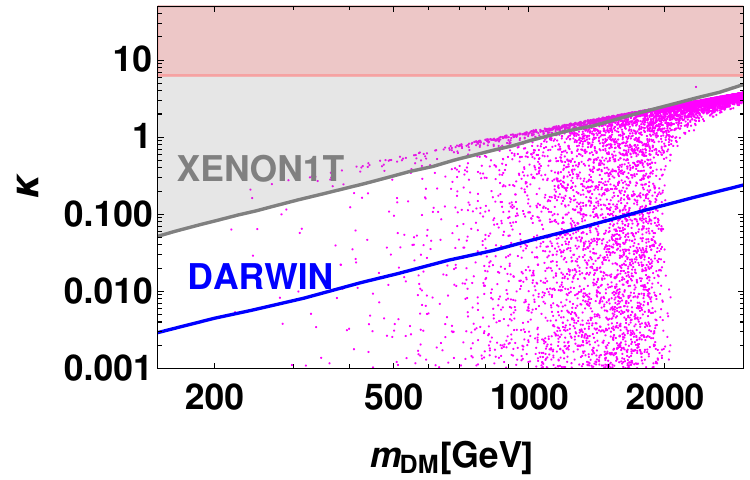}
\caption{
Allowed $m_{DM} - \kappa $ region leading to the correct scalar DM relic density (magenta points) in linear seesaw of Ref.~\cite{CarcamoHernandez:2023atk}.
  Here, $\kappa$ is the DM coupling to the SM Higgs boson. The pink band is disfavored by perturbativity.
  The gray region is excluded by XENON1T~\cite{Aprile:2018dbl} while the blue line corresponds to the sensitivity of DARWIN~\cite{DARWIN:2016hyl}.}
\label{DMplot}
\end{figure}

 In Fig.~\ref{DMplot}, we display the allowed region (magenta points) in the $m_{DM}-\kappa$ plane that reproduces the correct dark-matter relic abundance. 
In calculating the DM relic density, we have included its annihilation 
not only to SM states $WW$, $ZZ$, $H_1 H_1$, $t\overline{t}$, $b\overline{b}$, but also the components of the triplet scalar, which are the dominant channels. 
The masses of the charged and neutral components of $\Xi$ are varied in the range 0.2-5 TeV and the quartic coupling $\lambda_{11}$ is varied up to $0-4\pi$. 
 The pink band is disfavored by perturbativity, whereas the gray shaded region is disfavored by current limits. The region above the blue line lies within the sensitivity of the upcoming  DARWIN experiment~\cite{DARWIN:2016hyl}.  

\section{Charged lepton flavor violation}
\label{sec:cLFV}
By construction, in all the scenarios described here neutrino mass generation proceeds through a \textit{dark sector}.
Either the latter mediates neutrino mass generation through a radiative seesaw mechanism, as in Secs.~\ref{sec:simpl-scot-setup} or \ref{sec:singlet-triplet-scoto}, or it provides a radiative seed for neutrino mass generation that proceeds \textit{a la seesaw}, as in Sec.~\ref{sec:dark-low-scale-seesaw}.
As a result, in addition to the standard charged-current (CC) contributions associated to the seesaw sector~\cite{Valle:2015pba}, charged lepton flavor violation processes also receive dark-mediated contributions. 
These involve the new Yukawa couplings associated to the dark sector of the theory. Here lies a profound difference between our scotogenic-inspired approach to the dark-matter problem and the vanilla supersymmetric and/or IHDM WIMP cold dark-matter scenarios.
In the following subsections we discuss cLFV expectations for the various models we have discussed. These should be compared with Table~\ref{tab:LFV}.

\subsection{Simplest Scotogenic}   
\begin{figure}[!h]
\centering
\includegraphics[height=3.5cm,scale=0.3]{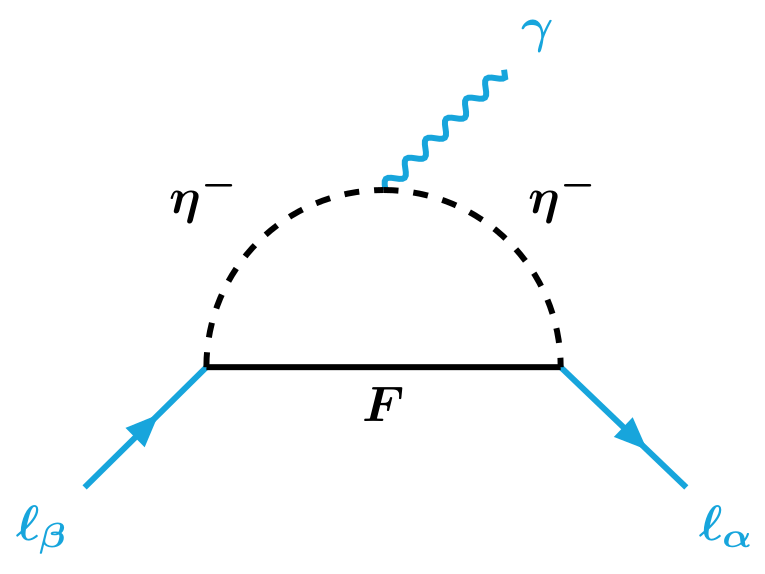}
\caption{
Feynman diagrams for charged lepton flavor violating processes $\ell_\beta\to\ell_\alpha\gamma$ in the scotogenic setup.} 
\label{fig:mutoegamma}
\end{figure}

In this section we present analytical results for the cLFV processes in the simplest scotogenic model, where dark loops involving $\eta^{\pm}-F$ exchange give a sizeable contribution to the cLFV rates. They involve the same Yukawa coupling matrix $ Y_F^{ij} \bar{L}_{i}\tilde{\eta} F_j$ which induces neutrino masses. The Feynman diagram for the cLFV process $\ell_\beta\to\ell_\alpha\gamma$ is shown in Fig.~\ref{fig:mutoegamma}. 

The branching ratio of $\ell_\beta\to\ell_\alpha\gamma$ in the original scotogenic model is given as~\cite{Toma:2013zsa}   
\begin{align}
\text{Br}(\ell_\beta \to \ell_\alpha\gamma)=\frac{3 \alpha_{\rm em}}{64\pi G_{F}^{2}m_{\eta^\pm}^{4}}\Big|\sum_{i=1}^{3}Y_{F}^{*\alpha i}Y_{F}^{\beta i}F_{2}(M_{F_{i}}^{2}/m_{\eta^\pm}^{2})\Big|^2 \text{Br}(\ell_\beta \to \ell_\alpha \nu_{\beta}\overline{\nu_{\alpha}}),
\label{eq:BRmue2gamma}
\end{align}
where $\alpha_{\rm em}=e^2/4\pi$ and $G_F$ is the Fermi constant. The loop function $F_{2}(x)$ is given by~\cite{Toma:2013zsa},
\begin{align}
F_{2}(x)=\frac{1-6x+3x^{2}+2x^{3}-6x^2\ln x}{6(1-x)^4}.
\label{eq:BRmue2gammaLoop}
\end{align}
Large $Y_F$ values are strongly constrained by the cLFV limits given in Table~\ref{tab:LFV}, since the diagrams in  Fig.~\ref{fig:mutoegamma} involve the Yukawa coupling matrix $Y_F$. 
The analytical expression for the $\mu - e$ conversion rate in nuclei reads as~\cite{Ibarra:2016dlb,Toma:2013zsa,Vicente:2014wga} 
\begin{align}
 \text{CR}(\mu - e, \text{Nucleus})\approx \frac{2 m_\mu^5 \alpha_{\rm em}^5 Z_{\rm eff}^4 F_p^2 Z}{(4\pi)^4\Gamma_{\rm capt} m_{\eta^\pm}^4}\Bigg|\sum_k Y_F^{*ek}Y_F^{\mu k} H_2\left(\frac{M_{F_k}^2}{m_{\eta^\pm}^2}\right)\Bigg|^2   ,
\end{align}
where $Z_{\rm eff}$, $Z$ and $F_p$ are the effective atomic charge, atomic number and nuclear matrix element. $\Gamma_{\rm capt}$ is the total relevant muon capture rate, given in Ref.~\cite{Kitano:2002mt} for various nuclei. The function $H_2(x)$ is given as 
\begin{align}
H_2(x)=1/3\, G_2(x)-F_2(x),~~~~\mathrm{with}~~~~ G_2(x)=\frac{2-9x+18x^2-11x^3+6x^3\text{log}\,x}{6(1-x)^4}    .
\end{align}
Stringent bounds from $\mu\to e\gamma$ and other cLFV processes, such as $\mu - e$ conversion in nuclei apply.  
Upcoming searches for $\mu - e$ conversion in nuclei are expected to reach very high sensitivities, e.~g. Mu2e~\cite{Glenzinski:2010zz,Mu2e:2012eea}, DeeMe~\cite{Aoki:2010zz,Natori:2014yba}, COMET~\cite{COMET:2009qeh,Kuno:2013mha} and PRISM/PRIME~\cite{prime}, with correspondingly higher restrictive power.
\begin{figure}[!h]
\centering
\includegraphics[height=5cm,width=0.45\textwidth]{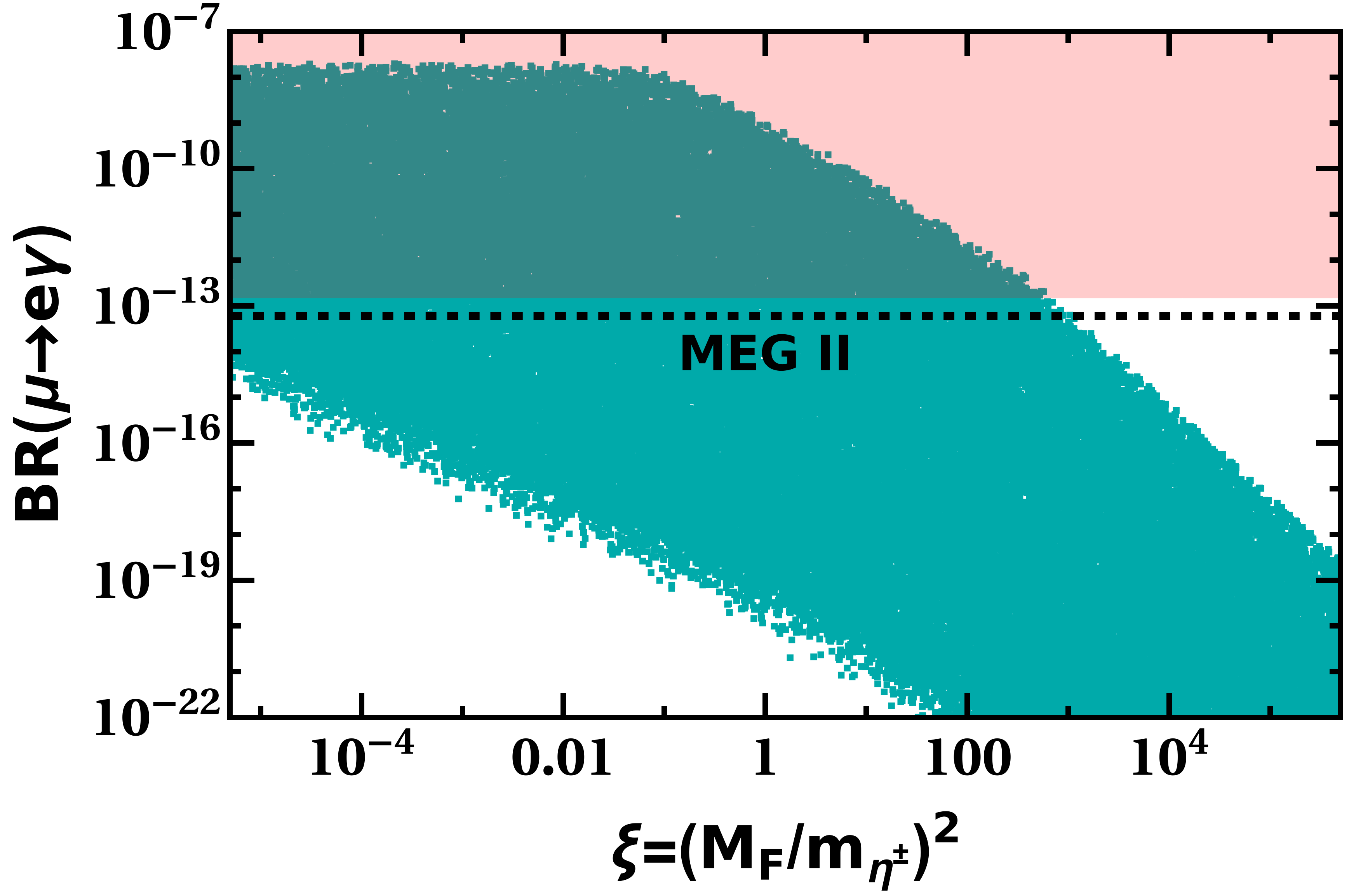}~~~
\includegraphics[height=5cm,width=0.45\textwidth]{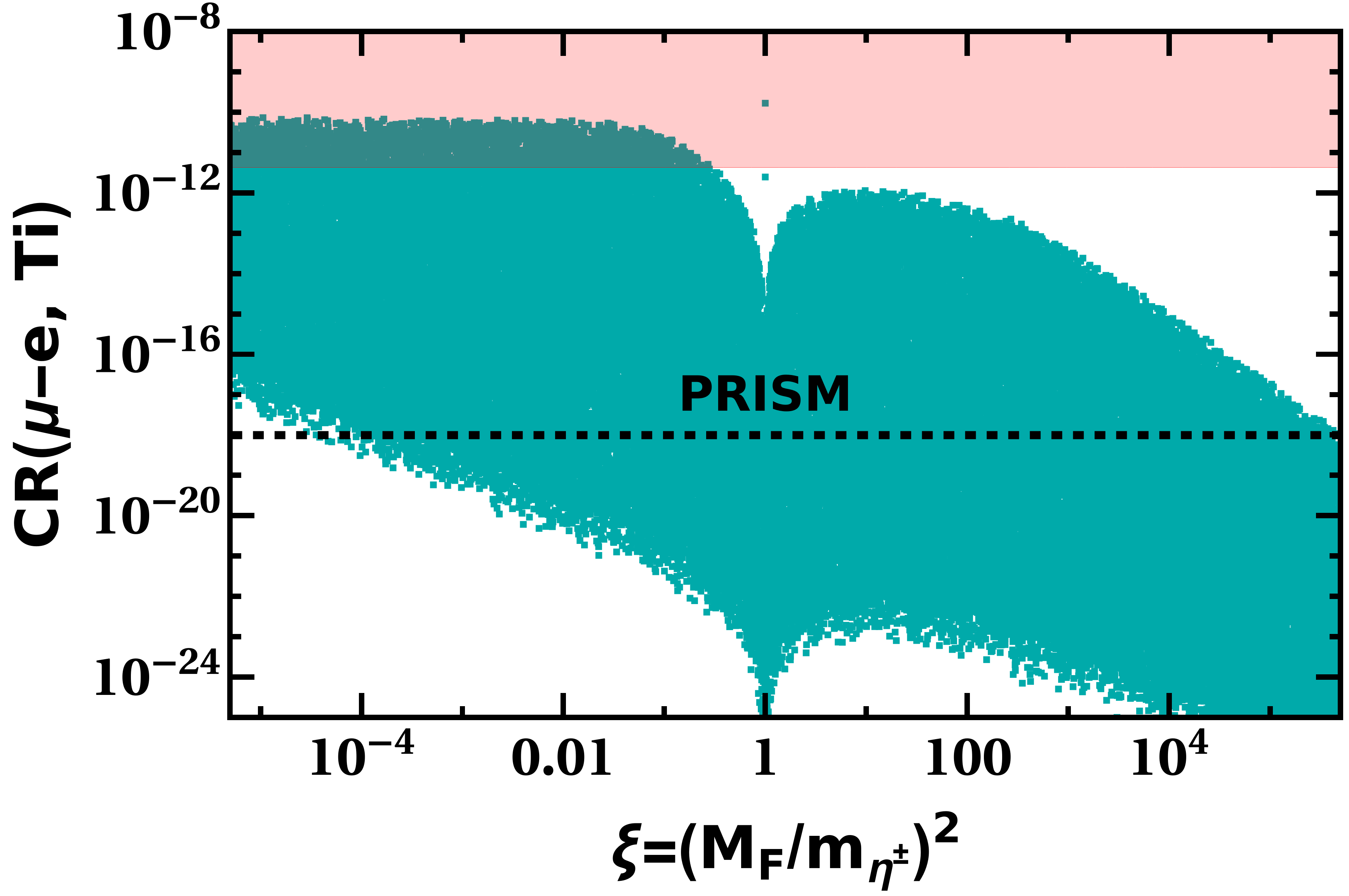}
\caption{
$\text{BR}(\mu\to e\gamma)$ and $\mu-e$ conversion in nuclei as a function of the mass squared ratio $\xi=(M_F/m_{\eta^{\pm}})^2$ in the original scotogenic model. The relevant parameters are taken as follows: $\lambda_3=\lambda_4=0.01$, $\lambda_5 \in [10^{-10}:10^{-5}]$, with oscillation parameters within $3\sigma$~\cite{deSalas:2020pgw,10.5281/zenodo.4726908}. The horizontal shaded bands are excluded by current limits in Table~\ref{tab:LFV}, while the black dashed line in the left and right panel are MEG II~\cite{MEGII:2018kmf} and PRISM~\cite{prism} projected sensitivities.}  
\label{fig:cLFV_Scoto} 
\end{figure}
\par Fig.~\ref{fig:cLFV_Scoto} displays the rates for the cLFV processes $\mu\to e\gamma$~(left panel) and $\mu - e$ conversion~(right panel) as a function of the dark mass squared ratio $\xi=(M_F/m_{\eta^{\pm}})^2$ in the
original scotogenic scheme. 
We assumed degenerate dark fermion masses, varying $\lambda_5$ in the range $10^{-10}$ to $10^{-5}$ and the oscillation parameters within their $3\sigma$ range~\cite{deSalas:2020pgw,10.5281/zenodo.4726908}. 
The shaded red band in each panel is excluded by current measurements. The pattern in each panel can be roughly understood from the fact that the functions $F_2(\xi)$ and $G_2(\xi)$ are almost constant when $\xi\ll 1$ and decrease when $\xi\gg 1$. 
Notice the suppression in the right panel for $\xi=1$. This pole is due to an exact cancellation between the contributions from the loop functions $G_2(\xi)$ and $F_2(\xi)$. 
The width of the region covered in each panel is mainly due to the variation of $\lambda_5$ which controls the variation of Yukawa coupling $Y_F$, see Eq.~\eqref{eq:Ynu}. 
The Yukawa coupling is large for small $\lambda_5$ and vice versa. For example, we find that for the fermionic DM case~($\xi<1$), the region with $\lambda_5<10^{-10}$ is mainly excluded from the current $\mu\to e\gamma$ upper limit. 
The allowed regions can be relaxed by taking into account the $F-\eta$ co-annihilation effects and/or by considering a hierarchical heavy neutrino spectrum. \par
On the other hand for the scalar DM case~($\xi>1$), we find that most of the parameter space is allowed, as there is no direct interconnection with cLFV processes, since these are mainly driven by Yukawa interactions.
Three-body cLFV processes, e.g. $\ell_\alpha\to 3\ell_\beta$ decays can also be competitive.
They mainly arise from the photonic contributions, coming from the $\gamma-$penguin diagram, and also the box diagrams~\cite{Vicente:2014wga}. 

\subsection{Revamped scotogenic}
\begin{figure}[h!]
	\vspace*{3mm}
 \includegraphics[height=3.5cm,scale=0.3]{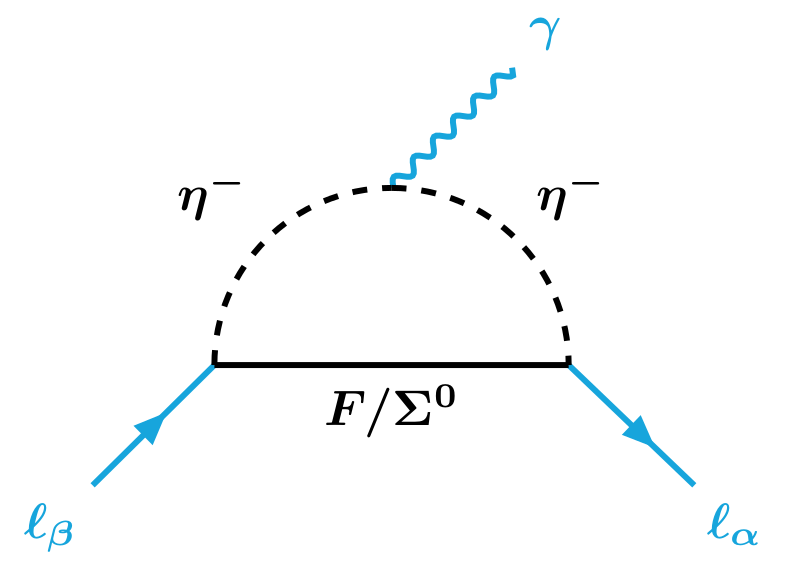}\hspace{10mm}
\includegraphics[height=3.5cm,scale=0.3]{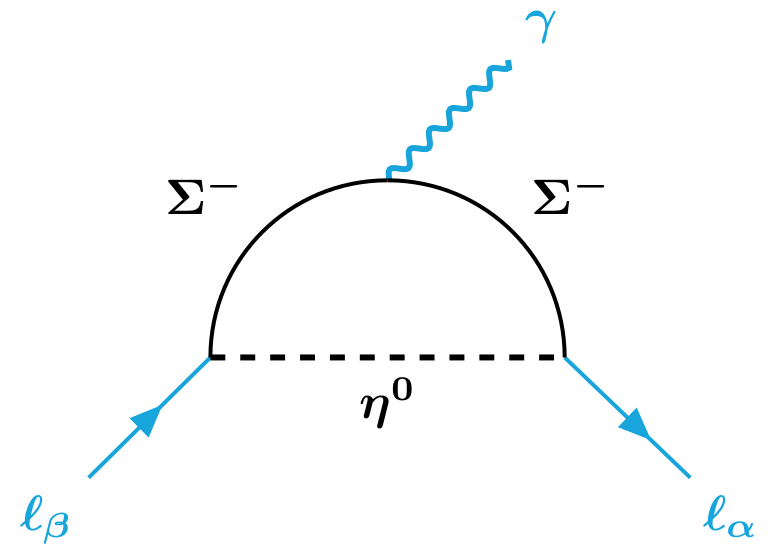}
	\caption{ 
     Leading contribution to $\mu \to e \gamma$ from photon penguin diagrams. There are also sub-leading box diagrams.}
	\label{fig:mutoe}
\end{figure}

Here we discuss the prospects for cLFV processes within the revamped singlet-triplet scotogenic model of Sec.~\ref{sec:singlet-triplet-scoto}. 
There are two \textit{dark mediated} Feynman diagrams which contribute to the decay rate $\mu \to e \gamma$~\cite{Restrepo:2019ilz,Rocha-Moran:2016enp,
Avila:2019hhv,
Karan:2023adm}, seen in Fig.~\ref{fig:mutoe}. 
The diagram with the neutral fermions $F_i$ running in the loop is common to scotogenic models, whereas the one with the charged $\Sigma^-$ mediator is present only in the singlet-triplet revamped extension. 
As we will see below, revamping impacts the expected cLFV phenomenology. \par

The analytic expressions for $\mu\to e\gamma$, $\mu\to 3e$ and $\mu - e$ conversion are given in Ref.~\cite{Rocha-Moran:2016enp} and we have confirmed them using the
FalvorKit~\cite{Porod:2014xia} of SARAH~\cite{Staub:2013tta} coupled to SPheno~\cite{Porod:2011nf} routines.  \\[-.3cm]
\begin{figure}[!h]
\centering
\includegraphics[height=6cm,width=0.48\textwidth]{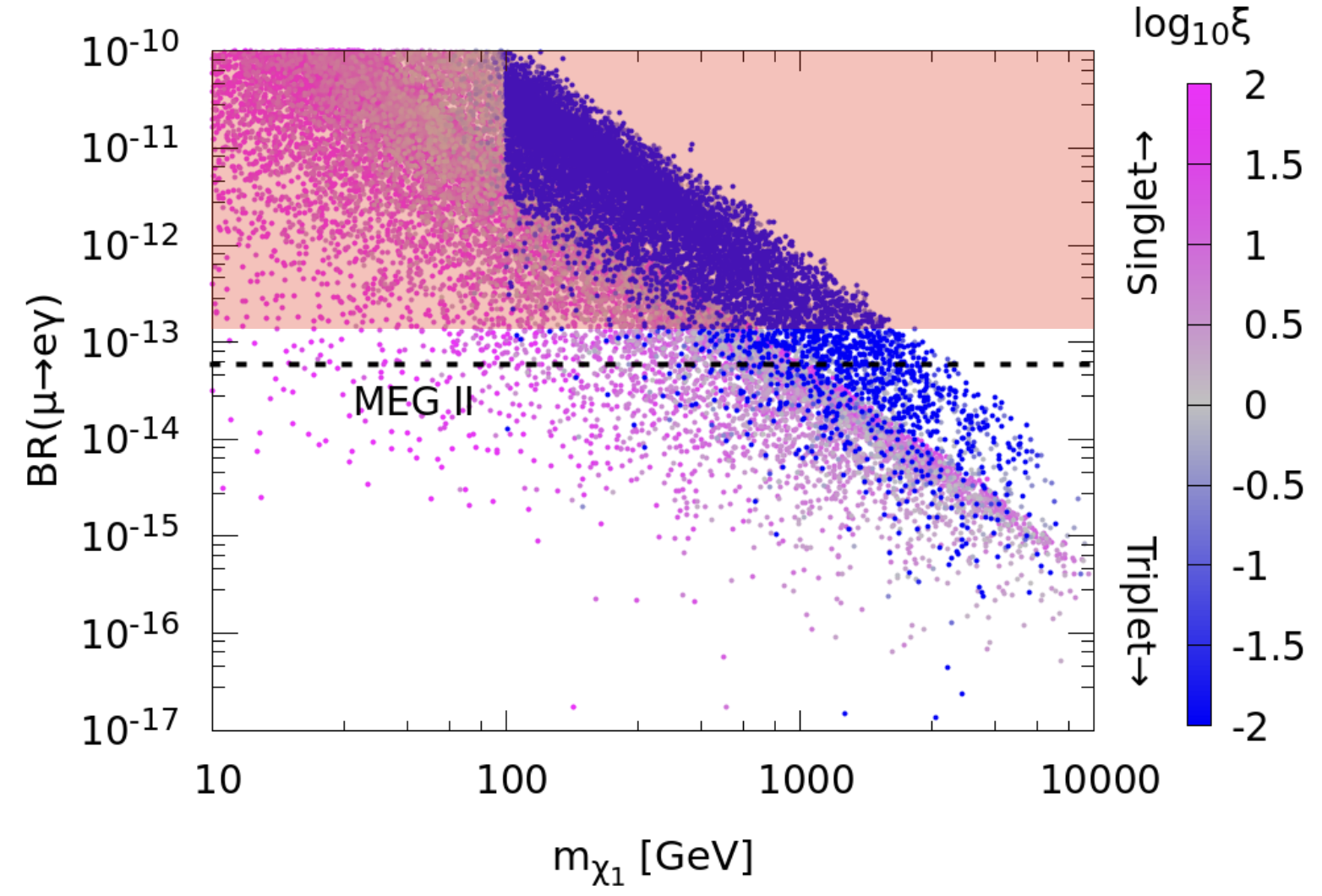}
\includegraphics[height=6cm,width=0.48\textwidth]{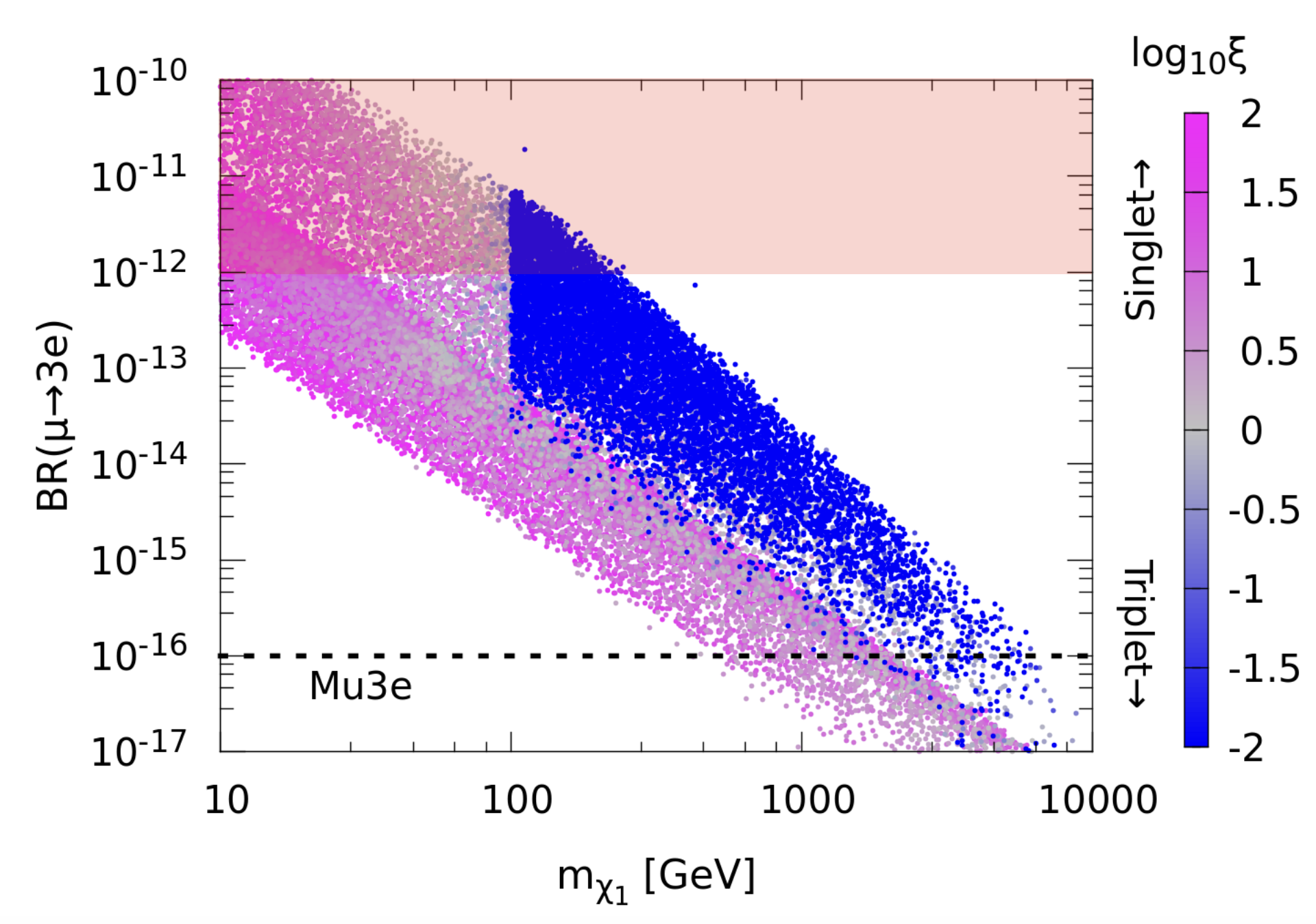}
\caption{
$\text{BR}(\mu\to e\gamma)$ and $\text{BR}(\mu\to 3e)$ as a function of the DM mass $m_{\chi_1}$ in the revamped scotogenic model. The color bar shows the triplet-singlet DM composition parameter $\xi$ in Eq.~\eqref{eq:dmness}. For high $\xi$ values we have mainly singlet
fermion DM ($F$), while vanishing $\xi$ corresponds to the limit of triplet fermion DM ($\Sigma^0$). The black dashed lines in both panels are the future projections from the MEG II~\cite{MEGII:2018kmf} and Mu3e~\cite{Blondel:2013ia,Mu3e:2020gyw} experiments, respectively.} 
\label{fig:cLFV_STM}
\end{figure}

\par 

When the revamped fermionic DM is mostly singlet, the phenomenology of dark-matter is specified by Yukawa interactions, hence one expects a correlation with cLFV phenomena, as in the simplest scotogenic model. 
On the other hand, for a mainly triplet fermionic dark-matter candidate, the DM phenomenology is determined by gauge interactions, with a strong interplay with cLFV physics. Notice the charged triplet contribution in Fig.~\ref{fig:mutoe}. 
Typically, one finds that cLFV rates decrease for large DM masses, as expected. 
For comparison, as we saw in Table~\ref{tab:LFV}, the MEG II  experiment excludes $\mu \to e \gamma$ branching ratio up to $1.5 \times 10^{-13}$~\cite{MEGII:2025gzr}. \\[-.3cm]

In Fig.~\ref{fig:cLFV_STM} we show $\text{BR}(\mu\to e\gamma)$~(left panel) and $\text{BR}(\mu\to 3e)$~(right panel) versus the fermionic DM mass $m_{\chi_1}$ in the full-fledged singlet-triplet revamped scotogenic model.
One sees that the correlation between dark-matter and cLFV phenomena is more involved. 
The color bar indicates whether the fermionic DM is triplet or singlet-like. High values of $\xi$ is the limit for singlet fermionic dark-matter, while low $\xi$-values correspond to triplet-like fermionic dark matter $\Sigma^0$. 
Notice that because of LEP constraints on new exotic charged leptons, as described in section~\ref{sec:fdm-Constraints}, triplet-like fermionic DM (shown by the blue points) must have mass higher than 102 GeV~\cite{OPAL:2003zpa,L3:2001xsz}.
 The horizontal black-dashed lines in the left and right panels are the future projected sensitivities expected at the MEG II~\cite{MEGII:2018kmf} and Mu3e~\cite{Blondel:2013ia,Mu3e:2020gyw} experiments, respectively.
\subsection{Dark linear seesaw}

We first note that in dark inverse seesaw schemes the cLFV phenomena will proceed through the usual charged-current (CC) weak interaction contribution, mediated by heavy singlet neutrino exchange~\cite{Bernabeu:1987gr,Gonzalez-Garcia:1991brm,Deppisch:2005zm},
a feature characteristic of all usual seesaw extensions~\cite{Minkowski:1977sc,Marciano:1977wx,Cheng:1980tp,Lim:1981kv,Langacker:1988up,Alonso:2012ji,Ilakovac:1994kj,DeRomeri:2016gum}.
In contrast to the original scotogenic model and most of its extensions, 
in the dark inverse seesaw the dark sector is \textit{hidden}, i.e. singlet under the \sm \SM gauge group, and therefore does not play a direct role in inducing cLFV processes. \\[-.2cm]

\begin{figure*}[tbh]
\begin{center}
\includegraphics[height=3.cm,scale=0.3]{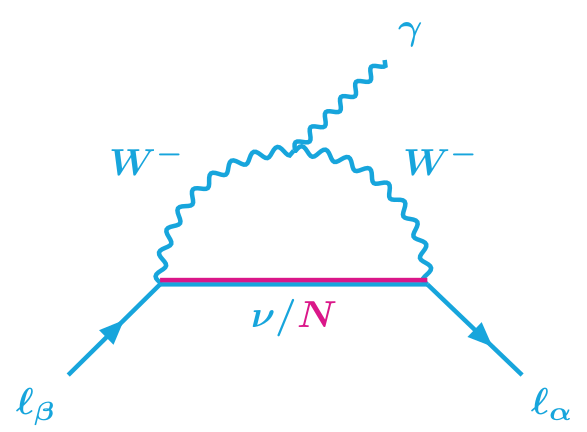}
\hspace{10mm}
\includegraphics[height=3.cm,scale=0.3]{Plots/mu_egam_scoto_new.png}
\end{center}
\caption{
  Feynman diagrams that contribute to $l_{\beta}\to l_{\alpha}\gamma$ processes in the dark linear seesaw.
  The left diagram shows the charged-current contribution, whereas the one in the right shows the dark-sector contribution.}
\label{fig:figmuegFeyn}
\end{figure*}

For this reason we move directly to discuss the cLFV implications within the dark linear seesaw approach. In particular, we focus on the radiative decays
$\ell_\beta\to\ell_\alpha\gamma$, the most sensitive of which is the $\mu \to e\gamma$ decay.
The relevant amplitudes inducing $\mu \to e\gamma $ decay involve the Feynman diagrams in Fig.~\ref{fig:figmuegFeyn}. 
Besides the usual CC contribution, characteristic of type-I seesaw (left panel) there is a contribution arising from the dark sector (right diagram). \\

In order to estimate the seesaw-type contribution to cLFV rates, it is convenient to parametrize the ``Dirac'' submatrix $m_D$ in terms
of the measured oscillation parameters. In the same spirit as~\cite{Casas:2001sr} one can give an approximate expression for extracting the $m_D$ parameters as follows~\cite{Forero:2011pc,Cordero-Carrion:2019qtu},
\begin{equation}
m_{D} = U_{\rm lep} \sqrt{m_\nu} A^{T} \sqrt{m_\nu} U_{lep}^{T}\left(
\varepsilon ^{T}\right) ^{-1}M^{T} , ~~~~~~~\textrm{with} ~~~ A =\left( 
\begin{array}{ccc}
\frac{1}{2} & a & b \\ 
-a & \frac{1}{2} & c \\ 
-b & -c & \frac{1}{2}%
\end{array}
\right) , \label{mdeqn}
\end{equation}
where $a$, $b$ and $c$  are taken to be real numbers, $m_\nu = {\rm diag}\left( {m_{1}},{m_{2}},{m_{3}}\right)$ is given by the light neutrino masses. We assume the basis in which the charged lepton mass matrix is diagonal.
From this expression it is clear that, given $m_\nu$ from oscillation measurements, one can fix $m_{D}$ in accordance with
  the corresponding mass scale ratio $\varepsilon / M$ so as to fit the oscillations. Likewise, concerning mixing parameters,
  one can choose the off-diagonal entries of $m_{D}$ to match the measured “solar” , “atmospheric”  and “reactor” angles in $U_{\rm lep}$.
  In fact, taking as a simple ansatz $M$ and $\varepsilon$ diagonal and proportional to identity, one sees that $m_{Dij}$
  can be chosen so as to fit the observed oscillation mixing angles. 

In order to determine the CC amplitude, the key ingredient is the full lepton mixing matrix, which has a rectangular form~\cite{Schechter:1980gr}. 
Such rectangular mixing matrix describes not only the CC contributions of the light neutrinos, 
but also those of the heavy mediators, weighted by a $\mathcal O \left(\frac{m_{D}}{M}\right)$ factor~\cite{Schechter:1981cv}. 
We find that the CC light-neutrino contribution to the $\mu \to e\gamma $ decay can be sizeable, thanks to effective unitarity violation of the relevant lepton-mixing sub-matrix. 
Moreover, one has potentially large contributions also due to the exchange of the six sub-dominantly coupled heavy quasi-Dirac states, that can lie at the TeV scale. \\

Concerning the dark-sector contributions, we stress that the same dark sector Yukawa couplings $Y^{(\eta)}$ generating neutrino masses radiatively via the linear seesaw mechanism, can also give rise to cLFV processes. 
Indeed, in the model of Ref.~\cite{CarcamoHernandez:2023atk} the dark charged scalar $\eta ^{\pm }$ and the dark fermions $F_{i}$ mediate cLFV through the couplings $Y^{(\eta)}$, as shown in Fig.~\ref{fig:figmuegFeyn}. 
The total branching ratio for the process $\mu \rightarrow e \gamma $ thus takes the form \cite{Langacker:1988up,Lavoura:2003xp},
\begin{align}
\text{Br}\left( \mu \to e \gamma \right) & =\frac{3\left( 4\pi \right)\alpha _{em}}{4G_{F}^{2}} \left\vert \sqrt{\frac{%
\alpha _{W}^{2}s_{W} }{m_{W}^{4}}}%
\sum_{k=1}^{9}K_{2k}^{\ast }K_{1k}\tilde{G}_{F}\left( \frac{M_{k}^{2}}{m_{W}^{2}}%
\right) \right.  
\left. + 
\sum_{k=1}^{3}\frac{Y_{2k}^{\left( \eta \right) }Y_{2k}^{\left( \eta \right)
}}{2 m_{\eta ^{\pm }}^{2}}F_{2}\left( 
\frac{m_{F_{k}}^{2}}{m_{\eta ^{\pm }}^{2}}\right) \right\vert ^{2},\hspace{%
0.5cm}\hspace{0.5cm}\hspace{0.5cm}\hspace{0.5cm}  \label{Brmutoegamma} \\
&\text{with, }  
\tilde{G}_{F}\left( x\right) =\frac{10-43x+78x^{2}-49x^{3}+18x^{3}\ln x+4x^{4}}{%
12\left( 1-x\right) ^{4}}.
\end{align}

\begin{figure*}[tbh]
  \includegraphics[height=6cm,scale=0.75]{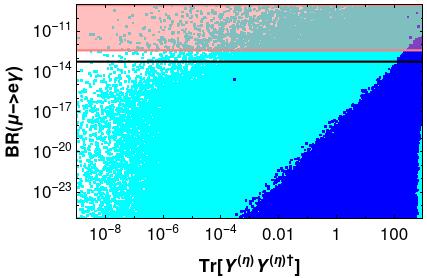}
\caption{
Correlations of $\text{Br}(\protect\mu \to e \protect\gamma)$ with the Yukawa coupling strengths Tr$[Y^{(\eta)} {Y^{(\eta)}}^\dag]$ in the dark linear seesaw. Cyan points give the CC contribution, whereas blue points show the dark-sector contribution. Note that for a strong perturbativity requirement on the dark Yukawa coupling the dark contributions always lie below the current limit.} 
\label{fig:figmueg}
\end{figure*}

In Eq. (\ref{Brmutoegamma}), the matrix $K$ is the $3\times 9$ rectangular mixing matrix describing the CC weak interaction
and includes the exchange of the three light active neutrinos with $k=1,2,3$ as well as the six mediators, with $k=4,5,..9$.
As mentioned earlier the latter form three quasi-Dirac heavy-neutrino pairs. The full lepton mixing matrix $K$ is given by 
\begin{equation}
K=\left( K_{L},K_{H}\right) ,
\end{equation}
where $K_{L}$ and $K_{H}$ are $3\times 3$ and $3\times 6$ matrices, respectively. These submatrices take the form:
\begin{eqnarray}
K_{L} &=&\left( 1_{3\times 3}-\frac{1}{2}m_{D}\left( M^{-1}\right)
^{T}M^{-1}m_{D}^{\dagger }\right) U_{\rm lep}=\left( 1_{3\times 3}-\frac{1}{2}%
VV^{\dagger }\right) U_{\rm lep}\,, \\
K_{H} &=&\left( -\frac{i}{\sqrt{2}}V,\frac{1}{\sqrt{2}}V\right) ,\hspace{1cm}%
~~~~~\mathrm{with}~~~~
V=m_{D}\left( M^{-1}\right) ^{T}.
\end{eqnarray}
 In Fig.~\ref{fig:figmueg}, we present the results for $\text{Br}(\protect\mu \to e \protect\gamma)$ versus  Tr$[Y^{(\eta)} {Y^{(\eta)}}^\dag]$. 
In order to optimize our parameter scan to generate these figures, ensuring that only viable solutions consistent with neutrino oscillation data are included, it is convenient to use the analytical approximation in Eq.~(\ref{mdeqn}). 
However, in presenting the numerical results we use the exact expressions for the diagonalization matrices.\\

\renewcommand{\arraystretch}{1.8}
\begin{table}[h]
\setlength\tabcolsep{0.25cm} \centering 
\begin{tabular}{|c||c|c|c|c|c||c|c|c|}
\hline
Parameters & $Y_{ij}^\eta$ & $ Y_{ij}^\xi = y\delta_{ij}$ & $m_{F_i}$ & $M_{ij} =M_N \delta_{ij}$ & $m_{D_1,D_2,D_{A_1},D_{A_2},\eta^{\pm}} $  & $\theta_{D}$ & $a,b,c$ \\ \hline
Range  & $[10^{-10}, 4\pi]$ & $[10^{-16},4\pi]$ & 
$[200,5000]$ GeV & $[200,5000]$ GeV  & $[200,5000]$ GeV  &0.01 & [-20,20] \\ \hline
\end{tabular}%
\caption{ 
The regions used in generating Fig.~\ref{fig:figmueg}.}
\label{tab:scan}
\end{table}

\par
In generating Fig.~\ref{fig:figmueg}, the neutrino oscillation parameters are varied in their $3\sigma $ ranges~\cite{deSalas:2020pgw,10.5281/zenodo.4726908},
the parameters $a$, $b$ and $c$ are varied in the range $ [-20,20]$ and the couplings $Y_{ij}^{(\eta)}$ are varied up to $4\pi$. 
For simplicity we took the heavy neutrino mediators as degenerate, varying their masses in the range $[200,5000]$ GeV. 
Concerning the dark sector parameters, $Y_{ij}^{(\xi)}$ is taken as $y \delta_{ij}$ with $y$ varied up to $4\pi$, while the dark mediator masses are varied in the range $[200,5000]$ GeV,  and the scalar mixing angle $\theta _{D}$ is fixed to be 0.01 implying $\theta _{D_A} = -0.01$.
The scan region is summarized in Table \ref{tab:scan}. 
The cyan and blue points in Fig.~\ref{fig:figmueg} show the CC and the dark-sector contributions to $\text{Br}(\protect\mu \to e \protect\gamma)$, respectively. 
The horizontal pink-shaded region corresponds to the bound obtained by the MEG experiment in~\cite{MEG:2016leq},
whereas the black line corresponds to the projected future sensitivity of $6 \times 10^{-14}$ for MEG-II~\cite{Baldini:2013ke,Meucci:2022qbh}.
For strongly perturbative $|Y^{(\eta)}|$ values, we find that the dark-sector contribution consistently lies below the current experimental limit, in contrast to the CC contribution that can exceed it.

\subsection{Scoto-seesaw}

\begin{figure}[ht!]
    \centering
    \includegraphics[height=3.5cm,scale=0.25]{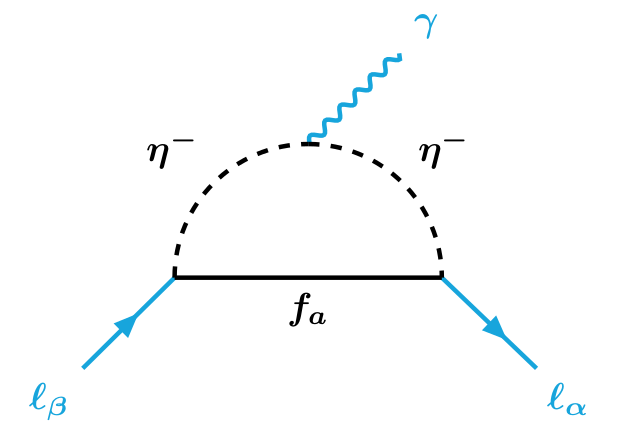}
    \includegraphics[height=3.5cm,scale=0.25]{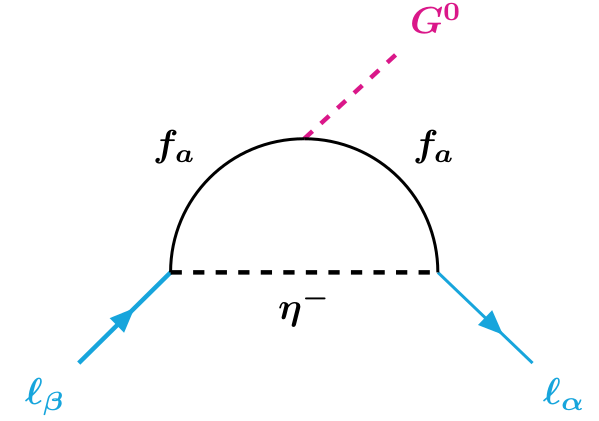}
    \caption{ 
    Leading contribution to $\mu \to e \gamma$ and to the novel cLFV decay $\mu \to e G^0$ through penguin diagrams~\cite{Leite:2023gzl}.}
    \label{fig:feyn_lfv_dyn_scoto}
\end{figure}

As we saw there is an interesting hybrid approach for putting together neutrino masses and dark matter, namely the scoto-seesaw mechanism.
Its main motivation is to reproduce the required value of the neutrino mass scales observed in oscillation experiments, Eq.~\eqref{eq:ratio}, by combining the seesaw and the scotogenic mechanism, Eq.~\eqref{eq:atm-sol-mass}.
Such hybrid constructions contain sources of lepton flavor  violation arising from the dark Yukawa couplings $Y_F$, as well as from the conventional Dirac Yukawa coupling $Y_N$. The first is associated to the \textit{dark} sector, while the second comes from the \textit{seesaw} mechanism.
For definiteness here we consider the dynamical scoto-seesaw model of Ref.~\cite{Leite:2023gzl}.
A novel feature of that scheme is the presence of a Nambu-Goldstone boson $G$, that can also lead to dark-mediated cLFV decays, in addition to the most standard ones, such as the $\mu \to e \gamma$ process, see Fig.~\ref{fig:feyn_lfv_dyn_scoto}.

\begin{figure}[ht!]
    \begin{minipage}{0.2\textwidth}
    \end{minipage}
    \begin{minipage}{0.58\textwidth}
    \includegraphics[height=5cm, width= 7cm]{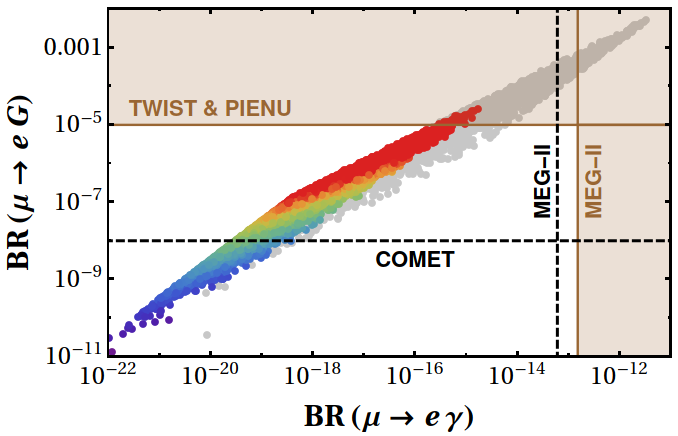}
    \end{minipage}
    \hspace{-3cm}
    \begin{minipage}{0.2\textwidth}
\vspace{-0.5cm}\includegraphics[trim={5mm 5mm  5mm 5mm},clip, height=4.5cm]{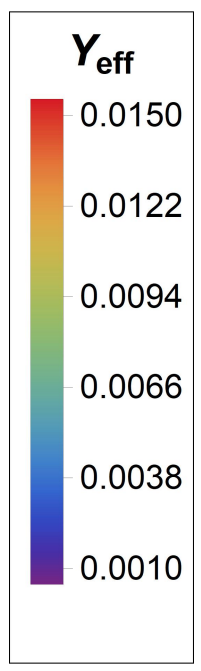}
\end{minipage}
    \caption{ 
    Correlations between $\mu \to e \gamma$ and $\mu \to e G$ decay branching ratios in dynamical secoto-seesaw~\cite{Leite:2023gzl}. Shaded bands are excluded by current experiments, while gray points are ruled out by stellar cooling bounds. The black-dashed lines are projected sensitivities, and colored dots represent varying Yukawa coupling values.}
    \label{fig:lfv_dyn_scoto}
\end{figure}

We compute these in the approximation $m_e/m_\mu\ll 1$.
The leading-order expressions for the branching ratios of the cLFV processes $\mu \to e \gamma$ and $\mu \to e G$ are given by: 
\begin{eqnarray}
    & \displaystyle \text{Br}(\mu \to e \gamma) \simeq \frac{\alpha_{\text{em}}\,m_\mu^5}{4\,m_{\eta^\pm}^4\,\Gamma_\mu}\,\bigg|\Big(Y^\eta \mathcal F_1 Y^{\eta\dagger}\Big)^{e\mu}\bigg|^2, \\
     & \displaystyle \text{Br}(\mu \to e G) \simeq \frac{m_\mu^3|\braket{G|\varphi_2}|}{32 \pi\, v_{\varphi_2}^2\, \Gamma_\mu}\,\bigg|\Big(Y^\eta \mathcal F_2 Y^{\eta\dagger}\Big)^{e\mu}\bigg|^2, 
\end{eqnarray}
where,
$ \displaystyle \mathcal{F}_1^{ab}= \frac{1-6x_a+3x_a^2+2x_a^3-6x_a^2 \log x_a}{96 \pi^2(1-x_a)^4} \,\delta_{ab}$~~~and~~~
$\mathcal F_2^{ab}= \frac{x_a(x_a-1-\log x_a)}{8\pi^2(x_a-1)^2}\,\delta_{ab}~$.\\[.2cm]
Here, $x_a=(m_{f_a}^2/m_{\eta^\pm}^2)$, where the indices $(a,b)$ take values 1 and 2, and $\braket{G|\varphi_2}$ denotes the projection of the Goldstone boson along $\varphi_2$, a singlet scalar present in the model.
Fig.~\ref{fig:lfv_dyn_scoto} illustrates how the expected rates for the $G$-emitting process are more promising than those of the 
standard $\mu \to e \gamma$  process and how they are correlated. 
These cLFV processes are mainly induced by TeV-scale dark particle loops.  
In order to display these  branching ratios, model parameters are fixed as: $\lambda_i=0.1$, $\mu_\eta=\mu_\sigma=1$ GeV, $\mu_1=5$ TeV, and $Y^N=Y^f_a=0.5$. 
Neutrino mass restrictions are incorporated assuming normal mass ordering. 
The VEV $v_\varphi$ is randomly varied between 1 TeV and 15 TeV, so the induced leptophilic $v_\Phi$ values range between 1 MeV to 280 MeV. 

As we saw in Table.~\ref{tab:LFV} the MEG-II experiment~\cite{MEGII:2025gzr} excludes the $\mu \to e \gamma$ branching ratio up to $1.5 \times 10^{-13}$, while the $\mu\to e G$ branching ratio is constrained to be below $10^{-5}$ by the TWIST~\cite{TWIST:2014ymv} and PIENU~\cite{PIENU:2020loi} experiments. 
These limits are shown as brown-shaded regions in the figure. 
Although Jodidio et al.~\cite{Jodidio:1986mz} report a stronger bound on $\text{Br}(\mu\to e G)$ of $2.6 \times 10^{-6}$, in the present scenario this constraint could be relaxed by an order of magnitude~\cite{Hirsch:2009ee}. \par 
On the other hand, stellar cooling constrains the diagonal Goldstone boson couplings to fermions~\cite{Bollig:2020xdr}, the corresponding excluded points being shown in gray. The color of different points represent the magnitude of Yukawa coupling combination, defined as: $Y_\text{eff}=\sqrt{|\sum_a Y_{ea}^\eta Y_{\mu a}^{\eta*}|}$.  The projected sensitivities of the MEG-II experiment for $\text{Br}(\mu\to e\gamma) \lsim 6 \times 10^{-14}$~\cite{MEGII:2021fah} and the COMET experiment~\cite{COMET:2018auw} for $\text{Br}(\mu\to eG) \lsim \mathcal O (10^{-8})$~\cite{Xing:2022rob} are indicated by the black dashed lines. One clearly sees from Fig. \ref{fig:lfv_dyn_scoto} that substantial improvement in the sensitivity of $\text{Br}(\mu\to e\gamma)$ is needed in order to comply with the current limit on $\text{Br}(\mu\to eG)$.
However, the predicted $\mu \to e \gamma$ and $\mu \to e G$ branching fraction correlations may depend on the parameter scan, see~\cite{Leite:2023gzl} for details.

All in all, this scheme gives rise to significant charged lepton flavor violating processes mediated by the dark sector particles.
Besides, due to the gauged B-L symmetry, there are also collider signatures associated to the Drell-Yan production of the seesaw messenger associated to the atmospheric mass generation, which can also lie at the TeV scale. We now turn to general collider implications of dark models of neutrino masses.


\section{Collider Implications }
\label{sec:collider}

 The new particles present in the models we have discussed above, such as those comprising the dark sector, can often produce signatures at $pp$ colliders, such as the LHC, the future FCC experiment now under discussion, or the
planned lepton collider proposals such as CEPC~\cite{CEPCStudyGroup:2018ghi,Antusch:2025lpm}, ILC~\cite{Barklow:2015tja}, CLIC~\cite{CEPCStudyGroup:2018ghi,FCC:2018evy,Bambade:2019fyw,CLIC:2018fvx,Antusch:2025lpm} or a muon collider~\cite{Delahaye:2019omf,Li:2023tbx,InternationalMuonCollider:2025sys,Stratakis:2025ofm,InternationalMuonCollider:2025sys}. 
 Indeed, the new TeV-scale dark states involved in generating neutrino masses could be produced at upcoming colliders. 
 
 In this section, we illustrate some of the related collider signatures for various model options. 
 We start with the production mechanisms and decay channels, and then discuss some of the associated phenomenological signatures across different facilities such as $e^+e^-/\mu^+\mu^-$ and $pp$ collisions. 
 Collider signatures will heavily depend on the nature of dark matter. Scalar production is dominated by the Drell-Yan mechanism but can be substantially enhanced by the t-channel exchange of dark fermions. Apart from usual dark matter signatures involving $\gamma+\slashed{E}_T$, $j+\slashed{E}_T$ or $\ell+\slashed{E}_T$, one can have interesting multi-lepton signatures with $\slashed{E}_T$. 
 More interestingly, for the case of a compressed dark sector mass spectrum or, alternatively, if the scotogenic dark matter is a FIMP, some of the dark scalars can be long-lived particles (LLPs). 
On the other hand, for the case of fermionic dark-matter in the revamped singlet-triplet scotogenic scenario 
 bears a close similarity with \textit{vanilla} supersymmetric neutralino dark-matter, except for the close inter-connection between dark-matter physics and neutrino phenomena, including cLFV processes, Table~\ref{tab:LFV}. 
\subsection{Simplest Scotogenic} 
\label{subsec:collider-scotogenic}

\begin{figure}[h]
\centering
\includegraphics[height=3.3cm, width=4.5cm]{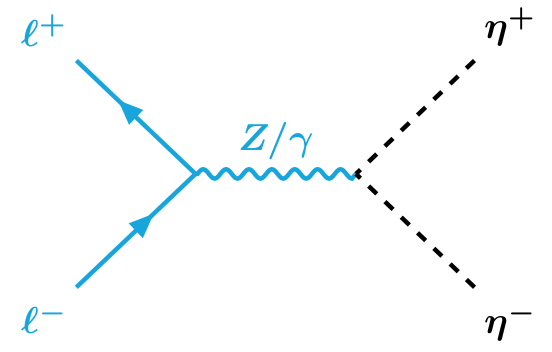}~~
\includegraphics[height=3.3cm, width=4.5cm]{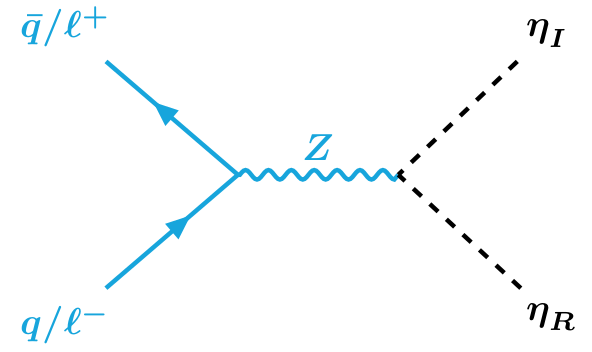}~~
\includegraphics[height=3.3cm, width=4.5cm]{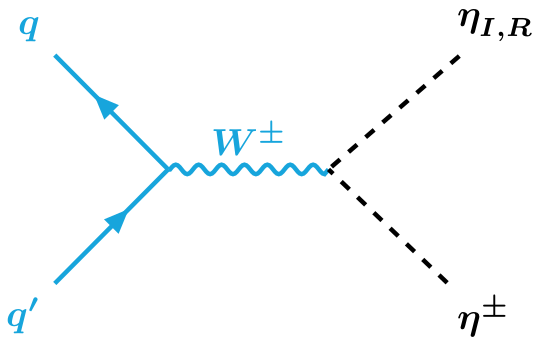}
\caption{
Dark scalar production channels. At $pp$ colliders $\eta^\pm$, $\eta_{R/I}$ are produced by neutral or charged current Drell-Yan mechanisms. 
At lepton colliders charged scalars can also be produced by t-channel  dark fermion exchange.} 
\label{fig:production-scotogenic-1}
\end{figure}

At proton-proton colliders, pair production of dark scalars proceeds via the neutral or charged-current Drell-Yan processes involving s-channel $\gamma/Z$ or $W^\pm$ exchange, as seen in Fig.~\ref{fig:production-scotogenic-1}.

\begin{figure}[!h]
\centering
\includegraphics[height=5.cm,width=0.45\textwidth]{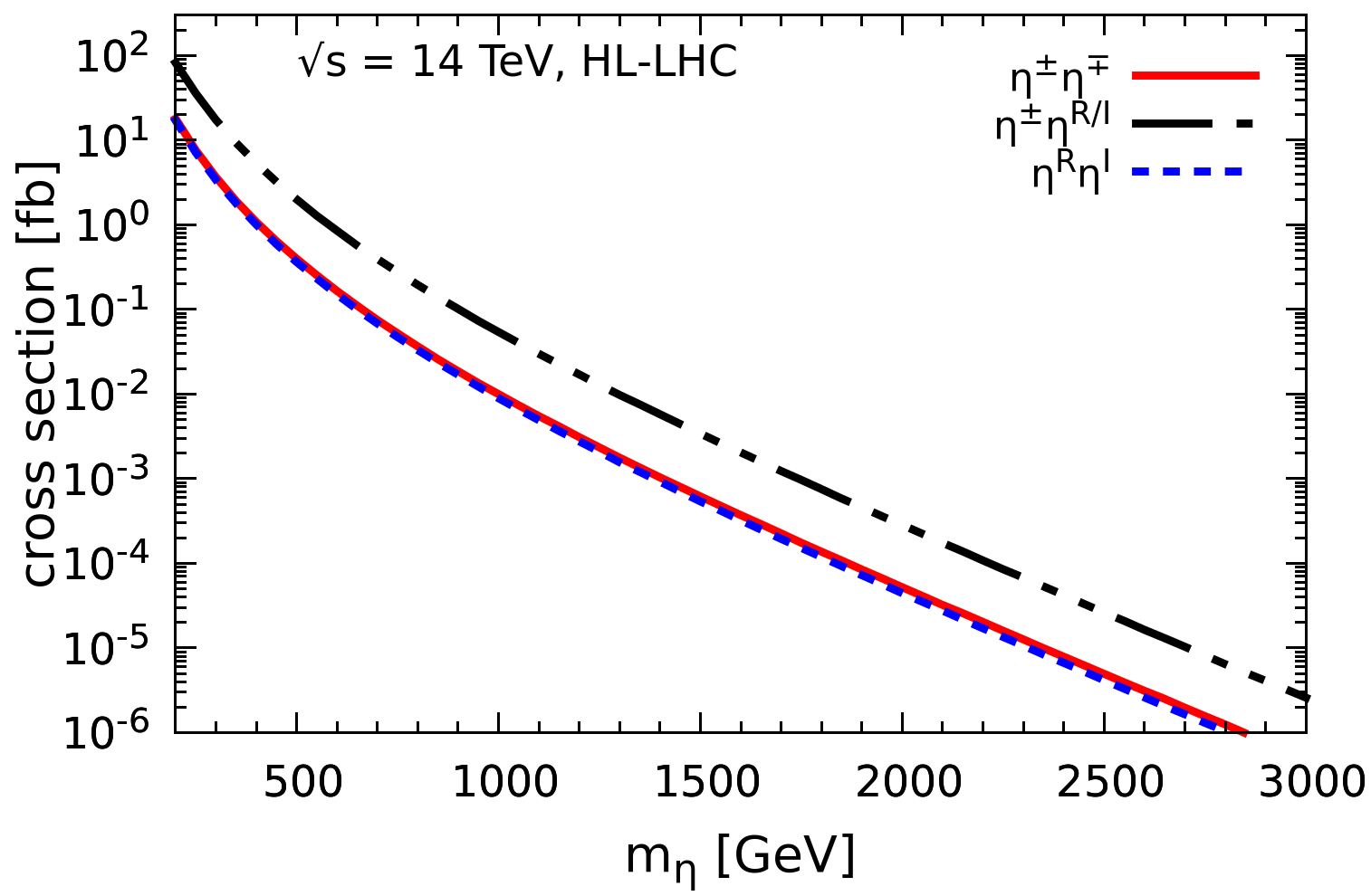}
\includegraphics[height=5.cm,width=0.45\textwidth]{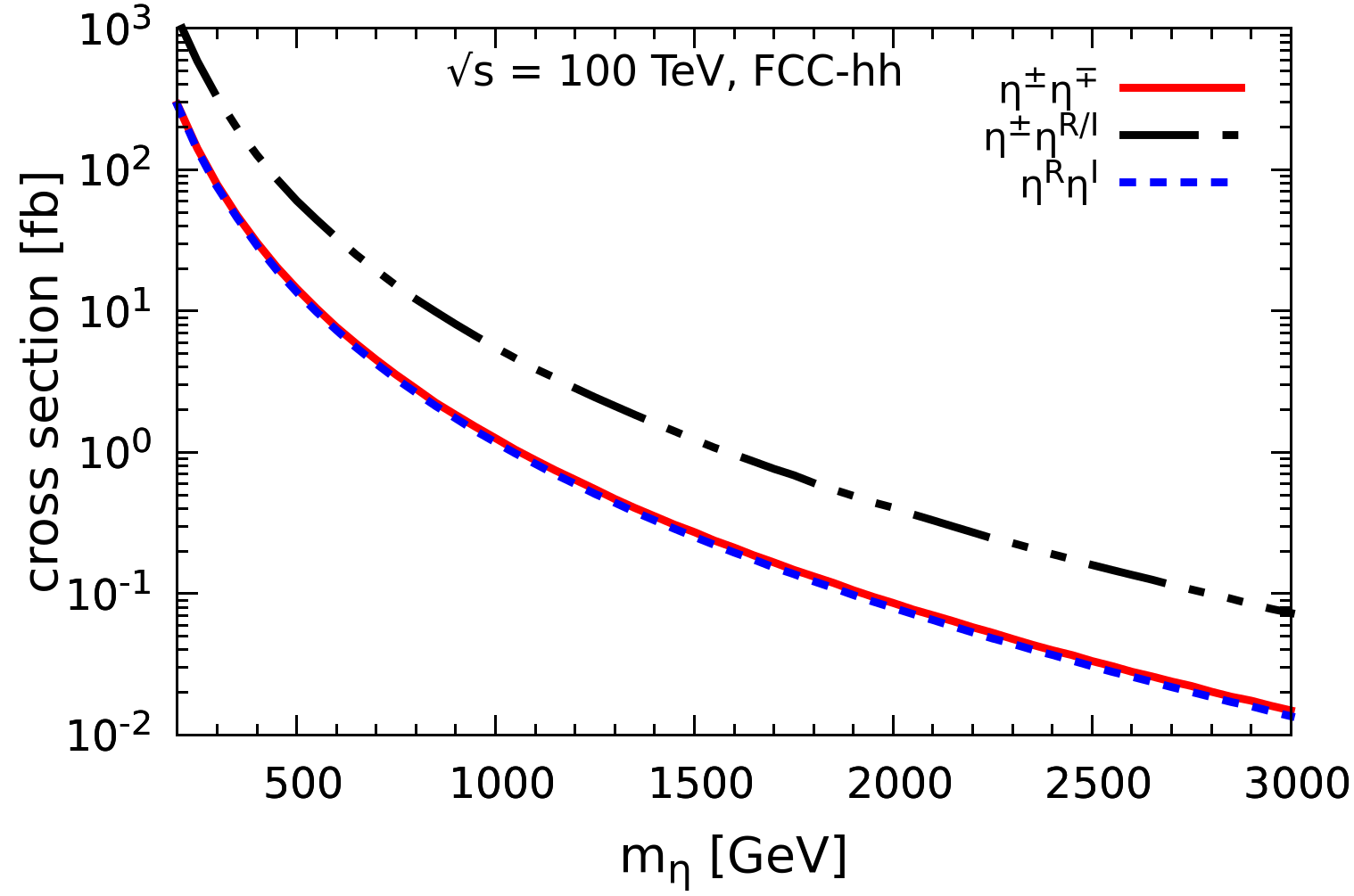}
\caption{
$\eta$-doublet scalar pair and associated Drell-Yan production cross-section versus the scalar mass $m_\eta$ for a $pp$ collider with $\sqrt{s}=14$ TeV~(left) and $\sqrt{s}=100$ TeV~(right).}
\label{fig:xs-scotogenic-pp}
\end{figure}

In Fig.~\ref{fig:xs-scotogenic-pp} we show the associated Drell-Yan cross sections versus the doublet scalar mass $m_\eta$ for $\sqrt{s}=14$ TeV~(HL-LHC~\cite{Apollinari:2015wtw}) and $\sqrt{s}=100$ TeV~(FCC-hh~\cite{FCC:2018byv}), respectively. 
For simplicity we have assumed a compressed mass spectrum in the dark scalar sector, $m_{\eta^\pm}\approx m_{\eta_R}\approx m_{\eta_I}$. \\

\begin{figure}
    \centering
\includegraphics[height=5.cm,width=0.49\textwidth]{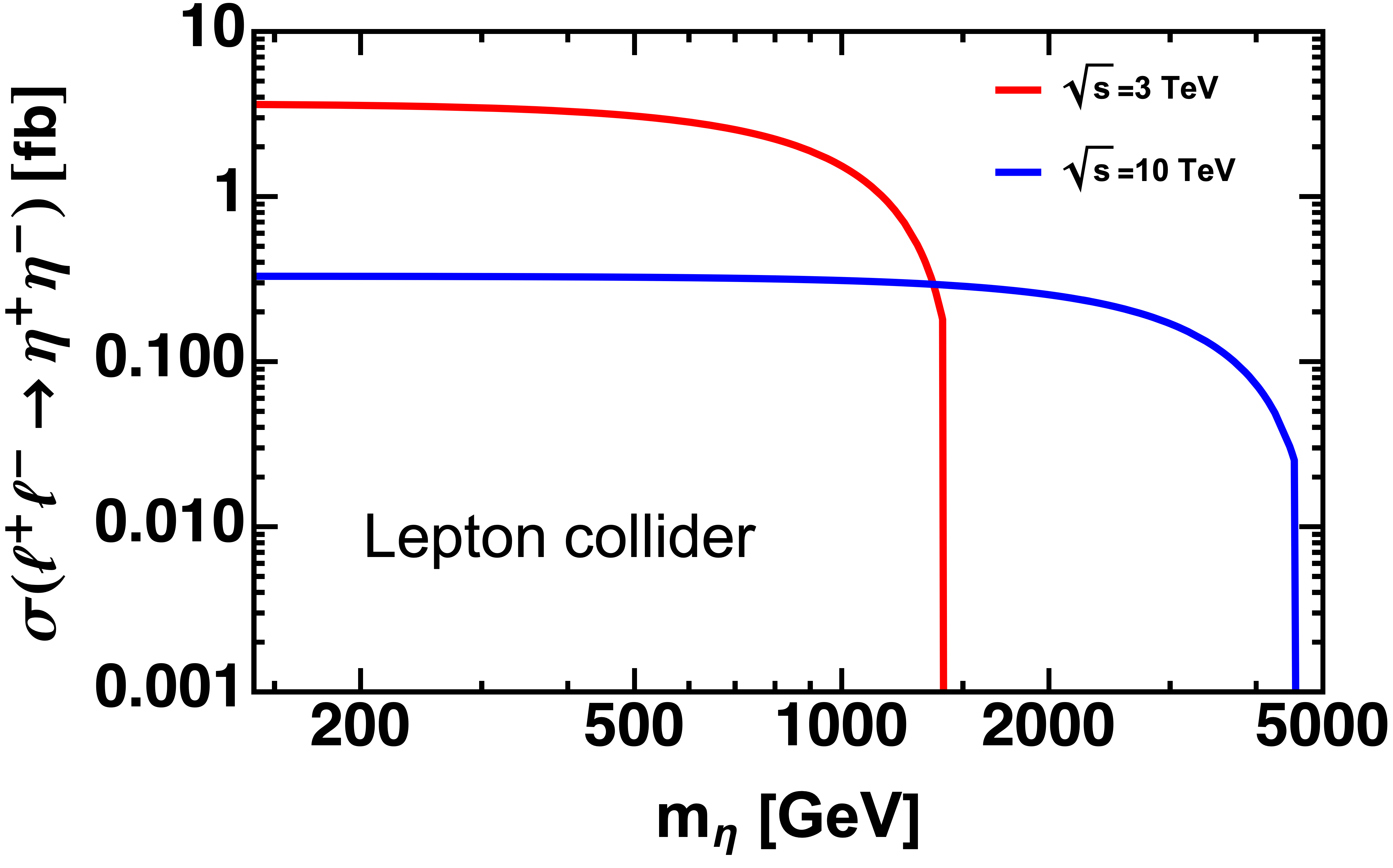}
\includegraphics[height=5.cm,width=0.49\textwidth]{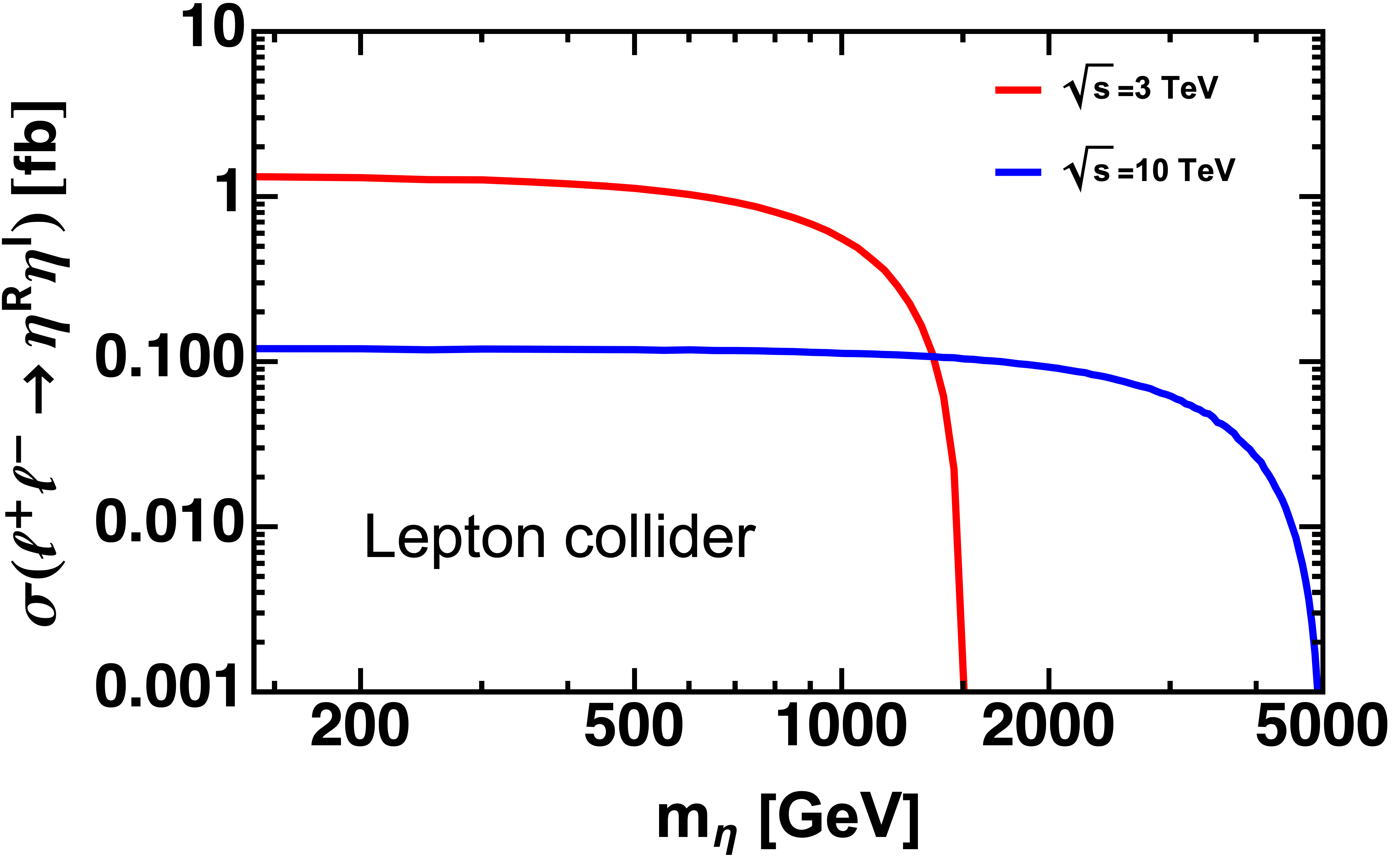}
\caption{
Dark doublet scalar pair and associated Drell-Yan production cross-section versus the scalar mass $m_\eta$ at a lepton collider~($\ell=e,\mu$) at $\sqrt{s}=3$ TeV~(red) and $\sqrt{s}=10$ TeV~(blue), respectively.}
\label{fig:xs-scotogenic-ll}
\end{figure}

Charged dark scalar pair production ($\eta^\pm\eta^\mp$) and the associated production of neutral dark scalars~($\eta_R\eta_I$) involves the s-channel exchange of $\gamma/Z$ and $Z$ boson, respectively, as seen in Fig.~\ref{fig:production-scotogenic-1}. 
The scalar pair and associated Drell-Yan production cross-sections 
of dark scalars at lepton colliders is shown in Fig.~\ref{fig:xs-scotogenic-ll} at center-of-mass energies $\sqrt{s}=3$ TeV~(red) and $\sqrt{s}=10$ TeV~(blue).  Here we have neglected the t-channel contribution to be discussed later.
One sees that production cross sections can also be sizable at planned lepton colliders such as ILC~\cite{Barklow:2015tja}, CLIC~\cite{CLIC:2018fvx,Aicheler:2012bya,CLIC:2016zwp,Abramowicz:2016zbo,CLICdp:2018cto}, CEPC~\cite{CEPCStudyGroup:2018ghi,Antusch:2025lpm}, FCC-ee~\cite{FCC:2018evy} and also at the muon collider~\cite{Delahaye:2019omf,Li:2023tbx,InternationalMuonCollider:2025sys,Stratakis:2025ofm,InternationalMuonCollider:2025sys}. 
\begin{figure}[h]
\centering
\includegraphics[height=3.0cm, width=5cm]{Plots/EpEm.png}~~~~~~~~
\includegraphics[height=3.0cm, width=3.0cm]{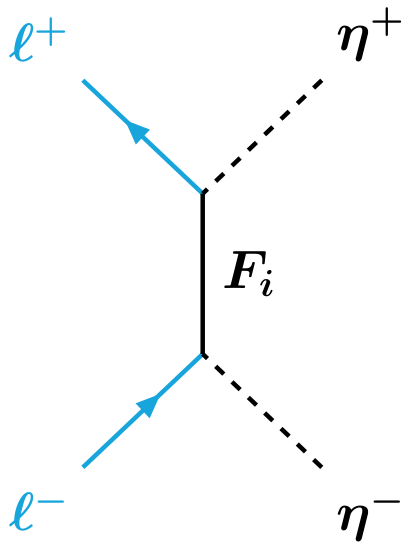}~~~~~~~
\includegraphics[height=3.0cm,width=3.0cm]{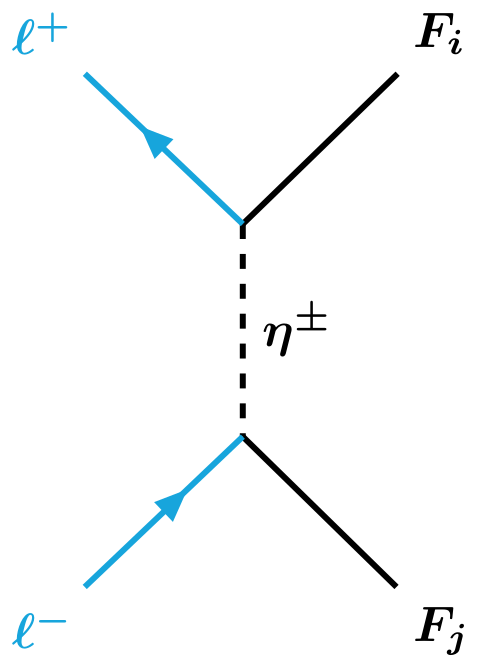}
\caption{
Feynman diagrams for dark fermion and scalar production at lepton colliders.
Left: Drell-Yan mechanism. 
Middle and right panels: t-channel diagrams for direct $\eta^\pm\eta^\mp$ production and $FF$ production, respectively.}

\label{fig:production-scotogenic-11}
\end{figure}
\begin{figure}[h]
\centering
\includegraphics[height=5.0cm,width=0.45\textwidth]{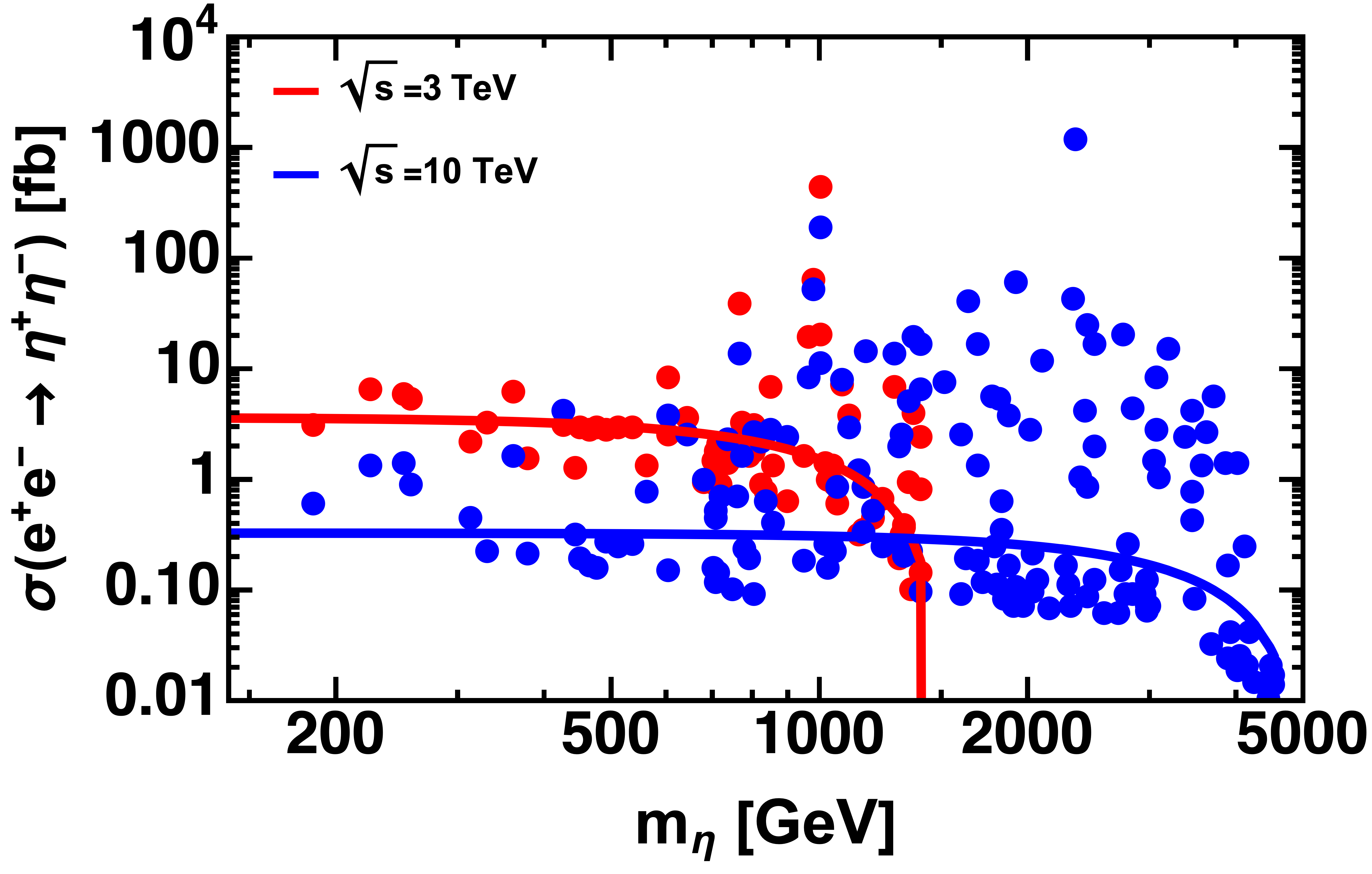}
\includegraphics[height=5.0cm,width=0.45\textwidth]{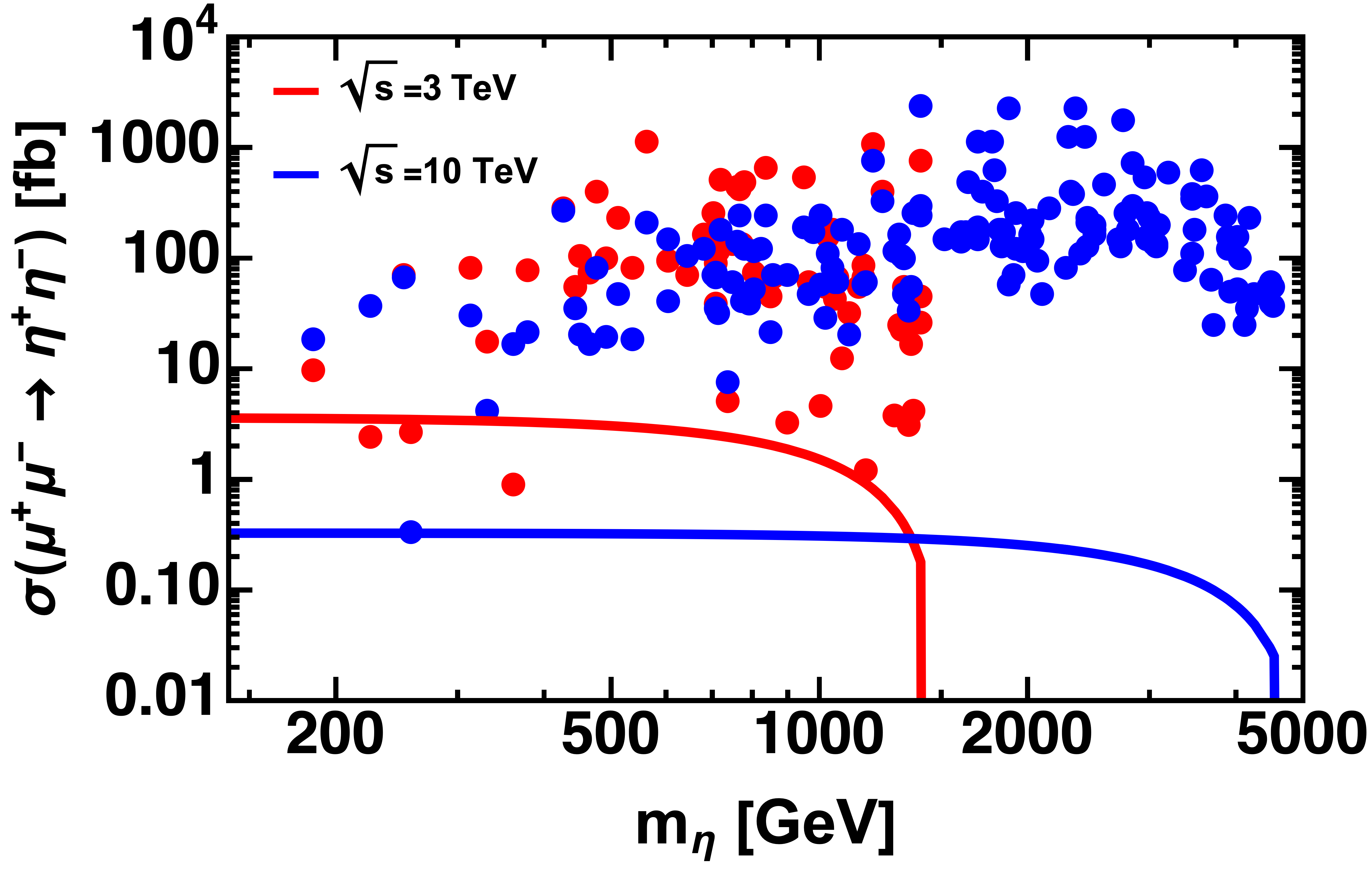}
\caption{ 
The red and blue solid lines are pure Drell-Yan cross-sections, while the red and blue points include also the t-channel contribution~for dark scalar production~(Fig.~\ref{fig:production-scotogenic-11}).}
\label{fig:xs-scotogenic-2}
\end{figure}
\par  Note that in addition to the Drell-Yan process, there can be an additional contribution to the charged dark scalar pair production at lepton colliders, via the t-channel, see Fig.~\ref{fig:production-scotogenic-11}. 
This new process is a characteristic feature of scotogenic-type models at lepton colliders, absent at $pp$-colliders. Such t-channel exchange contribution involves a new Yukawa $|Y_{F}^{\ell i}|^4$ that can be sizeable, so that the dark scalar production cross-section can even surpass the Drell-Yan contribution, similar to what happens for the leptophilic scalar in the linear seesaw scheme~\cite{Batra:2023mds,Batra:2023ssq} or in type-Ib seesaw schemes~\cite{Chianese:2021toe}. \par
In Fig.~\ref{fig:xs-scotogenic-2}, we show the cross-section for charged dark scalar production at electron~(left panel) and muon collider~(right panel) with $\sqrt{s}=3$ TeV~(red) and $\sqrt{s}=10$ TeV~(blue).  The solid red and blue lines correspond to the pure Drell-Yan contribution, whereas the red and blue points include the t-channel contributions with $Y_F^{e/\mu i}$ Yukawa coupling values consistent with the relevant constraints on the fermionic scotogenic DM candidate $F_1$.  Comparing the panels of Fig.~\ref{fig:xs-scotogenic-2} we find that the muon collider is more competitive than the $e^+e^-$ collider, for charged dark scalar pair production. 
This follows from the fact that for fermionic dark matter $F_1$, a hierarchical Yukawa structure $Y_F^{e1}\ll Y_F^{\mu 1}\sim Y_F^{\tau 1}\sim \mathcal{O}(1)$ is required in order to satisfy the cLFV constraints in Table~\ref{tab:LFV}. 
\par Having discussed the new dark scalar production modes, we now consider the associated collider signatures.  Once produced, dark scalars will decay to various possible  channels that depend on the nature of dark matter~(scalar or fermionic). For definiteness we take a compressed dark scalar mass spectrum $m_{\eta^\pm}\approx m_{\eta_R}\approx m_{\eta_I}$, so that SM gauge boson decay modes can be  neglected. 
 There are two different scenarios according to the nature of dark matter. \\ 

(1). {\bf Fermionic DM:} For the case of fermionic scotogenic dark matter $F_1$, the charged scalar $\eta^\pm$ can decay to all three families of dark fermions $F_i$. However for the mass hierarchy $M_{F_1} < m_{\eta^\pm}<M_{F_2}<M_{F_3}$ we have
\begin{align}
\Gamma(\eta^\pm\to F_1\ell_\alpha^\pm)=\frac{m_{\eta^\pm}}{16\pi}    \big|Y_F^{\alpha 1}\big|^2 \left(1-\frac{M_{F_1}^2}{m_{\eta^\pm}^2}\right)^2.
\label{eq:etaptolF}
\end{align}
\begin{figure}[h]
\centering
\includegraphics[height=5.cm,width=0.49\textwidth]{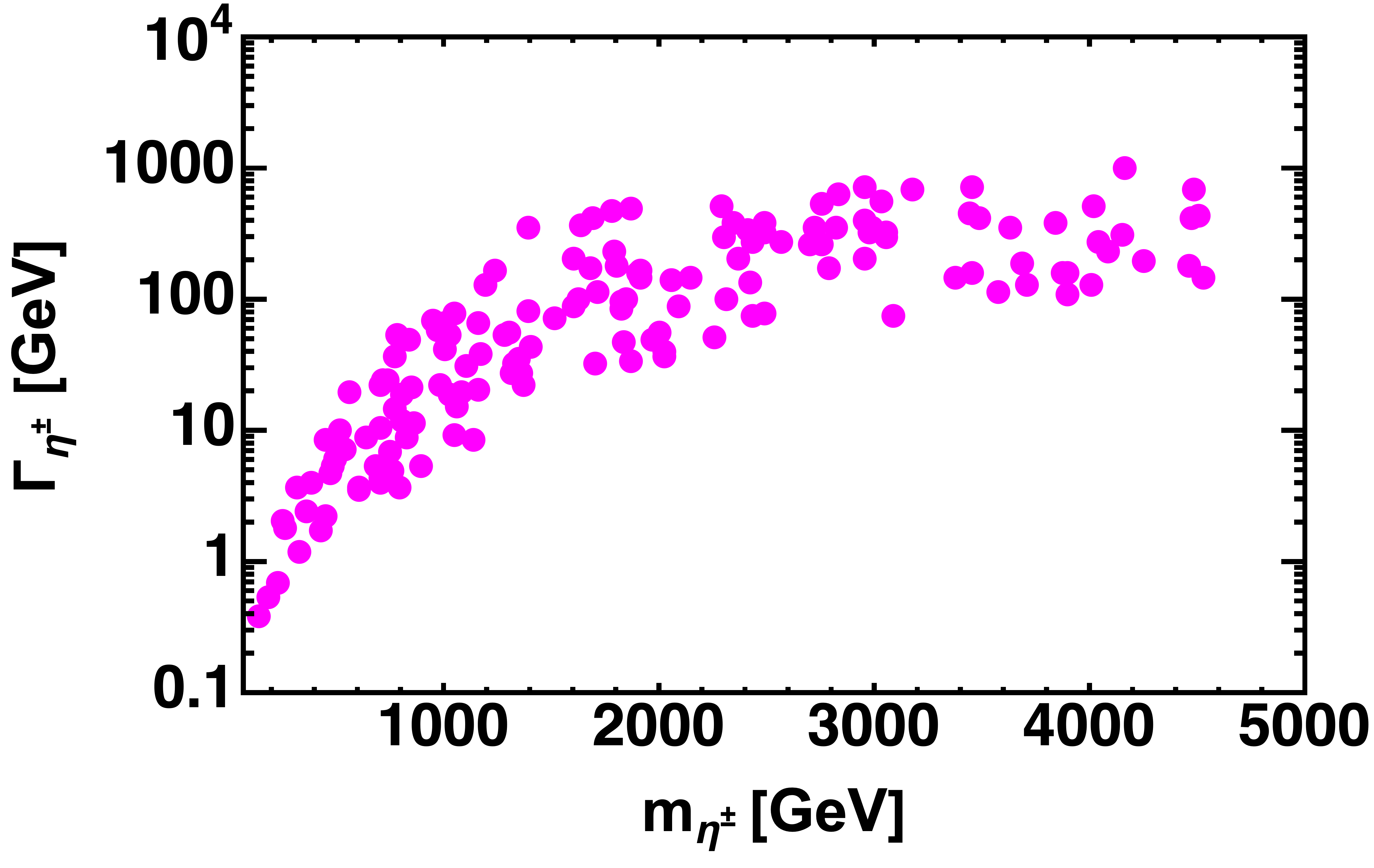}
\includegraphics[height=5.cm,width=0.49\textwidth]{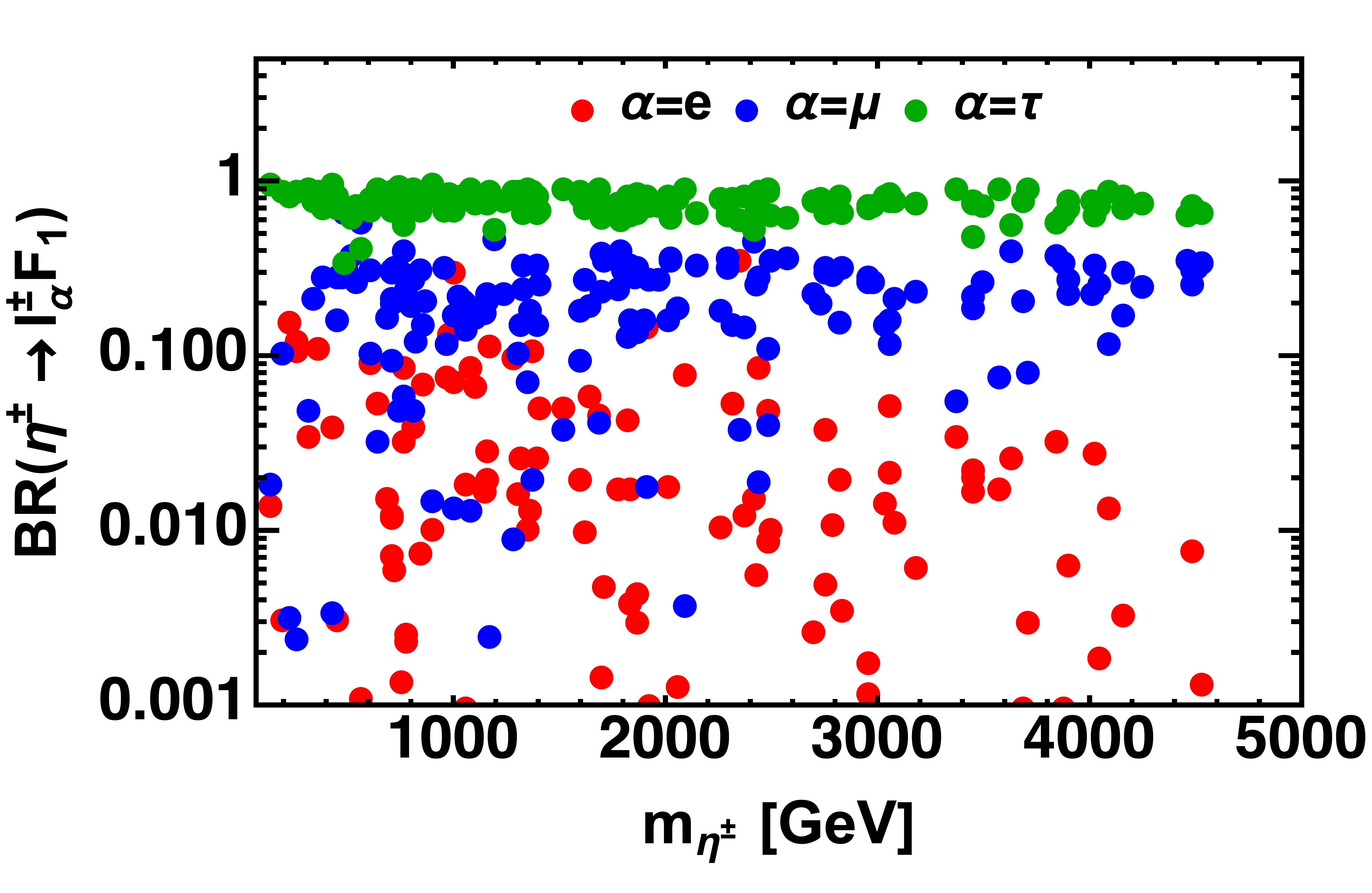}
\caption{
Charged dark scalar $\eta^\pm$ total decay width  (left) and branching ratios to leptonic modes $\ell_\alpha^\pm F_1$ (right).}
\label{fig:decay-scotogenic}
\end{figure}

In Fig.~\ref{fig:decay-scotogenic} we show the $\eta^\pm$ total decay width and branching ratios to $\ell_\alpha^\pm F_1$ for the parameter region consistent with $F_1$ dark matter. 
We see that the decay width $\Gamma_{\eta^\pm}$ is quite large and hence within this parameter region $\eta^\pm$ will always decay promptly. As the required fermionic DM Yukawa couplings follow the hierarchy $|Y_F^{e1}|\ll |Y_F^{\mu 1}|<|Y_F^{\tau 1}|$, the corresponding branching ratios also obeys the hierarchy $\text{BR}_{eF_1}\ll \text{BR}_{\mu F_1} < \text{BR}_{\tau F_1}$. Hence one can have, e.~g.
\begin{align} 
pp~~{\rm or}~~\ell^+\ell^-\to \eta^+\eta^-\to \mu^+\mu^-( \tau^+\tau^-) + \slashed{E}_T.
\end{align}

These signatures will be relevant to probe the fermionic dark matter $F_1$ and have been studied in the context of supersymmetry Refs.~\cite{ATLAS:2019lff,ATLAS:2019lng,CMS:2020bfa,CMS:2020bfa}.  
The di-tau signature is usually less promising than the di-muon channel, due to the poor tau reconstruction efficiency. Note also the possibility of lepton flavor violating di-lepton +
missing energy~($\slashed{E}_T$) signatures, i.e.
$\mu^+\tau^- +\slashed{E}_T$, $\mu^+ e^- +\slashed{E}_T$ and $\tau^+ e^- +\slashed{E}_T$. \par 
Besides the di-lepton signature, tere are other interesting signals. With a compressed mass spectrum $M_{F_1}\approx m_{\eta^\pm}$~(associated with co-annihilation effects for $F_1$ dark matter), the leptons from $\eta^\pm\to \ell_\alpha^\pm F_1$ decays in this scenario are relatively soft, hence hard to detect. 
Searches for the compressed mass region need the low-$p_T$ criterion, see Ref.~\cite{ATLAS:2019lng}. 
On the other hand, if the heavier singlet fermions $F_{j}$~($j=2,3$) are light enough, they can be singly and doubly produced at lepton collider via $\ell^+\ell^- \to F_j F_1,  F_j F_j$ as shown in the right panel of Fig.~\ref{fig:production-scotogenic-11}.  
\par The cascade decay of dark fermions $F_j\to\ell_\alpha^\pm\eta^\mp\to \ell_\alpha^\pm\ell_\beta^\mp F_1$ can lead to di-lepton and tetra-lepton signatures~\cite{Liu:2022byu}.  As this signature is mediated by the t-channel exchange of $\eta^\pm$, the cross-section will be enhanced by the relatively large value of $Y_F^{\ell 1}$.  This is why a muon collider is best suited to study these type of signals.  Another interesting channel is the mono-photon signature $\ell^+\ell^-\to F_1 F_1 \gamma\to \gamma + \slashed{E}_T$, where the photon will come from initial state radiation~\cite{Casarsa:2021rud,Habermehl:2020njb,Black:2022qlg,Liu:2022byu}. 
Associated production $pp\to \eta^\pm\eta_{R/I}$ can also give a mono-lepton+$\slashed{E}_T$ signature, resulting from the decay chain $\eta^\pm \to \ell_\alpha^\pm F_1$ and $\eta_{R/I}\to \nu F_1 $. 

\subsection{FIMP fermionic scotogenic dark matter} 
\label{subsec:fimp_col}
\par Notice that collider signatures would be completely different in the case of FIMP dark matter. The phenomenology crucially depends on the dark fermion masses relative to the dark scalars. We focus on three representative cases: 
\begin{itemize}
\item {\bf $M_{F_1}< m_{\eta^\pm},m_{\eta_{R/I}}<M_{F_{2,3}}$:} In this scenario, the charged scalar $\eta^\pm$ can only decay into the FIMP $F_1$ and a charged lepton, with a rate given by Eq.~\eqref{eq:etaptolF}. 
Using Eqs.~\eqref{eq:etaptolF} and (\ref{eq:FIMPy1}) the proper decay-length of $\eta^\pm$, $c\tau~(\eta^\pm)$, can be computed as,  
\begin{align}
c\tau(\eta^\pm)\approx 8.3\text{ m}\left(\frac{M_{F_1}}{10\,\text{KeV}}\right) \left(\frac{100\,\text{GeV}}{m_{\eta^\pm}}\right)^2,    
\end{align}
and can clearly exceed typical  detector sizes. Depending on the decay length, the dark scalar  $\eta^\pm\eta^\mp$production, followed by the decay $\eta^\pm\to\ell_\alpha^\pm F_1$ could lead to two displaced charged leptons associated with missing energy. 
\item {\bf $M_{F_1}<M_{F_2}<m_{\eta^\pm},m_{\eta_{R/I}}<M_{F_3}$:} In this case, the charged dark scalar can decay to both  $\ell_\alpha^\pm F_1$ and   $\ell_\alpha^\pm F_2$. The dominant decay mode is $\eta^\pm\to \ell_\alpha^\pm F_2$,  
 the Yukawa coupling $y_1=\left(\sum_{\alpha}|Y_F^{\alpha 1}|^2\right)^{1/2}$ obeys Eq.~\eqref{eq:FIMPy1} and $y_2=\left(\sum_{\alpha}|Y_F^{\alpha 2}|^2\right)^{1/2}$ lies in the range $[10^{-5}:10^{-2}]$~\cite{Hessler:2016kwm}. 
 Hence $\eta^\pm$ will dominantly decay into the next-to-lightest dark fermion $F_2$, with a decay-length which is typically below 1 mm. Subsequently, $F_2$ decays as $F_2\to \ell_\alpha^-\ell_\beta^+ F_1$ and $F_2\to\nu_\alpha\overline{\nu_\beta} F_1$. The corresponding rate is, 
 \begin{align}
 \Gamma(F_2\to \ell_\alpha^-\ell_\beta^+ F_1) \approx \Gamma(F_2\to\nu_\alpha\overline{\nu_\beta} F_1) =\frac{M_{F_2}^5}{6144\pi^3 m_{\eta^\pm}^4}\left(|Y_F^{\beta 1}|^2 |Y_{F}^{\alpha 2}|^2 + |Y_{F}^{\alpha 1}|^2 |Y_F^{\beta 2}|^2\right)  ,
 \end{align} 
Within the FIMP dark-matter region, the $F_2$ decay-length can be approximated as, 
\begin{align}
  c\tau(F_2)\;\approx\; 2\times 10^{13}\,\text{m}\,\left(\frac{M_{F_1}}{\text{10 keV}}\right)\left(\frac{m_{\eta^\pm}}{\text{500 GeV}}\right)^3\left(\frac{\text{100 GeV}}{M_{F_2}}\right)^5\left(\frac{10^{-3}}{y_2}\right)^2~,
  \label{eq:F2length}
\end{align} 
which is much larger than the detector size. Hence, the experimental signature of this scenario consists of two prompt charged leptons and missing energy, from the production of $\eta^\pm\eta^\mp$, followed by the dark scalar decay $\eta^\pm\to\ell_\alpha^\pm F_2$ with a long-lived $F_2$. 
\item {\bf $M_{F_1}<M_{F_2}<M_{F_3}<m_{\eta^\pm},m_{\eta_{R/I}}$:} This scenario produces two different signatures. The first one arises from the decay $\eta^\pm\to\ell_\alpha^\pm F_2$, which gives $\ell_\alpha^\pm\ell_\beta^\mp + \slashed{E}_T$, as $F_2$ has a long decay-length, see Eq.~\eqref{eq:F2length}. Hence this signal is identical to the previous case of $M_{F_1}<M_{F_2}<m_{\eta^\pm},m_{\eta_{R/I}}<M_{F_3}$.  The second signature arises from the decay $\eta^\pm\to\ell_\alpha^\pm F_3$ with $F_3$ decaying to $\ell_\alpha^-\ell_\beta^+ F_2$. In contrast to $F_2$, which is stable within the detector, $F_3$ can decay fast enough with the proper decay length, 
\begin{align}
  c\tau(F_3)\;\approx\; 0.4\,\text{m}\left(\frac{\text{100 GeV}}{M_{F_3}}\right) \left(\frac{m_{\eta^\pm}}{M_{F_3}}\right)^4\left(\frac{10^{-3}}{y_2}\right)^2\left(\frac{10^{-3}}{y_3}\right)^2.
  \label{eq:F3length}
\end{align}
Hence, $F_3$ can decay inside the detector at macroscopic distances for some parameter choices. This leads to a clean prompt-di-lepton + displaced-di-lepton signature, that can be searched for efficiently at the LHC~\cite{CMS:2014xnn,CMS:2014hka}. 
\end{itemize}
(2). {\bf Scalar dark matter:} The scalar DM case is characterized by  $m_{\eta^\pm},m_{\eta_{R/I}}<M_{F_i}$, so the generic signature to be searched for is missing energy~($\slashed{E}_T$), measured through the total transverse momentum recoil of the visible particles in the event. 
Typical signatures in the scotogenic model are $\slashed{E}_T+X$, where $X$ can be one or two jets~\cite{ATLAS:2016bek,Haisch:2018hbm}, two leptons~\cite{ATLAS:2018lkz} or one photon~\cite{ATLAS:2017nga}. 
Although these are all interesting, we find that the most promising one is $\slashed{E}_T+\text{jet}$~(mono-jet) arising from $pp\to\eta_{R/I}\eta_{R/I} + j$,  $\eta^\pm\eta_{R/I} + j$ and $\eta^\pm\eta^\mp + j$ processes. Here one looks for events with one high-$p_T$
jet~($\geq \mathcal{O}(100\,\text{GeV})$) with pseudo-rapidity $|\eta|<2.4$ and $\slashed{E}_T>200$ GeV for the ATLAS and CMS detectors~\cite{ATLAS:2016bek,ATLAS:2016neq}. 
\par If the mass of $F_1$, $M_{F_1}$, is larger than that of the dark scalar particles, then $F_1$ pair production by $\mu^+\mu^-$ annihilation through t-channel $\eta^\pm$ exchange can lead to additional signatures. In this scenario, $F_1$ can decay as $F_1\to\ell_\alpha^\pm \eta^\mp$ or $F_1\to\nu_j\eta^\alpha$ (with $j = 1, 2, 3$ and $\alpha=R,I$), with partial decay widths given by 
\begin{align} 
& \Gamma(F_1\to\ell_\alpha^\pm \eta^\mp) = \frac{M_{F_1}}{32\pi} |Y_F^{\alpha 1}|^2 \left(1-\frac{m_{\eta^\pm}^2}{M_{F_1}^2}\right)^2 ,\\
&\Gamma(F_1\to\nu_j \eta^\alpha) = \sum_j\frac{M_{F_1}}{32\pi} |Y_F^{j 1}|^2 \left(1-\frac{m_{\eta^\alpha}^2}{M_{F_1}^2}\right)^2 . 
\end{align}
$F_1$ can decay promptly or can be long-lived depending on the values of Yukawa couplings $Y_F$ and the mass differences $m_{\eta_{R/I}}-M_{F_1}$, $m_{\eta^\pm}-M_{F_1}$. If it decays promptly then we will have the usual di-lepton signature in association with missing energy, whereas for macroscopic decay lengths we have a displaced di-lepton signature. 

Notice also that the small mass difference among the new particles of the dark sector translates into possible long-lived particles (LLP) signatures. 
Two possible LLP scenarios were discussed in~\cite{Avila:2021mwg}, showing that when the charged dark scalar $\eta^{\pm}$ is the long-lived candidate, there is a chance for detecting disappearing charged tracks, when a dark-matter candidate $\eta_R$ of mass around $300-400$ GeV makes up the total relic density. 
\subsection{Revamped Scotogenic}
\label{Revamped-collider}
\begin{figure}[h]
\subfloat[]{\includegraphics[height=3.0cm,width=0.3\linewidth]{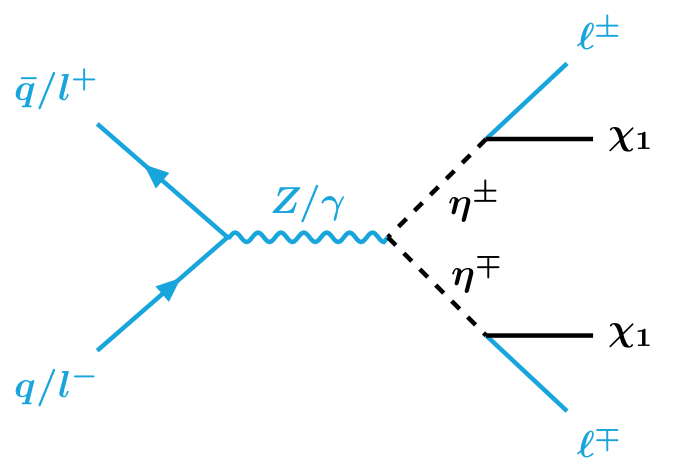}}\hfil
\subfloat[]{\includegraphics[height=3.0cm,width=0.3\linewidth]{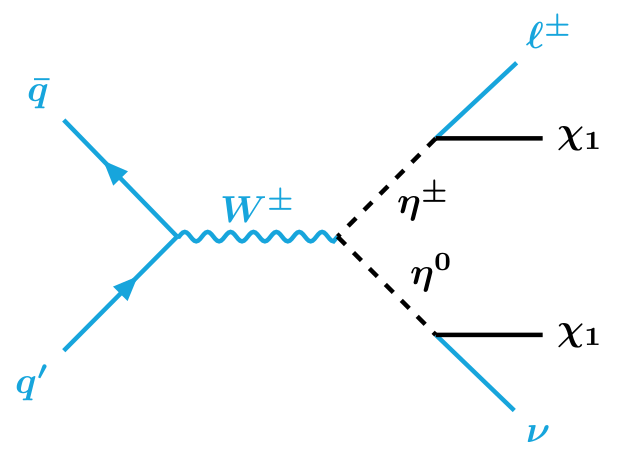}}\hfil \\
\subfloat[]{\includegraphics[height=3.0cm,width=0.3\linewidth]{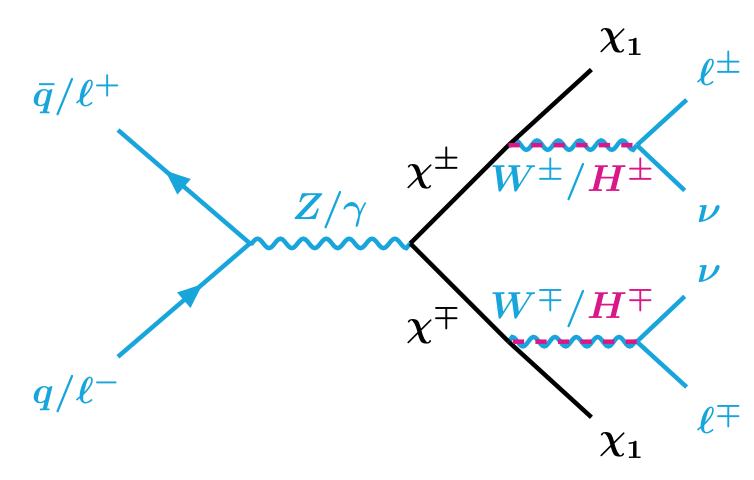}}\hfil
\subfloat[]{\includegraphics[height=3.0cm,width=0.3\linewidth]{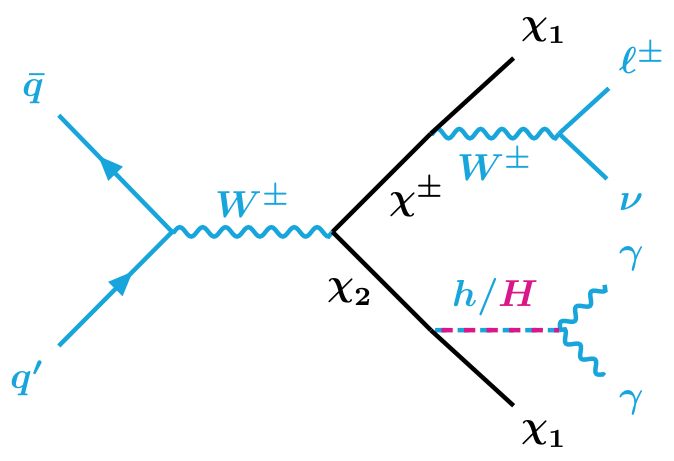}}
\caption{ 
Diagrams for dark fermion production in the revamped scotogenic setup. Diagrams (a) and (b) depict the production of the scalar doublet $\eta$, resulting in $2\ell+\slashed{E}_T$ and $\ell+\slashed{E}_T$. 
 Diagrams (c) and (d) show the dark fermion pair-production processes $\chi^\pm,\chi_2^0$, which lead to $2\ell+\slashed{E}_T$, and $\ell+2\gamma+\slashed{E}_T$ final states.} 
\label{fig:collider-singlet-triplet1}
\end{figure}

Here we focus on the
detection prospects of scalar/fermionic dark matter at various colliders within the revamped scotogenic scenario. 
The collider phenomenology for the scalar dark matter $\eta_R$ case is nearly the same as we discussed above for the simplest scotogenic model. 
The most promising signature is jet+$\slashed{E}_T$ arising from $pp\to\eta_R\eta_R+g(q)$ processes.
In contrast, the case of fermionic dark matter is much more interesting in the revamped singlet-triplet scotogenic model and bears a close analogy to supersymmetric neutralino
dark-matter~\cite{Jungman:1995df}. 
As before, in analogy to the supersymmetric case, we will refer here to the lightest scotogenic particle as LSP. 
For the case of pure triplet $\Sigma^0$ dark-matter, the relic density is satisfied around 2.5 TeV, 
Fig.~\ref{fig:relic-pure-triplet} left, so that the expected DM production cross-section at the HL-LHC is very small. 
Thanks to the mixing of singlet and triplet fermions $F$ and $\Sigma^0$ in the revamped scheme, however,  fermionic DM masses $\mathcal{O}$(100)~GeV or lower are possible, hence DM can be produced even in the 14 TeV $pp$ collider. 
Fermionic dark-matter is produced by the Drell-Yan processes associated to  $pp\to Z/\gamma\to \chi^+\chi^-$, $pp\to W^\pm\to \chi^\pm \chi_2^0$ and $pp\to h/H\to \chi_2^0\chi_2^0$, respectively~\cite{Restrepo:2019ilz,Choubey:2017yyn}, as seen in Fig.~\ref{fig:collider-singlet-triplet1}. 

\par
\begin{figure}[h]
\centering
\includegraphics[height=4.5cm,width=0.45\textwidth]{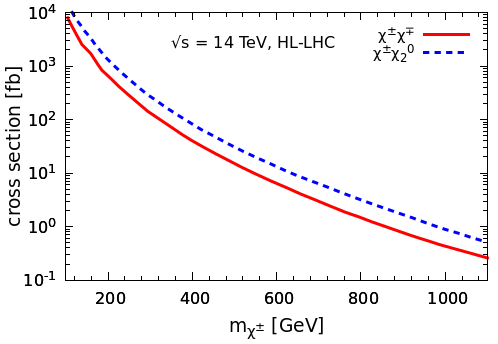}~~~~~~~
\caption{ 
 Cross-sections for $\chi^\pm\chi^\mp$~(red-solid) and $\chi^\pm\chi_2^0$~(blue-dashed) production for a fixed $m_{\chi_2^0}-m_{\chi_1^0}=$ 20 GeV. } 
\label{fig:collider-singlet-triplet-xs}
\end{figure}

 In Fig.~\ref{fig:collider-singlet-triplet-xs}, we show the production cross-sections for $pp\to \chi^+\chi^-$ and $pp\to \chi^\pm \chi_2^0$ for a fixed $m_{\chi_2^0}-m_{\chi_1^0}=$ 20 GeV.
Note that the $\chi^\pm\chi_2^0$ production cross-section is bigger than that for $\chi^\pm\chi^\mp$ pair-production, i.e. $\sigma(\chi^\pm \chi_2^0)> \sigma(\chi^\pm \chi^\mp)\gg \sigma(\chi_2^0\chi_2^0)$, the latter is not shown in Fig.~\ref{fig:collider-singlet-triplet-xs}.  
Note that fermionic dark matter can also be produced through dark scalar pair production e.g. $\eta^+\eta^-$ and $\eta^+\eta^0$ with cross-section estimates already given in the previous sub-section.  
As for the simplest scotogenic, in  lepton colliders there is also a t-channel contribution to $\eta^\pm\eta^\mp$ production. Here we focus on hadron colliders, LHC and HL-LHC.  
Having discussed the various possible production modes, we now discuss the associated collider signatures of fermionic scotogenic dark matter. 
Once produced, they will decay to various possible decay channels depending on kinematics, as shown in Fig.~\ref{fig:collider-singlet-triplet1}. We will examine three representative benchmarks, each characterized by their spectrum and hierarchy of Lagrangian parameters~\cite{Lozano:2025tst}: 
\begin{itemize}
\item[\textbf{A}.] \textbf{Heavy dark scalar doublet} $\boldsymbol{ (M_F < M_\Sigma \ll  m_{\eta}})$: In this case the scalar doublet decouples, so the dominant production channels are $pp\to\chi_2^0\chi^\pm$ and $pp\to\chi^\pm\chi^\mp$.
Given this mass hierarchy, the LHC phenomenology is primarily governed by the lower processes shown as diagrams (c) and (d) of Fig.~\ref{fig:collider-singlet-triplet1}. \par 
Charged dark states will decay through a $W/H$ boson giving $\chi^\pm\to \chi_1^0 W^{\pm *}/H^{\pm *}\to \chi_1^0\ell^\pm\nu$ and the neutral state $\chi_2^0$ will decay to $\chi_1^0$ along with a Higgs into various possible final states.  
  One can exploit one of the cleanest Higgs boson decay modes into two photons in order to explore this production mechanism.   One can require $105\text{ GeV}<m_{\gamma\gamma}<160\text{ GeV}$ for the invariant mass window of the photons, in order to isolate events originating from Higgs decays.  As a result, Fig.~\ref{fig:collider-singlet-triplet1}(c) and (d) will lead to di-lepton plus missing energy~($2\ell+\slashed{E}_T$) final state, and mono-lepton plus di-photon in association with missing energy~($\ell+2\gamma+\slashed{E}_T$), respectively. The search for $\ell+2\gamma+\slashed{E}_T$ was originally designed to probe the production and decay of charged and neutral electroweakinos in supersymmetry~\cite{ATLAS:2020qlk,CMS:2017moi,ATLAS:2016uwq}.   
\begin{figure}
    \centering
    \includegraphics[width=0.45\linewidth]{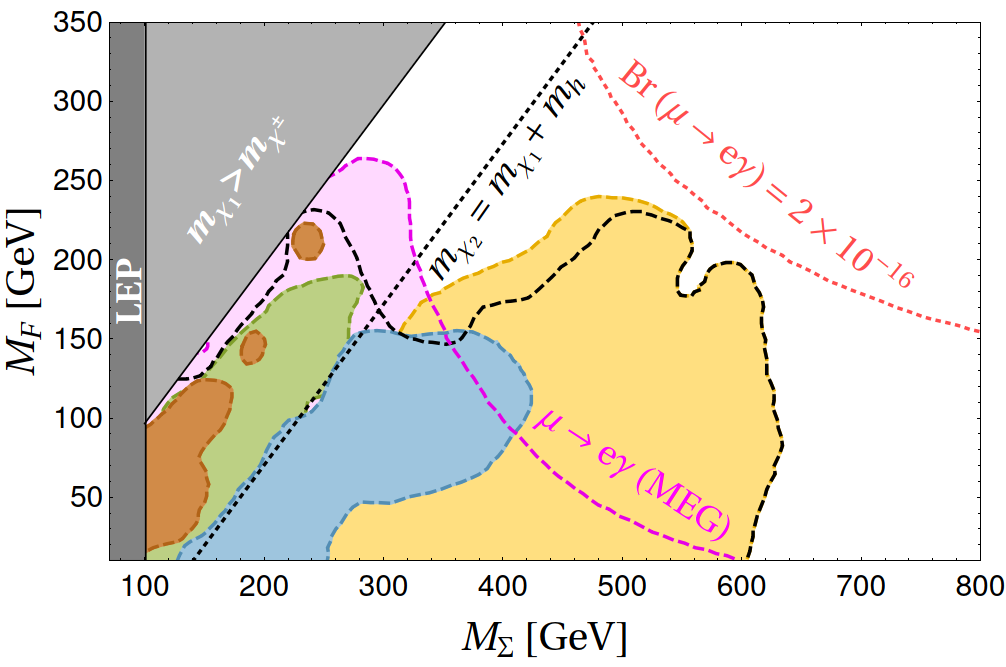}
    \hfil
    \includegraphics[width=0.45\linewidth]{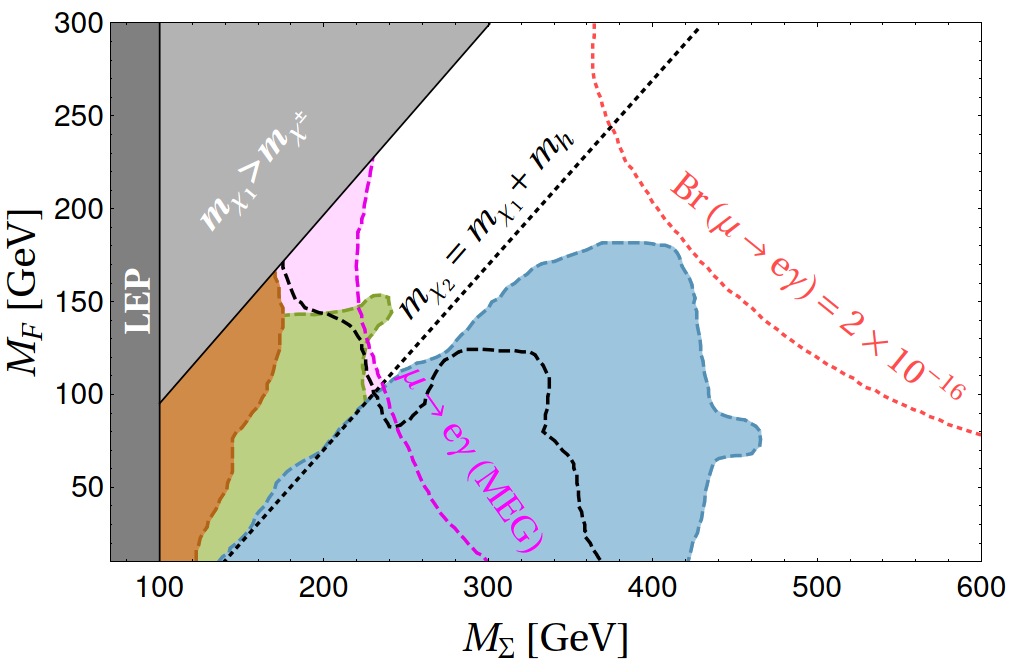}
    \caption{
    95\% C.L. excluded regions in the $M_\Sigma-M_F$ plane for $v_\Omega=1.5$ GeV. The panels have  $\mu_1=1.1$ GeV (left) and $\mu_1=1376.2$ GeV (right).  Colored areas correspond to the parameter regions covered by different LHC searches such as, $2\tau+\slashed{E}_T$(yellow)~\cite{ATLAS:2019gti}, $2\ell+\slashed{E}_T$~(green)~\cite{CMS:2017moi}, $\ell+2\gamma+\slashed{E}_T$~(blue)~\cite{ATLAS:2020qlk} and $2\ell~(\text{soft})+\slashed{E}_T$~(dark orange)~\cite{CMS:2016zvj}. Dashed magenta curves indicate cLFV limits, with the magenta region excluded by $\mu\to e \gamma$ searches at MEG.}
    \label{fig:MfMs1}
\end{figure}
Searches involving two leptons with low transverse momentum~\cite{CMS:2016zvj} arise in scenarios with small mass differences in the dark sector, resulting in a small transfer of momentum to the final-state particles. 
Dedicated searches for events with two tau leptons and large missing transverse energy~\cite{ATLAS:2019gti} can also play a significant role.  
These searches were originally designed to target stau pair production, with each stau decaying to a tau lepton. The same final-state topology also arises in the revamped scotogenic model.  
 In contrast to SUSY, where parameters are largely unconstrained, within the scotogenic approach the viable region is restricted, e.g. from neutrino oscillation and cLFV physics, making it both predictive and testable. 
\par In Fig.~\ref{fig:MfMs1}, we show the sensitivity of this scenario at HL-LHC, where in the left and right panels we fix $\mu_1=1.1$ GeV and $\mu_1=1376.2$ GeV, respectively. The details of the input parameter choices are given in Ref.~\cite{Lozano:2025tst}.  The vertical gray band in both panels is excluded by the LEP constraint $m_{\chi^\pm}< 100$~GeV~\cite{ALEPH:2002gap, DELPHI:2003uqw, L3:2000euy, OPAL:1997xnz}. The triangular gray region gives a charged LSP, not a viable dark matter candidate.  
Colored areas are the regions covered by different LHC searches such as, $2\tau+\slashed{E}_T$(yellow)~\cite{ATLAS:2019gti}, $2\ell+\slashed{E}_T$~(green)~\cite{CMS:2017moi}, $\ell+2\gamma+\slashed{E}_T$~(blue)~\cite{ATLAS:2020qlk} and $2\ell~(\text{soft})+\slashed{E}_T$~(dark orange)~\cite{CMS:2016zvj}. The black dashed line corresponds to the total exclusion region. 
\item[\textbf{B}.] \textbf{Heavy dark triplet fermion} $\boldsymbol{(M_F<m_\eta \ll M_\Sigma)}$: This benchmark has the fermionic triplet decoupled, resulting in $pp\to\eta^\pm\eta^\mp,\eta^\pm\eta^0$ as the dominant channels. 
Hence, in this scenario the processes relevant for the LHC phenomenology arise from diagrams (a) and (b) of Fig.~\ref{fig:collider-singlet-triplet1}.  Two dark neutral scalars can also be produced, but with a much smaller cross section than processes with one or two charged states.  
Once produced, the charged dark scalar decays into a charged lepton and the LSP~($\eta^\pm\to\ell^\pm \chi_1^0$), whereas the neutral state decays into a neutrino and the LSP~($\eta^0\to\chi_1^0\nu$). 
Hence this decay chain, Fig.~\ref{fig:collider-singlet-triplet1}(a), gives to two leptons plus ~$\slashed{E}_T$ ($2\ell+\slashed{E}_T$), whereas Fig.~\ref{fig:collider-singlet-triplet1}(b) leads to mono-lepton plus $\slashed{E}_T$~($\ell+\slashed{E}_T$). 
\begin{figure}
    \centering
    \includegraphics[height=4cm,width=0.35\linewidth]{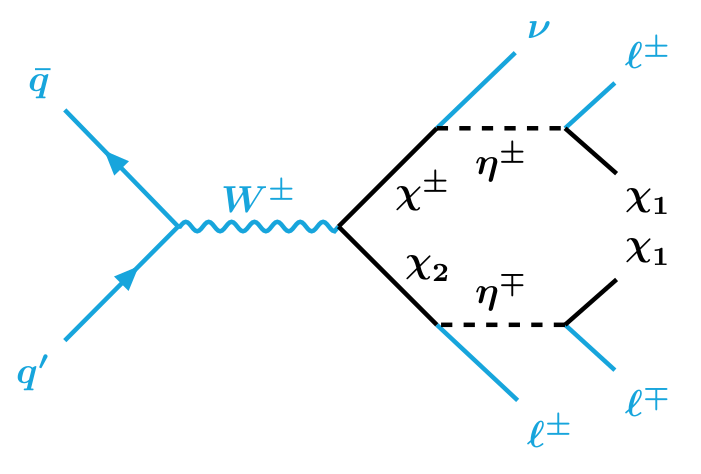}\hfil
 \includegraphics[height=4cm,width=0.35\linewidth]{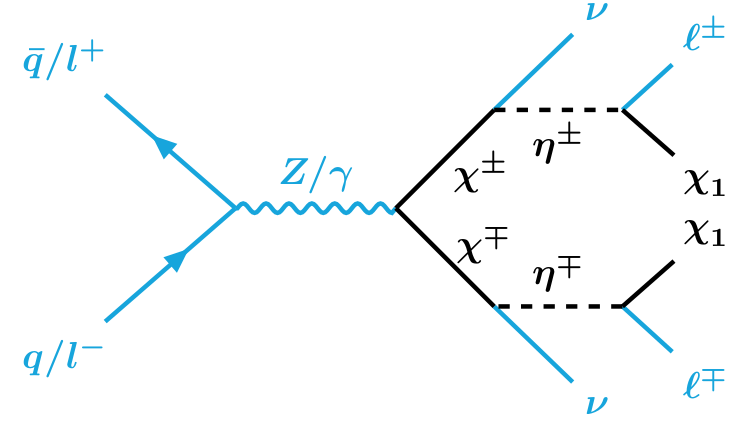}
\caption{ 
Diagrams for the production and subsequent decay of dark particles in the revamped scotogenic model with $M_\Sigma>m_\eta+M_F$. The signature here consists of 2–3 leptons accompanied by missing transverse energy.}
    \label{fig:collider-singlet-triplet2}
\end{figure}
 \item[\textbf{C}.] \textbf{Light dark sector} $\boldsymbol{(M_F <  M_\eta \simeq M_\Sigma)}$: This most agressive benchmark is characterized by low values for $M_F$, $M_\Sigma$ and $m_\eta$, with $M_F$ set as the smallest to ensure a viable fermionic LSP as dark matter candidate. This implies that all the production diagrams of Fig.~\ref{fig:collider-singlet-triplet1} are important. 
 If $M_\Sigma>m_\eta+M_F$, the relevant decays are $\chi^\pm\to \eta^\pm \nu$ and  $\chi_2^0\to \eta^\pm \ell^\mp$ followed by the subsequent decay $\eta^\pm\to \ell^\pm \chi_1^0$, see Fig.~\ref{fig:collider-singlet-triplet2}. This leads to 2-3 leptons plus $\slashed{E}_T$ as the main signal. 
\end{itemize}   

\par One sees that the fermionic nature of the LSP allows for various spectrum configurations, leading to a wide range of potential signals at collider 
 experiments. While the spirit of the above discussion closely follows the approach used in SUSY dark matter scenarios, in our case, the dark matter parameters are constrained in order to comply with the findings of neutrino oscillation experiments. 
 In particular, they can also be explored in synergy with searches for cLFV processes. Such studies offer complementary insights to conventional DM searches at collider experiments, providing a characteristic feature of our \textit{dark neutrino mass} approach.

\subsection{Collider signatures of linear seesaw}  

In this subsection we briefly discuss the collider signatures associated to the dark-matter candidates in the linear seesaw model.  
As usual, due to the remnant dark symmetry, the scalar and fermionic dark-matter candidates will be produced in pairs. 
Because of the small mixing between $\eta^0_{R}$ and $\xi^0_{R}$ as well as $\eta^0_{I}$ and $\xi^0_{I}$, we can take them aproximately as the mass eigenstates. 
The neutral components of the dark doublet can be produced in pairs via the Drell-Yan mechanism mediated by the $Z$-boson or through vector boson fusion.  This will correspond to a final state of 2 jets plus missing energy at the collider.
Detailed studies of collider signatures arising from pair production of neutral components of the dark doublet via vector boson fusion are provided in~\cite{Dutta:2017lny}.

\begin{figure}[tbh]
\centering
\includegraphics[width=0.4\textwidth,height=4.0cm]{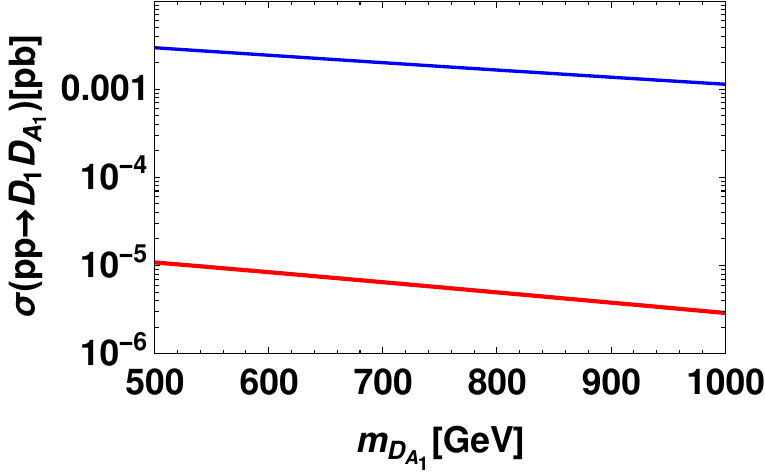}
\caption{
  Total Drell-Yan cross section for the CP-even/CP-odd scalar dark-matter production versus the CP-odd scalar dark-matter mass $m_{D_{A_1}}$
  at a proton-proton collider for $\protect\sqrt{s}=14$
TeV (red line)
and $\protect\sqrt{s}=100$ TeV (blue line). }
\label{fig:pptoD1Da1}
\end{figure}

Fig.~\ref{fig:pptoD1Da1} displays the total cross section for the pair production of $\eta^0_R$ and $\eta_0^I$ via the Drell-Yan mechanism at a proton-proton collider for $\protect\sqrt{s}=14$ TeV (red line)
and $\protect\sqrt{s}=100$ TeV (blue line) as a function of the CP-odd dark scalar mass $m_{D_{A_1}}$ varying in the range from $500$ GeV up to $1.0$ TeV. 
Here the mass of the CP even dark scalar, $m_{D_{1}}$, has been set to be equal to $1$ TeV. 
As shown in Figure \ref{fig:pptoD1Da1}, the total cross section for the CP-even and CP-odd dark scalar production at the LHC via the Drell-Yan mechanism reaches $\mathcal{O}(10^{-5})$pb  for $m_{D_{A_1}}=0.5$ TeV, and decreases as $m_{D_{A_1}}$ increases. 
The total Drell-Yan production cross section increases by two orders of magnitude by going to a $100$ TeV proton-proton collider, where the cross section reaches $3\times 10^{-3}$ pb when the CP-odd dark scalar mass is set to $0.5$ TeV. 

For the case of a fermionic DM candidate, the pair-production of the charged components of the dark doublet through the Drell-Yan mechanism and their subsequent decays will give rise to an opposite-sign di-leptons plus missing energy signature.  
The observation of an excess of such events with respect to the SM background could provide support of this model at the LHC. 
Quantitative results would require detailed collider simulations. 


\section{Implications for cosmology}
\label{sec:cosmology}

Models where neutrino masses arise radiatively through interactions involving dark sector particles, e.g. odd under a $\mathbb{Z}_2$-symmetry, can have rich implications for cosmology, concerning not only the issue of dark matter itself, but possibly also other aspects associated to baryogenesis, inflation, and structure formation.
Below we briefly illustrate this with some examples related to dark matter and baryogenesis.

\begin{center}
   {\bf \small Dark-matter in Scotogenic Models: WIMPs versus FIMPs}  
\end{center}

Concerning the core theme of this review, the leading option considered has been that of WIMP dark-matter produced by freeze-out in the early Universe. 
In the associated theories either neutrino masses are generated directly by loops involving the exchange of dark particles, or the latter provide a radiative seed for generating neutrino masses \textit{a la seesaw}.
The dark matter candidate, bosonic or fermionic, is in thermal equilibrium with the \sm plasma in the early Universe. As the Universe expands and cools, DM annihilation rates drop below the Hubble rate, and  freeze-out occurs at some relic abundance, presumably in agreement with the measured one at the Planck telescope~\cite{Planck:2018vyg}.

Alternatively, one may have a freeze-in dark-matter picture, characteristic of the FIMP scenario. Here the DM candidate never thermalizes with the \sm bath and is produced via decays or scatterings of thermal particles. The dark-matter yield builds up until the production terminates, typically, as the plasma cools below the mediator's mass.
Throughout the review we have studied the dark matter density in various theoretical schemes, as well as the corresponding direct, indirect and collider detection prospects.

\begin{center}
   {\bf \small Baryogenesis through Leptogenesis} 
\end{center}
    
The Standard Model fails to explain the observed imbalance between matter and antimatter in the Universe, known as the baryon asymmetry of the universe (BAU)~\cite{Kolb:1983ni,Riotto:1999yt}. Observational data from the CMB and primordial BBN reveal a small but significant excess of matter over antimatter, namely the baryon-to-photon ratio 
\begin{equation}
\label{eq:eta-BAU}
  \eta_B^{obs}\approx 6.1 \times 10^{-10}.  
\end{equation}
    The process of baryogenesis via leptogenesis \cite{Fukugita:1986hr}, in which a lepton asymmetry in the early universe is transformed into a baryon asymmetry through the electroweak sphaleron transition~\cite{Rubakov:1996vz}, is an elegant idea~\cite{Davidson:2008bu}. 
 The usual thermal type-I seesaw leptogenesis involves two or three families of heavy right-handed neutrinos (RHNs). These RHNs get produced in the thermal bath through scatterings and later the out-of-equilibrium CP-violating decay of the lightest of them creates a lepton asymmetry.   
 In the simplest type-I seesaw the lightest RHN mass obeys~\cite{Davidson:2002qv}
 \begin{equation}
      M_1 \gtrsim 10^9 \mathrm{GeV}, 
 \end{equation}
which may be translated into an upper bound on the light neutrino masses \cite{Buchmuller:2003gz}.
This limit on the lightest RHN is required in order to be compatible with $\eta_B^{obs}$. The inclusion of flavor effects in leptogenesis can reduce this bound by roughly an order of magnitude~\cite{Davidson:2008bu,Antusch:2010ms}, but cannot evade the limit on the scale, which still remains high. 
Clearly such heavy RHNs are too far from any possible direct observation at present or future collider experiments~\cite{Deppisch:2013jxa, ATLAS:2025uah, CMS:2024xdq}.
 Scotogenic scenarios can avoid this problem, while providing a dark matter completion of the Standard Model with new physics potentially accessible at colliders~\cite{Hugle:2018qbw,Singh:2023eye}.

The starting point of the discussion is the baryon-to-photon ratio $\eta_B$. This can be generally expressed as the product of three quantities~\cite{Buchmuller:2004nz,Fong:2012buy}:
\begin{equation}
	\eta_B=C\,\kappa\,\varepsilon\,.
\end{equation}
Here the parameter $C\approx0.01$ includes the sphaleron conversion factor,
and the dilution factor accounting for entropy release after leptogenesis. 
 The efficiency factor $\kappa$ captures the impact of washout processes on the generated lepton asymmetry, while $\epsilon$ is the CP-asymmetry produced in the  heavy neutrino decays.
 
 In the context of scotogenic scenarios, the CP-ymmetry is driven from the decays of the lightest dark fermion $F_1$, 
 $$\epsilon=\frac{\Gamma(F_1\to l\eta)-\Gamma(F_1\to \bar l\eta^\dagger)}{\Gamma(F_1\to l\eta)+\Gamma(F_1\to \bar l\eta^\dagger)}~.$$ 
 The efficiency $\kappa$ depends on the decay parameter $K=\frac{\Gamma(z\to\infty)}{H(z=1)}$, where $\Gamma$ is the decay width of the lightest dark fermion and $H$ is the Hubble expansion rate.

 Depending on the washout regime, $\kappa$ is approximated as~\cite{Buchmuller:2004nz,Hugle:2018qbw}: 
\begin{equation}
	\kappa\simeq 
    \begin{cases}
        \frac{9\pi^2}{64} K^2\; &(\text{for}\; K\lesssim1\implies\text{Weak washout})\\
        \frac{1}{1.2\,K \, (\ln K)^{0.8}}\; &(\text{for}\; K\gtrsim4 \implies\text{Strong washout})
    \end{cases}	
\end{equation}
with $z=M/T$, the ratio of the dark fermion mass to the photon bath temperature. Note that any $B-L$ asymmetry generated in these decays could be wiped out due to washout effects~\cite{Buchmuller:2004nz,Fong:2012buy}. In the strong washout regime, $\Delta L = 1$ processes such as the inverse decay $l \eta \to F_1$ remain the dominant channels responsible for washing out the generated lepton asymmetry. However, in the weak washout regime, $\Delta L = 2$ processes — such as $l l \to \eta \eta$ and $l \eta \to \bar{l} \eta^\dagger$ — also become relevant and can contribute significantly to the washout. It is worth mentioning that flavor effects~\cite{Blanchet:2008pw,Davidson:2008bu,Antusch:2010ms}, thermal corrections~\cite{Giudice:2003jh,Kiessig:2010pr}, quantum kinetic effects~\cite{Buchmuller:2000nd,DeSimone:2007gkc} as well as $\Delta L=1$ scatterings \cite{Luty:1992un,Plumacher:1996kc} might introduce non-negligible corrections.

\subsection{Scotogenic leptogenesis: Fermionic LSP}

While scotogenic models have the potential to evade the Davidson-Ibarra bound~\cite{Clarke:2015hta}, detailed analysis shows that achieving a successful thermal leptogenesis scenario at the TeV scale remains challenging~\cite{Kashiwase:2012xd}. 
This difficulty arises because it is very hard to reconcile the observed BAU with neutrino oscillation data at such low scales. 
We structure our next discussion into two distinct scenarios, fermionic LSP and and scalar LSP, depending on the LSP identity.\\[-.2cm]

For the case of fermionic scotogenic dark matter, achieving successful thermal leptogenesis at the TeV scale is highly challenging. 
 Typically, achieving the correct relic density for fermionic dark matter requires large Yukawa couplings, which in turn lead to significant washout of the generated lepton asymmetry, making the fermionic DM scenario problematic. 
 While a near-degeneracy among the dark fermion masses—akin to the so-called resonant leptogenesis~\cite{Pilaftsis:1997jf,Pilaftsis:2003gt,Suematsu:2011va,Kashiwase:2013uy}—can enhance the asymmetry~\cite{Suematsu:2011va}, the required Yukawa couplings become too small to ensure the correct relic abundance.

An alternative approach is explored in Ref~\cite{Mahanta:2019gfe}, where the heavier dark fermion $F_2$ generates the lepton asymmetry via out-of-equilibrium decays into SM leptons and the dark scalar $\eta$.
 In this scenario successful leptogenesis can occur with inverted neutrino mass ordering (IO) provided $M_1^{min}$ is around 
$$M_{1,IO}^{min}\sim20 \text{ TeV},$$
while the normal neutrino mass ordering (NO) still requires a much higher scale, approximately $$M_{1,NO}^{min}\sim10^4\text{TeV}.$$
 While these results soften the Davidson-Ibarra bound, they are still very far from providing hopes for a collider-testable thermal leptogenesis scenario in the scotogenic approach. We now turn to the alternative case of scalar LSP.

\subsection{Scotogenic leptogenesis: Scalar LSP}

The dynamics and viability of scalar LSP leptogenesis in the context of the simplest scotogenic scheme of Sec.~\ref{sec:simpl-scot-setup} can change significantly depending on the number of dark fermions present in the model. 
     \begin{itemize}
           \item \textbf{Two dark fermions:} 
           The same way as the type-I seesaw mechanism may be a \textit{missing partner} one~\cite{Schechter:1980gr}, the minimal scotogenic model may also harbor an incomplete set of dark fermion copies.
           In fact, as seen in Sec.~\ref{sec:neutrino-parameters}, current oscillation data can be explained by just two neutrino mass splittings, solar and atmospheric. Hence only two active neutrinos need to acquire mass, and hence only two dark fermion species are required. 
           
           Unfortunately in this case the situation does not improve compared to vanilla type-I seesaw leptogenesis. 
           In order to reproduce the observed BAU, Eq.~\eqref{eq:eta-BAU}, the lightest dark fermion must still be very heavy.  Ref.~\cite{Hugle:2018qbw} finds that, for the case of normal active neutrino mass ordering, the lower bound is approximately~$$M_{1,NO}^{min}\approx 10^{10}\text{ GeV},$$ while for inverted mass ordering, the bound becomes even more stringent, $$M_{1,IO}^{min}\approx 10^{12}\text{ GeV}.$$ 
           
           This outcome arises because, throughout the relevant parameter region this scenario is characterized by the decay parameter $K\gsim 10$, indicating a  strong washout regime. As a result, any generated lepton asymmetry is significantly suppressed.

           \item \textbf{Three dark fermions:}  
           The situation changes significantly when three dark fermions species are introduced.
           In this case, the complex orthogonal matrix $R$ in Eq. \eqref{eq:Ynu}, which determines the Yukawa couplings $Y_F$, becomes a $3\times3$ matrix and involves three complex angles. 
           In this scenario, the system enters the strong washout regime only if the lightest active neutrino mass satisfies $m_{\nu_l}\gsim 10^{-3}$ eV with $m_\eta \not\approx M_1$~\cite{Hugle:2018qbw}. 
           In this parameter region the results essentially mirror those of the two dark fermion case, and we obtain  $$M_1^{min}\approx10^{10} \text{ GeV},$$ a similar lower-bound as before. In this case, however, both normal and inverted mass orderings yield similar outcomes.  

\vspace{3mm}

          Taking $M_1\gg m_\eta$, as the lightest neutrino mass $m_{\nu_l}$ drops below $10^{-3}$ eV, the system no longer remains in the strong washout regime, and the corresponding minimum mass $M_1^{min}$ required for successful leptogenesis also starts to decrease.
          As $m_{\nu_l}$ falls below $10^{-6}-10^{-7}$ eV, the system enters the weak washout regime, where $\Delta L=2$ lepton-number-violating scattering processes become significant. By numerically solving the relevant Boltzmann equations, one finds that the minimum dark fermion mass required to generate the observed baryon asymmetry can be approximated as: $M_1^{min}\approx 4.6 \times 10^7(m_{\nu_l}/eV)^{0.3}$ GeV. 
           Moreover, the lepton asymmetry generated from the decay of $F_1$
  must be converted into a baryon asymmetry by electroweak sphalerons, which go out of equilibrium at a temperature of roughly $T_{sph}\approx 130$ GeV. 
  This requirement imposes a lower bound on the lightest neutrino  mass $m_{\nu_l}\gsim 10^{-12}$eV, and a corresponding lower bound on the dark fermion mass~\cite{Hugle:2018qbw,Borah:2018rca}: $$M_1^{min}\sim 10\text{ TeV}.$$

Let us also mention a recent claim~\cite{Racker:2024fpn} that low-scale leptogenesis in the scotogenic model is achievable with $$M_1^{min}=1.3 (2.5) \text{ TeV}$$  
 for thermal (zero) initial abundance of $F_1$. These benchmarks comply with cLFV bounds, neutrino data, and scalar dark matter limits. The presence of heavy dark scalars and spectator processes, though not generating a lepton asymmetry, acts during leptogenesis and influences the final BAU,
 exponentially suppressing washout effects, and enabling successful leptogenesis at the TeV scale.          
 
 \item \textbf{Revamped Scotogenic Model:} The scalar LSP scenario has been explored in the revamped scotogenic model~\cite{Singh:2023eye}.  The minimal scenario with two dark fermions—one singlet and one triplet—closely resembles the simplest scotogenic scheme with two dark fermions, requiring 
$$M_1^{min}\gsim 10^9\text{ GeV}$$ for successful leptogenesis.  
However, in the presence of three dark fermions (two singlets and one triplet), and with a mass hierarchy $M_1<M_\Sigma\ll M_2$, one can lower $$M_1^{min}\sim3.1 \text{ TeV},$$ provided the lightest neutrino mass obeys $m_{\nu_l}\leq 10^{-5}$ eV. 
While the simplest scotogenic scenario suggests $M_1^{min}\sim$10 TeV, here one finds that $M_1^{min}$ drops significantly to a few TeV.

   \end{itemize}

In summary, reconciling neutrino oscillation data with the BAU at the TeV scale remains highly challenging. This difficulty is more pronounced in fermionic LSP scenarios than in scalar ones. Among scalar LSP schemes, the revamped scotogenic model with three dark fermion species has a comparative advantage over the simplest scotogenic setup. Nonetheless,  all such scenarios escape detection at existing colliders, as the minimum scale exceeds a few TeV or so.
 However they may end up becoming testable at future facilities such as the muon collider or FCC, a situation substantially more satisfactory than high-scale type-I seesaw. Alternative mechanisms—such as resonant leptogenesis \cite{Pilaftsis:1997jf,Pilaftsis:2003gt,Suematsu:2011va,Kashiwase:2013uy}, Akhmedov-Rubakov-Smirnov (ARS) leptogenesis \cite{Akhmedov:1998qx,Baumholzer:2018sfb}, or CP-violating Higgs decays \cite{Hambye:2016sby,Hambye:2017elz,Baumholzer:2018sfb}—may offer more promising avenues for making the LSP detectable at the LHC, albeit at the cost of introducing mass degeneracies and/or additional symmetries.

 {Concerning thermal fermionic dark matter production in simple scotogenic scenarios, the detailed nature of the cosmological phase transition in the early universe can play an important role. Depending on the parameters, the scotogenic model can produce one-step and two-step first order phase transitions~\cite{Shibuya:2022xkj}. Both of these scenarios may generate detectable gravitational wave signatures at the future interferometers like LISA~\cite{Caprini:2015zlo,Caprini:2019egz}, BBO~\cite{Corbin:2005ny}, DECIGO~\cite{Seto:2001qf,Kawamura:2020pcg}, U-DECIGO~\cite{Kudoh:2005as}.}
{For a one-step phase-transition, dark matter can be produced through the standard freeze-out mechanism, with sufficiently low nucleation temperature and satisfying the observed relic abundance. However, for a two-step transition, dark matter should never be thermal, so as to prevent over-closing the universe. This is possible in a freeze-in scenario~\cite{Shibuya:2022xkj}. } 

{A non-standard cosmological history can also affect fermionic dark matter production in scotogenic schemes. In standard cosmology, reheating happens almost instantaneously after inflation, with a reheating temperature higher than the dark matter decoupling temperature, making the freeze-out to occur in the radiation-dominated era. However, in some regions of parameters, the reheating temperature can be lower than the decoupling temperature of dark matter, forcing the freeze-out to occur in the reheating era~\cite{Roy:2025moo}. In this case, entropy injection to the thermal bath due to the late decay of inflatons helps achieving the relic dark matter abundance.}


\section{Conclusion and outlook}
\label{sec:conclusion}

In this review a broad and novel particle physics approach to dark matter was described, based on the idea of having a radiative \textit{dark origin} for neutrino masses.
The relevant Standard Model extension involves a \textit{dark sector} that accounts for the origin of dark matter, which is stabilized by the presence of a dark symmetry.
These schemes are more satisfactory as theories of dark matter than the simplest inert Higgs doublet model or the vanilla supersymmetric dark matter scenario.
Although the three relevant DM search techniques follow the same general complementary scheme illustrated in Fig.~\ref{fig:dm-collider}, there are additional neutrino probes that make this approach more appealing. Indeed, by having the issues of neutrino mass and dark matter closely interconnected, there are synergies between the two sectors and opportunities for cross-testing the idea that neutrino masses are indeed mediated or \textit{seeded} by a dark matter sector. 

After basic preliminaries in Sec.~\ref{sec:dark-preliminaries} we discussed the simplest inert Higgs doublet model in Sec.~\ref{sec:Inert}. The latter provides a theory benchmark, useful as a prelude for what comes next. 
In this case dark matter is bosonic and we give results for its relic abundance as a function of the dark-matter mass in Fig.~\ref{fig:Relic_scoto_seesaw}, while direct detection of dark matter via nuclear recoil (Fig.~\ref{fig:DD_scoto}) leads to
limits and projected sensitivities on the spin-independent WIMP-nucleon elastic scattering cross section given in Fig.~\ref{fig:DD-constraints}. Collider restrictions on the inert Higgs doublet model are given in Fig.~\ref{fig:higgs-decay-scoto}.

The main features and limitations of the simplest \textit{scotogenic} dark-matter scheme are discussed in Sec.~\ref{sec:simpl-scot-setup}, along with variations on the same theme. We discussed neutrino masses (Fig.~\ref{fig:neutrino-loop-scotogenic}) as well as the constraints that follow from oscillations, the relic fermionic DM density and the direct detection limits, given in Fig.~\ref{fig:loop_DD}. The possibility of accommodating the FIMP scenario within the scotogenic approach is illustrated in Fig.~\ref{fig:FIMP_12}.

A drawback of the minimal scotogenic picture, namely the possibility of loosing the dark symmetry as a result of
renormalization group evolution is illustrated in Fig.~\ref{fig:z2parity-scotogenic}.
In Sec.~\ref{sec:singlet-triplet-scoto} we discuss the revamped singlet-triplet scotogenic model which bypasses such dark symmetry breakdown problem. Simple fermionic dark-matter relic density predictions are shown in 
Fig.~\ref{fig:relic-pure-triplet} while
the complete picture is given in Fig.~\ref{fig:relic-mixed}. Direct fermionic DM detection through nuclear recoil is characterized by the spin-independent scattering cross sections presented in Fig.~\ref{fig:DD_sc} for various benchmark scenarios and co-annihilation assumptions.
The indirect detection gamma-ray signal is illustrated in Fig.~\ref{fig:ID_st}. The predictions of the revamped scenario for neutrinoless double-beta decay are given in Fig.~\ref{fig:dbd2}, while its advantages concerning the  behavior of the dark-symmetry are illustrated in Fig.~\ref{fig:z2parity-singlet-triplet}.

The scoto-seesaw picture combining the seesaw paradigm to account for the atmospheric scale, and the scotogenic one to provide the solar neutrino scale is described in Sec.~\ref{sec:scoto-seesaw}, see for example Fig.~\ref{fig:scoto-seesaw}.
Dark low-scale seesaw mechanisms are described in Sec.~\ref{sec:dark-low-scale-seesaw}, for example Fig.~\ref{fig:scoto-loop} and Fig.~\ref{fig:Neutrinoloopdiagram}
for the inverse and linear seesaw schemes, respectively. Collider restrictions for the dark inverse seesaw are given in Fig.~\ref{fig:limit-from-inv-mu} and Table~\ref{tab:2}.

Scalar dark-matter may arise in any scotogenic scenario, such as the simplest scheme, where the predicted relic densities are shown in Fig.~\ref{fig:relicscoto}.
The spin-independent elastic scattering cross-section in the simplest scotogenic model versus the DM mass is shown in Fig.~\ref{fig:ddscoto} left, while the indirect detection prospects are presented in the right panel of the same figure.
One sees how the DM mass must exceed 300~GeV, in order to be in agreement with current direct detection limits from LZ.
Notice that co-annihilation effects can make scalar DM masses below 500~GeV viable, in contrast to  the inert Higgs doublet model.
Spin-independent DM-nucleon elastic scattering cross section versus DM mass is shown in Fig.~\ref{fig:ddscalar} in the revamped scotogenic model, while results for DM annihilation cross-sections into $\gamma$ rays coming from various channels are given in Fig.~\ref{fig:idscalar} and compared with experimental limits and sensitivity projections.
Scalar dark-matter detection prospects in the dark inverse and dark linear low-scale seesaw schemes are also discussed.

Charged lepton flavor violation processes constitute a precious tool to probe schemes where a dark sector mediates or sources neutrino mass generation, as they can also be dark-mediated, as we discuss in Sec.~\ref{sec:cLFV}, see Fig.~\ref{fig:mutoegamma} and Fig.~\ref{fig:cLFV_Scoto} for the simplest scotogenic scheme and  Fig.~\ref{fig:mutoe}  and
Fig.~\ref{fig:cLFV_STM} for the revamped one.
For the dark linear seesaw there is an interplay between the conventional charged current contribution and that which comes from the dark-sector, as seen in Fig.~\ref{fig:figmuegFeyn} and Fig.~\ref{fig:figmueg}.
Some theories bring in exotic cLFV processes involving the emission of a Nambu-Goldstone boson, for example $\mu \to e G$, in addition to the $\mu \to e \gamma$ process, both arising from dark-sector contributions, as seen in Fig.~\ref{fig:feyn_lfv_dyn_scoto}. 
Fig.~\ref{fig:lfv_dyn_scoto} shows correlations between $\mu \to e \gamma$ and $\mu \to e G$ branching ratios, and that searches for the latter can be far more promising.

Finally, we note that the new particles present in the dark neutrino models are expected to produce signatures at pp colliders, such as the LHC, or at future hadron and lepton colliders, such as the ILC, CLIC, CEPC, FCC or a muon collider.
All in all, the key message to draw from the collider searches for dark matter signatures discussed in Sec.~\ref{sec:collider} is that, not only dark matter can be probed in many ways within collider experiments, but also that there are  important synergies with neutrino experiments, that enable dark matter searches to be cross-checked, for example, through cLFV studies as seen, for example, in Fig.~\ref{fig:MfMs1}.

\section*{Acknowledgements}

This work is funded by Spanish grants PID2023-147306NB-I00 (AEI/10.13039/501100011033), PID2023-146220NB-I00 (MCIU/AEI/10.13039/501100011033), and by  CEX2023-001292-S (MCIU/AEI/10.13039/501100011033) [Severo Ochoa Excellence grant], and by Generalitat Valenciana: Prometeo CIPROM/2021/054 and Prometeo/2021/071.
The work of A.K is partially supported by research grant number PID2020-114473GB-I00 and  20227S3M3B under the program PRIN 2022 of the Italian Ministero dell’Università e Ricerca (MUR).
The work of S.M. is supported by KIAS Individual Grants (PG086002) at Korea Institute for Advanced Study. S.S. thanks SERB, DST, Govt. of India for the grants SIR/2022/000432 and TAR/2023/000288, supporting an academic visit to IFIC. 
\appendix
\small 
\bibliographystyle{utphys}
\bibliography{bibliography} 

\end{document}